%% file: main.tex
\documentclass[10pt, a4paper, oneside]{Thesis} 

\graphicspath{{./Pictures/}} 
\DeclareGraphicsExtensions{.pdf}

\usepackage[square,colon,sort&compress]{natbib}

\usepackage{verbatim}  
\usepackage{MnSymbol} 
\usepackage{array} 
\usepackage[usenames,dvipsnames]{color} 

\usepackage{mathtools} 

\usepackage[framemethod=tikz]{mdframed}
\usepackage{relsize,etoolbox,thmtools}

\colorlet{myred}{red!50!black}

\declaretheoremstyle[
    headformat=\NAME\;\NUMBER:\color{black}\;\NOTE\;\!.,
    headfont=\bfseries\color{myred}, 
    notefont=\bfseries,
    notebraces={}{},
    headpunct={\newline},
    postheadspace=0pt,
    postheadhook={},
    spaceabove=0pt,
    spacebelow=0pt,
    bodyfont=\relsize{-1},
    mdframed={
	 linecolor=myred,
            linewidth=1pt,
            leftmargin=1em,
            rightmargin=0em,
            innerleftmargin=0.5em,
            innerrightmargin=0.5em,
            innertopmargin=-0.75em,
            innerbottommargin=0.5em, 
            skipabove=-0.25em, 
            skipbelow=0.5em} 
]{notestyle}

\declaretheorem[
    style=notestyle,
    name=Note,
    numberwithin=chapter
]{note}

\makeatletter
\def\ll@note{%
  \protect\numberline{\csname the\thmt@envname\endcsname}%
  \ifx\@empty\thmt@shortoptarg
    \thmt@thmname
  \else
    \thmt@shortoptarg
  \fi}
\def\l@thmt@note{} 
\makeatother
\declaretheoremstyle[    
headformat=\NAME\;\NUMBER\color{black}\;(\!\!\NOTE).,
    headfont=\color{myred}\sffamily, 
    notefont={},
    notebraces={}{},
    headpunct={},
    postheadhook={},
    spaceabove=8pt,
    spacebelow=-1pt,
    bodyfont={}
]{mydefstyle}

\declaretheorem[
    style=mydefstyle,
    name=Definition,
    numberwithin=section
]{mydef}

\declaretheorem[
    style=mydefstyle,
    name=Theorem,
    numberlike=mydef
]{mythm}

\declaretheorem[
    style=mydefstyle,
    name=Criterion,
    numberlike=mydef
]{mycrit}

\declaretheoremstyle[    
headformat=\NAME:,
    headfont=\color{myred}\sffamily, 
    notefont={},
    notebraces={}{},
    headpunct={},
    postheadspace=\newline,
    postheadhook={},
    spaceabove=8pt,
    spacebelow=-1pt,
    bodyfont=\itshape
]{mytheoremstyle}

\declaretheorem[
    style=mytheoremstyle,
    name=Theorem,
    numberwithin=chapter
]{mytheorem}

\hypersetup{pdftitle={\ttitle}}
\hypersetup{pdfsubject=\subjectname}
\hypersetup{pdfauthor=\authornames}
\hypersetup{pdfkeywords=\keywordnames}

\newcommand{\noteref}[1]{Note~\hypersetup{linkcolor=myred}\ref{#1}\hypersetup{linkcolor=Blue}}
\newcommand{\defref}[1]{Def.~\hypersetup{linkcolor=myred}\ref{#1}\hypersetup{linkcolor=Blue}}
\newcommand{\thmref}[1]{Thm~\hypersetup{linkcolor=myred}\ref{#1}\hypersetup{linkcolor=Blue}}
\newcommand{\critref}[1]{Crit.~\hypersetup{linkcolor=myred}\ref{#1}\hypersetup{linkcolor=Blue}}
\newcommand{\refcol}[1]{\hypersetup{linkcolor=myred}\ref{#1}\hypersetup{linkcolor=Blue}}

\newcommand{\eqnref}[1]{Eq.~(\ref{#1})}
\newcommand{\eqnsref}[2]{Eqs.~(\ref{#1}) and (\ref{#2})}
\newcommand{\eref}[1]{(\ref{#1})}
\newcommand{\figref}[1]{Fig.~\ref{#1}}
\newcommand{\figsref}[2]{Figs.~\ref{#1} and \ref{#2}}
\newcommand{\tabref}[1]{Tab.~\ref{#1}}
\newcommand{\tabsref}[2]{Tabs.~\ref{#1} and \ref{#2}}
\newcommand{\secref}[1]{Sec.~\ref{#1}}
\newcommand{\secsref}[2]{Secs.~\ref{#1} and \ref{#2}}
\newcommand{\appref}[1]{App.~\ref{#1}}
\newcommand{\chapref}[1]{Chap.~\ref{#1}}

\newcommand{\footref}[1]{\textsuperscript{\ref{#1}}}

\newcommand{\bra}[1]{\langle #1|}
\newcommand{\ket}[1]{| #1 \rangle}
\newcommand{\braket}[2]{\langle #1 | #2 \rangle }
\newcommand{\e}{\mathrm{e}}
\newcommand{\ii}{\mathrm{i}}
\newcommand{\Tr}{\textrm{Tr}}
\renewcommand{\t}[1]{\textrm{#1}}

\newcolumntype{M}[1]{>{\centering\arraybackslash}m{#1}}
\newcolumntype{N}{@{}m{0pt}@{}}

\newcommand{\caps}[1]{{\normalfont\scshape #1}}

\newcommand{\HRule}{\rule{\linewidth}{0.5mm}}

\title{\ttitle} 

\begin{document}

\frontmatter 
\setstretch{1.3} 

\fancyhead{} 
\rhead{\thepage} 
\lhead{} 

\pagestyle{empty} 

\begin{titlepage}
\begin{center}
\setlength{\parskip}{0pt}
\textsc{\LARGE \univname}\\[1.25cm]
\textsc{\Large Doctoral Thesis}\\[0.5cm] 

\HRule \\[0.6cm] 
{\huge \bfseries {Precision bounds in \\[1.5ex] noisy quantum metrology}}\\[0.6cm] 
\HRule \\[1.5cm] 
 
\begin{minipage}{0.4\textwidth}
\begin{flushleft} \large
\emph{Author:}\\
\href{http://www.fuw.edu.pl/~jankolo}{\authornames} 
\end{flushleft}
\end{minipage}
\begin{minipage}{0.4\textwidth}
\begin{flushright} \large
\emph{Supervisor:} \\
\href{http://www.fuw.edu.pl/~demko}{\supname} 
\end{flushright}
\end{minipage}\\[1.5cm]

\begin{minipage}{0.8\textwidth}
\begin{flushright} \large
\emph{Referees:} \\
\href{http://www.fizyka.umk.pl/~darch}{Prof. dr hab.~Dariusz Chru\'{s}ci\'{n}ski}\\
(Nicolaus Copernicus University in Toru\'{n}, Poland)\\
\href{http://www.staff.amu.edu.pl/~agie}{Dr hab.~Andrzej Grudka}\\
(Adam Mickiewicz University in Pozna\'{n}, Poland)
\end{flushright}
\end{minipage}\\[1.5cm]

\large \textit{A thesis submitted in fulfilment of the requirements\\ for the degree of \degreename}\\[0.15cm] 
\textit{in the}\\[0.15cm]
\groupname\\ \deptname \\[1.0ex] \FACNAME \\[1.5cm]
 
{\large September 2014}
\end{center}
\end{titlepage}
\null\vfill
\begin{center}
\includegraphics[height=25mm]{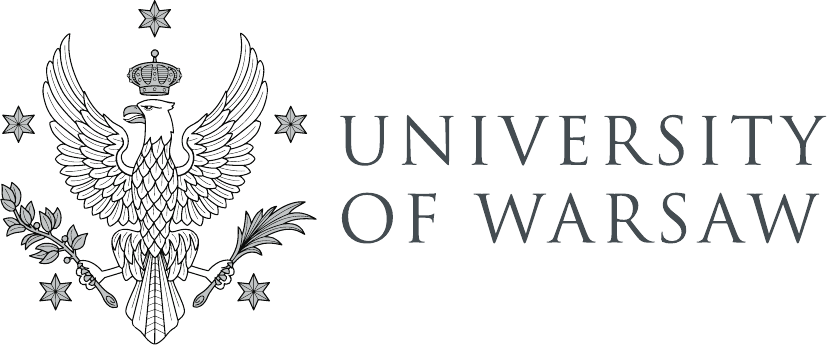}
\end{center}
\clearpage

\null\vfill 

\textit{``They say the first sentence in any speech is always the hardest. Well, that one's behind me, anyway."}

\begin{flushright}
Wis\l{}awa Szymborska \\ \small{(Nobel Lecture, 1996)}
\end{flushright}

\vfill\vfill\vfill\vfill\vfill\vfill\null 

\clearpage 

\null
\clearpage 


\addtotoc{Abstract} 

\abstract{\addtocontents{toc}{\vspace{1em}} 

\emph{Quantum metrology} is a vividly developing topic of current research in both
theoretical and experimental physics. Its main goal is to explore the capabilities of
quantum systems that, when employed as probes sensing physical parameters, 
allow to attain resolutions that are beyond the reach of classical protocols. 
Spectacularly, one may show that in an idealistic scenario,
by utilising the phenomena of \emph{quantum entanglement} and
super-classical correlations, a parameter of interest may be in principle determined
with mean squared error that scales as $1/N^2$ with the number of particles the 
system consists of---surpassing the $1/N$-scaling characteristic to classical statistics. 
However, a natural question arises, whether such 
an impressive quantum enhancement persists when one takes
into account the \emph{decoherence} effects, i.e.~the noise 
that distorts the system and is unavoidably present in any real-life implementation. 

In this thesis, we resolve a major part of this issue by describing general techniques 
that allow to quantify the attainable precision in metrological schemes, while accounting
for the impact of uncorrelated noise-types---ones that independently disturb each constituent 
particle (atom, photon) in the system. In particular, we show that the abstract geometrical
structure of a \emph{quantum channel} describing the noisy evolution of a single particle 
dictates critical bounds on the achievable quantum enhancement.
Importantly, our results prove that an infinitesimal amount of noise is enough to restrict the precision 
to scale classically in the asymptotic $N$ limit, what then constrains the maximal improvement 
to a constant factor. Although for relatively low numbers of particles 
the decoherence may be ignored, for large $N$ the presence of noise heavily alters the form 
of both states and measurements that should be employed to achieve the ultimate resolution. 
Crucially, however, the established bounds are then typically attainable with use of states 
and detection techniques that are natural to current experiments.

As we thoroughly introduce the necessary concepts and mathematical tools lying
behind the quantum metrological tasks, including the estimation theory techniques
in both \emph{frequentist} and \emph{Bayesian} frameworks, we hope that this work may 
be found attractive by researchers coming from the quantum information theory background
and willing to become more familiar with the current approaches to quantum metrology
problems. Throughout the work, we provide examples of applications of the methods 
presented to typical \emph{qubit noise models}, yet we also discuss in detail the
phase estimation task in \emph{Mach-Zehnder interferometry} both in the 
classical and quantum setting---with particular emphasis given to the photonic-losses 
model while analysing the impact of decoherence.

}
\clearpage 

\null
\clearpage 


\setstretch{1.3} 

\acknowledgements{
\addtocontents{toc}{\vspace{1em}} 
~\\
I would like to sorely thank my advisor Rafa\l{} Demkowicz-Dobrza\'{n}ski and Konrad Banaszek 
for their constant support throughout my graduate studies without which this thesis 
could have never come into existence.\\
\\
Rafa\l{} Demkowicz-Dobrza\'{n}ski -- most importantly for the patience in 
answering my many, never-ending questions and inspiration which has helped me to constantly progress 
in my research.\\
\\
Konrad Banaszek -- especially for the trust put in me four years ago when allowing me 
to join his research group despite my limited experience in the field of quantum information theory.\\
\\
My regards also go to Marcin Jarzyna with whom I have shared the pleasure of being a graduate 
student within the group. I hope that all the stimulating discussions we have had in the last years will also 
help him submit his thesis soon.\\
\\
Let me also greatly thank the referees of the thesis:~Dariusz Chrusci\'{n}ski and Andrzej Grudka, 
whose very accurate and pertinent remarks I have implemented in this version
of the manuscript.\\
\\
\\
Lastly, let me acknowledge the funding bodies that supported financially my graduate career:\\
Foundation for Polish Science TEAM project co-financed by the EU European Regional Development 
Fund, Polish NCBiR under the ERA-NET CHIST-ERA project QUASAR, FP7 IP projects Q-ESSENCE and SIQS.

}

\clearpage 


\begingroup

\renewcommand\bibname{List of Publications}
\addtotoc{\bibname}
\addtocontents{toc}{\vspace{1em}} 

\input{MyPubs/mypubs.bbl}
\let\clearpage\relax
\renewcommand\bibname{\Large{Beyond the scope of the thesis}}

\input{MyPubs/mypubs2.bbl}
\renewcommand\bibname{\Large{In print}}

\input{MyPubs/mypubs3.bbl}
\endgroup

\clearpage 


\lhead{\emph{Contents}} 
\tableofcontents 

\lhead{\emph{List of Figures}} 
\listoffiguresB 

\begingroup
\let\clearpage\relax

\lhead{\emph{List of Tables}} 
\listoftables 

\lhead{\emph{List of Notes}} 
\renewcommand{\listtheoremname}{List of Notes}
\listoftheorems[ignoreall,show={note}] 

\endgroup
\clearpage 

\pagestyle{fancy} 

\setstretch{1.3} 

\lhead{\emph{Abbreviations}} 
\listofsymbols{ll} 
{

\textbf{SQL} & \textbf{S}tandard \textbf{Q}uantum \textbf{L}imit:~$\Delta^2\tilde\varphi\!\sim\!1/N$\\
\textbf{HL} & \textbf{H}eisenberg \textbf{L}imit:~$\Delta^2\tilde\varphi\!\sim\!1/N^2$
\\[0.4cm]

\textbf{POVM} & \textbf{P}ositive \textbf{O}perator \textbf{V}alued \textbf{M}easure \\
\textbf{CPTP} & \textbf{C}ompletely \textbf{P}ositive \textbf{T}race \textbf{P}reserving map\\
\textbf{LGKS} & \textbf{L}indblad-\textbf{G}orini-\textbf{K}ossakowski-\textbf{S}udarshan form\\
\textbf{CJ} & \textbf{C}hoi-\textbf{J}amio\l{}kowski isomorphism/matrix
\\[0.4cm]

\textbf{PDF} & \textbf{P}robability \textbf{D}istribution \textbf{F}unction\\
\textbf{CLT} & \textbf{C}entral  \textbf{L}imit \textbf{T}heorem
\\[0.25cm]

\textbf{MSE} & \textbf{M}ean \textbf{S}quared \textbf{E}rror \\
\textbf{CRB} & \textbf{C}ram\'{e}r-\textbf{R}ao \textbf{B}ound \\
\textbf{FI} & Classical \textbf{F}isher \textbf{I}nformation \\
\textbf{ML} & \textbf{M}aximum \textbf{L}ikelihood estimator
\\[0.25cm]

$\overline{\textrm{\textbf{MSE}}}$ & Average \textbf{M}ean \textbf{S}quared \textbf{E}rror \\
\textbf{MMSE} & \textbf{M}inimum \textbf{M}ean \textbf{S}quared \textbf{E}rror estimator
\\[0.25cm]

\textbf{QCRB} & \textbf{Q}uantum \textbf{C}ram\'{e}r-\textbf{R}ao \textbf{B}ound \\
\textbf{QFI} & \textbf{Q}uantum \textbf{F}isher \textbf{I}nformation\\
\textbf{SLD} & \textbf{S}ymmetric \textbf{L}ogarithmic \textbf{D}erivative\\
\textbf{RLD} & \textbf{R}ight \textbf{L}ogarithmic \textbf{D}erivative\\
\textbf{NOON} & $\frac{1}{\sqrt{2}}(\ket{N,0}+\ket{0,N})$ state\\
\textbf{GHZ} & \textbf{G}reenberger-\textbf{H}orne-\textbf{Z}eilinger state\\
\textbf{BW} &  \textbf{B}erry-\textbf{W}iseman state
\\[0.4cm]

\textbf{SDP} & \textbf{S}emi-\textbf{D}efinite \textbf{P}rogram\\
\textbf{CS} & \textbf{C}lassical \textbf{S}imulation method\\
\textbf{QS} & \textbf{Q}uantum \textbf{S}imulation method\\
\textbf{CE} & \textbf{C}hannel \textbf{E}xtension method
\\[0.4cm]

\textbf{JS} & \textbf{J}ordan-\textbf{S}chwinger map

}

\clearpage 

\addtocontents{toc}{\vspace{2em}} 


\setstretch{1.1} 

\mainmatter 

\pagestyle{fancy} 


\input{./Chapters/intro}
\input{./Chapters/q_sys}
\input{./Chapters/est_theory}
\input{./Chapters/local_noisy_est}
\input{./Chapters/noisy_phase_est}
\input{./Chapters/conc}


\addtocontents{toc}{\vspace{2em}} 

\appendix 


\addtotoc{\emph{Appendices}} 

\input{./Appendices/AppendixA}
\input{./Appendices/AppendixB}
\input{./Appendices/AppendixC}
\input{./Appendices/AppendixD}
\input{./Appendices/AppendixE}
\input{./Appendices/AppendixF}
\input{./Appendices/AppendixG}
\input{./Appendices/AppendixH}
\input{./Appendices/AppendixI}
\input{./Appendices/AppendixJ}

\addtocontents{toc}{\vspace{2em}} 

\backmatter


\label{Bibliography}

\setstretch{1} 

\lhead{\emph{Bibliography}} 

\bibliographystyle{apsrmp4-1_custom} 

\bibliography{D:/DropBox/qi}

\end{document}

%% file: Chapters/intro.tex
\chapter{Introduction} 
\label{chap:intro} 
\lhead{Chapter 1. \emph{Introduction}} 

\section{Metrology -- the science of measurement}

The \emph{International Bureau of Weights and Measures} (BIPM),
located in S\`{e}vres (France) and serving since 1875 as one
of the primary 
guards ensuring uniformity of weights and measures
around the world, defines the term \emph{metrology} as%
\footnote{\url{http://www.bipm.org/en/convention/wmd/2004/}}: 
\begin{quote}
\begin{center}
"(...) the science of measurement, embracing both experimental and
theoretical determinations at any level of uncertainty in any field of science and technology. (...)"
\end{center}
\end{quote}
On the other hand, the ``\emph{Springer Handbook of Metrology and Testing}'' \citep{Czichos2011}
divides metrology into three sub-fields: \emph{scientific} (fundamental), \emph{technical} (industrial) 
and \emph{legal} (imposed by the national and international law).
Although the latter two stress the daily-life importance of precise measurements---which 
from the technological point of view play one of the main roles
in the rapidly developing industry, but also must be standardised
to ensure legal requirements vital to existence of modern society---their 
improvement can only be achieved owing to the first sub-field being constantly 
pursued by researchers around the globe. It is the development of metrology at 
the fundamental level that leads to the desired refinement of the 
up-to-date standards of quantities such as \emph{mass}, \emph{length} and \emph{time},
often requiring the research to be lead at the borderline of 
the current scientific state of the art.

One of the extensively contributing fields
is \emph{quantum metrology} that is a relatively young 
area of physics, currently intensively studied 
both at the theoretical and experimental levels. 
As the ultra-precise measurement 
schemes require the finest possible resolution of sensing, 
they are eventually condemned
to be limited by the fundamental building blocks describing the nature 
at the microscopic level, i.e.~by the laws of \emph{quantum mechanics}
which deals with physical phenomena at the nanoscopic scales.
Yet, the quantum theory has also to offer effects that
importantly allow to surpass the notions that may 
be naively inferred from classical statistics.
In particular, the ``spooky''---\citep{EPR1935}---feature 
of quantum theory known as \emph{entanglement} has been shown
to significantly enhance capabilities of precise-measurement techniques by 
exploring super-classical correlations between the system building blocks
(e.g.~individual atoms or photons) that sense the quantity of interest.
Furthermore, with the advent of new technologies allowing to 
control quantum systems by means of light-matter interactions,
such measurement precisions `beyond classical scaling-laws' have been
experimentally demonstrated. In fact, these achievements constituted 
one of the milestones accomplished in recent years within the field of 
quantum physics and motivated the Royal Swedish Academy of Sciences 
to present the \emph{Nobel Prize} in physics for 2012 to Serge Haroche and 
David J. Wineland ``\emph{for ground-breaking 
experimental methods that enable measuring 
and manipulation of individual quantum systems}"%
\footnote{\url{http://www.nobelprize.org/nobel_prizes/physics/laureates/2012/press.html}}
\citep{HarocheNobel,WinelandNobel}.
An important accomplishment of these experiments 
that has been crucial for their success, was the ability to defy 
the so-called quantum \emph{decoherence}
that typically disallows the quantum effects to be observed. 
Such a destructive phenomenon is a consequence of the inevitable 
interactions with the environment surrounding the quantum system 
examined, and leads to the presence of \emph{noise} 
affecting any measurements performed.
However, in the above experiments regimes have been 
spectacularly attained in which the impact of noise is negligible. 
On the other hand, this has imposed a novel open problem---which 
since then has been analysed by many researchers
also at the theoretical level---as to what extent the quantum enhancement 
of metrological protocols can be actually observed, but when dealing 
with real-life quantum systems in which the noise effects 
cannot be any more assumed to be small.

In this work, we would like to theoretically address the above issue and
discuss the consequences of the presence of \emph{decoherence},
which strongly affects the quantum system and thus delimits the accuracy of
the measurements performed. In particular, we employ the techniques 
developed in \emph{quantum information theory}, in order to establish 
general precision bounds that account for any sources of noise
that \emph{independently} affect each of the system building blocks, 
i.e.~the particles (atoms, photons) that a given quantum system consists of.

\section{Classical metrology}

Yet, before diving into the quantum mechanical framework designed to describe 
the quantum metrological tasks, one should acknowledge the immense field of \emph{estimation 
theory} \citep{Kay1993,Lehmann1998} that constitutes a major branch of statistics,
and has been developed to establish most efficient techniques
that allow to most accurately infer \emph{parameters} (representing
the quantities of interest) encoded in any randomly distributed data.
Thus, in any---potentially quantum-based---metrological problem one is bound
to use such techniques, as all that is always at hands of an experimentalist
are the statistics of the particular outcomes collected.
Typically, two philosophically differing approaches to such type of
problems are pursued, depending whether
one assumes the parameter being estimated
to be a fixed (deterministic) variable just parametrising
the physical model that predicts the outcome statistics, or 
accepts also the stochastic nature of the parameter and thus
its intrinsic random fluctuations.

Within the \emph{frequentist} approach to estimation, one
supposes the estimated parameter to be a deterministic variable which, 
if known, could in principle be stated up to any precision. Hence,
the inference process is limited only by the probabilistic nature of the 
measurement procedure, and one may thus utilise the well-established 
statistical techniques that allow to bound the effective \emph{Mean Squared Error} (MSE) 
of the estimation protocol. In particular, when restricting
to \emph{unbiased} estimators that on average output the correct parameter value, 
the so-called \emph{Fisher Information} (FI) can 
be evaluated and the \emph{Cram\'{e}r-Rao Bound} (CRB) may then 
be directly applied \citep{Kay1993,Lehmann1998}.

The \emph{Bayesian} approach is somehow complementary, yet by many
advocated to be the ``practically valid''  method of inference \citep[see e.g.][]{Jarzyna2014}. 
By assuming the parameter of
interest to also be randomly distributed, not only it explicitly accounts for the knowledge
one possesses about the quantity of interest before conducting the measurements, 
but also for the fact that the inference process should be interpreted as a 
data-updating routine. From the pragmatic point of view, the Bayesian techniques
turn out to be more approachable when dealing with parameters exhibiting \emph{symmetries},
what simplifies the form of the adequate Bayesian \emph{estimation costs} that 
must be minimised in order to establish the optimal inference strategy.
Furthermore, the structure of the Bayesian approach allows to naturally 
analyse the so-called \emph{adaptive measurement schemes} in which the knowledge about
the parameter is consecutively updated after each data-collection step,
so that the measurement settings may be gradually adjusted for the procedure 
to be more efficient.

Nevertheless, the distinction between the above two approaches
blurs and eventually vanishes in the limit of infinitely many experimental trials, 
so that it is often the case that both methods may be straightforwardly 
interrelated \citep{Vaart1998}.
However, let us already remark that in the quantum mechanical setting one 
of the key purposes of employing a quantum system consisting of many particles
is the ability to profit from its `stronger than classical' inter-particle correlations. 
As a result, even when restricting to single-particle
measurements, their outcomes are \emph{not} independently distributed, 
what opens doors to many novel interesting questions. In particular, as by raising
then the number of particles present in the setup one does not naturally increase 
the number of independent trials, no longer the classical intuitions relating 
the frequentist and Bayesian approaches apply \citep{Gill2013,Jarzyna2014}.

\section{Quantum metrology}

The classical \emph{problem of parameter estimation} 
has been explicitly restructured by \citet{Helstrom1976} and \citet{Holevo1982},
in order to incorporate the laws of quantum theory that dictate the 
measurement-outcome statistics.
In their seminal works, both frequentist and Bayesian 
tools known from classical estimation theory
have been adapted to apply in the quantum setting. 
Within the frequentist framework, the notions of \emph{Quantum Fisher Information} (QFI) 
and the \emph{Quantum  Cram\'{e}r-Rao Bound} (QCRB) 
have been established \citep{Nagaoka1989,Braunstein1994,Barndorff2000,Hayashi2005a},
whereas within the Bayesian paradigm the structure of the most general quantum 
measurements has been determined that naturally incorporates the estimated parameter symmetries.
In particular, for parameters exhibiting \emph{group symmetries}, the 
notion of the so-called \emph{covariant measurements} has been introduced,
which are provably always the optimal ones being also parametrised with 
the group elements induced by the estimation problem \citep{Chiribella2005,Hayashi2005a}.

On the other hand, without resorting to general \emph{quantum estimation} techniques,
but rather focusing on particular experimental scenarios and concrete measurement strategies,
it has been explicitly demonstrated that quantum enhancement of precision  
is theoretically possible in \emph{optical inteferometry} \citep{Caves1981,Bondurant1984,Yurke1986,Holland1993,Sanders1995,Dowling1998} 
and \emph{atomic spectroscopy} \citep{Wineland1992,Wineland1994,Bollinger1996}.
What is more, these pioneering results have manifested the ability of reaching 
super-classical accuracy already by utilising measurement schemes such as photon-counting, and 
by employing the well-known \emph{quantum-optical} techniques of \emph{squeezing}
applicable to both quantum states of atoms and light. 

Ten years later, the topic of quantum metrology has been extensively revisited
owing to the vividly developing field of \emph{quantum information theory}. This time,
the community has been stimulated to establish general frameworks allowing to predict ultimate, 
strategy-independent limits on the achievable parameter estimation precision, with particular interest 
in the role of \emph{quantum entanglement} enhancing the accuracy of parameter inference 
\citep{Giovannetti2001,Giovannetti2004,Giovannetti2006}.
As a result, the notions of the \emph{Standard Quantum Limit} (SQL) and the \emph{Heisenberg Limit}
(HL) have been grounded, which generally quantify the scaling of the estimated parameter 
\emph{Mean Squared Error} (MSE) with the number of particles employed, $N$. In the case of uncorrelated
particles, the $1/N$ \emph{SQL-like scaling} of the MSE is a natural consequence of the classical notion
of independent probing, or in other words the Central Limit Theorem (CLT). On the other hand, 
when quantum correlations in between the constituent particles are allowed,
the $1/N^2$ \emph{HL-like scaling} may be attained, which defines the ultimate quadratic
enhancement determined abstractly by the structure of the quantum theory.

At the beginning of XXIst century, with the advent of complex experimental techniques 
allowing to control light and matter, a vast number of experiments have been conducted
that demonstrated that the SQL can be indeed surpassed. These included the
ones achieving super-classical \emph{phase} resolution in \emph{optical interferometry} 
\citep{Mitchell2004,Nagata2007,Resch2007,Okamoto2008,Xiang2010,Kacprowicz2010}
with a spectacular implementation in the \emph{gravitational-wave detection} schemes 
\citep{LIGO2011,LIGO2013}.
Furthermore, the precision of estimation beyond SQL has also been attained in 
atomic-ensemble experiments in various configurations, in cases when one tries to
most accurately:~establish a time-duration reference -- \emph{atomic clocks} \citep{Appel2009,Louchet2010,Leroux2010,Ospelkaus2011},
register an atomic transition-frequency -- \emph{atomic spectroscopy} \citep{Leibfried2004,Schmidt2005,Roos2006}, 
or sense an external magnetic field -- \emph{atomic magnetometry} \citep{Wasilewski2010,Wolfgramm2010,Koschorreck2010,Napolitano2011,Sewell2012}.

\section{Noisy quantum metrology}

In all the above mentioned experiments, precision
beyond the SQL could have been attained owing to the spectacular control 
of the quantum systems involved, what allowed to diminish the amount of 
\emph{noise} inevitably present in the apparatus and nevertheless achieve the super-classical
accuracy. Yet, the issue of robustness of the above setups against 
various sources of \emph{decoherence} has soon been raised, in particular, questioning 
whether such experimental schemes could be utilised not only to surpass the SQL for a given 
number of particles (atoms, photons) employed, $N$, but also to observe a precision scaling that 
exceeds the classical $1/N$ dependence.

Such an issue, however, has long been left unanswered, as the original 
theoretical tools lacked the ability to efficiently incorporate the noise effects
into the metrological models considered. In fact, the above problem initiated a new topic 
of research aiming to establish general techniques that would allow to 
quantify the precision attained in quantum estimation protocols,
but would also explicitly account for the impact of noise either: by accurately
describing the destructive processes of \emph{decoherence} \citep{Zurek2003} 
and \emph{quantum fluctuations} \citep{Gardiner2000}; by adequately 
treating the overall ensemble of particles as an \emph{open quantum system} 
\citep{Breuer2002} constantly interacting with the environment;
or by utilising the abstract formalism of \emph{quantum channels} 
\citep{Nielsen2000,Bengtsson2006} to most generally model the 
noisy evolution of a given quantum system.

In a first step, the impact of decoherence on quantum metrological
protocols has been studied while considering particular
experimentally motivated measurement schemes and
models of noise \citep{Banaszek2009,Giovannetti2011,Maccone2011}.
In \emph{optical interferometry}, the primary attention has been paid
to the effect of \emph{photonic losses} that distorts the information
about the phase being estimated \citep{Dorner2009,Demkowicz2009,Kolodynski2010,Knysh2011},
but also to the impact of \emph{phase diffusion}
collectively affecting all the photons present in the setup \citep{Genoni2011,Genoni2012,Escher2012}.
In the \emph{atomic experiments}, on the other hand, particular focus 
has been given to the \emph{dephasing}-like noises affecting the atoms
in either correlated or uncorrelated manner 
\citep{Huelga1997,Ulam2001,Andre2002,Andre2004,Shaji2007,Dorner2012,Borregaard2013,
Macieszczak2014,Macieszczak2014a,Chaves2013}.

Nevertheless, the desired general frameworks have been soon proposed
that are capable of assessing the performance of quantum
metrological protocols in the presence of generic types of noise.
These have been possible due to the abstract mathematical description 
employed, which utilises the language of quantum channels originally
adapted to the metrological setting by \citet{Fujiwara2008} and \citet{Matsumoto2010}.
The resulting general approaches quantifying the precision bounds in the presence
of decoherence include: the \emph{purification-based} methods -- allowing
to consider a particular purified version of the system without
affecting its metrological properties \citep{Escher2011}; 
the \emph{simulation-based} methods -- allowing to simulate a given 
quantum channel representing the system evolution by means of
either independently distributed classical random variables \citep{Demkowicz2012}
or uncorrelated quantum states \citep{Kolodynski2013}; as well as 
the \emph{channel-extension--based} methods that despite being
numerical are always efficiently computable by means of 
\emph{semi-definite programming} \citep{Demkowicz2012,Kolodynski2013}.

Importantly, the above techniques have demonstrated that an 
\emph{infinitesimal} amount of generic noise, which independently affects each 
of the constituent particles of the system, is enough to force the asymptotic 
precision scaling with $N$ to be SQL-like. Thus, the maximal quantum enhancement 
is then always limited to reach at most a \emph{constant factor} improvement over 
the classical strategies. Moreover, the above results suggested that, due to 
the decoherence, the ultimate precision may always be attained in the asymptotic $N$ regime
without need of employing complex measurement schemes or preparing 
the system in exotic quantum states, but rather resorting to simple experimental
techniques, e.g.~the ones well-established in quantum optics such as the
photon-counting or light-squeezing.
Furthermore, the corresponding precision bounds, which obey the SQL-like scaling
imposed by the noise, may be typically saturated in a single experimental trial (shot), what indicates
the less prominent role of the inter-particle correlations \citep{Jarzyna2013,Jarzyna2014}, 
and contrasts the noiseless scenario in which maximally correlated states of particles
must be employed to attain the HL \citep{Giovannetti2006,Pezze2009}. Abstractly, the effects of uncorrelated noise 
may be seen as providing the necessary reason for the \emph{Quantum Local 
Asymptotic Normality} to hold \citep{Guta2006,Guta2007,Kahn2009}, so that 
the notion of the \emph{Central Limit Theorem} (CLT) of statistics can then be 
naturally generalised to the quantum setting. In particular, due to the 
impact of uncorrelated noise, one may thus effectively represent 
in the asymptotic $N$ limit any overall quantum state of the system by an 
``metrologically-equivalent'' Gaussian state. On the other hand, as some of the precision-bounding 
techniques allow to assess the accuracy of estimation not only in the 
asymptotic limit of many particles \citep{Kolodynski2013}, they may be also utilised 
to verify the performance of quantum estimation protocols in the regime of finite $N$,
in which it is generally not clear what type of measurements should be performed 
on a given system, and in what kind of states should it be prepared 
to achieve the optimal accuracy. Although they prove that for sufficiently low $N$ 
one may disregard the impact of decoherence, and thus assume the optimal noiseless solution
that generally employs highly non-classical states and measurements, it is an interesting 
open question how the structure of such optimal states and measurements varies with an increase 
in the number of particles---so that in the asymptotic $N$ limit, dominated by the noise, 
they eventually take a much simpler form.

\section{Outline of the thesis}

The main purpose of this thesis is to explicitly describe such general techniques
that allow to establish precision bounds in noisy quantum metrological schemes,
both in the asymptotic limit of infinitely many particles \citep{Demkowicz2012}
and for the quantum systems of finite size \citep{Kolodynski2013}. In particular,
we demonstrate in this work how the \emph{uncorrelated noise} modifies the effective
evolution of each \emph{single} constituent particle of the system, so that it may be 
effectively described by means of a \emph{quantum channel} which structure 
is sufficient to determine the ultimate-precision bounds of interest.

In order to do so, we introduce the mathematical language necessary
to describe the problems of noisy quantum metrology in \chapref{chap:q_sys},
where we also present the typical quantum estimation schemes of 
metrological relevance---the \emph{noisy-phase--estimation models} 
which we explicitly analyse throughout the work.
In \chapref{chap:est_theory}, we discuss in detail the tools of 
\emph{classical} and \emph{quantum estimation theory}, i.e.~the \emph{frequentist} 
and \emph{Bayesian} approaches to parameter inference, in order to 
explicitly apply them to the example of \emph{phase estimation} in 
\emph{Mach-Zehnder interferometry} both in the classical 
and quantum setting.

In \chapref{chap:local_noisy_est},
we move onto the main results of this work and demonstrate methods which allow us to
limit the ultimate precision in a general $N$-\emph{parallel--channel} estimation
protocol, stemming just from the structure of a \emph{single}
quantum channel describing the evolution of each constituent particle of the system.
We revisit the Mach-Zehnder interferometer and the phase estimation scenario in \chapref{chap:MZ_losses}
in order to analyse it in further detail, taking into account the impact
of \emph{photonic losses} both when following the frequentist and Bayesian approaches.

Finally, we summarise and conclude the discussion in \chapref{chap:conc}, where we
present some of  interesting problems that remain unresolved within the topic
of noisy quantum metrology, as well as indicate the relations to other methods
being complementary to our results. We also discuss the potential generalisations of
the established techniques to further metrologically relevant scenarios that are
beyond the scope of this work.

%% file: Chapters/q_sys.tex
\chapter{Metrology with realistic quantum systems} 
\label{chap:q_sys}
\lhead{Chapter 2. \emph{Metrology with realistic quantum systems}} 

In the following chapter, we discuss the mathematical framework
that is required in general to study the quantum metrological protocols, which
we analyse in further parts of this work. In particular,
we present the abstract language of quantum mechanics
allowing to describe a given \emph{quantum system} and
importantly the \emph{measurements}
performed on the system to investigate its features. We further discuss the \emph{evolution}
of such a system, which may be equivalently described with help of
\emph{quantum channels} or the time-differential \emph{master equation},
and give examples of models (i.e.~the \emph{noisy-phase--estimation} models
analysed in various configurations throughout this work) that are of particular
metrological relevance.
We argue that quantum metrology should be viewed as a problem
of determining the \emph{latent parameters} of the system evolution, which---in 
contrast to the \emph{observables} describing physical properties---are not directly measurable
and correspond to the variables that parametrise the \emph{model} assumed in the description. 
Thus, from maybe more philosophical perspective, they should \emph{not} be associated with the 
properties of the system per se. In the final part of this chapter,
we discuss in more detail the \emph{geometrical picture
of quantum channels}, which in the case of metrological problems 
not only describe the evolution, but also are importantly parametrised by
the latent variable, $\varphi$, being \emph{estimated}.
In particular, we explain the notion of \emph{$\varphi$-extremality} of a given channel
that will be later shown to be one of the key ingredients that forbid the
asymptotic HL-like precision scaling, $1/N^2$, with the number, $N$,
of the constituent particles.

\section{Mathematical description of a quantum system}

As the description of the subtleties of the \emph{quantum theory} is far beyond
the scope of this work, we can only point the reader to common textbooks introducing
quantum mechanics for explicit details \citep[see e.g.][]{Phillips2003,Griffiths2013},
and instead summarise---maybe brutally and iconoclastically---the
notion of quantum mechanics in a single sentence stating that:
\begin{quote}
The \emph{state} of a quantum system is described by its \emph{wave function},
whereas its measurable physical properties correspond to \emph{Hermitian operators} (\emph{observables}), 
which respectively are represented by \emph{vectors} and their adequate \emph{linear transformations},
so that the natural language of quantum mechanics is just the \emph{linear algebra}.
\end{quote}
However, we would like to explicitly deal with quantum systems subjected to \emph{noise},
which is manifested by the imperfect knowledge we possess about the system at a given time-instance.
Hence, we must further generalise the above framework of state vectors to \emph{density matrices},
or equivalently \emph{density operators}, in order 
to naturally incorporate in the description the classical notion of
\emph{statistical ensembles}. In what follows, we introduce the
necessary physical concepts and mathematical tools at a sufficient level 
from the quantum metrology point of view, 
yet their more detailed analysis may be found in \citep{Nielsen2000, Bengtsson2006}.

\subsection{Quantum states}
\label{sub:QSboundtates}

If a state of a given quantum system is perfectly known, it is described
by a vector, i.e.~a \emph{pure state}:~$\ket{\psi}\!\in\!\mathcal{H}$,
living in the Hilbert space $\mathcal{H}$. On the other hand,
if the system evolves according to some stochastic process, its state at a given moment
must be represented by an \emph{ensemble of pure states}:~$\{p_i,\ket{\psi_i}\}_i$
indexed by $i$ with $p_i$ representing the probability of the system being in a particular 
state $\ket{\psi_i}$. As a result, one then effectively denotes the state of the system
with help of a \emph{density operator/matrix}:
\begin{equation}
\varrho=\sum_{i}p_{i}\left|\psi_{i}\right\rangle \!\left\langle \psi_{i}\right|,
\label{eq:dens_op}
\end{equation}
which mathematically corresponds to a linear operator
on the Hilbert space $\mathcal{H}$ that is
\emph{non-negative} 
($\forall_{i}\!:p_{i}\!\ge\!0\Rightarrow\forall_{\left|\phi\right\rangle\in\mathcal{H}}\!:\left\langle \phi\right|\varrho\left|\phi\right\rangle \!\ge\!0$) 
and of \emph{unit trace} ($\sum_{i}p_{i}\!=\!1\Rightarrow\textrm{Tr}\!\left\{ \varrho\right\}\!=\!1$),
what we formally denote as%
\footnote{A density matrix is thus an element of the Banach space
of non-negative \emph{trace-class operators} of trace 1, which we denote as 
$\mathcal{T}(\mathcal{H})$ \citep{Reed1981}.}
$\varrho\!\in\!\mathcal{T}(\mathcal{H})$.
In other words, \eqnref{eq:dens_op} describes a \emph{probabilistic mixture} of state vectors,
and thus is normally termed to represent a \emph{mixed state} of the system. 

\begin{note}[Convexity of the space of density operators]
\label{note:conv_states}
Notice that consistently
a `mixture of mixed states', $\sum_i \!p_i\varrho_i$, is also a mixed state 
of the form \eref{eq:dens_op}, as it similarly corresponds to an ensemble  
$\left\{ p_{i}p_{i,j},\left|\psi_{i,j}\right\rangle \right\} _{i,j}$, which is built
after explicitly defining the ensembles for each $\varrho_i$ as $\varrho_i\!=\!\sum_{j}p_{i,j}\left|\psi_{i,j}\right\rangle \!\left\langle \psi_{i,j}\right|$.
More formally, this means that for any two mixed states $\varrho_{1/2}$ that belong to 
the space $\mathcal{T}(\mathcal{H})$, i.e.~the space of density matrices defined on the 
Hilbert space $\mathcal{H}$, their convex sum $\varrho\!=\!\lambda\varrho_1+(1-\lambda)\varrho_2$ with $0\!\le\!\lambda\!\le\!1$
also lies with within the space $\mathcal{T}(\mathcal{H})$.
Geometrically, such a fact proves that the space of density operators is \emph{convex}.
\end{note}

In order to study the quantum metrological scenarios, we 
must be able to correctly describe \emph{composite systems}; in 
particular, quantum systems consisting of many particles. 
Therefore, we very briefly 
review their representation within the language of density matrices.
Consider two distinct systems, labelled by A and B, which are isolated 
from one another (e.g.~not in physical contact
or space-like separated), so that they may be generally
treated as separate ensembles $\{p_i^\t{\tiny A/B},{\ket{\psi_i}}_\t{\tiny A/B}\}_i$
leading to $\varrho^\t{\tiny A/B}$ respectively.
Hence, we may intuitively construct the ensemble representing
the composite (bipartite) system AB as $\{p_i^\t{\tiny A}p_j^\t{\tiny B},{\ket{\psi_i}}_\t{\tiny A}{\ket{\psi_j}}_\t{\tiny B}\}_{i,j}$,
and thus the overall mixed state of the form \eref{eq:dens_op}:
\begin{equation}
\varrho^\t{\tiny AB}=\sum_{i,j}p_{i}^\t{\tiny A}p_{j}^\t{\tiny B}\left|\psi_{i}\right\rangle _\t{\tiny A}\left|\psi_{j}\right\rangle _\t{\tiny B}\,\prescript{}{\t{\tiny A}}{\left\langle \psi_{i}\right|}\prescript{}{\t{\tiny B}}{\left\langle \psi_{j}\right|}=\sum_{i}p_{i}^\t{\tiny A}\left|\psi_{i}\right\rangle _\t{\tiny A}\!\left\langle \psi_{i}\right|\;\otimes\;\sum_{j}p_{j}^\t{\tiny B}\left|\psi_{j}\right\rangle _\t{\tiny B}\!\left\langle \psi_{j}\right|=\varrho^\t{\tiny A}\otimes\varrho^\t{\tiny B},
\label{eq:tens_state}
\end{equation}
what proves that combining isolated quantum
systems results in their \emph{tensor-product} structure.
As a consequence, any state describing a quantum system consisting of $N$
particles that are \emph{uncorrelated} between one another---treated 
as isolated subsystems---most generally reads:
$\rho^N\!\!=\!\bigotimes_{n=1}^N\rho^{(n)}$, and in the case when each particle
is prepared in an identical state, say $\rho$, further simplifies to $\rho^N\!\!=\!\rho^{\otimes N}$.

On the other hand, we may ask the inverse question; 
how to describe the subsystem A given the most general combined 
state:~$\varrho^\t{\tiny AB}\!\in\!\mathcal{T}(\mathcal{H}_\t{\tiny A}\!\otimes\!\mathcal{H}_\t{\tiny B})$,
when we do not have access to the subsystem B (or equivalently vice versa). 
The state of the subsystem A corresponds then to the \emph{reduced density matrix},
$\varrho^\t{\tiny A}\!=\!\Tr_\t{\tiny B}\{\varrho^\t{\tiny AB}\}$,
obtained after performing the \emph{partial trace} operation over the system B,
which is most conveniently defined after choosing any orthonormal basis $\{{\ket{i}}_\t{\tiny B}\}_i$ in $\mathcal{H}_\t{\tiny B}$
for which:~$\varrho^\t{\tiny A}\!=\!\sum_i\prescript{}{\t{\tiny B}}{\bra{i}}\varrho^\t{\tiny AB}{\ket{i}}_\t{\tiny B}$.
One may easily prove that such an interpretation is the \emph{only one}
consistent with the quantum mechanical description (introduced in the following section) of the measurements,
which in such a situation are allowed to be performed \emph{only} 
on the part of the system that is at our disposal \citep{Nielsen2000}. 
Let us emphasise that the important consequence of the lack
of access to the whole system is the randomness
introduced while discarding (partial-tracing) some of its constituents.
Such a fact becomes most evident when considering any pure, and hence \emph{deterministic}, state
${\ket{\psi}}_\t{\tiny AB}$ of the composite system that is \emph{not separable}, i.e.~cannot be 
decomposed into a tensor product, ${\ket{\psi}}_\t{\tiny AB}\!\!\neq\!{\ket{\psi_1}}_\t{\tiny A}\!\!\otimes\!{\ket{\psi_2}}_\t{\tiny B}$.
It is so, as after tracing any of its subsystems one necessarily obtains a mixture
(e.g.~$\varrho^\t{\tiny A}\!=\!\Tr_\t{\tiny B}\{{\ket{\psi}}_\t{\tiny AB}\bra{\psi}\}\!\ne\!{\ket{\phi}}_\t{\tiny A}\!\bra{\phi}$
for A) and thus introduces \emph{stochasticity} into the description.
Yet, in case of composite systems possessing a (tensor-) product structure, 
by throwing away any of the subsystems we correctly do not affect the other ones, 
as most generally $\forall_{\varrho_{1/2}\in\mathcal{H}_\t{A/B}}\!:\Tr_\t{\tiny B}\{\varrho_1^\t{\tiny A}\!\otimes\!\varrho_2^\t{\tiny B}\}\!=\!\varrho_1^\t{\tiny A}$,
and similarly when tracing out over A.
For example, when considering quantum systems that comprise of particles 
that are uncorrelated with one another, after losing some of them 
the overall state of the rest will be unaffected, despite 
the total number of the constituent particles being diminished. 
In contrast, when particles are correlated---in fact \emph{entangled}
(see \secref{sub:ent} below)---by discarding any of them, we 
effectively introduce stochastic \emph{noise} that ``disturbs'' 
the surviving ones. Yet, let us emphasise that
such a ``disturbance"  should be understood as \emph{blurring
of the information} we possess about the system,
and \emph{not} as a process in which the particles are actually
physically affected \citep{Englert2013}.

\subsection{Quantum measurements}
\label{sub:Qmeas}

As mentioned above, the \emph{observables} of a quantum system
determine its physical properties and thus describe the quantities that
may be directly measured. Formally, they correspond to 
linear \emph{Hermitian}%
\footnote{Strictly speaking, observables correspond to \emph{self-adjoint} operators,
which apart from Hermiticity also demand the domains of $\hat O$
and $\hat O^\dagger$ to coincide \citep{Reed1981}. However, such an extra condition is 
naturally fulfilled by any \emph{bounded} operator---hence, the notation $\mathcal{B}(\mathcal{H})$ 
above---so that such mathematical subtleties may only play a role when 
dealing with infinite dimensional Hilbert spaces (e.g.~for position or momentum observables) \citep{Hall2013}.}
operators acting on the Hilbert space, 
i.e.\footnote{In this work, we primarily reserve the notation of a `hat', i.e.~$\hat O$,
only to operators that correspond to \emph{observables}. Yet, without loss of clarity, we also 
employ it in the final chapters to denote bosonic operators, while working within
the second-quantisation formalism utilised in the description of optical interferometry setups.}~%
$\hat O\!\in\!\mathcal{B}(\mathcal{H})$ such that $\hat O\!=\!\hat O^\dagger$,
what means that they can always be
expressed in the complete basis of projectors%
\footnote{If the eigenvalues of $\hat O\!=\!\sum_i \!o_i\ket{i}\!\bra{i}$ are non-degenerate, $P_i$ just correspond
to projections onto the eigenvectors:~$P_i\!=\!\ket{i}\!\bra{i}$.}:~$\{P_i\}_i$
satisfying $P_i P_j\!=\!P_i \delta_{ij}$ and $\sum_i \!P_i\!=\!\mathbb{I}$, such that
$\hat O\!=\!\sum_i \!o_i P_i$. Importantly, the real eigenvalues of $\hat O$, $o_i$,  
determine the possible values taken by the random variable $O$ 
that describes the outcomes obtained in a measurement of the observable. 
In particular, for a given state $\varrho$ of the system, 
in ``each shot'' one of the $o_i$-s is measured
with probability $p_i\!=\!\Tr\{\varrho \,P_i\}$,
so that on average $\left<O\right>\!=\!\sum_i \!p_i o_i\!=\!\sum_i \!\Tr\{\varrho \,P_i\} o_i\!=\!\Tr\{\varrho\,\hat O\}$
is obtained. Hence, such an observable-based 
\emph{projective measurement} is fully described by the operators $\{P_i\}_i$, each 
leading to an outcome labelled by $i$ occurring with probability $p_i$.

Let us note, however, that so-defined projective measurements are 
highly over-idealised, as by assuming all $P_i$ to be orthogonal 
with one another, we really demand ``by hand'' that we are able to 
accurately measure the observable of interest without heavily disturbing
the system. In such a peculiar setting, quantum mechanics predicts 
the system to be left in the relevant eigenstate of $\hat O$
after the measurement, what has maybe confusingly been termed
in the literature as a phenomenon of the ``wave-function collapse'' suggesting some dramatic
dynamical process \citep{Englert2013}. One should always bear in mind, 
that the quantum-state description should be just treated as a 
bookkeeping tool for the knowledge we possess about the system.
In reality, in order to precisely measure the observable, we must 
\emph{strongly} interact with the system. This, however, 
should not be seen as a problem, as we may still perfectly
register one of the eigenvalues $o_i$ while 
disrupting the system or even destroying it
(e.g.~consider optical experiments in which all the detection configurations
yielding various outcomes always lead to the same final state of the 
system---the vacuum---after absorbing all the photons) \citep{Steinberg2014}.

Furthermore, as in metrological problems we just seek measurement schemes
that lead to probability distributions from which the encoded evolution parameter 
may be most accurately inferred, the post-measurement state of 
the system is essentially of no interest. Hence, it is most appropriate to
consider the most general formalism of the so-called 
\emph{Postive Operator Valued Measure} elements (POVMs) \citep{Nielsen2000,Bengtsson2006},
which is designed to model all potential measurement-outcome statistics
at the price of being ambiguous in determining post-measurement state
of the system%
\footnote{%
For each POVM element $M_i$, one may construct an infinite number of \emph{generalised measurement
operators} $E_i$ satisfying $M_i\!=\!E_i^\dagger E_i$, each of which leads to a distinct
post-measurement state of the system:~$\varrho\!\to\!\frac{E_i\varrho E_i^\dagger}{\Tr\{E_i\varrho E_i^\dagger\}}$.
\label{footPOVM}
}. 
A POVM is generally represented by any set of \emph{quantum measurement operators} acting on
the system Hilbert space, $\{M_i\}_i$, that are \emph{non-negative}
$M_i\!\ge\!0$ and satisfy the \emph{completeness constraint} $\sum_i \!M_i\!=\!\mathbb{I}$.
Each outcome $i$ of the measurement is then obtained with 
probability $p_i\!=\!\Tr\{\varrho M_i\}$, yet (due to the ambiguity
of the measurement-action on the state\footref{footPOVM}) there exists 
an infinite number of experimental apparatuses that yield the measurement
statistics of a particular POVM. Thus, although the language
of POVMs is widely used and spectacularly successful in quantum information
theory---in particular, in the optimisation of protocols over
measurement strategies---it does not in principle provide a recipe how to 
construct its given implementation, which must be thus independently 
determined bearing in mind the experimental context considered.
Let us  note that nothing prevents 
us to also consider POVMs that are continuously parametrised
by a random variable $X$, and thus consist of an infinite number of non-negative elements.
Such ${\{M_x\}}_{X=x}$ yield then the outcome \emph{Probability Distribution Function}
(PDF):~$p(x)\!=\!\Tr\{\varrho M_x\}$,
with the completeness constraint now reading:~$\int\!\!\t{d}xM_x\!=\!\mathbb{I}$,
so that consistently $\int\!\!\t{d}x \,p(x)\!=\!1$.

\subsection{Entanglement of subsystems}
\label{sub:ent}

One of the most surprising phenomena that quantum theory has to offer
is its  ``spooky'' feature of \emph{entanglement}, recognised 
already at the beginning of XX century in the seminal papers
of \citet{Neumann1932,EPR1935} and \citet{Schrodinger1935}. 
Although it has been long considered rather as 
a peculiarity of quantum mechanics, after the advent
of rapidly developing field of quantum information theory
and modern experimental techniques, it has been explicitly shown to
be a genuine, tangible \emph{resource} being
at the heart of real-life quantum-based protocols in quantum:~cryptography,
communication, dense coding, teleportation and more \citep[see][for a review]{Horodecki2009}.

Importantly, from the quantum metrology perspective, it 
is exactly the presence of entanglement---or equivalently the \emph{non-classical 
correlations} in between the constituent particles of the 
system---that allows to surpass the standard limits imposed by classical statistics 
on the precision of parameter inference \citep{Giovannetti2001,Giovannetti2004,Giovannetti2006}.
That is why, we review the basic concepts of this phenomenon (with focus
on systems consisting of many, potentially indistinguishable particles), so
that our discussions of the quantum-enhanced---or really `entanglement-enhanced'---metrological protocols
in the further parts of this work may be clearer to the reader. 
Nevertheless, let us remark that a comprehensive explanation of the role 
of entanglement in quantum metrology has not yet been fully established,
being a vivid problem of current research \citep{Pezze2009,Hyllus2012,Toth2012}. 
Although the presence of entanglement is \emph{necessary}
for a super-classical precision to be observed in a metrological scenario,
there also exist highly particle-entangled states yielding no enhancement \citep{Hyllus2010a}. 
Yet, in the absence of any noise, one may establish a link between the so-called 
\emph{$N$-parallel--channel} schemes (discussed later in \chapref{chap:local_noisy_est})
that employ $N$ `maximally entangled' particles with each individually sensing the parameter of interest,
to the \emph{sequential} protocols  in which a \emph{single} particle
just senses the parameter $N$-times in a row \citep{Berry2009,Maccone2013}. 
Such an observation suggests
that the $N$-particle entangled states are beneficial, as they effectively 
simulate an $N$-fold amplification of the parameter sensitivity (see also \secref{sub:QEst_MZInter_HL}).
Such an interpretation, however, generally fails to be valid in the presence of noise
\citep{Maccone2013}, in which parallel strategies employing 
entangled particles seem to be more beneficial \citep{Demkowicz2014}.
On the other hand, it has been very recently shown that
(very noisy) states exhibiting \emph{bound-entanglement}---considered to 
be the weakest form entanglement possessing limited use in quantum protocols 
(not being distillable) \citep{Horodecki2009}---can also lead to the ultimate 
HL-like, $1/N^2$, precision scaling \citep{Czekaj2014}, what implies that even very cumbersome
types of inter-particle correlations may be \emph{witnessed} \citep{Horodecki2009}
by inspecting the enhancement of precision in an adequate metrological setting.

In general, a bipartite state consisting of subsystems A and B is called
\emph{separable}, if and only if it may be
written as a mixture of (tensor-)product states of the subsystems, i.e.
\begin{equation}
\varrho^\t{\tiny AB}=\sum_i p_i \;\varrho_i^\t{\tiny A}\otimes\varrho_i^\t{\tiny B}.
\label{eq:sep_state}
\end{equation}
On the other hand, $\varrho^\t{\tiny AB}$ is said to be \emph{entangled} if it is 
\emph{not} separable, i.e.~cannot be
written in the above form.
Notice that, due to the overall probabilistic distribution $\{p_i\}_i$, 
a separable state \eref{eq:sep_state} generally exhibits
\emph{classical}\footref{ft:discord} \emph{correlations} in between A and B, so that the subsystems
may be assumed to be uncorrelated only in the case of \eqnref{eq:tens_state}.
Crucially, the above definition of entanglement requires the
notion of a \emph{division into distinct parties} w.r.t.~which
the convex combination of product states \eref{eq:sep_state} may be defined.
Thus, for more than two constituent subsystems a given state 
may be considered entangled depending on the division chosen.

\begin{note}[Entanglement of a three-qubit state]
For instance,
consider a joint state of three qubits:~$\varrho^\t{\tiny ABC}\!=\!\frac{1}{2}\sum_{i,j=0}^1{\ket{i}}_\t{\tiny A}\bra{j}\otimes{\ket{i}}_\t{\tiny B}\bra{j}\otimes{\ket{0}}_\t{\tiny C}\bra{0}$ 
which is separable w.r.t.~the AB$|$C cut, but entangled w.r.t.~the A$|$BC cut.
Interestingly, making things even more complicated, one may find three-qubit states, 
$\varrho^\t{\tiny ABC}$, that
are separable w.r.t.~\emph{all} three bipartite cuts (AB$|$C, A$|$BC, AC$|$B),
but \emph{not} w.r.t.~the tripartite one (A$|$B$|$C) \citep{Acin2001a}.
Such states are in fact \emph{bound-entangled} \citep{Horodecki2009} and
naturally constructable with use of the formalism of
the so-called  \emph{unextendible product bases} \citep{Bennett1999,Bennett1999a}.
\end{note}

Nevertheless, we may generalise the notion of separability \eref{eq:sep_state}
to composite systems consisting of many particles, and define 
an $N$-particle quantum state to be \emph{fully separable}%
\footnote{Let us note that states \eref{eq:sep_state}/\eref{eq:sep_state_many_part},
despite being separable, may still possess \emph{non-classical correlations} typically quantified 
with use of the so-called \emph{quantum discord} \citep{Streltsov2015}, which
analysis is beyond the scope of this work.\label{ft:discord}}
if and only if it may be written in the form:
\begin{equation}
\rho^N\;=\;\sum_i p_i \;\bigotimes_{n=1}^N\rho_i^{(n)}\;=\quad\sum_i p_i \;\rho_i^{(1)}\otimes\rho_i^{(2)}\otimes\dots\otimes\rho_i^{(N)},
\label{eq:sep_state_many_part}
\end{equation}
so that it may be confidently termed \emph{not} to contain \emph{any} \emph{inter-particle entanglement}, 
being separable w.r.t.~to \emph{all} the possible 
cuts---particle groupings. In case we restrict only to pure states, 
the above full-separability condition simplifies to:
\begin{equation}
\ket{\psi^N}\;=\; \bigotimes_{n=1}^N\ket{\psi^{(n)}}\;=\quad\ket{\psi^{(1)}}\otimes\ket{\psi^{(2)}}\otimes\dots\otimes\ket{\psi^{(N)}},
\label{eq:sep_state_many_part_pure}
\end{equation}
showing that the particles must then be in a product state.
Let us emphasise that in order 
to apply the definitions \eref{eq:sep_state_many_part}/\eref{eq:sep_state_many_part_pure} and verify
the inter-particle entanglement of a given state $\rho^N$, we must be eligible 
to make the statement ``the $n$-th particle'', as
the notion of entanglement requires the subsystems to be unambiguously defined.
Hence, the decompositions \eref{eq:sep_state_many_part}/\eref{eq:sep_state_many_part_pure} are natural when 
dealing with systems consisting of \emph{distinguishable} particles
(e.g.~photons prepared in distinct time-bins, or atoms resident in different optical-lattice sites),
which in principle may thus be individually targeted.
However, we can also utilise \eqnsref{eq:sep_state_many_part}{eq:sep_state_many_part_pure}
when analysing systems consisting of a \emph{definite}%
\footnote{%
In fact (see \chapref{chap:MZ_losses}), we may also return to the first-quantisation picture in case of \emph{indefinite}
number of particles when we do \emph{not} possess a \emph{global phase reference} \citep{Molmer1997,Bartlett2007,Jarzyna2012}, as
we may then independently consider each $N$-particle sector \citep{Vaccaro2003}.
}
number of \emph{indistinguishable bosonic}%
\footnote{Similarly for \emph{fermionic} systems which, however, we do not consider within this work.}
particles. Yet, in order then to verify the inter-particle entanglement, we cannot use 
their natural description within the \emph{second-quantisation} formalism \citep{Scully1997,Schwabl2008,Altland2010}, 
but rather must return to the \emph{first-quantisation} picture \citep{Killoran2014},
in which the overall Hilbert space is unambiguously divided into a product
of the individual-particle Hilbert spaces, so that conditions \eref{eq:sep_state_many_part}/\eref{eq:sep_state_many_part_pure} 
apply. We discuss this issue in more detail below,
but let us also remark that, although such a procedure is appropriate from the point of 
view of the metrological protocols \citep{Demkowicz2015}---that (as later discussed) rely on a well-defined 
notion of distinct particles which number quantifies the resources---other approaches
to quantify entanglement of indentical-particle systems are also possible \citep{Shi2003,Stockton2003,Wiseman2003,Benatti2014}.

\subsection{Entanglement of indistinguishable particles}
\label{sub:sys_indist_part}

Consider a general pure state consisting of $N$ particles in two, 
\emph{bosonic} modes labelled by $a$ and $b$
\citep{Scully1997}:
\begin{equation}
\left|\psi^{N}_\t{\tiny bos}\right\rangle = \sum_{n=0}^N\, \alpha_n\; {\ket{n}}_a\,{\ket{N-n}}_b=\sum_{n=0}^N\, \alpha_n \; \ket{n,N-n},
\label{eq:Nph_state}
\end{equation}
written in the basis of vectors:~$\{\ket{n,N-n}\}_{n=0}^N$,
each representing ($n$,$N-n$) 
particles that occupy modes ($a$,$b$) respectively.
Hence, $\ket{\psi^{N}_\t{\tiny bos}}\!\in\!\mathcal{H}_\t{\tiny bos}$,
where $\mathcal{H}_\t{\tiny bos}$ is the bosonic subspace of the Hilbert space
with $\t{dim}\{\mathcal{H}_\t{\tiny bos}\}\!=\!N\!+\!1$.
Now, if we want to return to the first-quantisation picture
and treat particles as separate subsystems, we must associate
a 2-dimensional Hilbert space with each of them, so that each
corresponds to a \emph{qubit} with basis vectors $\{\ket{0(\equiv\!a)},\ket{1(\equiv\!b)}\}$
representing the particle being in either of the modes.
As a result, we are able to write every $\ket{n,N-n}$ vector in
the form:~$\ket{0}^{\otimes n}\otimes\ket{1}^{\otimes N-n}$, which, however,
must be \emph{symmetrised} over all particle permutations. 
Thus, adopting a binary notation in which an $N$-bit sequence $\boldsymbol{n}$
is utilised to represent a product state of $N$ \emph{distinguishable}
qubits:~$\left|\boldsymbol{n}\right\rangle\!=\!\left|n_{1}\right\rangle \otimes\dots\otimes\left|n_{N}\right\rangle$ 
with $\left|n_{i}\right\rangle\!\in\!\{\ket{0},\ket{1}\}$,
we can rewrite \eqnref{eq:Nph_state} as
\begin{eqnarray}
\ket{\psi_\t{\tiny bos}^N}
&=&
\sum_{n=0}^N \alpha_n\;
\frac{1}{\sqrt{|\Pi|}}
\sum_\Pi |\Pi[\underbrace{0,\dots,0}_n,\underbrace{1,\dots,1}_{N-n}]\rangle
\;=\;
\sum_{n=0}^N \alpha_n\;
\frac{1}{\sqrt{\binom{N}{n}}}\;
\sum_{\underset{\!\!|\boldsymbol{n}|=n}{\boldsymbol{n}=\boldsymbol{0}^N}}^{\boldsymbol{1}^N} \left|\boldsymbol{n}\right\rangle
\nonumber\\
&=&
\sum_{\boldsymbol{n}=\boldsymbol{0}^N}^{\boldsymbol{1}^N}
\left(
\sum_{n=0}^N \frac{\alpha_n\,\delta_{|\boldsymbol{n}|,n}}{\sqrt{\binom{N}{n}}}
\right)\;
\left|n_{1}\right\rangle \otimes \left|n_{2}\right\rangle \otimes\dots\otimes\left|n_{N}\right\rangle
\label{eq:NPh_perm_dist}.
\end{eqnarray}
We denote by $\sum_\Pi$ the sum over all permutations, $\Pi$,
which is also equivalently 
represented above in the binary notation by fixing the number of 1s
($n\!=\!|\boldsymbol{n}|$) appearing in a bit sequence.
We have explicitly written out the product state in \eqnref{eq:NPh_perm_dist}
to emphasise that indeed the notion of the ``$n$-th particle'' is now clearly defined,
so that the full-separability condition \eref{eq:sep_state_many_part_pure} directly applies.
In fact, \eqnref{eq:NPh_perm_dist} indicates that \eqnref{eq:sep_state_many_part_pure} can only be satisfied, 
if \eqnref{eq:NPh_perm_dist} may be rewritten as a tensor product 
$\ket{\xi}^{\otimes N}$ with $\ket{\xi}\!=\!\alpha\ket{a}\!+\!\beta\ket{b}$ being an arbitrary pure qubit state.

For comparison, notice that the most general pure state of $N$ \emph{distinguishable} particles
in two modes, or equivalently $N$ qubits, is supported by a vastly larger---$2^N\!$-dimensional---Hilbert 
space and reads:
\begin{equation}
\ket{\psi^N}
\;=\;
\sum_{\mathbf{n}=\boldsymbol{0}^N}^{\boldsymbol{1}^N}\alpha_{\boldsymbol{n}}\;\left|\boldsymbol{n}\right\rangle
\;=\;
\sum_{j=\{0,\frac{1}{2}\}}^{\frac{N}{2}}\,
\sum_{m=-j}^{j}
\tilde\alpha_{j,m}\;\ket{j,m}\,,
\label{eq:Nqubit_state}
\end{equation}
so that the bosonic state \eref{eq:NPh_perm_dist} 
may be interpreted as a special instance of $\ket{\psi^N}$ after matching the coefficients
$\alpha_{\boldsymbol{n}}$ with the ones appearing in the curly brackets in 
\eqnref{eq:NPh_perm_dist}. In order to make such a statement even clearer,
we have written $\ket{\psi^N}$ in \eqnref{eq:Nqubit_state} also in its \emph{angular-momentum}
representation \citep{Biedenharn1984,Devanathan1999}, in which the quantum numbers
($j$,$m$) respectively represent the eigenvalues of the total angular-momentum, 
$\hat J^2\!=\!\hat J_x^2\!+\!\hat J_y^2\!+\!\hat J_z^2$, and the 
${\hat J}_z\!=\!\frac{1}{2}\sum_{n=1}^N\hat\sigma_z^{(n)}$ operators%
\footnote{Throughout this work we adopt the standard notation $\hat \sigma_i$ with $i\!=\!\{x,y,z\}$ for Pauli spin-1/2 operators.}.
As a consequence, the bosonic (fully symmetric) subspace, $\mathcal{H}_\t{\tiny bos}$,
may then be associated with the one corresponding to the maximal total angular-momentum, $j\!=\!N/2$,
for which \eqnsref{eq:Nph_state}{eq:Nqubit_state} become equivalent after identifying all
$\ket{n,N-n}$ vectors with the so-called Dickes states \citep{Dicke1954}: $\ket{j\!=\!\frac{N}{2},m\!=\!n\!-\!\frac{N}{2}}$,
and thus $\alpha_n$ with $\tilde\alpha_{\frac{N}{2},{n-\frac{N}{2}}}$.

\begin{note}[Inter-particle entanglement of two-mode bosonic states]
\label{note:ent_Fock}
Let us consider two simple examples of $N$-particle two-mode bosonic 
states:~a \emph{Fock state} resident in one of the modes -- ${\ket{N}}_a{\ket{0}}_b$, 
and a \emph{twin-Fock state}%
\footnote{Assuming without loss of generality $N$ to be even.}
-- ${\ket{\frac{N}{2}}}_a{\ket{\frac{N}{2}}}_b$;
and investigate whether they contain any inter-particle entanglement. 
In case of a Fock state, the situation is completely straightforward, 
as it just corresponds to $N$ photons contained in mode $a$,
i.e.~$\left|N\right\rangle _{a}\left|0\right\rangle _{b}=\left|a\right\rangle ^{\otimes N}$,
that by definition are uncorrelated with one another.
Yet, such an obvious observation indicates that we need the 
indistinguishable particles to be somehow distributed between the two modes
for any inter-particle entanglement to be present.
On the other hand, in case of the twin-Fock state, as
only $\alpha_{n=\frac{N}{2}}\!=\!1$ is non-zero
in \eqnref{eq:Nph_state}, we obtain according to \eqnref{eq:NPh_perm_dist}:
\begin{equation}
\left|\frac{N}{2}\right\rangle _{a}\left|\frac{N}{2}\right\rangle _{b}=\frac{(N/2)!}{\sqrt{N!}}\sum_{\Pi}\Pi\left[\left|a\right\rangle ^{\otimes\frac{N}{2}}\left|b\right\rangle ^{\otimes\frac{N}{2}}\right],
\label{eq:TwinFock}
\end{equation}
where again $\sum_\Pi$ stands for a sum over permutations.
Importantly, as \eqnref{eq:TwinFock} 
cannot be rewritten into a product state of particles \eref{eq:sep_state_many_part_pure}, 
it proves that a twin-Fock state contains inter-particle entanglement%
\footnote{%
Actually, the states \eref{eq:Nph_state} that
do \emph{not} possess any inter-particle entanglement
are \emph{only} the ones that can  be obtained 
by impinging Fock and vacuum states respectively on the input ports of 
a general beam-splitter (see also \secref{sub:QEst_MZInter_HL}).
}.
As a consequence, in contrast to the Fock state (representing
the \emph{classical stategy} in a phase-estimation protocol
discussed in \secref{sub:ClEst_MZInter_SQL}), ${\ket{\frac{N}{2}}}_a{\ket{\frac{N}{2}}}_b$
may be utilised to achieve quantum enhancement 
in metrological protocols \citep{Holland1993}.
\end{note}

On the other hand, the first-quantisation--representation \eref{eq:NPh_perm_dist} of $\ket{\psi^N_\t{\tiny bos}}$ allows 
to extract from a general transformation of a bosonic $N$-particle system, $\ket{\psi_\t{\tiny bos}^N}\!\to\!\rho^N$,
the mathematical form of the evolution (see next section) of each constituent particle. This may be achieved by 
re-expressing also $\rho^N$ with help of \eqnref{eq:NPh_perm_dist} and identifying the effective
map according to which each \emph{single} particle evolves. Importantly, the properties of such a map 
will be shown later within this work to play a crucial role in quantum metrology, in principle determining the maximal
capabilities of a given metrological protocol. 

Moreover, let us also remark that, when considering \emph{noisy} 
quantum systems, the noise may often introduce extra degrees 
of freedom to the particles, opening thus doors to their potential distinguishability.
Mathematically, this means that the system is \emph{not} supported
by the bosonic subspace during the evolution, as it is ``taken out''
from $\mathcal{H}_\t{\tiny bos}$ by the action of the noise. 
For instance, consider $N$-photons
prepared in a state \eref{eq:Nph_state}, $\ket{\psi_\t{\tiny bos}^N}$, representing 
two modes of light impinged on the input ports of a beam-splitter. 
However, due to the imperfect (spatial/temporal) mode-matching
of the beams, some of the photons do not contribute to the splitting
process and lead to a non-coherent admixture additionally present in the outputted
light-modes (e.g.~yielding imperfect visibility of an interferometer) \citep{Demkowicz2015}.
Within the second-quantisation formalism, the non-contributing photons should be 
assumed to occupy extra distinct modes, which, however, are mistaken for
the output ones at future detection stages. Thus, the effective state describing 
all the outputted photons restricted to the two original modes 
is \emph{not} pure any more, despite still being permutation-invariant%
\footnote{%
The fact that all permutation-invariant states are bosonic is true 
for \emph{pure}, but \emph{not} for \emph{mixed} states. All
permutation-invariant mixed states may be written in a block-diagonal 
form:~$\rho_\t{\tiny p.i.}^N\!=\!\bigoplus_{j=0}^{N/2}\,\rho_j\!\otimes\!(\mathbb{I}_{d_j}/d_j)$,
where each subspace is parametrised by the total angular-momentum number $j$ and $d_j$ represents 
the dimension of each ``multiplicity'' space being proportional to $\mathbb{I}$ \citep{Bartlett2007}.
The \emph{bosonic mixed states} may lie only in the block corresponding to $j\!=\!N/2$.
}.
Furthermore, it is no longer supported only by the $j\!=\!N/2$ subspace in 
the angular-momentum representation,
and hence extends beyond the bosonic subspace.
As a result, the effective noise model does not preserve the 
\emph{bosonicity} of particles (cannot be described by means of just creation and annihilation 
operators of the output modes \citep{Scully1997,Schwabl2008,Altland2010}),
and in fact requires the first-quantisation picture of \eqnref{eq:NPh_perm_dist} to be employed.

\section{Quantum system dynamics}
\label{sec:Qevolution}

In quantum metrological problems, the parameter
of interest to be estimated from the measurements
performed on a given system is \emph{encoded} during
the system \emph{evolution}, so that---apart from 
the formalism describing the state of the system
at some time-instance---we must discuss more formally
how the system evolves in time. Let us remind the reader that
according to quantum mechanics the evolution of 
any state vector $\ket{\psi(t)}$ is governed by the Schr\"{o}dinger
equation \citep{Phillips2003,Griffiths2013}, which for a \emph{closed}
isolated system reads%
\footnote{%
Without loss of generality, throughout this 
work we assume the Planck's constant $\hbar$ to be equal to 1.
}:
\begin{equation}
\textrm{i}\frac{\textrm{d}}{\textrm{d}t}\left|\psi(t)\right\rangle =\hat{H}\left|\psi(t)\right\rangle
\quad\implies\quad
\textrm{i}\frac{\textrm{d}}{\textrm{d}t}\varrho(t)=\left[\hat{H},\varrho(t)\right],
\label{eq:schr_eq}
\end{equation}
and, as shown above, naturally generalises to the so-called von Neumann equation \citep{Breuer2002}
when considering the density matrix representation \eref{eq:dens_op} 
with $\varrho(t)\!=\!\sum_i \lambda_i(t)\,\ket{\psi_i(t)}\!\bra{\psi_i(t)}$.
As a result, a closed isolated system evolves in between times $t_0$ and $t$ 
under a \emph{unitary} transformation $U_\tau\!=\!\exp\!\left[-\ii\hat H\tau\right]$,
so that the final state of the system reads:~$\varrho(t)\!=\!U_{t-t_0}\varrho(t_0)U_{t-t_0}^\dagger$.

Now, as we would like to describe the evolution of noisy \emph{open} systems,
we must account for the degrees of freedom that are not under our control
and lead to the effect of \emph{decoherence} \citep{Zurek2003}, making
the overall evolution non-unitary. In such a case, \eqnref{eq:schr_eq}
still applies describing the evolution of the state
$\varrho^\t{\tiny SE}(t)\!\in\!\mathcal{H}_\t{\tiny S}\!\otimes\!\mathcal{H}_\t{\tiny E}$ 
containing the system (S) of interest, but \emph{also} the environment (E) which is beyond our reach.
However, we must ensure that at a given time instance $t_0\!=\!0$, 
from which we would like to describe the evolution and at which 
we importantly have full control of the system knowing its state 
$\varrho^\t{\tiny S}(0)\!\in\!\mathcal{H}_\t{\tiny S}$, the environment is isolated, 
i.e.~$\varrho^\t{\tiny SE}(0)\!=\!\varrho^\t{\tiny S}(0)\otimes\varrho^\t{\tiny E}$,
making the overall process physical.
Then, after the environment comes into contact with the system it introduces noise,
as the final state $\varrho^\t{\tiny SE}(t)$ must be traced-out (see \secref{sub:QSboundtates})
over the subspace $\mathcal{H}_\t{\tiny E}$ representing degrees of freedom 
we do not have access to.
As a result, after defining 
$U_\tau^\t{\tiny SE}\!=\!\exp\!\left[-\ii\hat H_\t{\tiny SE}\tau\right]$, we may most generally write:
\begin{equation}
\varrho^{\t{\tiny S}}(t)=
\Tr_{\t{\tiny E}}\!\left\{ U^{\t{\tiny SE}}_t\left(\varrho^{\t{\tiny S}}(0)\otimes\varrho^{\t{\tiny E}}\right)U^{\t{\tiny SE} \dagger}_t\right\}
=
\Lambda_t\!\left[\varrho^{\t{\tiny S}}(0)\right]
\qquad\Longleftrightarrow\qquad
\ii\frac{\t{d}}{\t{d}t}\varrho^{\t{\tiny S}}(t)
=
\Tr_{\t{\tiny E}}\!\left\{ \left[\hat{H}_\t{\tiny SE},\varrho^{\t{\tiny SE}}(t)\right]\right\},
\label{eq:evol_tr}
\end{equation}
where $\Lambda_t$ is thus the effective \emph{quantum channel} describing
the evolution of $\varrho^\t{\tiny S}(0)$ to $\varrho^\t{\tiny S}(t)$.
Notice already that $\Lambda_t\!:\mathcal{T}(\mathcal{H}_\t{\tiny S})\!\to\!\mathcal{T}(\mathcal{H}_\t{\tiny S})$ 
must be a linear map, as it is constructed by combining linear operations:~a unitary rotation
followed by the partial trace of the environmental subspace $\mathcal{H}_\t{\tiny E}$.

As from the perspective of quantum metrological protocols, we will be primarily interested 
(see \secref{sec:QEst} later)  in the overall transformation of the system from its `input' state 
at $t_0\!=\!0$ onto  its `output' state at $t$ for a \emph{fixed} evolution duration,
we drop in the next section the time-dependence of $\Lambda_t$, in order to 
discuss the general  structure of \emph{quantum channels}---\emph{Completely Positive Trace-Preserving} (CPTP)
maps---describing all possible physical `input-output transformations'
of the system. However, such an approach is not appropriate
when considering metrological schemes (e.g.~the frequency estimation scenarios 
studied later in \secref{sec:loc_freq_est}) in which the time duration $t$ is 
an extra degree of freedom that may be adjusted to purposefully vary the form of 
the effective quantum channel $\Lambda_t$. Thus, we also 
review the general description of an open system via the 
\emph{master equation}, i.e.~the differential equation on the r.h.s.~of \eqnref{eq:evol_tr}
(subject to the initial condition $\varrho^\t{\tiny SE}(0)\!=\!\varrho^\t{\tiny S}(0)\otimes\varrho^\t{\tiny E}$), 
which often is the one that phenomenologically specifies the dynamics of a given 
quantum system.

\subsection{\emph{Quantum channel} picture of system evolution}
\label{sub:Evo_QCh}

Let us consider a general \emph{quantum channel} defined via
\eqnref{eq:evol_tr}, i.e.~$\Lambda\!:\mathcal{T}(\mathcal{H}_\t{in})\!\to\!\mathcal{T}(\mathcal{H}_\t{out})$ \citep{Bruss2007},
which describes the system evolution from a given \emph{input} state $\varrho_\t{in}\!\in\!\mathcal{T}(\mathcal{H}_\t{in})$
(previously $\varrho^\t{\tiny S}(0)$) to the adequate \emph{output} state $\varrho_\t{out}\!=\!\Lambda\!\left[\varrho_\t{in}\right]
\!\in\!\mathcal{T}(\mathcal{H}_\t{out})$ (previously $\varrho^\t{\tiny S}(t)$). 
For generality, let us also assume that the system dimensions may vary during the evolution, i.e.~$d_\t{in}\!\ne\!d_\t{out}$ 
with $d_\t{in/out}\!=\!\dim\{\mathcal{H}_\t{in/out}\}$, what is still in agreement with \eqnref{eq:evol_tr},
in which the dimension of the traced-out environment E changes between $t_0$ and $t$.
Physically, this corresponds to the situation in which the noise either destroys or 
introduces new degrees of freedom to the system. For instance, one may consider
atoms which are perturbed by the interaction with an environment in a way, so that 
respectively either some of their energetic transitions become forbidden by the 
presence of decoherence, or they are excited into energetic levels that 
originally would not have been allowed while considering only the atomic free-evolution.

\paragraph{\emph{Complete Positivity} (CP) and other properties of quantum channels}~\\
Firstly, one should impose that the probability has to be conserved, so that
$\Tr\{\varrho_\t{in}\}\!=\!\Tr\{\varrho_\t{out}\}\!=\!1$ and thus any quantum map 
$\Lambda$ must be \emph{Trace-Preserving} (TP).
On the other hand, as \emph{none} of the eigenvalues (i.e.~mixing probabilities in \eqnref{eq:dens_op})
of the output state can be negative regardless of $\varrho_\t{in}$ chosen, 
any quantum channel must also be \emph{positive}, 
i.e.~$\forall_{\varrho_\t{in}\in\mathcal{T}(\mathcal{H}_\t{in})}\!:\Lambda\!\left[\varrho_\t{in}\right]\!\ge\!0$.

However, this is not the end of the story, as for a given quantum 
channel $\Lambda$ to be physical it must also describe a valid 
evolution of a system (S) being in contact with some ancilla (A), 
and thus constituting a part of an overall system+ancilla (SA) composite system. 
This is true only if for any initial bipartite state
($\varrho^\t{\tiny SA}\!\in\!\mathcal{T}(\mathcal{H}_\t{in}\otimes\mathcal{H}_\t{\tiny A})$
with arbitrary $d_\t{\tiny A}\!=\!\dim\{\mathcal{H_\t{\tiny A}}\}$)
the overall density matrix of the system SA remains positive after the action of $\Lambda$ on S.
Hence, this means that any quantum channel must also be \emph{Completely Positive} (CP), i.e.
\begin{equation}
\forall_{\varrho^\t{SA}\in\mathcal{T}(\mathcal{H}_\t{in}\otimes\mathcal{H}_\t{A})}:\quad\Lambda\otimes\mathcal{I}\left[\varrho^\t{\tiny SA}\right]\;\ge\;0,
\label{eq:CP}
\end{equation}
where $\mathcal{I}$ denotes the identity map that trivially acts on the subspace $\mathcal{H}_\t{\tiny A}$ without affecting it.
Notice that, if we restricted the bipartite inputs in the CP-definition \eref{eq:CP} to be separable (see \eqnref{eq:sep_state}),
\eqnref{eq:CP} would be trivially satisfied due to the positivity property of $\Lambda$. Thus, it
is really the ability of considering entangled states between the system and the ancilla, which strengthens 
the notion of positivity of a quantum channel to its complete positivity. 
Let us also note that, due to the linearity of $\Lambda$, we could have
equivalently restricted ourselves in the CP-definition \eref{eq:CP} only
to pure states ${\ket{\Psi}}_\t{\tiny SA}\!\in\!\mathcal{H}_\t{in}\otimes\mathcal{H}_\t{A}$.
As a result, one may directly see that it is sufficient to consider only
$d_\t{\tiny A}\!=\!d_\t{in}$
in order to ensure the CP property of $\Lambda$,
as any bipartite pure state ${\ket{\Psi}}_\t{\tiny SA}$ may always be expressed 
by utilising its \emph{Schmidt decomposition} as 
${\ket{\Psi}}_\t{\tiny SA}\!=\!\sum_{i=1}^{\min\{d_\t{in},d_\t{\tiny A}\}}\!\lambda_i\,{\ket{\psi_i}}_\t{\tiny S}{\ket{\phi_i}}_\t{\tiny A}$
with some $\lambda_i\!\ge\!0$ and orthonormal $\ket{\psi_i}\!\in\!\mathcal{H}_\t{in}$, $\ket{\phi_i}\!\in\!\mathcal{H}_\t{\tiny A}$ \citep{Nielsen2000,Bengtsson2006}.

Summarising, any physical quantum channel corresponds to a 
\emph{Completely Positive Trace-Preserving}
(CPTP) map that is defined to satisfy the above conditions.
We return once more to the analysis of CPTP maps in \secref{sec:qch_geom},
where we further parametrise them w.r.t.~the estimated parameter of interest in
quantum metrological protocols and discuss their \emph{geometrical}
properties.

\paragraph{\emph{Kraus representation} of a quantum channel and the \emph{Stinespring dilation theorem}}~\\
Furthermore, the construction
of \eqnref{eq:evol_tr} allows to naturally introduce the so-called 
\emph{Kraus representation} \citep{Kraus1983} of a given CPTP map
$\Lambda\!:\mathcal{T}(\mathcal{H}_\t{in})\!\to\!\mathcal{T}(\mathcal{H}_\t{out})$.
Notice that without loss of generality we may enlarge the subspace of
$\mathcal{H}_\t{\tiny E}\!\to\!\mathcal{H}_\t{\tiny E}\!\otimes\!\mathcal{H}_{\t{\tiny E}'}$,
and \emph{purify} $\varrho^\t{\tiny E}$ to ${\ket{\xi}}_{\t{\tiny EE}'}$ such that
$\varrho^\t{\tiny E}\!=\!\Tr_{\t{\tiny E}'}\{{\ket{\xi}}_{\t{\tiny EE}'}\bra{\xi}\}$.
Then, we may always rewrite \eqnref{eq:evol_tr} after relabelling the extended 
environment as the ``new effective"  environment, so that $\t{EE}'\!\to\!\t{E}$ and%
\footnote{%
Throughout this work, we represent by the \emph{calligraphic} fount
the \emph{super-operators}---maps acting on density matrices:~$\mathcal{T}(\mathcal{H}_\t{in})\!\to\!
\mathcal{T}(\mathcal{H}_\t{out})$---e.g.~$\mathcal{U}$ and $\mathcal{I}$ for a unitary and identity
channels reprectively, in contrast to $U$ and $\mathbb{I}$ representing operators acting on Hilbert spaces.
}
$\mathcal{U}^\t{\tiny SE}\otimes\mathcal{I}^{\t{\tiny E}'}\!\to\mathcal{U}^\t{\tiny SE}$:
\begin{equation}
\varrho_\t{out}=
\Lambda\!\left[\varrho_\t{in}\right]=
\Tr_{\t{\tiny E}}\!\left\{ U^\t{\tiny SE}\left(\varrho_\t{in}\otimes{\ket{\xi}}_\t{\tiny E}\bra{\xi}\right)U^{\t{\tiny SE} \dagger}\right\}
\;=\;
\sum_{i=1}^{d}\,K_i\,\varrho_\t{in}\,K_i^\dagger,
\label{eq:Kraus_decomp}
\end{equation}
where $K_i\!=\!\prescript{}{\t{\tiny E}}{\bra{i}}U^\t{\tiny SE}{\ket{\xi}}_\t{\tiny E}$
($K_i\!:\mathcal{H}_\t{in}\!\to\!\mathcal{H}_\t{out}$)
are the so-called \emph{Kraus operators} representing the action of $\Lambda$.
Note that $\{K_i\}_i$ are ambiguously defined after choosing
any set of vectors, $\{{\ket{i}}_\t{\tiny E}\}_{i=1}^{d}$, spanning the enlarged environmental subspace $\mathcal{H}_\t{\tiny E}$, 
such that $\sum_{i=1}^d{\ket{i}}_\t{\tiny E}\bra{i}\!=\!\mathbb{I}^\t{\tiny E}$. 
Thus, necessarily $d\!\ge\!\dim\{\mathcal{H}_\t{\tiny E}\}$, but the \emph{minimal} possible 
dimension of $\mathcal{H}_\t{\tiny E}$ which can be chosen to mimic the action of $\Lambda$
for any $\varrho_\t{in}$ defines the so-called \emph{rank}, $r$, of the quantum channel.
In general (see also \secref{sub:CJiso}), $1\!\le\! r \!\le\!d_\t{in}^2$
and $r\!=\!1$ corresponds to the case of a \emph{unitary} map ($\Lambda\!=\!\mathcal{U}$  with 
a trivial single Kraus operator $K\!=\!U$), whereas a quantum channel with $r\!=\!d_\t{in}^2$ is termed to be 
\emph{full-rank}---i.e.~it possesses the maximal possible number of linearly independent Kraus operators
for a given $\mathcal{H}_\t{in}$. 

\begin{figure}[!t]
\begin{center}
\includegraphics[width=0.85\columnwidth]{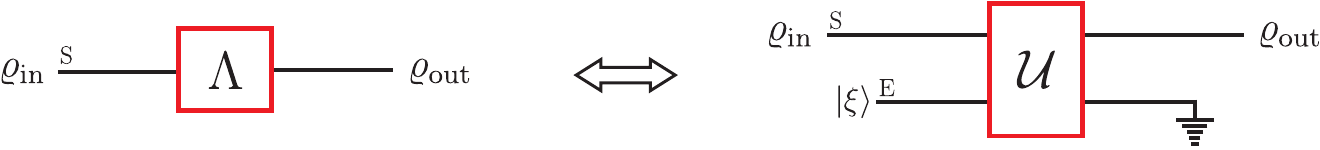}
\end{center}
\caption[Stinespring dilation theorem]{\textbf{Stinespring dilation theorem}
stating that \emph{any} quantum channel $\Lambda\!:\mathcal{T}(\mathcal{H}_\t{in})\to\mathcal{T}(\mathcal{H}_\t{out})$, 
i.e.~CPTP map that transforms all $\varrho_\t{in}\!\in\!\mathcal{T}(\mathcal{H}_\t{in})$ onto  some $\varrho_\t{out}\!\in\!\mathcal{T}(\mathcal{H}_\t{out})$,
may always be rewritten as a unitary transformation acting on an enlarged 
input space, $\varrho_\t{in}\otimes{\ket{\xi}}_\t{\tiny E}\bra{\xi}$,
with the environmental degrees of freedom eventually traced-out.}
\label{fig:stinespring}
\end{figure}

On the other hand, one may also show (see e.g.~\citep{Nielsen2000}) that, 
for a given Kraus representation of a quantum channel, one can always find all
$U^\t{\tiny SE}$, $\ket{\xi}_\t{\tiny E}$ and ${\ket{i}}_\t{\tiny E}$ in \eqnref{eq:Kraus_decomp},
which are defined in the enlarged space $\mathcal{H}_\t{\tiny S}\otimes\mathcal{H}_\t{\tiny E}$
and yield particular Kraus operators. Hence, the general system+environment interpretation 
\eref{eq:evol_tr} of the channel action is indeed always valid. Moreover, as any mapping $\Lambda$ expressed
in the \emph{Kraus form} \eref{eq:Kraus_decomp} must be CP (as $\Lambda$ admits then a 
positive Choi-Jamio\l{}kowski matrix representation \citep{Bengtsson2006}---see \secref{sub:CJiso}),
we arrive at the \emph{Stinespring dilation theorem} (see \figref{fig:stinespring}), which states 
that:

\begin{mythm}[Stinespring dilation theorem]
\label{thm:stinespring}
Any CPTP map may always be written according to \eqnref{eq:Kraus_decomp} 
as a unitary transformation acting on an enlarged space with the environmental degrees of 
freedom eventually traced-out.
\end{mythm}

Hence, the only constraint that must be generally satisfied by any Kraus operators 
reads:~$\sum_{i=1}^d K_i^\dagger K_i\!=\!\mathbb{I}^\t{\tiny S}$,
and it ensures that the TP-property is fulfilled by a given quantum channel.
As a consequence, given a set $\{K_i\}_{i=1}^d$ of Kraus operators of $\Lambda$, 
we may always construct another valid Kraus representation of the map with help
of any rectangular matrix%
\footnote{%
We denote the matrices $\mathsf{u}$ with a different fount 
to indicate that these do \emph{not} represent operators on Hilbert spaces, but 
should be treated as matrices of complex entries just specifying
linear transformations of Kraus operators. 
}
$\mathsf{u}$ of size $d'\!\times\!d$
that for any $1\!\le\!i\!\le\!d$ (and similarly for $j$) satisfies $\sum_{k=1}^{d'}\mathsf{u}_{ik}^\dagger\mathsf{u}_{kj}\!=\!\delta_{ij}$.
It is so, because we can then always write a new set of $d'$ Kraus operators:
\begin{equation}
\tilde K_i \;=\; \sum_{j=1}^d \mathsf{u}_{ij} K_j,
\label{eq:Kraus_trans}
\end{equation}
that also satisfy $\sum_{i=1}^{d'} \tilde K_i^\dagger \tilde K_i\!=\!\mathbb{I}^\t{\tiny S}$
and correctly represent the action of the channel, 
i.e.~$\Lambda[\bullet]\!=\!\sum_{i=1}^{d'}\tilde K_i\bullet\tilde K_i^\dagger$.
However, typically in problems which require optimisation over 
Kraus representations of a 
given quantum channel, it is enough to consider only the 
sets of linearly independent Kraus operators.
Thus, when searching for the optimal Kraus representation,
it is then sufficient to consider any $\{K_i\}_{i=1}^{r}$ with 
$r$ being the rank of the channel $\Lambda$
(e.g.~the unique \emph{canonical} Kraus representation defined with help 
of the Choi-Jamio\l{}kowski matrix---see \secref{sub:CJiso})
and generate all other Kraus representations of interest
via \eqnref{eq:Kraus_trans} after restricting to square---and 
thus \emph{unitary}---matrices $\mathsf{u}$.

\begin{note}[Partial trace as a CPTP map]
\label{note:partial_CPTP}
Lastly, let us show that the
\emph{partial trace} operation (previously introduced
in \secref{sub:QSboundtates} to represent the inability of accessing some
parts of a system) consistently corresponds to a valid CPTP map,
as it may be straightforwardly rewritten in the Kraus form \eref{eq:Kraus_decomp}.
Defining a general reduced density matrix as 
$\varrho^\t{\tiny A}\!=\!\Tr\{\varrho^\t{\tiny AB}\}
\!=\!
\sum_{i=1}^{d_\t{\tiny B}}\prescript{}{\t{\tiny B}}{\bra{i}}\varrho^\t{\tiny AB}{\ket{i}}_\t{\tiny B}$,
we may trivially construct the necessary Kraus operators $K_i\!:\mathcal{H}_\t{\tiny A}\!\otimes\!\mathcal{H}_\t{\tiny B}\!\to\!\mathcal{H}_\t{\tiny A}$,
such that each $K_i\!=\!\mathbb{I}^\t{\tiny A}\otimes\!\prescript{}{\t{\tiny B}\!}{\bra{i}}$ 
and thus $\varrho^\t{\tiny A}\!=\!
\sum_{i=1}^{d_\t{\tiny B}}K_i\,\varrho^\t{\tiny AB}K_i^\dagger$ as required.
\end{note}

\subsection{\emph{Master equation} picture of system evolution}
\label{sub:Evo_MEq}

From an alternative perspective, rather than analysing
the properties of a given quantum channel describing the overall evolution
between times $t_0$ and $t$, one may study
the form of the differential equation on the r.h.s.~of \eqnref{eq:evol_tr}.
Although, such a \emph{master equation} picture is equivalent to the quantum map 
representation given the initial conditions, i.e.~the input state $\varrho_\t{in}\!\equiv\!\varrho^\t{\tiny S}(t_0)$ at $t_0$,
it allows to determine the dynamical equations of motion at any time instance
for the reduced density matrix, and thus generalise the notion of 
classical stochastic processes into the quantum setting%
\footnote{In particular, the terminology of classical Markov master equations,
e.g.~the Chapman-Kolmogorov differential equation, \citep{Breuer2002}.} \citep{Breuer2002}.

Separating from the overall Hamiltonian $\hat H _\t{\tiny SE}$ in \eqnref{eq:evol_tr}
the Hamiltonian $\hat H_\t{\tiny S}$, which specifies the free---\emph{unitary}---evolution 
of the system in the absence of interactions with the environment,
we may write the master equation in the form:
\begin{equation}
\frac{\t{d}}{\t{d}t}\varrho(t)
\;=\;
-\ii\left[\hat{H},\varrho(t)\right]
\;+\;
\mathcal{L}\!\left[\varrho(t)\right]
\label{eq:MarkovMasterEq}
\end{equation}
where we have dropped the system-labelling S without loss of generality
and introduced the \emph{Louvillian} $\mathcal{L}$ (linear super-operator acting
on density matrices), which is responsible for the \emph{dissipative} part of the 
evolution---caused by the destructive impact of the environment. 
Notice that for the purpose of this work we have already assumed the 
Louvillian $\mathcal{L}$ to be \emph{time-independent}, so that 
\eqnref{eq:MarkovMasterEq} constitutes really the \emph{Markovian quantum master equation}
\citep{Breuer2002}. In general, in order to allow for non-Markovian effects \citep{Addis2014,Rivas2014}, 
i.e.~the emergence of memory in the evolution via the ``information back-flow'' from the environment
to the system, one must allow $\mathcal{L}$ to depend on time, yet other 
equivalent generalisations of \eqnref{eq:MarkovMasterEq} are also possible \citep{Chruscinski2010}.

The form of $\mathcal{L}$ may be determined by considering a particular 
physical model of the evolution under the Markovian assumptions of system-environment 
weak-coupling and infinitely small correlation time of the environment \citep{Breuer2002} 
(see e.g.~the \emph{quantum optical master equations} for the model of atom--electromagnetic-field 
interactions \citep{Gardiner2000}). 
However, one may generally show that for the Louvillian to be physical,
it must possess the so-called \emph{Lindblad-Gorini-Kossakowski-Sudarshan} (LGKS)
form, which is \emph{necessary and sufficient} for the corresponding effective quantum
map of \eqnref{eq:evol_tr}, i.e.~$\Lambda_t\!:\varrho(0)\!\to\!\varrho(t)$, to be CPTP 
(as defined in the previous section) for any $t$.

\paragraph{Lindblad-Gorini-Kossakowski-Sudarshan form}~\\
In their seminal works \citet{Gorini1976,Lindblad1976} have demonstrated
that for the Markovian master equation \eref{eq:MarkovMasterEq}
to always yield a valid CPTP map in the quantum channel picture of \secref{sub:Evo_QCh},
the Louvillian $\mathcal{L}$ must generally read:
\begin{equation}
\mathcal{L}\!\left[\varrho(t)\right]
\;=\;
\sum_{k=1}^{d^2-1} \,\gamma_k\,\left[L_k\,\varrho(t)\,L_k^\dagger\;-\;\frac{1}{2}\left\{L_k^\dagger L_k,\varrho(t)\right\}\right],
\label{eq:LGKS}
\end{equation}
where $\gamma_k\!\ge\!0$ are the non-negative `decay rates' of the dissipation (decoherence) process,
$\{A,B\}\!=\!AB\!+\!BA$ denotes the anti-commutator, $d$ is 
the dimension of the system Hilbert space, and $L_k$ are the so-called \emph{Lindblad 
operators}---that formally are \emph{not} constrained to possess any particular structure%
\footnote{Notice that, due to the master equation formulation \eref{eq:MarkovMasterEq}
and the LGKS form \eref{eq:LGKS},
the TP-property $\t{Tr}\{\varrho(t)\}\!=\!1$ is trivially preserved at all times,
as $\Tr\{\mathcal{L}[\varrho(t)]\}\!=\!0$ by construction.}.

Nevertheless, let us already remark that from the perspective of metrology
and, in particular, the results presented in this work, it is the quantum channel 
picture \eref{eq:Kraus_decomp} that turns out to be more appropriate, 
as it provides the geometric properties
of the evolution (discussed in \secref{sec:qch_geom}) that have a crucial
impact on the performance of metrological protocols.
Yet, when the time duration of the evolution, $t$, serves as an extra degree of freedom 
that one can control, or when the dynamics is simply specified 
at the differential equation level (e.g.~for Non-Markovian or non-commuting with $\hat H$
noise models \citep{Matsuzaki2011,Chin2012,Chaves2013}),
the master equation picture \eref{eq:MarkovMasterEq} must still be utilised 
in order to determine the form of the effective map $\Lambda_t$ in \eqnref{eq:evol_tr}, 
which properties will then importantly vary depending on the $t$ considered.
In what follows, we describe the qubit evolution models
of metrological relevance and determine the relation between their
quantum channel and master equation pictures---what happens to be 
straightforward due to the commutativity of the unitary, $\left[\hat H,\bullet\right]$, and 
the dissipative, $\mathcal{L}[\bullet]$, terms in \eqnref{eq:MarkovMasterEq} 
($\left[\hat H,\mathcal{L}[\bullet]\right]\!\equiv\!\mathcal{L}\!\left[\left[\hat H,\bullet\right]\right]$) for the evolutions considered.

\subsection{\caps{Example:} Noisy-phase--estimation channels of relevance to metrology}
\label{sub:noise_models}

\begin{figure}[!t]
\begin{center}
\includegraphics[width=1\columnwidth]{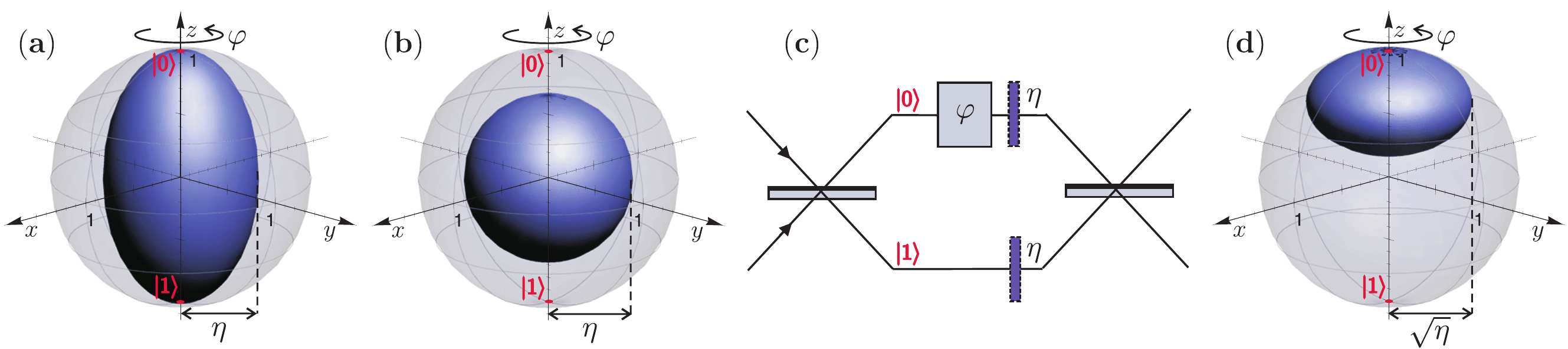}
\end{center}
\caption[Exemplary noisy-phase--estimation channels]
{\textbf{Noisy-phase--estimation channels}
of relevance to quantum metrological schemes.
In each case, a particle is represented by a \emph{qubit}
in the basis $\{\ket{0},\ket{1}\}$, which in the absence of noise solely rotates
around the $z$ axis by an angle specified by the parameter $\varphi$.
The noise is modelled by the quantum maps $\mathcal{D}_\eta$ of strength
$\eta$ (see \tabref{tab:noise_models}) that commute with such a rotation: (\textbf{a}) \emph{dephasing},
(\textbf{b}) \emph{depolarisation}, (\textbf{c}) \emph{loss}, (\textbf{d}) \emph{spontaneous emission}
(\emph{amplitude damping}). In case of the \emph{loss} noise model, 
the overall transformation does not possess a Bloch ball representation as it
corresponds to a qubit$\to$qutrit map. Yet, as shown in (\textbf{c}), the input qubit has then a 
natural interpretation of a photon propagating through an interferometer with 
equal power-transmittances of its arms.
\label{fig:noise_models}}
\end{figure}

Throughout this work, while analysing various tools introduced
to study the noise effects in quantum metrology, we consider the natural
examples of \emph{qubit} evolutions that are of particular relevance to metrological protocols.
As depicted in \figref{fig:noise_models}, these correspond to: 
\emph{dephasing}, \emph{depolarisation}, \emph{loss}, and \emph{spontaneous emission} (\emph{amplitude damping}) 
noise models; that are typically utilised when accounting for decoherence 
in optical interferometry \citep{Demkowicz2015} and atomic experiments \citep{Leibfried2003,Hammerer2010}.

\paragraph{Quantum channel picture}~\\
In order to describe these exemplary evolution models in the quantum channel picture \eref{eq:Kraus_decomp},
we consider a qubit representing a particle (two-level atom, photon in two modes) prepared in a state 
$\varrho_\t{in}$ which undergoes an overall CPTP map $\Lambda_\varphi$ according to
\begin{equation}
\varrho_\t{out}\;=\;\Lambda_\varphi\!\left[\varrho_\t{in}\right]\;=\;\mathcal{D}_\eta\!\left[\mathcal{U}_\varphi\!\left[\varrho_\t{in}\right]\right]\;=\;\sum_{i=1}^rK_i(\varphi)\,\varrho_\t{in}\,K_i^\dagger(\varphi)
\label{eq:ChPic}
\end{equation}
that is composed 
of consecutively a unitary $\mathcal{U}_\varphi$ and a pure-noise $\mathcal{D}_\eta$ channel.
$\mathcal{U}_\varphi\!\left[\varrho\right]\!=\!U_{\varphi}\varrho U_{\varphi}^\dagger$ 
with $U_{\varphi}\!=\!\exp\!\left[-\textrm{i}\hat\sigma_{z}\varphi/2\right]$,
so that it generates a rotation by an angle $\varphi$ around the $z$ axis in the Bloch ball representation (see \figref{fig:noise_models}),
whereas $\mathcal{D}_\eta$ models one of the noise-types depicted in 
\figref{fig:noise_models}, and specified in \tabref{tab:noise_models} with the 
effective noise strength set to $\eta$.
For convenience, we have chosen in \eqnref{eq:ChPic} $\mathcal{U}_{\varphi}$ to act before the decoherence,
yet the ordering of the two in principle plays no role,
as the unitary rotation commutes with the noise models considered. However, such a choice allows us to
similarly define the Kraus representation \eref{eq:Kraus_decomp} of $\Lambda_\varphi$ for all
four cases depicted in \figref{fig:noise_models}, i.e.~as $K_i(\varphi)\!=\!K_i U_\varphi$
with $K_i$ representing any valid (see \eqnref{eq:Kraus_trans}) set of Kraus operators 
for corresponding pure-noise maps $\mathcal{D}_\eta$.
Yet, for each above noise-type, we explicitly specify in \tabref{tab:noise_models} the canonical
Kraus operators of the adequate channel $\mathcal{D}_\eta$, which number then 
also determines (see \secref{sub:CJiso}) the rank%
\footnote{Note that $1\!\le\!r\!\le\!4$ for qubit-input channels, as $d_\t{in}^2\!=\!4$.}
$r$ of both $\mathcal{D}_\eta$ and%
\footnote{Also for $\Lambda_\varphi$, as the concatenation with a unitary map in \eqnref{eq:ChPic} does not vary the rank of a channel.}
$\Lambda_\varphi$.
Notice that the above-introduced formulation correctly applies also to 
the \emph{loss} noise model depicted in \figref{fig:noise_models}(\textbf{c}),
in which case the effective pure-noise map, $\mathcal{D}_\eta$, transforms a qubit into a qutrit 
with the third basis state (vacuum) representing the particle being lost, i.e.~$\{\ket{0},\ket{1}\}\to\{\ket{0},\ket{1},\ket{\t{vac}}\}$. 

\begin{table}[t!]
\begin{center}
\begin{tabular}{|M{2.25cm}|M{5.5cm}|M{5cm}|M{0.75cm}|N}
\hline 
\textbf{Noise model} 
& 
\textbf{Kraus representation of} $\mathcal{D}_{\eta}$
& 
\textbf{Liouvillian} $\mathcal{L}[\varrho]$
& 
$\eta(t)$
&\\[12pt]
\hline 
\hline
\emph{Dephasing}:
\relsize{-1}{($r\!=\!2$)}
&
\relsize{-1}{$K_{1}\!=\!\sqrt{\frac{1+\eta}{2}}\,\mathbb{I}$\;,\;\;$K_{2}\!=\!\sqrt{\frac{1-\eta}{2}}\,\hat\sigma_{z}$}
&
\relsize{-1}{$\frac{\gamma}{2}\left(\hat\sigma_{z}\varrho\,\hat\sigma_{z}-\varrho\right)$}
&
$\textrm{e}^{-\gamma t}$
&\\[16pt]
\hline 
\emph{Depolarization}:
\relsize{-1}{($r\!=\!4$)}
& 
\relsize{-1}{$K_{1}\!=\!\sqrt{\frac{1+3\eta}{4}}\,\mathbb{I}$\;,\;\;$\left\{ K_{i}\!=\!\sqrt{\frac{1-\eta}{4}}\,\hat\sigma_{i-1}\right\}_{i=2}^4$}
& 
\relsize{-1}{$\frac{\gamma}{2}\left(\frac{1}{3}\underset{i=1}{\overset{3}{\sum}}\hat\sigma_{i}\varrho\,\hat\sigma_{i}-\varrho\right)$}
& 
$\textrm{e}^{-\frac{2\gamma}{3}t}$
&\\[20pt]
\hline 
\quad\emph{Loss}:\newline
\relsize{-1}{($r\!=\!3$)}
& 
\relsize{-1}{
$K_{1}\!=\!
\left(\!\!\!\!\begin{array}{cc}
\sqrt{\eta} & 0\\
0 & \sqrt{\eta}\\
0 & 0
\end{array}\!\!\!\right)$\,,\;
$K_{2}\!=\!\left(\!\!\!\!\begin{array}{cc}
0 & 0\\
0 & 0\\
\sqrt{1-\eta} & 0
\end{array}\!\!\!\right)$\,,
$K_{3}\!=\!
\left(\!\!\!\begin{array}{cc}
0 & 0\\
0 & 0\\
0 & \sqrt{1-\eta}
\end{array}\!\!\!\right)\!$ 
}
&
\relsize{-1}{\[\gamma\underset{m=0}{\overset{1}{\sum}}\left(\sigma_{m,+}\varrho\,\sigma_{m,-}-\frac{1}{2}\left\{ \sigma_{m,-}\sigma_{m,+},\varrho\right\} \right)\]}
&
$\textrm{e}^{-\gamma t}$
&\\[60pt]
\hline 
\emph{Spontaneous emission}: 
\relsize{-1}{($r\!=\!2$)}
& 
\relsize{-1}{
$K_{1}\!=\!\left(\!\!\begin{array}{cc}
1 & 0\\
0 & \!\!\sqrt{\eta}
\end{array}\!\!\right)$\,,\;
$K_{2}\!=\!\left(\!\!\begin{array}{cc}
0 & \sqrt{1-\eta}\\
0 & 0
\end{array}\!\!\!\right)$
}
&
\relsize{-1}{$\gamma\left(\sigma_{+}\varrho\,\sigma_{-}-\frac{1}{2}\left\{ \sigma_{-}\sigma_{+},\varrho\right\} \right)$
\relsize{-2}{$\t{~}\quad\qquad\qquad\qquad\qquad\sigma_{\pm}\!=\!\frac{1}{2}\!\left(\hat\sigma_{x}\!\pm\!\textrm{i}\hat\sigma_{y}\right)$}}
& 
$\textrm{e}^{-\gamma t}$ 
&\\[26pt]
\hline 
\end{tabular}
\end{center}
\caption[Characteristics of the exemplary noisy-phase--estimation channels]{%
\textbf{Characteristics of the noisy-phase--estimation channels of \figref{fig:noise_models}}. 
For each qubit-evolution--model considered,
the canonical Kraus representation
of the pure-noise map $\mathcal{D}_{\eta}$
is specified, which also determines the rank, $r$, of the effective map
$\Lambda_\varphi$ in the quantum channel picture \eref{eq:ChPic}. 
Moreover, the corresponding 
Liouvillians, $\mathcal{L}$, of the LGKS form \eref{eq:LGKS}
are presented that apply at the level 
of the master equation \eref{eq:MEPic}.
In case of the \emph{loss} model, $\sigma_{m,+}\!=\!\ket{\textrm{vac}}\!\bra{m}$ 
are the generators of transitions into the vacuum state from the qubit basis vectors $\ket{m\!=\!\{0,1\}}$, 
such that $\sigma_{m,-}\!=\!\sigma_{m,+}^\dagger$.
As all pure-noise maps commute with the phase encoding, in order to construct the 
effective quantum channel, $\Lambda_{\omega,t}$, from the master equation \eref{eq:MEPic}, it is 
sufficient to set $\varphi\!\to\!\omega t$ in $\Lambda_\varphi$ and account for the time-dependence of 
the decoherence-strength parameters, $\eta\!\to\!\eta(t)$, in accordance with the last column.
}
\label{tab:noise_models}
\end{table}

\paragraph{Master equation picture}~\\
On the other hand, we may also determine the equivalent description
of the qubit evolution models of \figref{fig:noise_models} in the master 
equation picture \eref{eq:MarkovMasterEq}
by defining $\varrho(t_0)\!=\!\varrho_\t{in}$ at $t_0$ and specifying:
\begin{equation}
\frac{\t{d}}{\t{d}t}\varrho(t)
\;=\;
-\ii\,\frac{\omega}{2}\left[\hat\sigma_z,\varrho(t)\right]
\;+\;
\mathcal{L}\!\left[\varrho(t)\right],
\label{eq:MEPic}
\end{equation}
with the Liouvillians $\mathcal{L}$ (of the LGKS form \eref{eq:LGKS}) 
listed in \tabref{tab:noise_models} for each of the noise-types considered.
As a result, by integrating \eqnref{eq:MEPic} over the interval $[t_0,t]$
we can construct the effective map%
\footnote{We have on purpose chosen to denote the overall map between $t_0$ and $t$
as $\Lambda_{\omega,t}$, in order to match the convention later used in \secref{sec:loc_freq_est},
where $\omega$ of \eqnref{eq:MEPic} represents then the estimated detuning \emph{frequency} in atomic spectroscopy setups.}
such that $\varrho(t)\!=\!\Lambda_{\omega,t}\!\left[\varrho(t_0)\right]$, i.e.~the equivalent
of $\Lambda_t$ in \eqnref{eq:evol_tr}.
However, for the above qubit models, $\Lambda_{\omega,t}$ may also be
determined with help of the quantum channel picture
by simply setting $\varphi\!\to\!\omega t$ in $\Lambda_\varphi$
of \eqnref{eq:ChPic} and letting the decoherence strength be also a time-dependent function $\eta\!\to\!\eta(t)$---that 
we specify for each of the noise-types in the last column of \tabref{tab:noise_models}.
Importantly, such a straightforward relation between \eqnref{eq:ChPic} -- in which we have 
\emph{artificially} treated the noise and unitary maps to be separate, and \eqnref{eq:MEPic} -- that 
\emph{appropriately} models the noise to occur simultaneously to phase acquisition at all times, 
is only valid due to the commutativity of the unitary $\mathcal{U}_\varphi$ and pure-noise $\mathcal{D}_\eta$ 
channels in \eqnref{eq:ChPic}. In case of more general noise models, e.g. non-Markovian  \citep{Matsuzaki2011,Chin2012} or 
non-commuting \citep{Chaves2013}, we cannot naively separate the decoherence from the free evolution, but must rather
start with the phenomenologically determined equivalent of the master equation \eref{eq:MEPic} (e.g.~with $\mathcal{L}_t$ 
being now $t$-dependent), and by integrating it over $[t_0,t]$
construct the effective quantum map $\Lambda_t$ of \eqnref{eq:evol_tr}
such that $\varrho(t)\!=\!\Lambda_t\!\left[\varrho(t_0)\right]$ for a given $t$. 
On one hand, such a construction is always possible once the expressions for both $\varrho(t_0)$ and $\varrho(t)$ are known%
\footnote{As one may then construct the `dynamical' matrix transforming $\varrho(t_0)\!\to\!\varrho(t)$ and 
`reshuffle' it to obtain the Choi-Jamio\l{}kowski matrix \eref{eq:CJmatrix} \citep{Bengtsson2006}, from which 
the canonical Kraus representation may be directly determined (see \secref{sub:CJiso}).} \citep{Bengtsson2006},
yet one may also follow a general recipe and often construct the Kraus representation of 
the channel $\Lambda_t$ directly from the master equation,
i.e.~without need of explicit integration \citep{Andersson2007}.

\section{Quantum metrology -- estimation of \emph{latent} parameters of a quantum system}
\label{sec:qmet_latpars}

In statistics \citep{Everitt2010}, while analysing classical experiments
that yield random data, one divides 
the variables used in the description of the investigated process into two%
\footnote{In fact, one could also define a third category
that is beyond the scope of this work, i.e.~the 
\emph{hidden} variables -- ones that could in principle be measured and observed
but are not accessible for practical reasons. Notice that these also have
a direct application in the quantum setting, being naturally utilised 
when describing the phenomenon of \emph{quantum non-locality} and, 
in particular, the \emph{local hidden-variable models} \citep{Brunner2014}.}
natural categories of:
the \emph{observable} (manifest) variables -- ones representing
the \emph{properties} of the system undergoing the process 
that may be directly measured; and the \emph{latent} variables -- ones 
corresponding to the \emph{parameters} of the mathematical model assumed
to describe the stochastic process itself. Importantly, the latent parameters
are by definition intrinsic to the description and cannot be measured, so that 
for their determination one must resort to the techniques of \emph{statistical inference} 
\citep{Wasserman2004} and, in particular, the \emph{estimation theory} \citep{Kay1993}.

Such a formalism naturally carries over into the quantum setting, in 
which the observable variables correspond exactly to the 
\emph{quantum observables}---Hermitian operators $\hat O$
introduced in \secref{sub:Qmeas}---that, as explained before, 
indeed represent the physical properties (e.g.~position, momentum, spin, energy etc.)
of a given quantum system. The \emph{quantum latent parameters}, 
on the other hand, represent then the quantities that characterise the
system evolution (e.g.~the time $t$ in \eqnref{eq:schr_eq}, decay rates $\gamma_k$ in \eqnref{eq:LGKS},
$\varphi$ and $\eta$ parameters in \eqnref{eq:ChPic} or $\omega$ in \eqnref{eq:MEPic} etc.),
and should be interpreted as classical (possibly random) variables 
inbuilt in the description, extrinsic to the quantum system.

Importantly, given a fixed state of the system, $\varrho$, the
latent parameters cannot be even defined unless 
an explicit dynamical model is provided that explains 
how the parameters of interest have been \emph{encoded}.
Formally, when we label the state as $\varrho_\varphi$ with 
$\varphi$ representing a latent parameter, we 
also implicitly have in mind the $\varphi$-encoding process and thus 
the knowledge of how the form of 
$\varrho_\varphi$ varies with $\varphi$. On the other hand, 
as the most general quantum measurements
that may be performed on such a state,
i.e.~the POVMs $\{M_i\}_i$ introduced in \secref{sub:Qmeas},
just yield classical outcomes distributed with ($\varphi$-dependent)
probabilities $p_i(\varphi)\!=\!\Tr\{\varrho_\varphi M_i\}$,
in order to most accurately determine the value of $\varphi$
we must still utilise the classical estimation techniques \citep{Kay1993},
so that the parameter can be most efficiently inferred from the sampled data.
Hence, the extra freedom that the quantum framework really gives us is 
the choice of the \emph{optimal input state} on which the parameter $\varphi$ 
may be encoded,
but also the choice 
of the most effective measurement strategy---the \emph{optimal POVM}---that
leads to the distribution of outcomes 
from which $\varphi$ can then be most accurately deduced.
As a consequence, one may neatly conclude that 
the \emph{purpose of quantum metrology} simply boils down to
\begin{quote}
\begin{center}
``the most precise determination of the latent
parameters of a quantum system'',
\end{center}
\end{quote}
what, in practice, combines the optimisation of both states and 
measurements at the quantum mechanical level with classical
estimation tools that must be utilised at the statistical-data 
interpretation stage.

\subsection{Uncertainty relations for quantum observables}
Before focussing on the problem of quantum latent-parameters determination,
let us briefly discuss the consequences of the physical properties of a quantum
system being described by means of the observable formalism.
Crucially, an observable $\hat O$ fully determines the statistics
of its measurement, as every moment
of the random variable $O$ 
representing the measurement outcomes (being the 
eigenvalues, $o_i$, of $\hat O$, see \secref{sub:Qmeas}),
can be generally written as
\begin{equation}
\forall_{k\in\mathbb{N}}:\quad\left<O^k\right>=\sum_i p_i o_i^k=\sum_i\left<P_i\right>o_i^k=\left<\sum_iP_i o_i^k \right>=\left<\left(\sum_i P_i o_i\right)^k \right>=\left<\hat O^k \right>,
\label{eq:mom_obs}
\end{equation}
where we have acknowledged that the projectors $P_i$ satisfy $P_iP_j\!=\!P_i\delta_{ij}$, 
and naturally generalised the averaging operation, $\left<\bullet\right>\!=\!\sum_i \!p_i\,\bullet$, 
to the quantum setting, so that for a given state $\varrho$:
$\left<\bullet\right>\!=\!\Tr\{\varrho\,\bullet\}$. Importantly,
\eqnref{eq:mom_obs} proves that all the statistical measures 
of the random variable $O$ are just represented by their 
quantum mechanical (``hatted'') equivalents evaluated for the observable $\hat O$.
For instance, the \emph{variance} $\t{Var}\!\left[O\right]\!=\!\left<(O-\left<O\right>)^2\right>$
is directly translated onto%
\footnote{Following the standard convention,
we label by $\Delta^2\hat O$ the \emph{variance} of an observable.
Yet, as for random variables $\Delta^2 O$ is typically utilised to express the \emph{Mean Squared Error} \eref{eq:MSE} 
in estimation theory, we use the $\t{Var}[O]$ notation instead.} 
$\Delta^2 \hat O\!=\!\left<(\hat O-\left<\hat O\right>)^2\right>$,
which thus equivalently quantifies the spread of the outcomes distribution, 
when the observable of interest is measured.

One of the spectacular features of the quantum 
theory is the phenomenon of \emph{incompatibility of the observables} \citep{Griffiths2013,Phillips2003}, 
which states 
that if two observables, say $\hat A$ and $\hat B$, do not commute with one another,
i.e.~$\left[\hat A,\hat B\right]\!\ne\!0$,
there does not exist a quantum system which possesses both physical properties
associated with $\hat A$ and $\hat B$ precisely defined.
Such a fact is a direct consequence of the so-called \emph{Heisenberg uncertainty relation}
ensuring that one can only decrease one of the observable variances
at the expense of the other, i.e.
\begin{equation} 
\Delta^2\hat A\; \Delta^2\hat B 
\quad\geq\quad
\frac{1}{4}\left|\langle[\hat{A},\hat{B}]\rangle\right|^{2}+\left| \t{cov}(\hat A,\hat B) \right|^{2}
\quad\ge\quad
\frac{1}{4}\left|\langle[\hat{A},\hat{B}]\rangle\right|^{2},
\label{eq:un_rel}
\end{equation}
where the above stronger and weaker bounds have been discovered by (and are named
after) \citet{Schrodinger1930} and \citet{Robertson1934} respectively.
Similarly to random variables, $\t{cov}(\bullet,\bullet)$ denotes the \emph{covariance} of 
quantum observables, 
i.e.~$\t{cov}(\hat A,\hat B)\!=\!\frac{1}{2}\left<\left\{\hat A,\hat B\right\}\right>-\left<\hat A\right>\!\left<\hat B\right>$.
Operationally, \eqnref{eq:un_rel} means that given \emph{infinitely many} copies 
of the same system, if we measure $\hat A$ on half of them, while
$\hat B$ on the rest, the variances determined by the two data sets
collected must satisfy the above Robertson-Schr\"{o}dinger
inequalities. For completeness, let us note that this contrasts the scenario 
of the recently vividly researched topic of the 
\emph{noise-disturbance uncertainty relations} \citep{Busch2013,Ozawa2004,Branciard2013} 
which aim to relate the precision with which one measures
one of the observables, say $\hat A$, on a \emph{single} copy of the system to the 
error in measurement of the other observable, $\hat B$, performed afterwards.

\subsection{Inferring a latent parameter from a quantum observable}
At this preliminary stage of presentation, let us consider
a simple strategy of latent parameter inference,
in which we measure some  observable $\hat O$ of a system 
in a state $\varrho_\varphi$, in order to most accurately determine 
a fixed, \emph{deterministic} latent parameter $\varphi$. 
We achieve this by constructing an estimate%
\footnote{In fact, $\tilde \varphi$ is formally termed an \emph{estimator}, as defined later in \secref{sec:ClEst}.}
of $\varphi$, call it $\tilde \varphi$, which is built on a single outcome and
thus corresponds to a function of the random variable $O$:~$\tilde \varphi (O)$. 
Hence, if we assume that $O$ fluctuates in a very narrow region
around its mean%
\footnote{%
What (owing to the Central Limit Theorem) can always be assured, if we conduct 
a large enough number of procedure repetitions.
In fact, this is exactly the assumption of \emph{locality}
that is essential when pursuing the frequentist approach to 
parameter estimation (see \secref{sub:LocConseq}).
}, so that
 $\t{Var[O]}\!=\!\Delta^2\hat O\ll1$,
we may \emph{locally} Taylor-expand $\tilde \varphi$ up to first order around $\left\langle\hat O\right\rangle$ \citep{Barlow2013}:
\begin{equation}
\tilde{\varphi}(O)\;=\;\tilde{\varphi}\!\left(\left\langle \hat{O}\right\rangle \right)+\left.\frac{\textrm{d}\tilde{\varphi}}{\textrm{d}O}\right|_{\left\langle \hat{O}\right\rangle }\left(O-\left\langle \hat{O}\right\rangle \right)+\;\dots
\label{eq:Taylor_est}
\end{equation}
and by squaring and averaging \eqnref{eq:Taylor_est} obtain the \emph{error-propagation formula} (e.g. \citep{Wineland1992}):
\begin{equation} 
\Delta^{2}\tilde{\varphi}\;\approx\;\frac{\Delta^{2}\hat{O}}{\left|\frac{\textrm{d}\left\langle \hat{O}\right\rangle }{\textrm{d}\varphi}\right|^{2}},
\label{eq:ErrPropForm}
\end{equation}
where we have defined%
\footnote{$\Delta^2\tilde\varphi$ actually corresponds 
to the \emph{Mean Squared Error} \eref{eq:MSE} introduced later in \secref{sec:ClEst}.}
$\Delta^{2}\tilde{\varphi}\!=\!\left<\left(\tilde{\varphi}(O)-\varphi\right)^2\right>$ 
and assumed that at the mean value $\left<\hat O\right>$ we precisely
estimate the true parameter, i.e.~$\tilde{\varphi}\!\left(\left\langle \hat{O}\right\rangle \right)\!=\!\varphi
\implies \frac{\textrm{d}\tilde{\varphi}(\left\langle \hat{O}\right\rangle )}{\textrm{d}\varphi}\!=\!1$,
so that
$\left.\frac{\textrm{d}\tilde{\varphi}}{\textrm{d}O}\right|_{\left\langle \hat{O}\right\rangle }\!=\!\frac{\textrm{d}\tilde{\varphi}(\left\langle \hat{O}\right\rangle )}{\textrm{d}\varphi}\left[\frac{\textrm{d}\left\langle \hat{O}\right\rangle }{\textrm{d}\varphi}\right]^{-1}\!=\!\left[\frac{\textrm{d}\left\langle \hat{O}\right\rangle }{\textrm{d}\varphi}\right]^{-1}$.

Importantly, the error-propagation formula \eref{eq:ErrPropForm} 
quantifies the fluctuations of our latent-parameter--estimate $\tilde \varphi$
around the true value $\varphi$, which are unavoidable due to the
stochasticity of the observable measurement. As a result, 
\eqnref{eq:ErrPropForm} also specifies the ultimate sensitivity
of $\tilde \varphi$ to any variations of $\varphi$ (still being deterministic!) 
in such a ``small fluctuations of $O$'' (local---see later \secref{sub:ClEstLocal}) regime,
for a particular observable that is assumed to be measured. This
makes it a powerful tool that may be utilised to quantify performance 
of quantum metrological protocols employing a specific observable-based 
measurement strategy. In fact, it was \eqnref{eq:ErrPropForm} that has been 
utilised in the pioneering works of 
\citep{Caves1981,Bondurant1984,Yurke1986,Holland1993,Sanders1995,Dowling1998} to 
determine the maximal precision with which \emph{phase} may be resolved in optical interferometry,
and similarly for the estimation of atomic transition \emph{frequency} in atomic spectroscopy 
\citep{Wineland1992,Wineland1994,Bollinger1996}.

\begin{note}[Mandelstam-Tamm inequality -- time-energy uncertainty relation]
\label{note:time_energy}
A remarkable consequence of both the Robertson inequality \eref{eq:un_rel}
and the error-propagation formula \eref{eq:ErrPropForm} is the \emph{time-energy 
uncertainty relation} originally discovered by \citet{Mandelstam1945}%
\footnote{For an alternative version of the time-energy uncertainty relation based
on different principles, see \citep{Margolus1998}.}.
In the special case of \emph{time} being the parameter estimated, i.e.~$\varphi\!\equiv\!t$,
the derivative of the operator mean in \eqnref{eq:ErrPropForm}, may be 
generally rewritten utilising the dynamical von Neumann equation \eref{eq:schr_eq}, 
as $\frac{\textrm{d}\left\langle \hat{O}\right\rangle }{\textrm{d}t}\!=\!\ii\left\langle \left[\hat{H},\hat{O}\right]\right\rangle$.
As a result, after substituting $\hat A\!=\!\hat H$ and $\hat B\!=\!\hat O$ into
\eqnref{eq:un_rel}, as well as for $\Delta^2\hat O$ with help the error-propagation formula,
we obtain (independently of $\hat O$ assumed) the so-called \emph{Mandelstam-Tamm inequality}:
\begin{equation}
\Delta^2 \hat H\;\Delta^2 \tilde{t}\;\ge\;\frac{1}{4},
\label{eq:MTineq}
\end{equation}
which states that the variance of the Hamiltonian $\hat H$
sets a lower limit on the magnitude of fluctuations of the time-estimate $\tilde t$,
and thus the resolution with which one can sense the variations of the elapsed time $t$.
Notice that $\Delta^2 \hat H$ is really determined by the energy spectrum
of a given system, so that, for instance, assuming the system  to be in 
a pure state $\ket{\psi}$ written in the basis of the energy-eigenstates as
$\ket{\psi}\!=\!\sum_i\alpha_i\ket{E_i}$, where $\forall_i\!:\hat H\ket{E_i}\!=\!E_i\ket{E_i}$, 
the Hamiltionian variance just 
reads: $\Delta^2 \hat H\!=\!\sum_i \!E_i^2\,|\alpha_i|^2-\left(\sum_j\!E_j|\alpha_j|^2\right)^2$.
\end{note}

\section{Geometry of $\varphi$-parametrised quantum channels}
\label{sec:qch_geom}

Lastly, in order to describe in more detail the evolution of a quantum system 
employed in a metrological scenario,
we follow \secref{sub:Evo_QCh} and apply the language of quantum channels
that are then crucially responsible for the encoding of the latent parameter to be determined.
Hence, after identifying $\varrho_\varphi$ as the output state (previously 
labelled as $\varrho_\t{out}$ in \secref{sub:Evo_QCh}), we generally write 
$\varrho_\varphi\!=\!\Lambda_\varphi\!\left[\varrho_\t{in}\right]$,
where the evolution CPTP map,
$\Lambda_\varphi\!:\mathcal{T}(\mathcal{H}_\t{in})\!\to\!\mathcal{T}(\mathcal{H}_\t{out})$,
is now explicitly parametrised by the estimated parameter. Importantly,
we can thus effectively treat the variations of $\varphi$ as changes 
in the form of $\Lambda_\varphi$,
and---by defining a \emph{family} of CPTP maps $\{\Lambda_\varphi\}_\varphi$
parametrised by the latent parameter $\varphi$---interpret the $\varphi$-estimation
task as a problem of determining which of the channels from the 
family has acted on the input state $\varrho_\t{in}$.

As it is thus the family $\{\Lambda_\varphi\}_\varphi$ that contains 
all the information about the latent parameter, it is necessary to describe 
in more detail its mathematical form. That is why, in what follows, we introduce yet another tool
of quantum-channel formalism, i.e.~the \emph{Choi-Jamio\l{}kowski} (CJ)
\emph{isomorphism}, which allows us to establish 
the geometrical structure of the space of quantum channels.
As result, we are able to study the \emph{geometry}
of $\varphi$-parametrised CPTP maps and, in particular,
define the notion of \emph{$\varphi$-extremality} of
a given channel $\Lambda_\varphi$. 
We later show that the $\varphi$-extremality property
plays an important role when analysing metrological scenarios,
in which $\Lambda_\varphi$ is responsible for the parameter encoding. 
In the last part of this section, in order to make these ideas clear, 
we discuss in detail the geometrical properties of the 
exemplary qubit noisy-phase--estimation channels 
introduced in \secref{sub:noise_models}.

\subsection{Choi-Jamio\l{}kowski isomorphism}
\label{sub:CJiso}

Given a general quantum channel, $\Lambda\!:\mathcal{T}(\mathcal{H}_\t{in})\!\to\!\mathcal{T}(\mathcal{H}_\t{out})$,
introduced in \secref{sub:Evo_QCh}, we define its \emph{Choi-Jamio\l{}kowski} (CJ) \emph{matrix}
\citep{Choi1975,Jamiolkowski1972}, $\Omega_\Lambda$,
as a state supported by an enlarged space $\mathcal{T}\!\left(\mathcal{H}_\t{out}\!\otimes\!\mathcal{H}_\t{\tiny A}\right)$
(as in \eqnref{eq:CP}) with $\dim{\mathcal{H}_\t{\tiny A}}\!=\!d_\t{in}$, such that%
\footnote{Notice the similarity to the CP-property definition \eref{eq:CP}. In general, the CJ-isomorphism 
\eref{eq:CJmatrix} is valid for \emph{any} linear transformation $\Lambda$,
but $\Omega_\Lambda\!\ge\!0$ \emph{if and only if} $\Lambda$ is CP.\label{ft:CJ-CP}}
\begin{equation}
\Omega_\Lambda=\Lambda\otimes\mathcal{I}\!\left[\left|\mathbb{I}\right\rangle \!\left\langle \mathbb{I}\right|\right]\,,
\label{eq:CJmatrix}
\end{equation}
where $\left|\mathbb{I}\right\rangle \!=\!\sum_{i=1}^{d_\t{in}}\!\left|i\right\rangle_\t{\tiny S} \left|i\right\rangle_\t{\tiny A}$
is an (unnormalised) maximally entangled state defined on $\mathcal{\mathcal{H}_\t{in}}\!\otimes\!\mathcal{H}_\t{\tiny A}$---the 
system (S) input Hilbert space and the one of the ancilla (A).
We have without loss of generality defined $\left|\mathbb{I}\right\rangle$ to be unnormalised and thus $\Tr\{\Omega_\Lambda\}\!=\!d_\t{in}$,
in order to benefit from a concise notation for bipartite states, in which $\left|\phi\right\rangle \!=\!\sum_{i,j=1}^{d_\t{in}}\!\left\langle i\right|\!\phi\!\left|j\right\rangle \left|i\right\rangle_\t{\tiny S} \left|j\right\rangle_\t{\tiny A} \!=\!\phi\otimes\mathbb{I}\left|\mathbb{I}\right\rangle \!=\!\mathbb{I}\otimes\phi^{T}\left|\mathbb{I}\right\rangle$. For instance, for any three 
operators $A,B,C\!\in\!\mathcal{T}(\mathcal{H})$ with $\dim(\mathcal{H})\!=\!d_\t{in}$, we may then
equivalently write $A\!\otimes\!C\ket{B}\!=\!\ket{ABC^T}\!=\!ABC^T\!\!\otimes\!\mathbb{I}\,\ket{\mathbb{I}}\!=\!\mathbb{I}\!\otimes\!CB^T\!A^T\ket{\mathbb{I}}$.

As a result, given \emph{any} Kraus representation of $\Lambda$, e.g.~$\left\{ K_{i}\right\} _{i=1}^{d}$ of \eqnref{eq:Kraus_decomp}, 
we can always write the CJ matrix \eref{eq:CJmatrix} as $\Omega_\Lambda\!=\!\sum_{i=1}^d\ket{K_{i}}\!\bra{K_{i}}$, so that 
it may be interpreted as a mixture of states $\ket{K_{i}}$ and thus must be \emph{positive semi-definite}\footref{ft:CJ-CP}. 
On the other hand, assuming that $\Omega_\Lambda\!\ge\!0$, we may always perform its
eigendecomposition to obtain $\Omega_\Lambda\!=\!\sum_{i=1}^r \lambda_i\ket{\psi_i}\!\bra{\psi_i}$ with $\lambda_i\!>\!0$,
and by utilising the above bipartite notation \emph{unambiguously}
construct the \emph{canonical Kraus operators} of the corresponding quantum channel $\Lambda$ 
as $\{K_i\!=\!\sqrt{\lambda_i}\psi_i\}_{i=1}^r$ 
that satisfy $K_i\!\otimes\!\mathbb{I}\,\ket{\mathbb{I}}
\!=\!\sqrt{\lambda_i}\ket{\psi_i}$.
Importantly, the \emph{rank} $r$ (number of $\lambda_i\!>\!0$) of the CJ matrix thus
represents exactly to the \emph{rank} $r$ (defined in \secref{sub:Evo_QCh}) of its corresponding quantum channel.
Furthermore, by the above argumentation \emph{the map $\Lambda$ is CP---admits a Kraus representation---if 
and only if $\Omega_\Lambda\!\ge\!0$}. The TP
property of $\Lambda$, on the other hand, is assured by a constraint on
the CJ matrix:~$\Tr_\t{\tiny S}\{\Omega_\Lambda\}\!=\!\mathbb{I}^\t{\tiny A}\Leftrightarrow\sum_i\!K_i^\dagger K_i\!=\!\mathbb{I}^\t{\tiny S}$.

Formally, \eqnref{eq:CJmatrix} defines a linear mapping from the \emph{space of quantum channels} 
$\Lambda\!:\mathcal{T}(\mathcal{H}_\t{in})\!\to\!\mathcal{T}(\mathcal{H}_\t{out})$ onto the \emph{space
of their CJ matrices}  $\Omega_\Lambda\!\in\!\mathcal{T}\!\left(\mathcal{H}_\t{out}\!\otimes\!\mathcal{H}_\t{\tiny A}\right)$. 
On the other hand,
an inverse linear mapping may also be constructed,
by realising that the action of any channel $\Lambda$ may be written 
with use of its CJ matrix \eref{eq:CJmatrix} (see e.g.~\citep{Keyl2007,Bengtsson2006}) via:
\begin{equation}
\forall_{\varrho_\t{in}\in\mathcal{H}_\t{in}}\!: \quad
\Lambda\!\left[\varrho_\t{in}\right]=\Tr\!\left\{\Omega_\Lambda\!\left(\mathbb{I}\otimes\varrho_\t{in}^T\right)\right\}.
\label{eq:ch_action_via_CJ}
\end{equation}
Thus, the mapping between the two spaces actually 
corresponds to an \emph{isomorphism} \citep{Jamiolkowski1972}, so that importantly 
the space of all quantum channels $\Lambda\!:\mathcal{T}(\mathcal{H}_\t{in})\!\to\!\mathcal{T}(\mathcal{H}_\t{out})$
possesses the same \emph{geometric properties} as the 
space of the quantum states%
\footnote{%
Here, normalised to $\Tr\{\Omega\}\!=\!d_\t{in}$ rather to unity, what, however, is irrelevant.
}
$\Omega\!\in\!\mathcal{T}\!\left(\mathcal{H}_\t{out}\!\otimes\!\mathcal{H}_\t{\tiny A}\right)$.

\begin{note}[Convexity of the space of quantum channels]
\label{note:conv_channels}
In particular, as the space of density operators is \emph{convex} (see \noteref{note:conv_states}), so is the
space of quantum channels. In order to prove such a fact explicitly, note 
that if we construct a convex combination of two CPTP maps $\Lambda_{1/2}$,
i.e.~$\Lambda\!=\!\lambda\,\Lambda_1\!+\!(1\!-\!\lambda)\Lambda_2$ with any $0\!\le\!\lambda\!\le\!1$,
owing to the linearity of \eqnref{eq:CJmatrix} the corresponding CJ matrix
is also just a convex sum of $\Omega_{\Lambda_{1/2}}$, i.e.~$\Omega_\Lambda\!=\!\lambda\,\Omega_{\Lambda_1}\!+\!(1\!-\!\lambda)\Omega_{\Lambda_2}$.
As $\Omega_\Lambda$ is thus trivially positive semi-definite and satisfies $\Tr_\t{\tiny S}\{\Omega_\Lambda\}\!=\!\mathbb{I}^\t{\tiny A}$, 
its equivalent channel $\Lambda$ must respectively fulfil the CP and TP properties, and hence belong to the space of valid quantum maps.
\end{note}

\subsection{Extremal and $\varphi$-extremal quantum channels}
\label{sub:ExtremCh}

\paragraph{Extremal channels}~\\
Having shown that the space of all quantum channels can be equivalently 
interpreted as a \emph{convex} space of the corresponding CJ matrices \eref{eq:CJmatrix},
we define a given channel to be \emph{extremal}, if it cannot be
decomposed into a convex sum of other CPTP maps \citep{Bengtsson2006}:

\begin{mydef}[Extremality of a quantum channel]
\label{def:ch_extrem}
A CPTP map $\Lambda$ is \emph{extremal} if and only if there do \emph{not} exist
any distinct CPTP maps $\Lambda_{1/2}$, such that $\Lambda\!=\!\lambda \,\Lambda_1\!+\!(1\!-\!\lambda)\Lambda_2$ for some $0\!<\!\lambda\!<\!1$.
\end{mydef}

\begin{figure}[!t]
\begin{center}
\includegraphics[width=1\columnwidth]{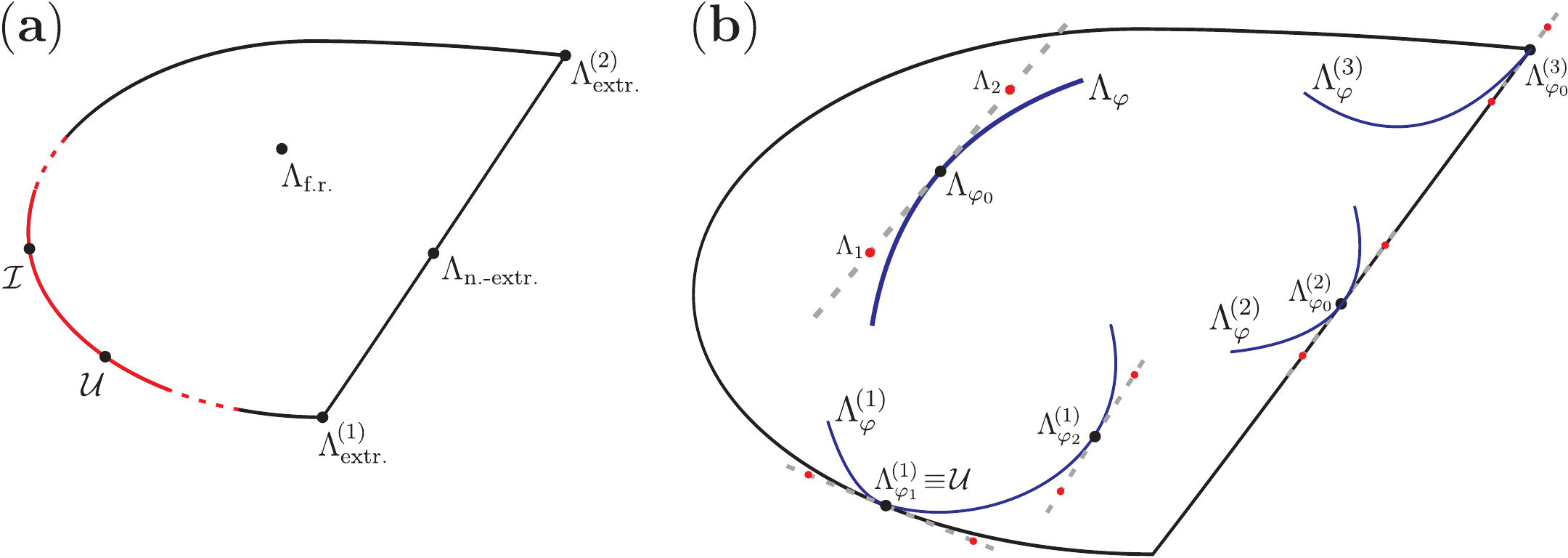}
\end{center}
\caption[Geometry of quantum channels -- extremality and $\varphi$-extremality]{
\textbf{Geometry of quantum channels -- extremality and $\varphi$-extremality}.\\
The \emph{convex} space of quantum channels is depicted as an oval shape 
with one of the sides cut to schematically represent the consequences
of the TP constraint selecting a hyperplane of CPTP maps in a larger space of CP maps.
\\
(\textbf{a}) \textbf{Extremality} property analysed for various channels.\\
$\mathcal{U}$ -- \emph{unitary} channel, being an \emph{extremal} map, lies 
within a convex surface of unitaries (\emph{red}) that also contain the identity map $\mathcal{I}$;
$\Lambda_\t{f.r.}$ --  exemplary \emph{full-rank} channel is naturally \emph{non-extremal},
as it lies strictly inside the channel space;
$\Lambda_\t{extr.}^{\!(1/2)}$ -- non-unitary channels that are
\emph{extremal}---not being decomposable into a mixture of CPTP maps;
$\Lambda_\t{n.-extr.}$ -- \emph{non-full-rank} channel which is located at 
a ``flat'' boundary, what still makes the map \emph{non-extremal}
(e.g.~may be expressed as a mixture of $\Lambda_\t{extr.}^{\!(1/2)}$).
\\
(\textbf{b}) \textbf{$\varphi$-extremality} property studied for 
four $\varphi$-parametrised families of CPTP maps: $\Lambda_\varphi$, $\{\Lambda_\varphi^{\!(i)}\}_{i=1}^3$.\\
$\Lambda_\varphi$ (\emph{thick blue}) -- presented to explicitly show the
tangential decomposition specified in \defref{def:ch_phi_extrem} that
yields the channel to be \emph{$\varphi$-non-extremal} at $\varphi_0$. As 
$\Lambda_{\varphi_0}$ is \emph{full-rank} such a decomposition is possible
irrespectively of the direction at which the curve crosses it;
$\Lambda_\varphi^{(1)}$ -- at $\varphi_1$ the channel is unitary 
and thus extremal, what implies \emph{$\varphi$-extremality}.
Yet, for any other parameter value, e.g.~$\varphi_2$, it is \emph{$\varphi$-non-extremal};
$\Lambda_\varphi^{(2)}$ -- at $\varphi_0$ the channel ceases
to be full-rank, but as the boundary is ``flat'' in the direction of 
$\partial_\varphi \Lambda_\varphi|_{\varphi_0}$ $\Lambda_{\varphi_0}^{(2)}$
is \emph{$\varphi$-non-extremal};
$\Lambda_\varphi^{(3)}$ -- at $\varphi_0$ the channel becomes extremal
and hence $\varphi$\emph{-extremal}.
}
\label{fig:ExtremCh}
\end{figure}

Geometrically, a natural consequence of \defref{def:ch_extrem} is the statement
that in the space of all CPTP maps there cannot exist a ball of valid quantum channels 
surrounding an extremal map, as then the above decomposition could always be constructed. 
Hence, as schematically depicted in \figref{fig:ExtremCh}(\textbf{a}), extremal channels can lie 
within the space of all quantum channels only at the \emph{boundaries}, which further cannot be ``flat'', as 
this would still allow for a decomposition into a mixture of other CPTP maps.

Following the same argumentation as in \noteref{note:conv_channels}, \defref{def:ch_extrem}
can be directly translated onto the space of CJ matrices and redefined in terms
of their convex combinations. Hence, a natural class of extremal CPTP maps 
may be specified by considering the ones that
yield CJ matrices of rank one, i.e.~lead to pure
$\Omega_\Lambda\!=\!\ket{\psi}\!\bra{\psi}$ in \eqnref{eq:CJmatrix}. As 
this may only occur when $\ket{\psi}\!=\!U\!\otimes\!\mathbb{I}\,\ket{\mathbb{I}}$ (pure 
CJ matrices \eref{eq:CJmatrix} correspond to unitary channels and vice versa \citep{Bengtsson2006}), 
all \emph{unitary} quantum channels serve as examples of extremal maps.
From the geometrical perspective (see \figref{fig:ExtremCh}(\textbf{a})), 
the unitary maps $\mathcal{U}$ being continuously parametrisable must not only lie at the 
boundary of the channel space, but also should intuitively form a ``smooth convex surface''
containing the (trivial unitary) identity map $\mathcal{I}$.

Notice that if we (incorrectly) assumed the CP condition to be sufficient
for a quantum channel to be physical, there would not exist any other extremal
channels, as in the space of positive semi-definite  density operators, $\Omega_\Lambda\!\ge\!0$, 
pure states are the only extremal points \citep{Bengtsson2006}. 
Yet, one must not forget also to impose the TP property
corresponding to an extra linear constraint on the CJ matrices: $\Tr_\t{\tiny S}\{\Omega_\Lambda\}\!=\!\mathbb{I}^\t{\tiny A}$,
which geometrically selects a hyperplane of CPTP maps in a larger space of CP maps. 
This may lead to non-trivial facets of the space of quantum channels,
what we schematically represent in \figref{fig:ExtremCh}
by a flat fragment of the boundary.
Importantly, it is thus the TP-property that is responsible for the existence
of non-unitary extremal CPTP maps.
In general, such non-trivial extremal channels may be identified by 
utilising the \emph{Choi criterion} that---as proven in \appref{chap:appChoiCrit}
following \citep{Choi1975}---is equivalent to the channel-extremality definition 
(\defref{def:ch_extrem}), but also provides a recipe of how to verify
if a given channel is extremal basing on its Kraus representation:

\begin{mycrit}[\textsf{Choi criterion for channel extremality}]
\label{crit:Choi}
Given a quantum channel $\Lambda$ of rank $r$ and a set of its linearly independent Kraus 
operators $\{K_i\}_{i=1}^r$, $\Lambda$ is \emph{extremal}
if and only if $\left\{K_i^\dagger K_j\right\}_{ij}$ is a set of $r^2$ 
\emph{linearly independent} matrices.
\end{mycrit}

On other hand, we term all quantum channels that 
do \emph{not} fulfil the above Choi criterion---or equivalently,
by \defref{def:ch_extrem}, \emph{are} decomposable into a convex sum of 
CPTP maps---to be \emph{non-extremal}. 

\paragraph{$\varphi$-extremal channels}~\\
A family of CPTP maps $\{\Lambda_\varphi\}_\varphi$,
where $\varphi$ is the latent parameter to be estimated
in a metrological scenario, geometrically corresponds 
to a \emph{curve} in the space of quantum channels
(\emph{solid blue} line(s)  in \figref{fig:ExtremCh}(\textbf{b})).
Crucially, the nature of a quantum metrological problem 
intrinsically defines a ``sense of direction'' along which 
the parameter of interest varies. Hence, we naturally adapt 
the concept of channel extremality
specified in \defref{def:ch_extrem}, so that
it encapsulates such notion of parameter-induced direction:

\begin{mydef}[$\varphi$-extremality of a quantum channel]
\label{def:ch_phi_extrem}
An element of a family of CPTP maps $\{\Lambda_\varphi\}_\varphi$ is 
\emph{$\varphi$-extremal} at a given $\varphi_0$ if and only if there do \emph{not} exist
any distinct CPTP maps $\Lambda_{1/2}$ that lie in the space of quantum channels
along the line \emph{tangential} at $\varphi_0$ to the curve representing the family,
and yield $\Lambda\!=\!\lambda \,\Lambda_1\!+\!(1\!-\!\lambda)\Lambda_2$  for some $0\!<\!\lambda\!<\!1$.
\end{mydef}

Although the above definition possesses a neat geometrical interpretation,
which we explicitly depict in \figref{fig:ExtremCh}(\textbf{b})
(\emph{thick blue line}), it can be formalised with help of the CJ-matrix representation \eref{eq:CJmatrix}.
Notice that the tangential direction in the space of quantum channels (and equivalently
CJ matrices)  is specified at a given $\varphi_0$ by the derivative%
\footnote{%
Throughout this work, by derivative of a matrix/vector we 
simply mean the matrix/vector obtained after 
computing the derivatives of all the entries.}
of the CJ matrix:~$\dot\Omega_{\Lambda_{\varphi_0}}\!\!\equiv\!\left.\partial_\varphi\Omega_{\Lambda_\varphi}\right|_{\varphi=\varphi_0}$
(which then by \eqnref{eq:ch_action_via_CJ} also defines the derivative of 
the channel $\left.\partial_\varphi\Lambda_\varphi\right|_{\varphi=\varphi_0}$).
Thus, all the maps lying along the tangent (i.e.~along 
a given \emph{dashed grey line} in \figref{fig:ExtremCh}(\textbf{b})) 
may be defined as the ones with CJ 
matrices reading $\Omega_{\Lambda_{\varphi_0}}\!\!+\!\epsilon\,\dot\Omega_{\Lambda_{\varphi_0}}$
for any $\epsilon\!\in\!\mathbb{R}$. 
On the other hand, the construction of the 
convex decomposition in \defref{def:ch_phi_extrem} 
is possible, if one may ``follow'' the tangential curve by \emph{any} (even infinitesimally small) 
distances in \emph{both} directions away from $\Lambda_{\varphi_0}$,
while remaining within the space of CPTP maps.
Hence, for this \emph{not} to be true, so that $\Lambda_\varphi$ is
$\varphi$-extremal at $\varphi_0$ (and \emph{red dots} in \figref{fig:ExtremCh}(\textbf{b}) lie outside the channel space), 
there must not exist $\epsilon\!>\!0$ such that both $\Omega_{\Lambda_{\varphi_0}}\!\!\pm\!\epsilon\,\dot\Omega_{\Lambda_{\varphi_0}}$
are positive semi-definite. As explicitly shown in \appref{chap:appPhiExtremCond},
such a statement may be further reformulated to 
define the criterion for $\varphi$-extremality as:

\begin{mycrit}[Criterion for channel $\varphi$-extremality]
\label{crit:ch_phi_extrem}
An element of a family of CPTP maps $\{\Lambda_\varphi\}_\varphi$ is 
\emph{$\varphi$-extremal} at a given $\varphi_0$
if and only if the derivative of its CJ matrix, $\dot{\Omega}_{\Lambda_{\varphi_0}}$,
is \emph{not} contained within the support of the CJ matrix $\Omega_{\Lambda_{\varphi_0}}$.
\end{mycrit}

Similarly to non-extramal channels, we may also define
the class of \emph{$\varphi$-non-extremal} CPTP maps, i.e.~all
that do \emph{not} satisfy \critref{crit:ch_phi_extrem}.
Then, as $\varphi$-non-extremality guarantees existence of a 
valid decomposition (one along the tangent), any $\varphi$-non-extremal
map must naturally be non-extremal. Equivalently, an extremal
channel is by definition $\varphi$-extremal. We prove explicitly these
statements in \appref{chap:appPhiExtremCond} by showing
that \defref{def:ch_extrem} and \critref{crit:Choi} indeed imply 
respectively \defref{def:ch_phi_extrem} and \critref{crit:ch_phi_extrem}.
However, let us clearly emphasise that, as the notion of $\varphi$-extremality
strongly depends on the geometry of a particular channel family,
\emph{there exist $\varphi$-extremal channels which are non-extremal} (see
e.g.~the loss noise model analysed in the following section).

\begin{note}[Geometry of full-rank quantum channels]
Lastly, let us discuss the \emph{full-rank} quantum channels that, 
as defined in \secref{sub:Evo_QCh}, possess a maximal number of linearly independent 
Kraus operators ($r\!=\!d_\t{in}^2$).
In the CJ-matrix picture \eref{eq:CJmatrix}, such condition translates onto the
statement that \emph{all} the eigenvalues of $\Omega_\Lambda$ are strictly greater than zero. 
Geometrically, this corresponds exactly to the situation in which one
may construct a ball of valid CPTP maps surrounding $\Lambda$ of interest, or
loosely speaking, one may ``move away'' from $\Lambda$ in any direction 
without crossing any boundary of the channel space (what occurs when one of the 
eigenvalues of $\Omega_\Lambda$ changes sign). Hence, 
as indicated in \figref{fig:ExtremCh}(\textbf{a}), \emph{the full-rank channels are 
the ones that lie strictly inside the space of CPTP maps},
and thus are naturally \emph{non-extremal}.
Furthermore, considering a family $\{\Lambda_\varphi\}_\varphi$
and a full-rank channel $\Lambda_{\varphi_0}$ at some $\varphi_0$, it must be 
also $\varphi$\emph{-non-extremal} irrespectively of the parameter-induced 
geometry. Due to the presence of valid CPTP maps in any direction away from $\Lambda_{\varphi_0}$,
no matter what form the derivative $\left.\partial_\varphi\Lambda_\varphi\right|_{\varphi=\varphi_0}$
takes, the convex decomposition along the tangent is always possible. Formally,
one may directly see that \critref{crit:ch_phi_extrem} is indeed fulfilled in the CJ-matrix picture, 
as $\Omega_{\Lambda_{\varphi_0}}$ is full-rank (its eigenvectors span the whole space) so that it 
definitely supports $\dot\Omega_{\Lambda_{\varphi_0}}$.
\end{note}

\subsection{\caps{Example:} Noisy-phase--estimation channels}
\label{sub:noise_models_geom}

\begin{table}[!t]
\begin{center}
\begin{tabular}{|M{2.2cm}||M{2.9cm}|M{2.9cm}|M{2.9cm}|M{2.9cm}|N}
\hline 
\textbf{Noise model:} & \emph{Dephasing} & \emph{Depolarization} & \emph{Loss} & \emph{Spontaneous emission}  
&\\[20pt]
\hline
{\relsize{-1}{\bfseries Rank:}} & $2$ & $4$ {\relsize{-1}{(full-rank)}} & $3$  & $2$ 
&\\[10pt]
\hline 
{\relsize{-1}{\bfseries Extremal:}} & no & no & no & \textbf{yes} 
&\\[12pt]
\hline 
{\relsize{-1}{\bfseries $\varphi$-extremal:}} & no & no & \textbf{yes} & \textbf{yes}
&\\[12pt]
\hline
{\relsize{-1}{\bfseries Geometric interpretation:}}
& 
\includegraphics[width=2.9cm]{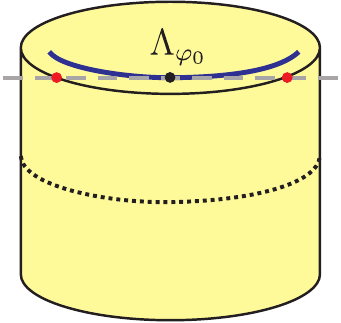}
& 
\includegraphics[width=2.9cm]{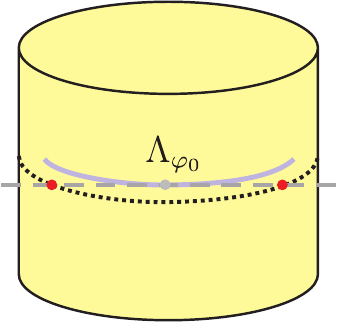} 
& 
\includegraphics[width=2.9cm]{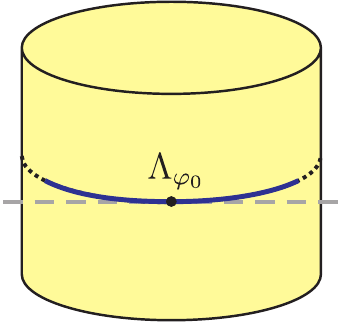}
& 
\includegraphics[width=2.9cm]{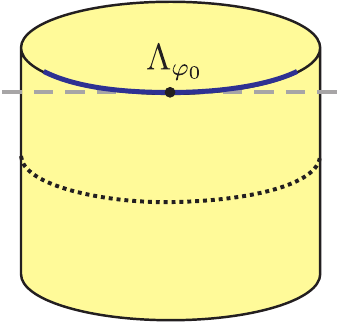}
&\\[88pt]
\hline
\end{tabular}
\end{center}
\caption[Geometrical properties of the noisy-phase--estimation models]{%
\textbf{Geometrical properties of the noisy-phase--estimation models} depicted in \figref{fig:noise_models}.
The channel-ranks are dictated by the pure-noise maps specified in \tabref{tab:noise_models},
so that only the \emph{depolarisation} channel corresponds to a full-rank CPTP map.
The \emph{loss} and \emph{spontaneous emission} models yield $\varphi$-\emph{extremal} channels, 
where in the latter case this is a consequence of the noise-map being \emph{extremal}.
In the last raw, we present an intuitive representation depicting geometry of the curve 
dictated by each channel family, i.e.~its schematic location in the space all CPTP maps.
}
\label{tab:noise_models_extrem}
\end{table}

As an example, let us return to the qubit evolution models 
introduced in \secref{sub:noise_models} (i.e.~the noisy-phase-estimation channels
depicted in \figref{fig:noise_models}) and discuss their geometrical properties, 
in particular, verifying the notions of extremality and $\varphi$-extremality.
For each model, we consider the overall channel $\Lambda_\varphi\!=\!\mathcal{U}_\varphi\!\circ\!\mathcal{D}_\eta$
defined in \eqnref{eq:ChPic}, with an adequate pure-noise map $\mathcal{D}_\eta$ 
specified in \tabref{tab:noise_models}.
We summarise the results in \tabref{tab:noise_models_extrem},
but let us already note that due to the \emph{unitary} parameter-encoding
the channels enjoy a circular symmetry%
\footnote{In fact, the \emph{U(1) symmetry of phase} (which we discuss in more detail 
later in \secref{sub:AvCostCircSymm}), as the parameter $\varphi$ appears in the CJ matrices of all four channels 
only through the $\e^{\pm\ii\varphi}$ factors.},
so that their geometrical properties are $\varphi$-independent. That is why, we are
able to \emph{schematically} represent the space of CPTP maps containing
all%
\footnote{To be precise, the loss channel is a qubit-qutrit map and thus belongs
to a different space of qubit-qutrit channels. Yet, this does not stop us 
to \emph{intuitively} interpret such a space in \tabref{tab:noise_models_extrem} 
in the same manner as for the other models.}
four channels of interest as a cylinder,
where the curve $\{\Lambda_\varphi\}_\varphi$ in all cases
corresponds to a circle that lies in a horizontal plane 
being symmetric around the vertical axis.
In the last row of \tabref{tab:noise_models_extrem}, we
present such a metaphorical ``geometric interpretation'' for each noise-type,
where in each case we intuitively draw a segment of the 
curve $\{\Lambda_\varphi\}_\varphi$
in agreement with the geometric features we obtain.

In order to verify extremality of the channels, we apply 
the Choi criterion (\critref{crit:Choi}) to each set of 
Kraus operators specified in \tabref{tab:noise_models}, and come to conclusion that only 
the \emph{spontaneous emission} (\emph{amplitude damping}) model yields 
an \emph{extremal} CPTP map \citep{Bengtsson2006}, which therefore is also 
$\varphi$\emph{-extremal}.
In the ``cylinder-representation'', we thus choose its corresponding curve
to coincide with the upper edge of the cylinder, what adequately disallows
any convex decomposition of $\Lambda_{\varphi_0}$ to be legal.
Secondly, we verify the $\varphi$-extremality criterion 
(\critref{crit:ch_phi_extrem}) to find out that also 
the \emph{loss} noise-type fulfils it, what thus proves that the loss model
yields a \emph{$\varphi$-extremal} but \emph{non-extremal} channel. 
Hence, we interpret such a fact by drawing its curve on the cylinder 
in a way, so that it is decomposable in the vertical direction, but not along the 
parametrisation-induced tangent. 
The \emph{depolarisation} channel is the simplest one to analyse, as it constitutes
a full-rank map. Thus, for all $\varphi$ the corresponding 
$\Lambda_\varphi$ lie strictly inside the cylinder,
making the map $\varphi$\emph{-non-extremal} (and hence \emph{non-extremal}). 
On the other hand, the \emph{dephasing} noise model is \emph{not} full-rank and 
therefore must be located on the boundary of the channel space.
Yet, as we verify that it is $\varphi$\emph{-non-extremal}---its 
$\Omega_{\Lambda_{\varphi_0}}$ supports $\dot\Omega_{\Lambda_{\varphi_0}}$
in accordance with \critref{crit:ch_phi_extrem}---such 
boundary must be flat in the tangential direction,
i.e.~we draw its curve on the top facet of the cylinder.

%% file: Chapters/est_theory.tex
\chapter{Limits to precise estimation of latent parameters} 
\label{chap:est_theory} 
\lhead{Chapter 3. \emph{Limits to precise estimation of latent parameters}} 

\section{Classical estimation theory \label{sec:ClEst}}

Before going into details of quantum mechanical aspects of metrology,
we review the fundamentals of estimation theory that lies at the heart
of any, also classical, metrological problem. The essential question
that has been addressed by statisticians long before the invention
of quantum mechanics is how to most efficiently extract information
from a given data set, which is determined by some non-deterministic
process. In particular, if there exists a (global) latent parameter
that affects the measurement outcomes collected, e.g.~temperature at
which an experiment is performed or strength of a magnetic field distorting
the electromagnetic signal measured, to what extent is one able to
determine its value basing on the gathered sample of data. This issue
is normally termed as the \emph{problem of parameter estimation}
\citep{Kay1993,Lehmann1998}.

\subsection{The parameter estimation problem \label{sub:ClEstProb}}

Mathematically, the parameter estimation problem corresponds to the
situation, in which we are given an $N$-point data set $\mathbf{x}\!=\!\left\{ x_{1},x_{2},\dots,x_{N}\right\} $
representing a realisation of $N$ \emph{independent} identically-distributed
random variables, $X^{N}$, each distributed according to a common
\emph{Probability Density Function} (PDF), $p_{\varphi}(X)$, that
depends on an unknown parameter $\varphi$ we wish to determine. Our
goal is to construct an \emph{estimator} $\tilde{\varphi}_{N}(\mathbf{x})$
which should be interpreted as a function that outputs the most accurate
estimate of the parameter $\varphi$ based on a given data set. Importantly,
as the estimator $\tilde{\varphi}_{N}$ is built on random data, it
is a random variable itself and its statistical properties, such as
the mean or the variance, are dictated by the data statistics, i.e.~%
the collective factorisable PDF:~$p_{\varphi}(\mathbf{x})\!=\!\prod_{i=1}^{N}p_{\varphi}(x_{i})$. 

Typically, two approaches to the above problem are undertaken depending
on the probabilistic nature of the estimated parameter. In the so
called \emph{frequentist} approach, $\varphi$ is assumed to be a
\emph{deterministic} variable with a fixed value that, if known, could
in principle be stated to any precision. Moreover, as the sample size
is taken to be large enough that the frequencies of the outcomes approximate
well their probabilities---hence, the name of the approach---it
is presumed that without loss of generality any estimation protocol
may be taken to be \emph{local}, i.e.~designed for a particular value
of $\varphi$. In contrast, when following the \emph{Bayesian} paradigm,
the estimated parameter is a \emph{random} variable itself, so that
the estimation protocol has to apply \emph{globally}, i.e.~it must
be optimised for a given range of values the parameter may take. In
such a case, the intrinsic fluctuations of $\varphi$ account for
the lack of knowledge about the parameter we possess \emph{prior} to
performing the estimation. We describe both approaches in detail below.

\subsection{Frequentist approach -- \emph{local} estimation of a\emph{ deterministic} parameter}
\label{sub:ClEstLocal}

\subsubsection{Imposing the unbiasedness and minimising the Mean Squared Error}

As within the \emph{frequentist} approach the estimated parameter
is assumed to be a deterministic variable of fixed value, we may write
the variance of any given estimator $\tilde{\varphi}_{N}(\mathbf{x})$
built on an $N$-point data sample as 
\begin{equation}
\left.\textrm{Var}\!\left[\tilde{\varphi}_{N}\right]\right|_{\varphi}=\left\langle \left(\tilde{\varphi}_{N}(\mathbf{x})-\left\langle \tilde{\varphi}_{N}\right\rangle _{\varphi}\right)^{2}\right\rangle _{\varphi}=\int\!\!\textrm{d}^{N}\! x\; p_{\varphi}(\mathbf{x})\left(\tilde{\varphi}_{N}(\mathbf{x})-\left\langle \tilde{\varphi}_{N}\right\rangle _{\varphi}\right)^{2},
\label{eq:EstVar}
\end{equation}
where $\varphi$ is the true value of the parameter and $\left\langle \tilde{\varphi}_{N}\right\rangle _{\varphi}\!=\!\int\!\textrm{d}^{N}\! x\; p_{\varphi}(\mathbf{x})\,\tilde{\varphi}_{N}(\mathbf{x})$
is the mean value of the estimator. We term an estimator to be
\emph{consistent}, if it outputs with certainty the correct value
of the estimated parameter when the sample size is infinitely increased,
i.e.
\begin{equation}
\lim_{N\rightarrow\infty}\tilde{\varphi}_{N}\!\left(\mathbf{x}\right)=\varphi,
\label{eq:EstCons}
\end{equation}
where taking the asymptotic $N$ limit should be understood as convergence
in probability, so that the distribution of $\tilde{\varphi}_{N}$
becomes infinite-narrowly peaked around the nominal value $\varphi$.
In particular, the consistency of the estimator implies that its mean
and variance must converge asymptotically to $\underset{N\rightarrow\infty}{\lim}\!\left\langle \tilde{\varphi}_{N}\right\rangle _{\varphi}\!=\!\varphi$
and $\underset{N\rightarrow\infty}{\lim}\!\!\left.\textrm{Var}\!\left[\tilde{\varphi}_{N}\right]\right|_{\varphi}\!=\!0$
respectively. On the other hand, the performance of any estimator
is quantified by the \emph{Mean Squared Error} (MSE), i.e.~the average \emph{squared distance} of $\tilde{\varphi}_{N}$ from the true
value $\varphi$:
\begin{equation}
\left.\Delta^{2}\tilde{\varphi}_{N}\right|_{\varphi}=\left\langle \left(\tilde{\varphi}_{N}(\mathbf{x})-\varphi\right)^{2}\right\rangle _{\varphi}=\int\!\!\textrm{d}^{N}\! x\; p_{\varphi}(\mathbf{x})\left(\tilde{\varphi}_{N}(\mathbf{x})-\varphi\right)^{2},
\label{eq:MSE}
\end{equation}
which is then minimised by an \emph{optimal }estimator. 

Let us emphasise that such an optimal estimator may turn out to be
\emph{local}---optimised for a particular value of the parameter---as 
in general there may not exist a \emph{single} estimator minimising
the MSE \eref{eq:MSE} for \emph{all} values of $\varphi$. As a
matter of fact, such a local estimator seems to be useless from
the practical point of view, as it requires the estimated parameter
to be exactly known before the estimation procedure! However, as any
realistic, or in other words \emph{global}, estimator
can perform only worse at a given $\varphi$ than the estimator specially
optimised for that parameter value, we may limit the performance of
any potential strategy by establishing the ultimate bounds on precision
achieved by the local estimators. Moreover, the issue of locality
becomes less and less important with growth of the sample size and
in particular may be fully ignored when investigating protocols in
the asymptotic $N$ limit, in which the local---frequentist---precision
bounds are guaranteed to be saturable \citep{Vaart1998}. We discuss
the consequences of this counter-intuitive paradigm in \secref{sub:LocConseq}
below, but one should already bear in mind that such \emph{local estimation
regime} is always presumed within the frequentist approach, what has
then strong implications on its applicability to any real-life problems.
On the other hand, if luckily a \emph{global} estimator may be constructed
that minimises the MSE for any parameter value, then trivially it
is also \emph{locally }optimal for any $\varphi$.

The minimisation of \eqnref{eq:MSE} is addressed within
the frequentist framework in two steps. When considering \emph{global}
estimators, they are firstly restricted to be \emph{unbiased},
so that they always output on average the true parameter value by
fulfilling for any $\varphi$ the condition
\begin{equation}
\left\langle \tilde{\varphi}_{N}\right\rangle _{\varphi}=\int\!\!\textrm{d}^{N}\! x\; p_{\varphi}(\mathbf{x})\,\tilde{\varphi}_{N}(\mathbf{x})=\varphi.
\label{eq:UnbiasConstr}
\end{equation}
As a result, the \eqnsref{eq:EstVar}{eq:MSE}
become equivalent and the minimisation of the MSE is then tantamount
to minimising the variance of a given unbiased estimator. Yet again,
as we are allowed to consider and focus on the case of local estimators,
we may relax the \emph{unbiasedness constraint} \eref{eq:UnbiasConstr},
so that it holds only up to $O(\delta\varphi^{2})$ in the vicinity
of particular parameter true value chosen, say $\varphi_{0}$ with
$\varphi\!=\!\varphi_{0}\!+\!\delta\varphi$. \eqnref{eq:UnbiasConstr}
then fixes only the mean value of the estimator and its differential
w.r.t.~$\varphi$:
\begin{equation}
\left\langle \tilde{\varphi}_{N}\right\rangle_{\varphi_{0}}\!\!=\varphi_{0}\,,\qquad\left.\frac{\partial\left\langle \tilde{\varphi}_{N}\right\rangle}{\partial\varphi} \right|_{\varphi_{0}}\!\!=\;1.
\label{eq:LocUnbiasConstr}
\end{equation}
However, out of the above constraints 
only the second one has a non-trivial meaning, because knowing the
true value $\varphi_{0}$ while estimating locally we can always
adequately shift the estimator and satisfy the first one. Thus effectively,
the \emph{local unbiasedness constraints} \eref{eq:LocUnbiasConstr}
fix only the ``speed'' of change of the estimator mean with the parameter at 
its true value. Importantly, any \emph{global} unbiased
estimator satisfying \eqnref{eq:UnbiasConstr} trivially
satisfies the \emph{local} conditions \eref{eq:LocUnbiasConstr} at any $\varphi_0$,
so that we may utilise \eqnref{eq:LocUnbiasConstr} in the 
following section to bound the precision 
of \emph{any} unbiased estimator.

Yet, before doing so, let us remark that by imposing either of the
unbiasedness constraints, \eref{eq:UnbiasConstr} or \eref{eq:LocUnbiasConstr},
the frequentist approach naturally excludes all the \emph{biased}
estimators that potentially may not only be more accurate but also
be the only ones minimising the MSE \eref{eq:MSE}. However, as we
require any estimator to be \emph{consistent}, by averaging \eqnref{eq:EstCons}
over the outcomes one should realise that such a practical restriction
also forces any estimator to be \emph{unbiased} in the asymptotic
$N$ limit. Hence, although the biased estimators may lead to improved
precision beyond the scope of the frequentist approach for finite
sample sizes, they may be ignored in the $N\!\rightarrow\!\infty$
limit for which the frequentist approach is really designed for.

\subsubsection{Ultimate precision and the Cram\'{e}r-Rao Bound}
\label{sub:CRB}

Stemming from the local unbiasedness condition on the ``speed'' of
change of the estimator mean \eref{eq:LocUnbiasConstr} 
and assuming the single-outcome PDF, $p_{\varphi}(X)$,
to be \emph{regular} at a given $\varphi_{0}$, i.e.%
\footnote{%
Crucially, the regularity assumption allows to interchange 
the order of $\int\!\textrm{d}x$ and $\partial/\partial\varphi$
in any expression averaged over the outcomes of $X$ at
a particular parameter value $\varphi_0$.
}
\begin{equation}
\left\langle \frac{\partial}{\partial\varphi}\ln p_{\varphi}\right\rangle_{\!\varphi_{0}}\!\!\!=\,0\quad\implies\quad\!\!\int\!\!\textrm{d}x\left[p_{\varphi}(x)\frac{\partial\, \ln p_{\varphi}(x)}{\partial\varphi}\right]_{\varphi_{0}}\!\!\!=\int\!\!\textrm{d}x\left.\frac{\partial\, p_{\varphi}(x)}{\partial\varphi}\right|_{\varphi_{0}}\!\!\!=\!\left[\frac{\partial}{\partial\varphi}\int\!\!\textrm{d}x\, p_{\varphi}(x)\right]_{\varphi_{0}}\!\!\!=\,0,
\label{eq:LocRegCond}
\end{equation}
one may construct by means of a Cauchy-Schwarz inequality---see e.g.~\citep{Kay1993}
for the derivation---the so-called \emph{Cram\'{e}r-Rao Bound} (CRB)
that lower-limits the MSE of any unbiased estimator:
\begin{equation}
\left.\Delta^{2}\tilde{\varphi}_{N}\right|_{\varphi_{0}}\ge\frac{1}{N\left.F_{\textrm{cl}}\!\left[p_{\varphi}\right]\right|_{\varphi_{0}}}.
\label{eq:CRB}
\end{equation}
Although we have specified the CRB as a bound on the MSE \eref{eq:MSE} which is 
the adequate figure of merit, one should bear in mind that due to unbiasedness
$\left.\Delta^{2}\tilde{\varphi}_{N}\right|_{\varphi_{0}}\!\!=\!\!\left.\t{Var}\!\left[\tilde{\varphi}_{N}\right]\right|_{\varphi_{0}}$
and the CRB equivalently lower-limits the variance \eref{eq:EstVar} of the estimator, which
is the more experimentally relevant quantity that may be determined for an unknown $\varphi$ basing 
only on the data gathered. We have also kept the notation $\left.\dots\right|_{\varphi_0}$ in 
\eqnref{eq:CRB}, in order to stress that both its l.h.s.~and 
r.h.s.~may in principle depend on $\varphi_0$, as the bound 
is derived fixing a particular value of the estimated parameter. 
Nevertheless, the CRB applies also to \emph{global}
unbiased estimators which form does not vary with $\varphi$. The crucial quantity limiting
ultimately the MSE in \eqnref{eq:CRB} is the so-called (classical)
\emph{Fisher Information} (FI), $F_{\textrm{cl}}$, that can be expressed
using one of the equivalent formulae below:
\begin{equation}
F_{\textrm{cl}}\!\left[p_{\varphi}\right]=\int\!\textrm{d}x\;\frac{1}{p_{\varphi}(x)}\left[\frac{\partial\, p_{\varphi}(x)}{\partial\varphi}\right]^{2}=\left\langle \left(\frac{\partial}{\partial\varphi}\ln p_{\varphi}\right)^{2}\right\rangle =-\left\langle \frac{\partial^{2}}{\partial\varphi^{2}}\ln p_{\varphi}\right\rangle.
\label{eq:FI}
\end{equation}
Being \emph{non-negative} and \emph{additive} the FI has the interpretation of an
\emph{information measure} \citep{Arndt2001} which increase indicates a higher precision potentially
achievable by estimation protocols. In particular, at a given $\varphi_{0}$,
$\left.F_{\textrm{cl}}\left[p_{\varphi}\right]\right|_{\varphi_{0}}\!=\!0$
proves that one cannot extract any information about the parameter
from a sample, whereas divergent $\left.F_{\textrm{cl}}\left[p_{\varphi}\right]\right|_{\varphi_{0}}\!=\!\infty$
implies that the true value $\varphi_{0}$ can in principle be perfectly determined,
what, however, is guaranteed only in the asymptotic $N$ limit, as
discussed in \secsref{sub:CRBSat}{sub:MLE} below.
The non-negativity of the FI follows naturally from the second expression
in \eqnref{eq:FI} and its additivity can be easily
verified by using the last expression of \eqnref{eq:FI}, in order to prove
that $F_{\textrm{cl}}\!\left[p_{\varphi}^{(1,2)}\right]\!=\! F_{\textrm{cl}}\!\left[p_{\varphi}^{(1)}\right]\!+\! F_{\textrm{cl}}\!\left[p_{\varphi}^{(2)}\right]$
for any factorisable $p_{\varphi}^{(1,2)}(X_{1},X_{2})\!=\! p_{\varphi}^{(1)}(X_{1})\, p_{\varphi}^{(2)}(X_{2})$.
Importantly, when dealing with \emph{independently} distributed
samples $F_{\textrm{cl}}\!\left[p_{\varphi}^{N}\right]\!=\! N\, F_{\textrm{cl}}\!\left[p_{\varphi}\right]$,
what indeed leads to the CRB \eref{eq:CRB} being fully determined
by the distribution of the single random variable $X$ and, most importantly,
to the \emph{SQL-like} \emph{scaling} of $1/N$ for the MSE.

\begin{note}[Central Limit Theorem -- asymptotic estimation of the PDF mean]
\label{note:CLT}
As aside, let us note that the additivity property of FI is consistent
with the natural intuition one may infer from the \emph{Central Limit
Theorem} (CLT). One may look at the CLT as a special kind of an estimation
problem in which the mean, $\mu$, of a distribution (of a random variable $X$) 
is treated as the parameter to be determined, $\varphi\!\equiv\!\mu$,
whereas the sample average, $\tilde{\varphi}_{N}(\mathbf{x})=\frac{1}{N}\sum_{i=1}^{N}x_{i}$,
constitutes an example of a global unbiased estimator that always
saturates the CRB \eref{eq:CRB} in the asymptotic $N$ limit. According
to the CLT, the PDF of such an estimator converges to a Gaussian distribution
with $\Delta^{2}\tilde{\varphi}_{N}\!\!\overset{N\rightarrow\infty}{=}\t{Var}\!\left[X\right]/N$.
Hence, $\Delta^{2}\tilde{\varphi}_{N}$ saturates asymptotically
the CRB, as the FI \eref{eq:FI} calculated with
respect to the mean of any Gaussian distribution corresponds to the
inverse of its variance:
\begin{equation}
F_{\textrm{cl}}\left[\thicksim\exp\!\left(-\frac{\left(x-\varphi\right)^{2}}{2\,\t{Var}\!\left[X\right]}\right)\right]=-\left\langle \frac{\partial^{2}}{\partial\varphi^{2}}\left[-\frac{\left(x-\varphi\right)^{2}}{2\,\t{Var}\!\left[X\right]}\right]\right\rangle =\frac{1}{\t{Var}\!\left[X\right]}.
\label{eq:FImeanGauss}
\end{equation}
\end{note}

Lastly, let us remark that the FI is a \emph{local quantity},
what should be expected acknowledging the fact that it was derived
basing on the local unbiasedness conditions \eref{eq:LocUnbiasConstr}
valid up to $O\!\left(\delta\varphi^{2}\right)$ around a given $\varphi_{0}$.
Consequently, as explicitly
stated in the first expression of the definition \eref{eq:FI}, the
FI at a given parameter value, $\left.F_{\textrm{cl}}\left[p_{\varphi}\right]\right|_{\varphi_{0}}$, 
is dependent only on $p_{\varphi_0}(X)$ and $\left.(\partial\,p_{\varphi}(X)/\partial\varphi)\right|_{\varphi_0}$,
so that it is fully specified by just fixing at $\varphi_0$:~the PDF and its ``speed''
of change with $\varphi$. Crucially, this means that \emph{all} parametrised PDFs of $X$ that coincide up to 
$O\!\left(\delta\varphi^{2}\right)$ with one another at $\varphi_{0}$ are 
\emph{equivalent} from the point of view of their FI.
In fact, the locality is also a consequence of the FI possessing a neat geometric interpretation \citep{Amari2007}.
Expanding the \emph{angular distance}%
\footnote{%
Also known as the Bhattacharyya distance, for which the fidelity is termed as the Bhattacharyya coefficient \citep{Bengtsson2006}.
}
\citep{Wootters1981,Amari1987} between 
the neighbouring PDFs $p_{\varphi_0}(X)$ and $p_{\varphi_0+\delta\varphi}(X)$ around $\varphi_0$, which 
is defined as $D_\t{cl}(p_1,p_2)\!=\!\arccos[\mathsf{Fid}_\t{cl}(p_1,p_2)]$ with
$\mathsf{Fid}_\t{cl}(p_1,p_2)\!=\!\int\!\!\t{d}x\sqrt{p_1(x)\, p_2(x)}$ being 
the \emph{fidelity} \citep{Nielsen2000} of PDFs $p_{1/2}$, we obtain for $\delta\varphi\!\ge\!0$:
\begin{equation}
D_\t{cl}\!\left(p_{\varphi_0},p_{\varphi_0+\delta\varphi}\right)=\frac{1}{2}\sqrt{\left.F_{\textrm{cl}}\!\left[p_{\varphi}\right]\right|_{\varphi_0}}\,\delta\varphi+O\!\left(\delta\varphi^{2}\right).
\label{eq:FIviaStDist}
\end{equation}
Hence, the FI \eref{eq:FI} may be equivalently interpreted as
the square of the speed, $F_{\textrm{cl}}\!\left[p_{\varphi}\right]\!=\!4\left(\t{d}D_\t{cl}/\t{d}\varphi\right)^2$,
with which $p_{\varphi}$ is ``moving'' along the path of $\varphi$ at a particular parameter value.

\subsubsection{Locality of the frequentist approach and its consequences}
\label{sub:LocConseq}

When analysing performance of realistic estimation protocols, two
issues may arise when utilizing the CRB to quantify the accuracy of
any practical, hence \emph{global}, estimator chosen. Firstly, in
a more general case than the Gaussian distribution mean estimation
\eref{eq:FImeanGauss}, the FI \eref{eq:FI} and hence the CRB \eref{eq:CRB}
may depend on the true value $\varphi_{0}$. Thus, it is then ambiguous
which parameter value to substitute into the CRB to most accurately
bound the precision achieved given particular sample of data and a
global unbiased estimator. Secondly, even when the evaluated FI turns out to
be parameter-independent, one must always verify if the CRB actually
corresponds to an inequality which may be saturated by a \emph{single}
estimator at any $\varphi$. In the following section, we show that 
for any finite $N$ a global unbiased estimator, which saturates the CRB
irrespectively of the parameter value, exists \emph{only} for PDFs 
belonging to a subclass of the \emph{exponential probability distributions}.
Hence, for a general $p_{\varphi}(X)$, unless we 
are dealing with very large samples, the CRB should be treated with caution.

On the other hand, we show in \secref{sub:CRBSat} that \emph{locally}
one can always%
\footnote{%
Yet, one must be careful when dealing with parameters
that do not carry a standard \emph{topology} of a real line,
e.g.~in phase estimation problems in which the parameter is an element
of the circle group (see for instance \secref{sub:ClEst_MZInter_SQL_Freq}).
}
construct an estimator that saturates the CRB \eref{eq:CRB}.
From the physical perspective, this means that the CRB is always meaningful
when the goal is really to design an estimator that is most sensitive
to small deviations from a known value of the parameter. This corresponds
to the situation when one has investigated perfectly all the properties
of a given physical system that afterwards is subjected to some external
fluctuations, which vary the parameter of interest by a small amount.
Intuitively, this must be the most optimistic scenario of estimation
that one may consider, so it is consistent that the CRB defines the
ultimate bound on the achievable precision. Moreover, this fact also 
explains why in the situation when the parameter is unknown one may
attain the CRB by infinitely increasing the sample size $N$. Then, 
one may always use a fraction of the outcomes in order
to learn the value of the estimated parameter sufficiently enough
to enter the local estimation regime, so that the CRB becomes more
and more accurate as $N\!\rightarrow\!\infty$. 

Unfortunately, the frequentist approach does not give a recipe how
to quantify $N$ for which the local regime of estimation may
already be assured, and thus the accuracy of the CRB as a bound on precision. 
Yet, as shown in \eqnref{eq:FImeanGauss} with the example
of application of the CLT to estimation, the CRB \eref{eq:CRB} represents
a tight inequality when the independently distributed data statistics
approach their corresponding asymptotic Gaussian distribution. Such
phenomenon is true in general for any (asymptotically) unbiased estimator
defined on \emph{independently} distributed data and is known under the name 
of the Local Asymptotic Normality \citep{Vaart1998}.
Hence, one way to quantify the speed at which the precision of the
optimal strategy approaches the CRB with $N$ is to determine the
rate at which the overall PDF of the sampled
data converges to its asymptotic Gaussian form. Nevertheless, the method 
which bases on and is constructed to account for the progressive
improvement of knowledge about the parameter with growth of $N$ is
really the Bayesian inference which we discuss in \secref{sub:ClEstGlobal}.

\subsubsection{\label{sub:CRBSat}Saturability of the CRB}

An unbiased estimator that saturates the CRB \eref{eq:CRB} is said
to be \emph{efficient} in that it efficiently uses the sampled data.
As the CRB is derived by means of the Cauchy-Schwarz inequality, the
sufficient and necessary condition for its saturability, which is
imposed on the PDF and a \emph{global} unbiased estimator, corresponds to
the statement \citep{Kay1993}:
\begin{equation}
\frac{\partial}{\partial\varphi}\ln\, p_{\varphi}(\mathbf{x})= F_{\textrm{cl}}\!\left[p_{\varphi}^N\right]\,\left(\tilde{\varphi}_{N}(\mathbf{x})-\varphi\right),\label{eq:CRBSatCond}
\end{equation}
where the multiplicative factor on the r.h.s.~becomes fixed to $F_{\textrm{cl}}\!\left[p_{\varphi}^N\right]\!=\!N\, F_{\textrm{cl}}\!\left[p_{\varphi}\right]$,
so that:~by differentiating \eqnref{eq:CRBSatCond} w.r.t.~$\varphi$, averaging it
over the outcomes,  and applying the unbiasedness condition \eref{eq:UnbiasConstr};~we
recover the FI definition \eref{eq:FI} for $p_{\varphi}^N$. One should note
that the above requirement ceases to make mathematical sense when
$N\!\rightarrow\!\infty$, so the fact of \eref{eq:CRBSatCond}
not being satisfiable for any finite $N$ does not stop the CRB from
being potentially tight in the asymptotic $N$ limit. 

Defining%
\footnote{%
In order to shorten the notation, we represent the \emph{derivatives w.r.t.~the estimated parameter} throughout
this work with `\emph{overdots}', so that e.g.~$\dot{\eta}(\varphi)\equiv\partial_{\varphi}\, \eta(\varphi)$,
$\dot{p}_{\varphi}(x)\equiv\partial_{\varphi}\, p_{\varphi}(x)$,
$\dot{\varrho}_{\varphi}\equiv\partial_{\varphi}\,\varrho_{\varphi}$,
$\left|\dot{\psi}_{\varphi}\right>\equiv\partial_{\varphi}\!\left|\psi_{\varphi}\right\rangle $
and $\dot{K}(\varphi)\equiv\partial_{\varphi}\, K(\varphi)$ etc.} 
$\ddot\lambda(\varphi)\!=\!\frac{\partial^{2}\lambda(\varphi)}{\partial\varphi^{2}}\!=\!F_{\textrm{cl}}\!\left[p_{\varphi}\right]$,
where $\lambda(\varphi)$ is some general, outcome-independent function,
we may write the most general form of the PDF of $X^{N}$ that satisfies
\eqnref{eq:CRBSatCond} for \emph{any} $\varphi$ as
\begin{equation}
p_{\varphi}(\mathbf{x})=\exp\left\{ N\left[\dot\lambda(\varphi)\left(\tilde{\varphi}_{N}(\mathbf{x})-\varphi\right)+\lambda(\varphi)+c_{N}(\mathbf{x})\right]\right\} 
\label{eq:PDFEffEst}
\end{equation}
with $c_{N}(\mathbf{x})$ being an arbitrary, parameter-independent
function. Moreover, as the data distribution is described with independently distributed
random variables, the l.h.s.~of \eqnref{eq:PDFEffEst} factorizes,
so that $p_{\varphi}(\mathbf{x})\!=\!\prod_{i=1}^{N}p_{\varphi}(x_{i})$.
Therefore, if there exists an efficient estimator for a single outcome,
$\tilde{\varphi}(x)$ for $N\!=\!1$, then the overall efficient estimator
may be simply taken to be the mean of such individual estimators,
$\tilde{\varphi}_{N}(\mathbf{x})\!=\!\frac{1}{N}\sum_{i=1}^{N}\tilde{\varphi}(x_{i})$,
where after choosing $c_{N}(\mathbf{x})\!=\!\frac{1}{N}\sum_{i=1}^{N}c(x_{i})$
the individual PDF must satisfy
\begin{equation}
p_{\varphi}(x)=\exp\left\{ \dot\lambda(\varphi)\left(\tilde{\varphi}(x)-\varphi\right)+\lambda(\varphi)+c(x)\right\} \label{eq:PDFEffEstN1}
\end{equation}
that is consistent with the general expression \eref{eq:PDFEffEst}
after substituting $N\!=\!1$.

The PDFs that possess the form \eref{eq:PDFEffEst}, and hence allow
for a global unbiased estimator that satisfies the criterion \eref{eq:CRBSatCond},
belong to the so-called \emph{exponential family} which
is a well established class of PDFs in probability theory that encapsulates
most common distributions such as:~Gaussian, Bernoulli, gamma, chi-squared,
binomial, Poissonian and many others \citep{Kay1993}. In general,
an exponential PDF reads:
\begin{equation}
p_{\varphi}^\t{\tiny exp}(\mathbf{x})=h(\mathbf{x})\;\exp\left\{ N\!\left[\eta(\varphi)\,\frac{T(\mathbf{x})}{N}-A(\varphi)\right]\right\},
\label{eq:PDFExpFam}
\end{equation}
where $\eta,\,A$ and $T,\,h$ are some standard parameters that
characterise a given distribution and depend on the estimated parameter
and the sample outcomes respectively%
\footnote{%
In statistics, the parameters of an exponential PDF are normally termed
\citep{Lehmann1998}:~$h(\mathbf{x})$--base measure, $T(\mathbf{x})$--sufficient
statistic, $\eta(\varphi)$--natural parameter, $A(\varphi)$--log-partition
function.}.
Then, for the single-outcome, exponential PDF 
$p_{\varphi}^\t{\tiny exp}(X)$ the FI \eref{eq:FI} 
may be written as 
\begin{equation}
F_{\textrm{cl}}\!\left[p_{\varphi}^\t{\tiny exp}\right]=\dot\eta(\varphi)\;\frac{\partial}{\partial\varphi}\!\left[\frac{\dot A(\varphi)}{\dot\eta(\varphi)}\right].\label{eq:FIExpFam}
\end{equation}
Comparing \eqnsref{eq:PDFEffEst}{eq:PDFExpFam}
it becomes evident that a legal global estimator for $p_{\varphi}^\t{\tiny exp}(\mathbf{x})$
reads $\tilde{\varphi}^\t{\tiny exp}_{N}(\mathbf{x})\!=\!T(\mathbf{x})/N$, but for the validity of 
the CRB \eref{eq:CRB} it must also be assured to satisfy the unbiasedness constraint \eref{eq:UnbiasConstr},
which then naturally guarantees the CRB-saturability requirement \eref{eq:CRBSatCond}
to be fulfilled. It is easy to show that for an exponential PDF of
the form \eref{eq:PDFExpFam} the unbiasedness and CRB-saturability
conditions are met by such an estimator \emph{if and only if}
\begin{equation}
\left\langle\tilde{\varphi}^\t{\tiny exp}_{N}\right\rangle _{\varphi}=\,\int\!\! \t{d}^{N}\!x\; p_{\varphi}^\t{\tiny exp}(\mathbf{x})\,\frac{T(\mathbf{x})}{N}\,=\,\frac{\dot A(\varphi)}{\dot\eta(\varphi)}\,=\,\varphi.
\label{eq:UnbiasedConstrExpFam}
\end{equation}
Furthermore, the expression \eref{eq:FIExpFam} for the FI of $p_{\varphi}^\t{\tiny exp}(X)$
simplifies then to $F_{\textrm{cl}}\!\left[p_{\varphi}^\t{\tiny exp}\right]\!=\!\dot\eta(\varphi)$
what becomes also clear when comparing \eqnsref{eq:PDFEffEst}{eq:PDFExpFam}
and noting that $\eta(\varphi)\!=\!\dot\lambda(\varphi)$.

On the other hand, by just solving the CRB-saturability condition \eref{eq:CRBSatCond}
for $\tilde{\varphi}_{N}(\mathbf{x})$, a \emph{local} efficient estimator may always 
be constructed for \emph{any}%
\footnote{%
Yet, the estimator \eref{eq:LocEffEst} may lead to inconclusive answers when estimating 
parameters \emph{not defined on a real line}, i.e.~\emph{not} of standard topology, e.g.
see the case of a circularly symmetric parameter discussed in \secref{sub:ClEst_MZInter_SQL_Freq}.
} PDF $p_{\varphi}(X)$ at a given true
value $\varphi_{0}$, so that
\begin{equation}
\tilde{\varphi}_{N,\varphi_{0}}(\mathbf{x})\,=\,\varphi_{0}+\frac{1}{N\left.F_{\textrm{cl}}\!\left[p_{\varphi}\right]\right|_{\varphi_{0}}}\left.\frac{\partial\ln p_{\varphi}(\mathbf{x})}{\partial\varphi}\right|_{\varphi_{0}},
\label{eq:LocEffEst}
\end{equation}
which then trivially satisfies \eqnref{eq:CRBSatCond} and
as necessary fulfils the local unbiasedness conditions \eref{eq:LocUnbiasConstr}.
Although the local estimator $\tilde{\varphi}_{N,\varphi_{0}}$ is not constructable
when the second term in \eqnref{eq:LocEffEst} is divergent, this may 
occur only for pathological parameter values for which the estimation problem is 
ill-defined, i.e.:~when $\left.\partial\ln p_{\varphi}(\mathbf{x})/\partial\varphi\,\right|_{\varphi_{0}}$
is infinite, what makes the PDF regularity assumption \eref{eq:LocRegCond}
invalid and the CRB \eref{eq:CRB} not applicable; or 
when $\left.F_{\textrm{cl}}\!\left[p_{\varphi}\right]\right|_{\varphi_{0}}\!\!=\!0$
and no information about the parameter is extractable from the PDF.
Lastly, while returning to the special case of exponential PDFs and
substituting $p_{\varphi}^\t{\tiny exp}(\mathbf{x})$ \eref{eq:PDFExpFam}
into \eqnref{eq:LocEffEst}, we obtain the corresponding general form of 
a local efficient estimator:
\begin{equation}
\tilde{\varphi}_{N,\varphi_{0}}^\t{\tiny exp}(\mathbf{x})\,=\,\varphi_{0}+\left[\frac{T(\mathbf{x})}{N}-\frac{\dot A(\varphi_{0})}{\dot\eta(\varphi_{0})}\right]\left[\!\left.\frac{\partial}{\partial\varphi}\!\left[\frac{\dot A(\varphi)}{\dot\eta(\varphi)}\right]\right|_{\varphi_{0}}\right]^{-1},\label{eq:LocEffEstExpFam}
\end{equation}
which may be verified to satisfy the local unbiasedness constraints
\eref{eq:LocUnbiasConstr} as required.

\subsubsection{\label{sub:MLE}Maximum Likelihood estimator}

As remarked in the previous section, although a global unbiased estimator may not exist that
satisfies the saturability condition \eref{eq:CRBSatCond} for a
given $p_{\varphi}(\mathbf{x})$, \eqnref{eq:CRBSatCond}
loses its mathematical validity when $N\!\rightarrow\!\infty$ opening
doors to the CRB \eref{eq:CRB} being potentially saturable. As a
matter of fact, there exists an estimator, i.e.~the \emph{Maximal
Likelihood} (ML) estimator, which always turns out to be efficient
in the asymptotic $N$ limit. The ML estimator is formally defined as
\begin{equation}
\tilde{\varphi}_{N}^\t{\tiny ML}(\mathbf{x})\,=\,\underset{\varphi}{\textrm{argmax}}\, p_{\varphi}(\mathbf{x})\,=\,\underset{\varphi}{\textrm{argmax}}\,\ln p_{\varphi}(\mathbf{x}),
\label{eq:MLE}
\end{equation}
but intuitively should be understood as a function that for a given
instance of outcomes, $\mathbf{x}$, outputs the value of parameter
for which this data sample is most probable, i.e.~the \emph{likelihood
function} $l_{\mathbf{x}_{0}}(\varphi)\equiv p_{\varphi}(\mathbf{x}\!=\!\mathbf{x}_{0})$
is maximal. Although $\tilde{\varphi}_{N}^\t{\tiny ML}$ is generally biased for finite $N$,
it is unbiased asymptotically for any PDF:~$\left\langle \tilde{\varphi}_{N}^\t{\tiny ML}\right\rangle_\varphi\!\!\!\overset{N\rightarrow\infty}{=}\!\varphi$,
so that the CRB then applies and, crucially, is always saturated, as also
$\left.\Delta^{2}\tilde{\varphi}_{N}^\t{\tiny ML}\right|_\varphi\!\!\!\overset{N\rightarrow\infty}{=}\!1/(N F_{\textrm{cl}}\!\left[p_{\varphi}\right])$
\citep{Kay1993,Lehmann1998,Vaart1998}. However, similarly to the
problem discussed in \secref{sub:LocConseq} of certifying sufficient sample size that assures 
locality of estimation,
the frequentist approach does not give a recipe how to quantify $N$
for which the ML estimator attains the CRB up to a certain accuracy.

As the logarithm is a monotonic function, in the last expression of
\eqnref{eq:MLE} we equivalently consider the logarithm of
the likelihood (the log-likelihood) to deduce the parameter value
that maximizes the probability. Thus, in a typical situation when the log-likelihood
possesses a single maximum, the ML estimator may be then interpreted as a solution
to the equation 
\begin{equation}
\frac{\partial}{\partial\varphi}\,\ln p_{\varphi}(\mathbf{x})=0,\label{eq:MLECond}
\end{equation}
which is warranted to indicate the maximum if also the condition $\frac{\partial^{2}}{\partial\varphi^{2}}\ln p_{\varphi}\!\left(\mathbf{x}\right)<0$
is fulfilled at the $\varphi$ considered.

Looking back at the CRB-saturability requirement \eref{eq:CRBSatCond},
one should note that there always exists a parameter value---independently 
of the particular sampled data $\mathbf{x}$ obtained---$\varphi\!=\!\tilde{\varphi}_{N}(\mathbf{x})$
for which the r.h.s.~of \eqnref{eq:CRBSatCond} vanishes.
Importantly, at such $\varphi$, condition \eref{eq:CRBSatCond}
becomes equivalent to the requirement \eref{eq:MLECond} 
determining the ML estimator.
Furthermore, the differential of \eqnref{eq:CRBSatCond} w.r.t.~$\varphi$ there reads
\begin{equation}
\left.\frac{\partial^{2}}{\partial\varphi^{2}}\,\ln p_{\varphi}\!\left(\mathbf{x}\right)\right|_{\varphi=\tilde{\varphi}_{N}(\mathbf{x})}=-\, N\left.F_{\textrm{cl}}\!\left[p_{\varphi}\right]\right|_{\varphi=\tilde{\varphi}_{N}(\mathbf{x})},\label{eq:MLECondMaxAss}
\end{equation}
and, due to the non-negativity of the FI \eref{eq:FI}, guarantees
that $\varphi\!=\!\tilde{\varphi}_{N}(\mathbf{x})$ corresponds
to a maximum of the likelihood function. Hence, for a PDF that leads to
$l_{\mathbf{x}}(\varphi)$ possessing a single maximum in $\varphi$, the ML estimator constructed 
by satisfying \eref{eq:MLECond} must always be identical to the global efficient estimator 
satisfying \eqnref{eq:CRBSatCond} if such one exists. 
$\tilde{\varphi}_{N}^\t{\tiny ML}$
fulfils then the condition \eref{eq:CRBSatCond} after choosing without loss
of generality $\varphi\!=\!\tilde{\varphi}_{N}(\mathbf{x})$,
which also is the parameter value outputted by the ML estimator \eref{eq:MLE}.
Therefore, in such a special case, the ML estimator is not only always unbiased, but also saturates the
CRB for \emph{any} $N$ and not just in the asymptotic limit.

\begin{note}[ML estimator for the mean of a Gaussian PDF]
For instance, let us consider the mean estimation problem of $N$ independent
variables, $X^{N}$, distributed according to a Gaussian 
PDF:~$p_{\varphi}(x)\!\sim\!\t{exp}\!\left[\frac{-(x-\varphi)^{2}}{2\,\t{Var}[X]}\right]$,
for which the condition \eref{eq:MLECond} reads
\begin{equation}
\frac{\partial}{\partial\varphi}\,\ln p_{\varphi}(\mathbf{x})=\frac{\partial}{\partial\varphi}\sum_{i=1}^{N}\frac{-\left(x_{i}-\varphi\right)^{2}}{2\,\textrm{Var}[X]}=\frac{N}{\textrm{Var}[X]}\left(\frac{1}{N}\sum_{i=1}^{N}x_{i}-\varphi\right)=0,
\label{eq:MLECondGauss}
\end{equation}
so that indeed the ML estimator \eref{eq:MLE} corresponds to the average of a sample, 
$\tilde{\varphi}_{N}^\t{\tiny ML}(\mathbf{x})\!=\!\frac{1}{N}\sum_{i=1}^{N}x_{i}$,
which is the global unbiased estimator saturating the CRB 
\eref{eq:CRB}---as discussed in \noteref{note:CLT}.
\end{note}

Considering the case of $N$ independent variables, $X^{N}$, each being distributed according 
to a PDF within the \emph{exponential} family of distributions \eref{eq:PDFExpFam} satisfying
the saturability constraint \eref{eq:UnbiasedConstrExpFam}, the
ML estimator must then correspond to $\tilde{\varphi}_{N}^\t{\tiny ML}(\mathbf{x})\!=\! T(\mathbf{x})/N$
that is the corresponding global efficient estimator.

\subsubsection{Estimating a transformed parameter}

As shown in \secref{sub:CRBSat}, a given estimation
problem may not allow for a global estimator to exist that fulfils
the CRB-saturability criterion \eref{eq:CRBSatCond} unless the
asymptotic $N$ limit is considered, in which the ML estimator \eref{eq:MLE}
is guaranteed to be efficient. However, it may happen that the same estimation problem
may still be solvable efficiently with a global estimator regardless of $N$ when
estimating a \emph{transformed parameter} $g(\varphi)$. One may explicitly
prove, see e.g.~\citep{Kay1993}, that for a general $g(\varphi)$
the CRB \eref{eq:CRB} transforms to 
\begin{equation}
\left.\Delta^{2}\tilde{g}_{N}\right|_{\varphi_0}\ge\frac{\left[\dot{g}(\varphi_0)\right]^{2}}{N\left.F_{\textrm{cl}}\!\left[p_{\varphi}\right]\right|_{\varphi_0}},\label{eq:CRBFunPar}
\end{equation}
because the FI \eref{eq:FI} under such parameter change just rescales,
$F_{\textrm{cl}}\!\left[p_{g(\varphi)}\right]\!=\! F_{\textrm{cl}}\!\left[p_{\varphi}\right]/\left[\dot{g}(\varphi)\right]^{2}$,
due to the chain rule $\partial/\partial\varphi\!\equiv\! \dot{g}(\varphi)\,\partial/\partial g$.
As in the case of the CRB, the \emph{transformed CRB}
\eref{eq:CRBFunPar} lower-bounds the MSE, and hence the variance, of any unbiased estimator
$\tilde{g}_{N}$ and may be saturated by a single estimator if
the transformed version of criterion \eref{eq:CRBSatCond} is
satisfied for any $\varphi$:
\begin{equation}
\frac{\partial}{\partial\varphi}\ln\, p_{\varphi}(\mathbf{x})=N\,\frac{F_{\textrm{cl}}\!\left[p_{\varphi}\right]}{\dot{g}(\varphi)}\,\left(\tilde{g}_{N}(\mathbf{x})-g(\varphi)\right).\label{eq:CRBSatCondFunPar}
\end{equation}
Thus, in particular, we may always fulfil the corresponding requirement \eref{eq:UnbiasedConstrExpFam}
applicable to the exponential family of PDFs \eref{eq:PDFExpFam} and 
the estimator $\tilde{g}^\t{\tiny exp}_{N}(\mathbf{x})=T(\mathbf{x})/N$ by choosing
the transformed parameter such that
\begin{equation}
\left\langle\tilde{g}^\t{\tiny exp}_{N}\right\rangle _{\varphi}\,=\,\frac{\dot{A}(\varphi)}{\dot{\eta}(\varphi)}\,=\,g(\varphi),
\label{eq:CRBSatCondExpFamFunPar}
\end{equation}
what guarantees%
\footnote{%
Yet, by transforming the parameter we may introduce more pathological
parameter values for which $\dot{g}(\varphi)\!=\!\pm\infty$ or $\dot{g}(\varphi)\!=\!0$, as then
$F_{\textrm{cl}}\!\left[p_{g(\varphi)}\right]\!=\!F_{\textrm{cl}}\!\left[p_{\varphi}\right]/\left[\dot{g}(\varphi)\right]^{2}\!=\!0$
or $\frac{\partial}{\partial g} \ln p_{g(\varphi)}\!=\!\frac{1}{\dot{g}(\varphi)}\frac{\partial}{\partial \varphi} \ln p_{\varphi}\!=\!\pm\infty$ 
invalidating the CRB.}
$\tilde{g}^\t{\tiny exp}_{N}$ to be a global estimator saturating the transformed CRB \eref{eq:CRBFunPar} for any $N$.

\subsection{Bayesian approach -- \emph{global} estimation of a \emph{stochastic} parameter}
\label{sub:ClEstGlobal}

\subsubsection{Average Mean Squared Error}
\label{sub:AvMSE}

Within the \emph{Bayesian} approach, the estimated parameter 
$\varphi$ is assumed to be a random variable that is distributed
according to a \emph{prior} PDF, $p(\varphi)$, representing the knowledge
about $\varphi$ one possesses before performing the estimation. Therefore,
in contrast to the frequentist philosophy of \secref{sub:ClEstLocal},
where the estimated parameter was assumed to have a fixed,
well defined value, it is a particular \emph{realisation} of the
parameter that is really estimated in a real-life experiment. As a
consequence, an optimal estimator must not only be \emph{global} 
and minimise the MSE \eref{eq:MSE}, but also has to take into account which 
values of $\varphi$ are more probable according to $p(\varphi)$. 
Hence, such an estimator must minimise the \emph{Average
Mean Squared Error} ($\overline{\textrm{MSE}}$):
\begin{equation}
\left\langle \Delta^{2}\tilde{\varphi}_{N}\right\rangle=\int\!\! \t{d}\varphi\; p(\varphi)\left.\Delta^{2}\tilde{\varphi}_{N}\right|_{\varphi}=\!\int\!\!\textrm{d}\varphi\; p(\varphi)\!\!\int\!\!\textrm{d}^{N}\! x\; p(\mathbf{x}|\varphi)\left(\tilde{\varphi}_{N}(\mathbf{x})-\varphi\right)^{2},
\label{eq:AvMSE}
\end{equation}
which is the MSE \eref{eq:MSE} averaged over the possible values
of the estimated parameter. $p(\mathbf{x}|\varphi)$ is the PDF previously 
labelled as $p_\varphi(\mathbf{x})$ within the frequentist approach,
which due to stochastic character of the parameter now represents
a conditional probability predicting for a given realisation of $\varphi$ the outcome statistics.
Consequently, the $\overline{\textrm{MSE}}$ may be interpreted also as the
mean \emph{squared distance} between the estimator and parameter realisations averaged 
according to the joined PDF $p(\mathbf{x},\varphi)$, which is defined 
via the \emph{Bayes' theorem}---hence the name of the approach---in two equivalent ways:
\begin{equation}
p(\mathbf{x},\varphi)=p(\mathbf{x}|\varphi)\, p(\varphi)=p(\varphi|\mathbf{x})\, p(\mathbf{x}).
\label{eq:BayesTh}
\end{equation}
In general, the conditional probabilities satisfy $\int\! \t{d}^{N}\!x\, p(\mathbf{x}|\varphi)\!=\!\int\! \t{d}\varphi\, p(\varphi|\mathbf{x})\!=\!1$
and the probability of a particular sample corresponds to the marginal $p(\mathbf{x})\!=\!\int\! \t{d}\varphi\, p(\mathbf{x},\varphi)$.
By utilizing the last expression in \eqnref{eq:BayesTh}, one can rewrite \eqnref{eq:AvMSE} as 
\begin{equation}
\left<\Delta^{2}\tilde{\varphi}_{N}\right>=\int\!\! \t{d}^{N}\!x\; p(\mathbf{x})\left[\int\!\! \t{d}\varphi\; p(\varphi|\mathbf{x})\left(\tilde{\varphi}_{N}(\mathbf{x})-\varphi\right)^{2}\right]
\end{equation}
to show that the $\overline{\textrm{MSE}}$ is minimal for an estimator
which minimises the term in square brackets for each $\mathbf{x}$.
Hence, we may determine the form of the optimal \emph{Minimum Mean
Squared Error} (MMSE) estimator:%
\footnote{%
In accordance with the convention introduced in \secref{sub:ClEstLocal}, 
where the subscript in $\left\langle \dots\right\rangle _{\varphi}$ denoted the 
fixed parameter value, we generalise this notation here, so that it
explicitly specifies the PDF w.r.t.~which the averaging should be performed,
e.g.~$\left\langle \dots\right\rangle _{p(\phi|\bullet)}\!=\!\int\!\textrm{d}\phi\, p(\phi|\bullet)(\dots)$.}
\begin{equation}
\frac{\partial}{\partial\tilde{\varphi}_{N}(\mathbf{x})}\int\!\! \t{d}\varphi\; p(\varphi|\mathbf{x})\left(\tilde{\varphi}_{N}(\mathbf{x})-\varphi\right)^{2}=0\quad\implies\quad\tilde{\varphi}_{N}^\t{\tiny MMSE}(\mathbf{x})=\!\int\!\! \t{d}\varphi\; p(\varphi|\mathbf{x})\,\varphi=\left\langle \varphi\right\rangle _{p(\varphi|\mathbf{x})},
\label{eq:MMSE_Est}
\end{equation}
which simply corresponds to the average parameter value computed with
respect to the \emph{posterior} PDF,
$p(\varphi|\mathbf{x})$, that in principle may always be computed 
due to the Bayes' theorem \eref{eq:BayesTh} via
\begin{equation}
p(\varphi|\mathbf{x})=\frac{p(\mathbf{x}|\varphi)\, p(\varphi)}{\int\!\! \t{d}\varphi\, p(\mathbf{x}|\varphi)\, p(\varphi)}.
\label{eq:PostPDF}
\end{equation}
Within the Bayesian framework, one should view the process of data inference
as a procedure in which the effective PDF of the estimated parameter $\varphi$
becomes updated. Hence, the posterior PDF $p(\varphi|\mathbf{x})$ represents
the prior $p(\varphi)$ that has been reshaped and narrowed-down after learning
the sample $\mathbf{x}$, whereas the MMSE estimator \eref{eq:MMSE_Est} just outputs the mean of 
such an effective distribution. Moreover, the \emph{minimal} $\overline{\textrm{MSE}}$ \eref{eq:AvMSE} then reads
\begin{equation}
\left<\Delta^{2}\tilde{\varphi}_{N}^\t{\tiny MMSE}\right>=\int\!\! \t{d}^N\!x\; p(\mathbf{x})\left[\int\!\! \t{d}\varphi\; p(\varphi|\mathbf{x})\left(\varphi-\left\langle \varphi\right\rangle _{p(\varphi|\mathbf{x})}\right)^{2}\right]=\int\!\! \t{d}^N\!x\; p(\mathbf{x})\left.\textrm{Var}\!\left[\varphi\right]\right|_{p(\varphi|\mathbf{x})},
\label{eq:MinAvMSE}
\end{equation}
so that  it represents the variance of the parameter $\varphi$
computed also with respect to $p(\varphi|\mathbf{x})$ and
averaged over all the possible outcomes.

Firstly, let us emphasise that, in contrast to the local approach of \secref{sub:ClEstLocal},
in order to establish the optimal estimator, i.e.~the MMSE estimator \eref{eq:MMSE_Est}, 
we did \emph{not} have to force it to be \emph{unbiased}. In fact, assuming a 
particular parameter realisation $\varphi_0$, the ``local mean'' of the MMSE estimator reads
\begin{equation}
\left\langle \tilde{\varphi}_{N}^\t{\tiny MMSE}\right\rangle_{\varphi_0}=\int\!\! \t{d}^N\!x\; p(\mathbf{x}|\varphi_0)\,\tilde{\varphi}_{N}^\t{\tiny MMSE}(\mathbf{x})=\iint\!\! \t{d}^N\!x\,\t{d}\varphi\;p(\varphi|\mathbf{x})\,p(\mathbf{x}|\varphi_0)\,\varphi=\left<\varphi\right>_{p^N\!(\varphi|\varphi_0)},
\label{eq:MMSE_EstLocMean}
\end{equation}
where $p^N\!(\varphi|\varphi_0)\!=\!\int\!\! \t{d}^N\!x\,p(\varphi|\mathbf{x})\,p(\mathbf{x}|\varphi_0)$ is 
the probability of inferring the parameter value $\varphi$ on average given the true value $\varphi_0$.
Thus, $\tilde{\varphi}_{N}^\t{\tiny MMSE}$ is unbiased from the local perspective only
if $\left<\varphi\right>_{p^N\!(\varphi|\varphi_0)}\!=\!\varphi_0$ what is not true in general.

For instance, when the prior PDF $p(\varphi)$ is a distribution much more sensitive to any parameter changes 
than the distribution $p(\mathbf{x}|\varphi)$ dictating the outcomes collected, the variations of the posterior PDF \eref{eq:PostPDF} with $\varphi$ are 
predominantly determined by the prior PDF with the sampled data playing a marginal role, so that 
$p(\varphi|\mathbf{x})\!\approx\!p(\mathbf{x}|\varphi)p(\varphi)\!\approx\!p(\varphi)$. As a result, also $p^N\!(\varphi|\varphi_0)\!\approx\!p(\varphi)$
and the prior PDF is then responsible for the bias in \eqnref{eq:MMSE_EstLocMean}. Moreover,
in such a prior-dominant case, the minimal $\overline{\textrm{MSE}}$ \eref{eq:MinAvMSE}
approximately equals the variance of the prior PDF and the distribution of the
MMSE estimator 
becomes narrowly peaked around the prior mean.
Therefore, it is really important within the Bayesian approach to choose an appropriate $p(\varphi)$ 
such that, on one hand, it adequately represents the knowledge about the parameter 
before the estimation, but, on the other, it does not significantly
overshadow the information obtained from the data collected.

\begin{note}[Bayesian approach with a Dirac delta prior distribution]
\label{note:deltaprior}
An extremal but instructive example of a prior PDF is the Dirac delta distribution, 
$p_\delta(\varphi)\!=\!\delta\!\left(\varphi\!-\!\varphi_{0}\right)$,
which represents the case when we perfectly know the estimated parameter
before performing the estimation. Importantly, such
situation is \emph{not} equivalent to the local estimation regime
of the frequentist approach, in which (as explained in \secref{sub:LocConseq}) one
seeks for an estimator most sensitive to parameter fluctuations
and the prior knowledge, even being complete, is thus irrelevant. 
In contrast, when we substitute $p_\delta(\varphi)$
into \eqnref{eq:PostPDF} to compute the MMSE estimator \eref{eq:MMSE_Est}, 
we simply obtain $\tilde{\varphi}_{N}^\t{\tiny MMSE}(\mathbf{x})\!=\!\varphi_{0}$
yielding $\left<\Delta^{2}\tilde{\varphi}_{N}^\t{\tiny MMSE}\right>\!=\!0$, 
so that the Bayesian solution to the estimation problem correctly suggests just to output 
the perfectly known value of $\varphi$ and fully ignore any data collected.
\end{note}

Secondly, let us stress that so far we did \emph{not} require at any stage the sampled
data to be \emph{independently} distributed. Such property, which
previously guaranteed locality in the asymptotic $N$ limit within the frequentist
approach (see \secref{sub:LocConseq}) is \emph{not necessary} in the derivation
of the optimal Bayesian estimator, which relies only on the form of the posterior PDF
\eref{eq:PostPDF}. In fact, as independently distributed data may be 
interpreted as if it was collected carrying out consecutive \emph{repetitions} of 
the estimation protocol, the Bayesian results in such a case may be understood
as a \emph{progressive} updating of the knowledge we possess about the parameter.
Notice that the data may then be freely split into parts treated as separate samples, e.g.~$\mathbf{x}\!=\!\{\mathbf{x}_1,\mathbf{x}_2\}$.
Due to the independence property $p(\mathbf{x}\,|\varphi)\!=\!p(\mathbf{x}_1|\varphi) p(\mathbf{x}_2|\varphi)$ and
one may rewrite the posterior PDF by dividing the numerator and denominator in \eref{eq:PostPDF} by $p(\mathbf{x}_1)$ as:
\begin{equation}
p(\varphi|\mathbf{x})=\frac{p(\mathbf{x}_2|\varphi)\, p(\varphi|\mathbf{x}_1)}{\int\!\! \t{d}\varphi\, p(\mathbf{x}_2|\varphi)\, p(\varphi|\mathbf{x}_1)},
\label{eq:PostPDFprogress}
\end{equation}
so that $p(\varphi|\mathbf{x})$ may be reinterpreted as if it was calculated basing only on the 
outcomes $\mathbf{x}_2$ but for the prior already updated with the results $\mathbf{x}_1$. 
Hence, for independently distributed data, we could equivalently arrive 
at the MMSE estimator \eref{eq:MMSE_Est} evaluated for $p(\varphi|\mathbf{x})$, 
if we constructed it progressively by repeating the protocol and varying the 
prior PDF, while including more and more outcomes in each round,
what effectively narrows down the spread of each consecutively obtained estimator 
that eventually coincides with $\tilde{\varphi}_{N}^\t{\tiny MMSE}(\mathbf{x})$ when all the data is finally employed.

On the other hand, the independent character of the sampled data allows to establish
a link between the global and local results and, in particular, give an operational 
meaning to the $\overline{\textrm{MSE}}$ \eref{eq:AvMSE}, so that it is
not just a figure of merit with respect to which the estimator is derived, but is also
related to the estimator variance that, as remarked before, is of experimental significance
being determinable basing only on the outcomes gathered. For independent data, one may 
prove that the role of any well-behaved prior distribution becomes negligible in the 
asymptotic $N$ limit. This is so, because the Local Asymptotic Normality%
\footnote{%
In particular, the Bernstein--von Mises theorem \citep{Vaart1998}.} \citep{Vaart1998} 
then assures that for \emph{any} regular%
\footnote{%
By regular prior distribution we mean that it possesses ``finite information'' about the parameter,
so that it is smooth and for all $\varphi$:~$F_{\textrm{cl}}\!\left[p(\varphi)\right]\!<\!\infty$,
what excludes e.g.~the Dirac delta distribution $p_\delta(\varphi)\!=\!\delta\!\left(\varphi\!-\!\varphi_{0}\right)$ discussed in \noteref{note:deltaprior}.}
$p(\varphi)$, 
the posterior $p(\varphi|\mathbf{x})$ \eref{eq:PostPDF}---treated as 
a distribution of $\varphi$ for a given $\mathbf{x}$ and the parameter true value $\varphi_0$---always
becomes equivalent to a Gaussian PDF as $N\!\rightarrow\!\infty$, with mean that may be viewed
as a random variable accounting for the fluctuations of $\mathbf{x}$, which
is also normally distributed with mean and variance respectively equal to $\varphi_0$ and 
the inverse of the FI:~$1/(N\!\left.F_{\textrm{cl}}\!\left[p(X|\varphi)\right]\right|_{\varphi_0})$. 
As the mean of the posterior PDF corresponds to the MMSE estimator \eref{eq:MMSE_Est},
such observation proves that  for regular priors $\tilde{\varphi}_{N}^\t{\tiny MMSE}$ 
is always asymptotically unbiased and locally saturates the CRB \eref{eq:CRB}.
Hence, it converges to the ML estimator \eref{eq:MLE} with
$\left\langle\tilde{\varphi}_{N}^\t{\tiny MMSE}\right\rangle_{\varphi_0}\!\!=\!\left<\varphi\right>_{p^N\!(\varphi|\varphi_0)}\!\!\!\!\!\overset{N\rightarrow\infty}{=}\!\varphi_0$ and
$\left.\Delta^{2}\tilde{\varphi}_{N}^\t{\tiny MMSE}\right|_{\varphi_0}\!\!\!\!\!\overset{N\rightarrow\infty}{=}\!1/(N\!\left.F_{\textrm{cl}}\!\left[p(X|\varphi)\right]\right|_{\varphi_0})$%
\footnote{%
In fact, any other reasonable estimator built on the posterior PDF \eref{eq:PostPDF} will also converge asymptotically to the local results, 
e.g.~estimators corresponding to the median and the mode of $p(\varphi|\mathbf{x})$ mentioned in the following section \citep{Vaart1998}.}.
As a consequence, one may replace for each realisation of $\varphi$ the MSE, $\left.\Delta^{2}\tilde{\varphi}_{N}^\t{\tiny MMSE}\right|_\varphi$, in \eqnref{eq:MinAvMSE}
by its corresponding CRB, so that the minimal $\overline{\t{MSE}}$ \eref{eq:MinAvMSE}
may always be rewritten in the asymptotic $N$ limit as
\begin{equation}
\left<\Delta^{2}\tilde{\varphi}_{N}^\t{\tiny MMSE}\right>= \left<\left.\Delta^{2}\tilde{\varphi}_{N}^\t{\tiny MMSE}\right|_\varphi\right>_{p(\varphi)}\overset{N\rightarrow\infty}{=}\left<\frac{1}{N F_{\t{cl}}\!\left[p(X|\varphi)\right]}\right>_{p(\varphi)}\ge\frac{1}{N\left<F_{\t{cl}}\!\left[p(X|\varphi)\right]\right>_{p(\varphi)}},
\label{eq:MinAvMSE_Jensen}
\end{equation}
where the last expression follows from the Jensen inequality \citep{Jensen1906}
stating that for any concave function $f(X)$:~$\left<f(X)\right>\!\ge\!f(\left<X\right>)$.
Most importantly, although the lower bound \eref{eq:MinAvMSE_Jensen} becomes trivial when 
the FI \eref{eq:FI} is independent of the estimated parameter, 
i.e.~$\forall_\varphi\!:\,F_{\t{cl}}\!\left[p(X|\varphi)\right]\!=\!F_\t{cl}$,
the minimal $\overline{\t{MSE}}$ \eref{eq:MinAvMSE} coincides then asymptotically with the CRB, as 
$\left<\Delta^{2}\tilde{\varphi}_{N}^\t{\tiny MMSE}\right>\!\overset{N\rightarrow\infty}{=}1/F_\t{cl}$,
and hence by \eqnref{eq:CRB} constitutes for $N\!\to\!\infty$ a lower bound on the variance of any
(locally) unbiased estimator.

Lastly, let us remark that \eqnref{eq:MinAvMSE_Jensen}
can also be derived by means of the so-called
\emph{Bayesian CRB} \citep{Trees1968,Gill1995}, 
which applies \emph{regardless} of $N$ and lower-limits the $\overline{\t{MSE}}$ 
\eref{eq:AvMSE} for any Bayesian estimator at a price of
requiring the prior PDF not only to be regular, 
but also to vanish at the end-points, i.e.~$p(a)\!=\!p(b)\!=\!0$ for $\varphi\!\in\![a,b]$:
\begin{equation}
\left\langle \Delta^{2}\tilde{\varphi}_{N}\right\rangle \ge \frac{1}{F_{\textrm{cl}}\!\left[p(\varphi)\right]+N \left\langle F_{\textrm{cl}}\!\left[p(X|\varphi)\right]\right\rangle _{p(\varphi)}}\overset{N\rightarrow\infty}{=}\frac{1}{N \left\langle F_{\textrm{cl}}\!\left[p(X|\varphi)\right]\right\rangle _{p(\varphi)}}\,,
\label{eq:BayesCRB}
\end{equation}
where $F_{\textrm{cl}}\!\left[p(\varphi)\right]$ is the FI of the prior distribution.
As a result, \eqnref{eq:BayesCRB} is unfortunately not valid when considering uniform
prior PDFs, i.e.~$p(\varphi)\!\simeq\!1$ such that $\int\!\!\t{d}\varphi\, p(\varphi)\!=\!1$, which we focus on in this work. Thus, in the following sections
when considering Bayesian estimation problems, we always utilise \eqnref{eq:MinAvMSE_Jensen} 
to establish connection with the complementary frequentist results for independently distributed samples.

\subsubsection{Average \emph{cost}}
\label{sub:AvCost}

Within the frequentist approach, one is restricted to use the MSE \eref{eq:MSE},
$\Delta^{2}\tilde{\varphi}_{N}$, which figure of merit is the \emph{squared
distance} between the estimator and the true parameter value:~$\left(\tilde{\varphi}-\varphi\right)^{2}$,
as only then by imposing the unbiasedness the CRB \eref{eq:CRB}
may be derived and utilized. On the other hand, within the Bayesian
framework, nothing prevents us to consider other figures of merit,
i.e.~\emph{cost functions} $C(\tilde{\varphi},\varphi)$, in order
to generalise the $\overline{\textrm{MSE}}$,
$\left\langle \Delta^{2}\tilde{\varphi}_{N}\right\rangle $, and define
the \emph{average cost}, $\left\langle \mathcal{C}(\tilde{\varphi}_N)\right\rangle $
\citep{Kay1993}:
\begin{equation}
\left\langle \mathcal{C}(\tilde{\varphi}_N)\right\rangle=\int\!\! \t{d}\varphi\; p(\varphi)\left.\mathcal{C}(\tilde{\varphi}_N)\right|_{\varphi}=\!\int\!\!\textrm{d}\varphi\; p(\varphi)\int\!\!\textrm{d}^{N}\! x\; p(\mathbf{x}|\varphi)\;C\!\left(\tilde{\varphi}_{N}(\mathbf{x}),\varphi\right),
\label{eq:AvCost}
\end{equation}
which in some situations may turn out to be more appropriate than
the $\overline{\textrm{MSE}}$ \eref{eq:AvMSE} corresponding
to the special case of $C\!\left(\tilde{\varphi},\varphi\right)\!=\!\left(\tilde{\varphi}-\varphi\right)^{2}$.
For example, other common cost functions that are often considered
include the Absolute Error (AE) and the Hit-or-Miss Error (HME) \citep{Kay1993}:
\begin{equation}
C_{\textrm{AE}}(\tilde{\varphi},\varphi)=\left|\tilde{\varphi}-\varphi\right|,\qquad C_{\textrm{HME}}(\tilde{\varphi},\varphi)\!=\!\begin{cases}
0, & \left|\tilde{\varphi}-\varphi\right|\le\delta\\
1, & \left|\tilde{\varphi}-\varphi\right|>\delta
\end{cases}\textrm{ with }\delta\!\ll\!1.
\label{eq:CostFuns}
\end{equation}
The average cost in the case of AE, $\left\langle \mathcal{C}_{\textrm{AE}}(\tilde{\varphi}_N)\right\rangle $,
does not differ significantly from the $\left\langle \Delta^{2}\tilde{\varphi}_{N}\right\rangle$,
but due to the cost function being linear and not
quadratic it does not penalise that much for estimates being far
from the true parameter value. As a consequence, the optimal Bayesian
estimator does not correspond then to the mean of the posterior
PDF, $p(\varphi|\mathbf{x})$, as in the case of the $\overline{\textrm{MSE}}$
and the MMSE estimator \eref{eq:MMSE_Est}, but rather to its \emph{median}.
The HME, on the other hand, is most restrictive rewarding only the estimates being
approximately the real value of the parameter, so that the optimal
estimator minimising the average cost, $\left\langle \mathcal{C}_{\textrm{HME}}(\tilde{\varphi}_N)\right\rangle $,
corresponds to choosing the most probable value of $\varphi$ with 
respect to the posterior PDF, i.e.~its \emph{maximum} (mode)%
\footnote{%
For explicit derivations of the optimal estimators for the AE and HME
cost functions see e.g.~\citep{Kay1993}.}.
In particular, such an estimator is equivalent \emph{regardless of} $N$ to the ML estimator
\eref{eq:MLE} considered within the frequentist approach when the
prior PDF is assumed to be uniform, i.e.~$p(\varphi)\!\simeq\!const$, \citep{Vaart1998}.

However, we present the cost functions \eref{eq:CostFuns} to the reader
only as instructive examples, as we do not utilise them explicitly
within this work. This is because, we restrict ourselves to average costs \eref{eq:AvCost}
which converge in the asymptotic $N$ limit to the $\overline{\textrm{MSE}}$
\eref{eq:AvMSE}, i.e.~$\left\langle \mathcal{C}\!\left(\tilde{\varphi}_{N}\right)\right\rangle\!\!\overset{N\rightarrow\infty}{=}\!\!\left\langle \Delta^{2}\tilde{\varphi}_{N}\right\rangle$. 
As any consistent estimator (see \eqnref{eq:EstCons}) attains the parameter true value with $N\!\rightarrow\!\infty$,
so that $\tilde\varphi$ approaches $\varphi$ with $N$, such constraint is fulfilled by
considering only cost functions that satisfy $C(\tilde\varphi,\varphi)\!=\!({\tilde\varphi}-\varphi)^2\!+\!O\!\left[({\tilde\varphi}-\varphi)^3\right]$.
As a result, the \emph{optimal} Bayesian estimator, $\tilde{\varphi}_{N}^\t{\tiny opt}$, \emph{always} 
converges asymptotically to the MMSE estimator, $\tilde{\varphi}_{N}^\t{\tiny MMSE}$, and the average cost
$\left\langle \mathcal{C}\!\left(\tilde{\varphi}_{N}^\t{\tiny opt}\right)\right\rangle$ 
attains with $N\!\rightarrow\!\infty$  the \emph{minimal} $\overline{\textrm{MSE}}$ \eref{eq:MinAvMSE}.
Furthermore, if the data is \emph{independently} distributed, $\tilde{\varphi}_{N}^\t{\tiny opt}$ 
also approaches the ML estimator \eref{eq:MLE} and 
$\left\langle \mathcal{C}\!\left(\tilde{\varphi}_{N}^\t{\tiny opt}\right)\right\rangle$ may be related 
to the frequentist results via \eqnsref{eq:MinAvMSE_Jensen}{eq:BayesCRB}. 
In principle, $\left\langle \mathcal{C}\!\left(\tilde{\varphi}_{N}^\t{\tiny opt}\right)\right\rangle$ may 
then be even utilised to derive the local precision bounds and the form of the CRB \eref{eq:CRB}.

\subsubsection{Average \emph{cost} with circular symmetry}
\label{sub:AvCostCircSymm}

We have introduced the concept of general average cost $\left\langle \mathcal{C}(\tilde{\varphi}_{N})\right\rangle$
\eref{eq:AvCost}, in order to be able to \emph{globally} solve \emph{phase estimation
problems}, in which $\varphi$ is a circular parameter or more formally
an element of the circle group%
\footnote{%
Typically denoted by $\mathbb{T}$ or $U(1)$---being also the group
of unitary $1\!\!\times\!\!1$ matrices.}
$U(1)$ satisfying $\varphi\!\equiv\!\varphi+2\pi n$ for any $n\!\in\!\mathbb{Z}$.
In such a case, the squared-distance cost function employed in the
$\overline{\textrm{MSE}}$ \eref{eq:AvMSE} is not valid, as it does
not respect the parameter \emph{topology} which must be taken into
account when performing the integral over all parameter values contained
in the prior PDF in \eqnref{eq:AvCost} \citep{Holevo1982}.
This contrasts the frequentist case of \secref{sub:ClEstLocal},
where the strategy was optimised to sense only small variations of $\varphi$
and such issues were completely ignored. Such ignorance, however, can lead
within the local approach to estimators that give inconclusive answers, 
as they disregard any parameter symmetry (see e.g.~the later discussed estimator 
of \eqnref{eq:LocEffEst_BinPhi}).
On the other hand, when pursuing the Bayesian approach, 
we may \emph{correctly} account for the parameter topology, what
in the case of the circular parameter is achieved by requiring the cost function to
be:~\emph{symmetric} -- $C(\tilde{\varphi},\varphi)\!=\! C(\varphi,\tilde{\varphi})$,
\emph{group invariant} -- $\forall_{\phi\in U(1)}\!:\,C(\tilde{\varphi}\!+\!\phi,\varphi\!+\!\phi)\!=\! C(\tilde{\varphi},\varphi)$
and \emph{periodic} -- $\forall_{n\in\mathbb{Z}}\!:\, C(\tilde{\varphi}\!+\!2\pi n,\varphi)\!=\! C(\tilde{\varphi},\varphi)$,
so that most generally we may rewrite it as a function of the difference
between the estimator value and the parameter realisation, $\theta\!=\!\tilde{\varphi}-\varphi$,
which reads:
\begin{equation}
C(\tilde{\varphi},\varphi)=C(\theta)=\sum_{k=0}^{\infty}\, c_{k}\,\cos\!\left(k\,\theta\right).
\label{eq:CostFun_gen}
\end{equation}
Furthermore, as $C(\theta)$ must rise monotonically from $C(0)\!=\!0$
at $\theta\!=\!0$ to some $C(\pi)\!=\! C_{\textrm{max}}$ at $\theta\!=\!\pi$,
so that $C'(\theta)\!\ge\!0$, the coefficients $c_{k}$ must fulfil
following constraints:
\begin{equation}
\sum_{k=0}^{\infty}c_{k}=0,\qquad\sum_{k=0}^{\infty}(-1)^{k}c_{k}=C_{\max},\qquad\sum_{k=1}^{\infty}k^{2}c_{k}\le0,\qquad\sum_{k=1}^{\infty}k^{2}(-1)^{k}c_{k}\ge0,
\end{equation}
which may be satisfied by imposing $\forall_{k>0}\!:\, c_{k}\le0$
and taking $c_{k}$ to decay at least quadratically with $k$. Lastly,
as mentioned in the previous section, we require the average cost
$\left\langle \mathcal{C}\!\left(\tilde{\varphi}_{N}\right)\right\rangle$
to converge to the $\overline{\textrm{MSE}}$
$\left\langle \Delta^{2}\tilde{\varphi}_{N}\right\rangle$
in the asymptotic $N$ limit, in which the estimator 
approaches the true
parameter and $\theta\!\rightarrow\!0$. Such constraint
imposes that for small $\theta$:~$C(\theta)\!=\!\theta^{2}\!+\! O\!\left(\theta^{4}\right)$
and hence $\sum_{k=1}^{\infty}k^{2}c_{k}\!=\!-2$.

In all the Bayesian estimation problems considered in our work that
deal with a circularly symmetric parameter, we will consider the simplest
cost function introduced by \citet{Holevo1982}, which satisfies 
all the above-mentioned conditions with
$c_{0}\!=\!-c_{1}\!=\!2$, $\forall_{k>1}\!:\, c_{k}\!=\!0$ 
and explicitly reads:
\begin{equation}
C_{\textrm{H}}(\tilde{\varphi},\varphi)=C_{\textrm{H}}(\tilde{\varphi}-\varphi)=4\,\sin^{2}\!\left(\frac{\tilde{\varphi}-\varphi}{2}\right).
\label{eq:CostFun_H}
\end{equation}
Consequently, it leads to the average cost \eref{eq:AvCost} of the form:
\begin{equation}
\left\langle \mathcal{C}_{\textrm{H}}(\tilde{\varphi}_{N})\right\rangle =4\int\!\! d\varphi\; p(\varphi)\int\!\! d\mathbf{x}\; p(\mathbf{x}|\varphi)\;\sin^{2}\!\left(\frac{\tilde{\varphi}_{N}(\mathbf{x})-\varphi}{2}\right).
\label{eq:AvCost_H}
\end{equation}
Following the same argumentation as in \eqnref{eq:MMSE_Est}
describing the derivation of the MMSE estimator optimal 
for the $\overline{\textrm{MSE}}$ \eref{eq:AvMSE}, one may prove
that $\left\langle \mathcal{C}_{\textrm{H}}(\tilde{\varphi}_N)\right\rangle$ is minimised
if an estimator, $\tilde{\varphi}_{N}^\t{\tiny H}$,
can be found that for any possible data sample $\mathbf{x}$ collected
satisfies the following condition:
\begin{equation}
\int\!\! d\varphi\;\, p(\varphi|\mathbf{x})\;\sin\!\left(\tilde{\varphi}_{N}^\t{\tiny H}(\mathbf{x})-\varphi\right)=0.\label{eq:H_Est}
\end{equation}

\begin{figure}[!t]
\begin{center}
\includegraphics[width=0.9\columnwidth]{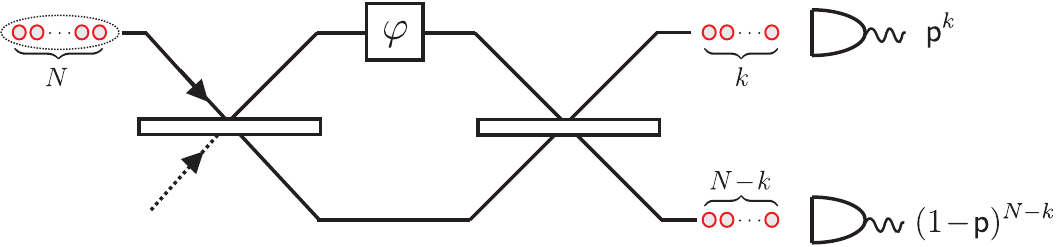}
\end{center}
\caption[Mach-Zehnder interferometry with uncorrelated photons]
{\textbf{Schematic description of a Mach-Zehnder interferometer}
which introduces a relative \emph{phase} delay $\varphi$ in between 
the light beams travelling in its arms. A light state of $N$ \emph{uncorrelated}
photons, i.e.~a Fock state $\ket{N}$,  
is impinged onto one of the input ports. As a result,
each of the constituent photons may be treated independently and
is detected with probabilities $\mathsf{p}\!=\!\cos^{2}\!\frac{\varphi}{2}$ 
and $1\!-\!\mathsf{p}\!=\!\sin^{2}\!\frac{\varphi}{2}$ 
at the output ports. Hence, the overall distribution of registering $k$ 
photons in one of the ports (and thus $N\!-\!k$ in the other) is binomal.
\label{fig:MZinter_cl}}
\end{figure}

\subsection{\caps{Example:} Mach-Zehnder interferometry with uncorrelated photons}
\label{sub:ClEst_MZInter_SQL}

Let us consider in detail a physically motivated example of a \emph{Mach-Zehnder interferometer}
depicted in \figref{fig:MZinter_cl}, for which the estimated
$\varphi$ represents the relative \emph{phase}
delay in between the interferometer arms \citep{Demkowicz2015}, so that the parameter
naturally exhibits a circular symmetry, and thus we choose $\varphi\!\in\!\left[-\pi,\pi\right]$. When
a state of light consisting of $N$ uncorrelated photons%
\footnote{%
In particular, a Fock state $\ket{N}$, which may be also simulated 
by impinging a coherent state of light and post-selecting only events when the total 
number of photons registered at the output ports turns out to be $N$ \citep{Demkowicz2015}.
Importantly, such a state does not possess any correlations (even classical) in between the 
constituent photons, as explicitly shown in \noteref{note:ent_Fock}.}
is shone
on the input port, each individual photon may be detected in one of
the two output ports with probabilities $\mathsf{p}\!=\!\cos^{2}\!\frac{\varphi}{2}$
and $1\!-\!\mathsf{p}\!=\!\sin^{2}\!\frac{\varphi}{2}$ respectively,
what resembles a coin-tossing experiment with binary probabilities
$\mathsf{p}$ and $1\!-\!\mathsf{p}$ dictating the `heads'/`tails'
outcomes. Hence, the overall probability distribution of registering
$k$ photons in one port and $N\!-\!k$ in the other is binomial and reads
\begin{equation}
p_{\varphi}^{N}(k)=\binom{N}{k}\,\mathsf{p}^{k}\,\left(1-\mathsf{p}\right)^{N-k}=\binom{N}{k}\left(\cos^{2}\!\frac{\varphi}{2}\right)^{k}\left(\sin^{2}\!\frac{\varphi}{2}\right)^{N-k},
\label{eq:PDF_BinPhi}
\end{equation}
naturally belonging to exponential family of PDFs \eref{eq:PDFExpFam}
with $h(k)\!=\!\binom{N}{k}$, $T(k)\!=\! k$, $\eta(\varphi)\!=\!\ln\!\left(\cot^{2}\!\frac{\varphi}{2}\right)$
and $A(\varphi)\!=\!\ln\!\left(\textrm{cosec}^{2}\frac{\varphi}{2}\right)$.
We study the problem of $\varphi$-estimation in such a scenario from both
frequentist and Bayesian perspectives below.

\subsubsection{Frequentist approach \label{sub:ClEst_MZInter_SQL_Freq}}

Calculating the FI \eref{eq:FI} with the integral $\int\!\textrm{d}\mathbf{x}$
replaced accordingly by the sum $\sum_{k=0}^{N}$ due to discreteness
of the distribution \eref{eq:PDF_BinPhi} or equivalently utilizing
\eqnref{eq:FIExpFam} for the exponential family of PDFs,
we derive the CRB \eref{eq:CRB} via
\begin{equation}
F_{\textrm{cl}}\!\left[p_{\varphi}^{N}\right]=N\, F_{\textrm{cl}}\!\left[p_{\varphi}\right]=N\quad\implies\quad\Delta^{2}\tilde{\varphi}_{N}\ge\frac{1}{N}\;.
\label{eq:FI&CRB_BinPhi}
\end{equation}
Although the above CRB is independent of the actual parameter value,
investigating the CRB-saturability condition \eref{eq:CRBSatCond}, we
realise that the binomial PDF \eref{eq:PDF_BinPhi} with such a parametrisation
does not satisfy the corresponding requirement \eref{eq:UnbiasedConstrExpFam}
for a global efficient estimator to exist, because 
\begin{equation}
\frac{\dot{A}(\varphi)}{\dot{\eta}(\varphi)}=\cos^{2}\!\frac{\varphi}{2}\ne\varphi.
\label{eq:PDF_SatCond_BinPhi}
\end{equation}
However, for a given parameter true value $\varphi_{0}$, we may
always construct a local efficient estimator following the prescription
of \eqnref{eq:LocEffEstExpFam}:
\begin{equation}
\tilde{\varphi}_{N,\varphi_{0}}\!(k)=\varphi_{0}+\cot\!\left(\frac{\varphi_{0}}{2}\right)-\frac{2k}{\sin(\varphi_{0})\, N},
\label{eq:LocEffEst_BinPhi}
\end{equation}
which is correctly unbiased:~$\left<\tilde{\varphi}_{N,\varphi_{0}}\right>_{\varphi_0}\!\!=\!\varphi_0$, 
and its MSE indeed saturates the CRB \eref{eq:FI&CRB_BinPhi}:
$\!\left.\Delta^2\tilde{\varphi}_{N,\varphi_{0}}\right|_{\varphi_0}\!=\!1/N$.
Yet, for this to be generally true, one must let in the calculation $\tilde{\varphi}_{N,\varphi_{0}}\!\in\![-\infty,\infty]$ 
and completely ignore the fact that $\tilde{\varphi}_{N,\varphi_{0}}$ is utilised to 
determine a circularly symmetric parameter. 
Hence, the applicability of the estimator \eref{eq:LocEffEst_BinPhi} becomes doubtful 
for the true values $\varphi_0$, for which $\tilde{\varphi}_{N,\varphi_{0}}$ exits out of the 
range of a single period, i.e.~$[-\pi,\pi]$, as it does not associate then equivalent parameter values 
with one another (does not respect the fact that e.g.~$\varphi\!=\!3\pi/2\!\equiv\!-\pi/2$). 
On the other hand, for the parameter values $\varphi_{0}\!=\!\{0,\pm\pi\}$, 
the estimator \eref{eq:LocEffEst_BinPhi} diverges 
even if one allows it to be defining points on the whole real line, 
but so actually does $\left.\left(\textrm{d}\ln p_{\varphi}^{N}(k)/\textrm{d}\varphi\right)\right|_{\varphi_{0}}$,
so that the regularity assumption \eref{eq:LocRegCond} fails and the
CRB is not valid there at all. As a matter of fact, these pathological points correspond
to the cases of $\mathsf{p}\!=\!\{0,1\}$, for which
the estimation problem becomes deterministic, as by learning then
whether respectively $k\!=\!\{0,N\}$ we may deduce the parameter value without
any error.

\begin{figure}[!t]
\includegraphics[width=1\columnwidth]{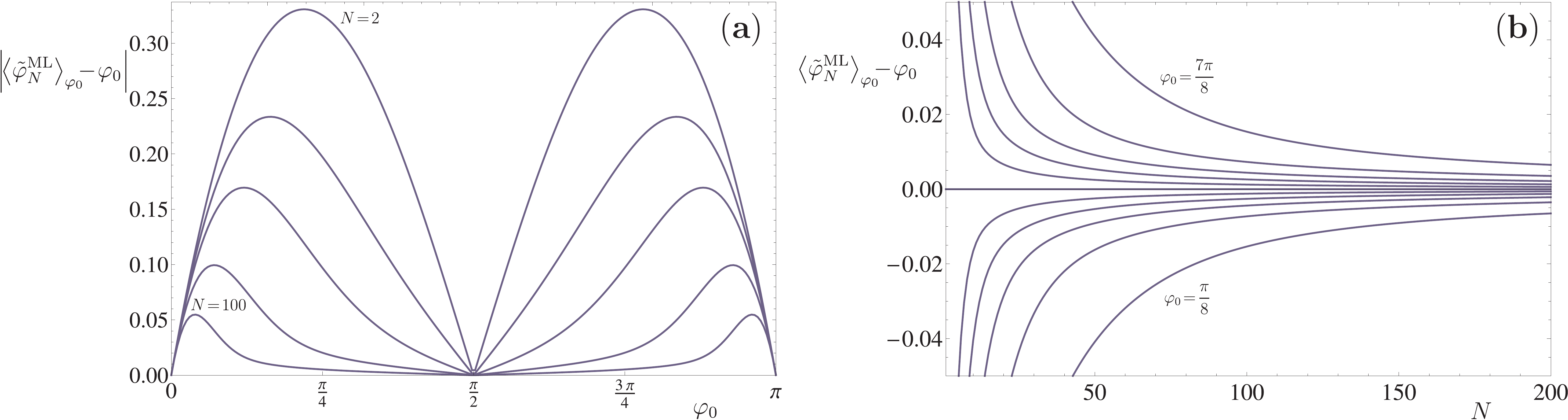}
\caption[Bias of the ML estimator]{%
\textbf{Bias of the ML estimator} \eref{eq:MLE_BinPhi} as a function of:~(\textbf{a})
-- the parameter true value $\varphi_{0}$,
and (\textbf{b}) -- the sample size $N$.
The curves in (\textbf{a}) are depicted for $N\!=\!\!\left\{ 2,5,10,30,100\right\} $, whereas
the ones in (\textbf{b}) correspond to the equally-distributed values:
$\varphi_{0}\!=\!\{\frac{\pi}{8},\dots,\frac{\pi}{2},\dots,\frac{7\pi}{8}\}$.
Both plots clearly indicate a decrease of the bias with an increase
in $N$ and the special parameter values $\varphi_{0}\!=\!\{0,\frac{\pi}{2},\pi\}$,
for which the bias vanishes irrespectively of the sample size.
Nevertheless, the ML estimator is \emph{locally unbiased} according to \eqnref{eq:LocUnbiasConstr} 
\emph{only} in the asymptotic $N$ limit.
\label{fig:BiasMLE_ClEst_BinPhi}
}
\end{figure}

\begin{figure}[!t]
\includegraphics[width=1\columnwidth]{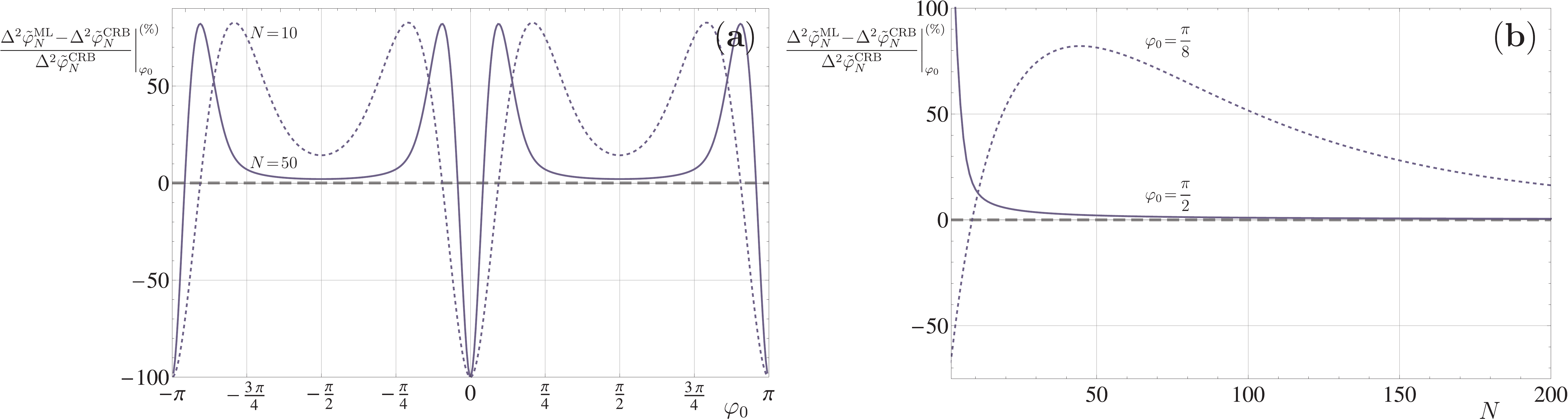}
\caption[MSE of the ML estimator]{%
\textbf{MSE of the ML estimator} \eref{eq:MLE_BinPhi}
in comparison with the CRB \eref{eq:FI&CRB_BinPhi} as a function of:
(\textbf{a}) -- the parameter true value $\varphi_{0}$, and (\textbf{b})
-- the sample size $N$. In both plots the CRB is represented by 
the horizontal line (\emph{grey dashed}). The curves in (\textbf{a}) are depicted for $N\!=\!\!\left\{10,50\right\} $
and show:~the pathological points $\varphi_{0}\!=\!\{0,\pm\pi\}$
at which $\forall_N\!:\Delta^2\tilde{\varphi}_{N}^\t{\tiny ML}\!=\!0$,
and the optimal points $\varphi_{0}\!=\!\pm\frac{\pi}{2}$
for which 
the CRB is most rapidly attained. The MSE dependence
on $N$, plotted in (\textbf{b}), indicates that for parameter
values close to the pathological points, e.g.~$\varphi_{0}\!=\!\frac{\pi}{8}$,
$\Delta^2\tilde{\varphi}_{N}^\t{\tiny ML}$ surpasses
the CRB at low $N$ to still attain it asymptotically from above,
whereas at the optimal $\varphi_{0}\!=\!\pm\frac{\pi}{2}$ the CRB
is attained within negligible margin of error already for $N\!\approx\!50$.
\label{fig:MSEMLE_ClEst_BinPhi}}
\end{figure}

Secondly, we consider the ML estimator \eref{eq:MLE} that is guaranteed
to saturate the CRB \eref{eq:FI&CRB_BinPhi} independently of $\varphi$
in the asymptotic $N$ limit. As we have just shown the estimation
problem not to allow for a global efficient estimator to exist, the conditions
\eref{eq:CRBSatCond} and \eref{eq:MLECond} are not equivalent and thus
$\tilde{\varphi}_{N}^\t{\tiny ML}$ must be found by explicitly solving
\eqnref{eq:MLECond} for the PDF \eref{eq:PDF_BinPhi}:
\begin{equation}
\tilde{\varphi}_{N}^\t{\tiny ML}(k)=\underset{\varphi}{\textrm{argmax}}\,\ln p_{\varphi}^{N}(k)=\pm2\;\textrm{arccot}\!\left(\sqrt{\frac{k}{N-k}}\,\right).
\label{eq:MLE_BinPhi}
\end{equation}
The two equivalent maxima arise due to the ambiguity in the sign
of $\varphi$, as $\forall_{N,k}\!:p_{\varphi}^{N}(k)\!=\! p_{-\varphi}^{N}(k)$.
As a consequence, before applying $\tilde{\varphi}_{N}^\t{\tiny ML}$
as a global estimator, we must possess extra information that allows
us to deduce whether the parameter true value $\varphi_{0}$ is positive.
The necessary division of the whole range $\left[-\pi,\pi\right]$
into halves is dictated by the parametrisation $\mathsf{p}\!=\!\cos^{2}\!\frac{\varphi}{2}$,
as $\varphi$ may be unambiguously inferred from $\mathsf{p}$ only after restricting
to parameter sub-ranges in which $\mathsf{p}$ is monotonic in $\varphi$,
i.e.~$\left[-\pi,0\right]$ and $\left[0,\pi\right]$. Thus,
the ML estimator \eref{eq:MLE_BinPhi} is non-smooth at $\varphi_{0}\!=\!\left\{0,\pm \pi\right\} $, 
which consistently are the pathological points at which the regularity
condition \eref{eq:LocRegCond}---and hence the CRB \eref{eq:FI&CRB_BinPhi}---fails%
\footnote{%
Notice that, in contrast to the local efficient estimator \eref{eq:LocEffEst_BinPhi}, the ML estimator \eref{eq:MLE_BinPhi}
does not exit the parameter range for which it is defined, i.e.~$\left[-\pi,0\right]$ or $\left[0,\pi\right]$,
so that it does \emph{not} violate the circular symmetry of the parameter.
}. 
Taking $\varphi_{0}\!\ge\!0$ we depict the bias of the ML estimator in \figref{fig:BiasMLE_ClEst_BinPhi}
as a function of consecutively $\varphi_{0}$ and $N$. As required
by the asymptotic unbiasedness property, the bias diminishes gradually
with $N$ for any $\varphi_{0}$. Yet, it vanishes regardless of $N$ at
the special value $\varphi_{0}\!=\!\frac{\pi}{2}$ and the pathological
$\varphi_{0}\!=\!\{0,\pm\pi\}$ for all of which thus 
$\textrm{Var}\!\left[\tilde{\varphi}_{N}^\t{\tiny ML}\right]\!=\!\Delta^2\tilde{\varphi}_{N}^\t{\tiny ML}$.
However, one should bear in mind that vanishing of the estimator bias is not sufficient
for the fulfilment of the local unbiasedness conditions \eref{eq:LocUnbiasConstr}, which
importantly also constrain the ``speed'' of change of the estimator mean.
Hence, the ML estimator, being locally unbiased only in the asymptotic $N$ limit, 
can in principle surpass the CRB \eref{eq:FI&CRB_BinPhi} for any finite $N$ and $\varphi_0$.
Yet, this is not the case for the special parameter value $\varphi_{0}\!=\!\frac{\pi}{2}$, which intuitively should be the optimal one 
as it corresponds to the point of the steepest variation of $\mathsf{p}$ with $\varphi$, 
i.e.~$\textrm{argmax}_{\varphi}\!\left|\textrm{d}\mathsf{p}/\textrm{d}\varphi\right|$,
leading to the highest parameter sensitivity. The MSE plots%
\footnote{%
We have chosen to plot the MSE, $\Delta^2\tilde{\varphi}_{N}$, that adequately quantifies the local performance
of any estimator, but one should be aware that the estimator variance  $\textrm{Var}\!\left[\tilde{\varphi}_{N}\right]$
(for which \figref{fig:MSEMLE_ClEst_BinPhi} in the case of $\tilde{\varphi}_{N}^\t{\tiny ML}$ changes insignificantly) 
is the more experimentally relevant quantity, being
determinable basing only on a large data sample collected for an unknown $\varphi$. 
}
in \figref{fig:MSEMLE_ClEst_BinPhi}
confirm these facts by indicating that for $\varphi_{0}\!=\!\frac{\pi}{2}$,
represented by the saddle points in \figref{fig:MSEMLE_ClEst_BinPhi}(\textbf{a}), $\Delta^2\tilde{\varphi}_{N}^\t{\tiny ML}$ 
never surpasses and most rapidly attains the CRB---up to 5\% already at $N\!\approx\!50$,
as shown in \figref{fig:MSEMLE_ClEst_BinPhi}(\textbf{b}).
On the other hand, for the pathological values $\varphi_{0}\!=\!\left\{0,\pm \pi\right\} $ at which
CRB is not valid even asymptotically, $\tilde{\varphi}_{N}^\t{\tiny ML}$
is errorless with $\Delta^2\tilde{\varphi}_{N}^\t{\tiny ML}\!=\!0$
for any $N$. Furthermore, for low $N$ and the parameter values 
close to these pathological points, for which the ML estimator 
significantly violates the local unbiasedness constraints \eref{eq:LocUnbiasConstr},
$\Delta^2\tilde{\varphi}_{N}^\t{\tiny ML}$ indeed surpasses the CRB.
As a result, the curve in \figref{fig:MSEMLE_ClEst_BinPhi}(\textbf{b})
representing relative percentage excess over the CRB as a function of
$N$ for $\varphi_{0}\!=\!\frac{\pi}{8}$ starts negative below the
horizontal line representing the CRB. Nevertheless, the sub-CRB regions
lying below the CRB-threshold in \figref{fig:MSEMLE_ClEst_BinPhi}(\textbf{a})
narrow down as the bias evanesces with $N\!\rightarrow\!\infty$ and,
as required, $\Delta^2\tilde{\varphi}_{N}^\t{\tiny ML}$
attains asymptotically the CRB from above for any $\varphi_{0}\!\ne\!\{0,\pm\pi\}$.

Lastly, we seek for the optimal form of a transformed parameter, i.e.~$g(\varphi)$ in \eqnref{eq:CRBFunPar}, such that a global
efficient estimator exists. Rewriting the necessary condition \eref{eq:CRBSatCondExpFamFunPar}
applicable to exponential PDFs \eref{eq:PDFExpFam}, we deduce that
\begin{equation}
\frac{\dot{A}(\varphi)}{\dot{\eta}(\varphi)}=\cos^{2}\!\frac{\varphi}{2}=\mathsf{p},
\end{equation}
so that the necessary parameter to be efficiently estimated corresponds simply to
the probability of each binary outcome:~$\mathsf{p}$. The corresponding 
transformed CRB \eref{eq:CRBFunPar} limiting the MSE and variance of 
any unbiased estimator of $\mathsf{p}$ then reads
\begin{equation}
\Delta^{2}\tilde{\mathsf{p}}_{N}\ge\frac{\mathsf{p}(1-\mathsf{p})}{N}
\label{eq:CRBp}
\end{equation}
and is assured to be saturated for any $\mathsf{p}$ by the global estimator $\tilde{\mathsf{p}}_{N}(k)\!=\! k/N$, 
which thus coincides with the ML estimator
\eref{eq:MLE} that is now efficient \emph{regardless of} $N$
and not only in the asymptotic $N$ limit. In contrast to the CRB \eref{eq:FI&CRB_BinPhi}
on $\Delta^2\tilde{\varphi}_{N}$, the PDF \eref{eq:PDF_BinPhi} fulfils the regularity 
condition \eref{eq:LocRegCond} with respect to $\mathsf{p}$ for the previously 
pathological values $\mathsf{p}\!=\!\{0,1\}$, so that the transformed CRB \eref{eq:CRBp} 
still applies to $\Delta^{2}\tilde{\mathsf{p}}_{N}$ at these 
extremal points, where it actually vanishes---correctly indicating the possibility of perfect parameter determination.
However, one should note that although the transformed CRB \eref{eq:CRBp}
corresponds now to a tight inequality, it is parameter dependent.
Hence, as discussed in \secref{sub:LocConseq}, in any real-life
estimation protocol in order to actually compute the CRB, one requires
some prior knowledge of the estimated binary probability $\mathsf{p}$ that
is assured again only in the local estimation regime.

\subsubsection{Bayesian approach \label{sub:ClEst_MZInter_SQL_Bay}}

When solving the above estimation problem within
the Bayesian approach, we treat $\varphi$ as a random variable and
thus alter the notation of \eqnref{eq:PDF_BinPhi}, so that
$p^{N}\!(k|\varphi)\!\equiv\! p_{\varphi}^{N}(k)$ represents now the
conditional PDF. As the estimated parameter describes the phase delay of the
Mach-Zehnder interferometer depicted in \figref{fig:MZinter_cl},
we must account for its circular symmetry and therefore
minimise the average cost \eref{eq:AvCost_H}, $\left\langle \mathcal{C}_{\textrm{H}}(\tilde{\varphi}_{N})\right\rangle $,
as the figure of merit rather than the $\overline{\textrm{MSE}}$
\eref{eq:AvMSE}, $\left\langle \Delta^{2}\tilde{\varphi}_{N}\right\rangle $.
Furthermore, as we ideally do \emph{not assume any prior knowledge} about the estimated
phase, we firstly take the prior PDF to correspond to a uniform distribution
over the full parameter period, i.e.~$p(\varphi)\!=\!1/(2\pi)$
for $\varphi\!\in\![-\pi,\pi]$. However, in such a case, when we
compute the posterior PDF $p^{N}\!(\varphi|k)$ with help of \eqnref{eq:PostPDF}
after substituting for the binomial PDF \eref{eq:PDF_BinPhi}, we find
the necessary condition \eref{eq:H_Est} for the optimal estimator
to be trivially satisfied by $\tilde{\varphi}_{N}^\t{\tiny H}(k)\!=\!0$
which completely disregards the data collected:~$k$. Because of choosing
the whole parameter range $\varphi\!\in\![-\pi,\pi]$, we are again
not able to resolve the sign ambiguity $\pm\varphi$, which previously
lead to two distinct solutions of the ML estimator \eref{eq:MLE_BinPhi}
within the local approach. Within the Bayesian framework the consequences
are even more serious, as it is more beneficial to always output $\varphi\!=\!0$, which is
the only non-ambiguous parameter value, rather than infer from the
data any other realisation of $\varphi$ that, due to its equally probable counterpart
with opposite sign, introduces on average an overwhelming error. Hence, similarly
to the frequentist solution, for the estimation problem not to be ``ill-defined", 
we must possess extra information that allows us to restrict 
the parameter search to a sub-range in
which $\varphi$ is unambiguously determined by the outcomes, i.e.~in
which $\mathsf{p}\!=\!\cos^{2}\frac{\varphi}{2}$ is monotonic in
$\varphi$. 

\begin{figure}[!t]
\includegraphics[width=1\columnwidth]{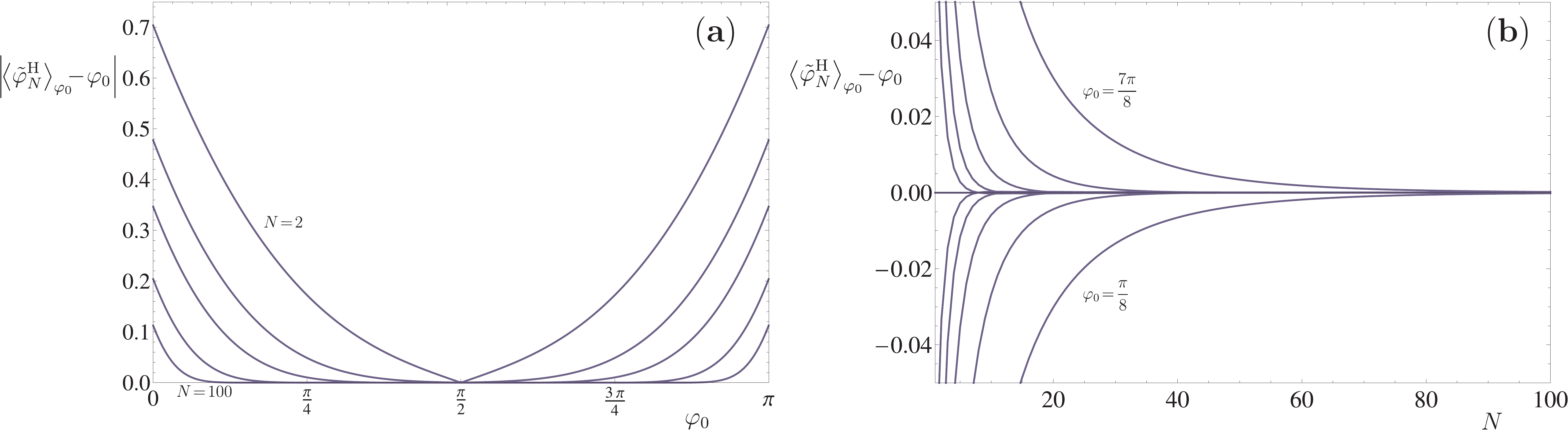}
\caption[Bias of the Bayesian estimator]{%
\textbf{Bias of the Bayesian estimator} \eref{eq:HE_BinPhi} as a function of:~(\textbf{a}) 
-- the parameter true value $\varphi_{0}$,
and (\textbf{b}) -- the sample size $N$.
The curves in (\textbf{a}) are depicted for $N\!=\!\!\left\{ 2,5,10,30,100\right\} $, whereas
the ones in (\textbf{b}) correspond to the equally-distributed values:
$\varphi_{0}\!=\!\{\frac{\pi}{8},\dots,\frac{\pi}{2},\dots,\frac{7\pi}{8}\}$.
Both plots clearly indicate a rapid decrease of the bias with an increase
in $N$. At the special parameter value $\varphi_{0}\!=\!\frac{\pi}{2}$
the bias vanishes irrespectively of the sample size, what however
does \emph{not} guarantee the Bayesian estimator to be \emph{locally unbiased}
unless the asymptotic $N$ limit is considered.
\label{fig:BiasHE_ClEst_BinPhi}
}
\end{figure}

\begin{figure}[!t]
\includegraphics[width=1\columnwidth]{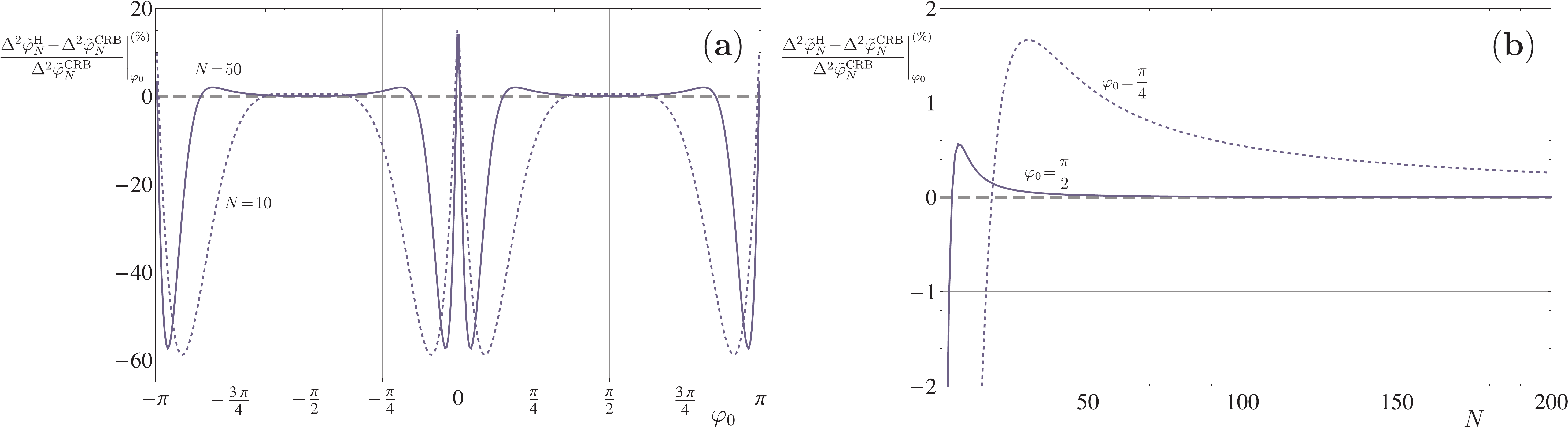}
\caption[MSE of the Bayesian estimator]{%
\textbf{MSE of the Bayesian estimator} \eref{eq:HE_BinPhi}
in comparison with the CRB \eref{eq:FI&CRB_BinPhi} as a function of:
(\textbf{a}) -- the parameter true value $\varphi_{0}$, and (\textbf{b})
-- the sample size $N$. In both plots the CRB is represented by 
the horizontal line (\emph{grey dashed}). The curves in (\textbf{a}) are depicted for $N\!=\!\!\left\{10,50\right\} $
and show the optimal points $\varphi_{0}\!=\!\pm\frac{\pi}{2}$
for which 
the CRB is most rapidly attained. The MSE dependence
on $N$, plotted in (\textbf{b}), indicates that 
$\Delta^2\tilde{\varphi}_{N}^\t{\tiny ML}$ surpasses
the CRB at low $N$ to still attain it asymptotically from above.
For the optimal $\varphi_{0}\!=\!\pm\frac{\pi}{2}$ the CRB
is attained to negligible precision already for $N\!\approx\!50$.
\label{fig:MSEHE_ClEst_BinPhi}}
\end{figure}

That is why, we choose $\varphi\!\in\![0,\pi]$ and the adequate prior
distribution $p(\varphi)\!=\!1/\pi$, so that \eqnref{eq:H_Est} now
yields a non-trivial form of the optimal Bayesian estimator:
\begin{equation}
\tilde{\varphi}_{N}^\t{\tiny H}(k)=\textrm{arccot}\!\left[\frac{\left(k-\frac{N}{2}\right)\Gamma\!\left(k+\frac{1}{2}\right)\Gamma\!\left(N-k+\frac{1}{2}\right)}{k!(N-k)!}\right],
\label{eq:HE_BinPhi}
\end{equation}
where $\Gamma\!\left(x\right)$ stands for the Euler Gamma function. 
Consistently with $\left\langle \mathcal{C}_{\textrm{H}}(\tilde{\varphi}_{N})\right\rangle \!\overset{N\rightarrow\infty}{=}\!\left\langle \Delta^{2}\tilde{\varphi}_{N}\right\rangle$
and discussions of \secref{sub:AvCost},
the estimator \eref{eq:HE_BinPhi} converges in the asymptotic $N$ limit to the ML
estimator \eref{eq:MLE_BinPhi}, i.e.~$\forall_k\!:\tilde{\varphi}_{N}^\t{\tiny H}(k)\!\overset{N\rightarrow\infty}{=}\!\tilde{\varphi}_{N}^\t{\tiny ML}(k)$. 
However, one should note that for any finite $N$ due to averaging over a 
uniform prior distribution the optimal Bayesian estimator---in 
contrast to the ML estimator \eref{eq:MLE_BinPhi}---does not output with certainty the previously pathological values $\varphi\!=\!\{0,\pm\pi\}$
for the extremal outcomes $k\!=\!\{0,N\}$, e.g.~$\tilde{\varphi}_{N}^\t{\tiny H}(N)\!=\!\t{arccot}\!\left\{\sqrt{\pi}\,\Gamma(N+1/2)/[2(N-1)!]\right\}\overset{N\rightarrow\infty}{=}0$
in comparison to $\tilde{\varphi}_{N}^\t{\tiny ML}(N)\!=\!0$ for any $N$. 
Such behaviour is clearly depicted in \figref{fig:BiasHE_ClEst_BinPhi}(\textbf{a}) illustrating the bias
of the Bayesian estimator
to be non-zero at $\varphi\!=\!\{0,\pm\pi\}$, which
leads to a significant MSE shown in \figref{fig:MSEHE_ClEst_BinPhi}(\textbf{a}) at these values.
Note that \figsref{fig:BiasHE_ClEst_BinPhi}{fig:MSEHE_ClEst_BinPhi}
illustrate the local performance of $\tilde{\varphi}_{N}^\t{\tiny H}$ for a given realisation 
$\varphi_0$, so that the MSE \eref{eq:MSE} is indeed the adequate figure of merit.
As a result, \figsref{fig:BiasHE_ClEst_BinPhi}{fig:MSEHE_ClEst_BinPhi}
are directly comparable to \figsref{fig:BiasMLE_ClEst_BinPhi}{fig:MSEMLE_ClEst_BinPhi}
describing the performance of $\tilde{\varphi}_{N}^\t{\tiny ML}$.
Surprisingly, the Bayesian estimator, derived with the global approach in mind,
turns out to be more effective than the ML estimator even from the local perspective.
In particular, its bias diminishes much more rapidly with the sample size $N$ and its MSE attains the 
CRB \eref{eq:FI&CRB_BinPhi} much faster. This can be seen when comparing the sub-plots (\textbf{b}) 
of \figsref{fig:BiasHE_ClEst_BinPhi}{fig:MSEHE_ClEst_BinPhi} with respectively 
\figsref{fig:BiasMLE_ClEst_BinPhi}{fig:MSEMLE_ClEst_BinPhi} after noticing
a significant change in the corresponding plot scales.
Again, as shown in \figref{fig:MSEHE_ClEst_BinPhi}, the optimal parameter value for which the Bayesian estimator $\tilde{\varphi}_{N}^\t{\tiny H}$ most 
quickly saturates the CRB is $\varphi_0\!=\!\frac{\pi}{2}$. Yet|in contrast 
to $\tilde{\varphi}_{N}^\t{\tiny ML}$|$\tilde{\varphi}_{N}^\t{\tiny H}$ at 
this special point surpasses the CRB at low $N$, so that
the corresponding curve in \figref{fig:MSEHE_ClEst_BinPhi}(\textbf{b}) depicting
the relative percentage excess over the CRB starts below the
horizontal line representing the CRB. This is however consistent,
as the Bayesian estimator, similarly to the ML estimator, only asymptotically 
fulfils the local unbiasedness conditions \eref{eq:LocUnbiasConstr}, so that in principle it may
surpass the CRB \eref{eq:FI&CRB_BinPhi} for any finite $N$ and $\varphi_0$.

Nevertheless, one should not forget that, although \figsref{fig:BiasHE_ClEst_BinPhi}{fig:MSEHE_ClEst_BinPhi}
describe the local performance of the Bayesian estimator \eref{eq:HE_BinPhi} for a known true value $\varphi_0$,
$\tilde{\varphi}_{N}^\t{\tiny H}$ does not neglect the circular symmetry of the problem
being derived basing on the average cost $\left\langle \mathcal{C}_{\textrm{H}}(\tilde{\varphi}_{N})\right\rangle$.
In fact, substituting the optimal Bayesian estimator
$\tilde{\varphi}_{N}^\t{\tiny H}$  into \eqnref{eq:AvCost_H}, we obtain
the expression for the overall average cost:
\begin{equation}
\left\langle \mathcal{C}_{\textrm{H}}(\tilde{\varphi}_{N}^\t{\tiny H})\right\rangle =2\left[1-\frac{1}{\pi\left(N+1\right)}\sum_{k=0}^{N}\sqrt{4+\left(\frac{(N-2k)\,\Gamma(k+\frac{1}{2})\,\Gamma(N-k+\frac{1}{2})}{k!(N-k)!}\right)^{2}}\right],
\label{eq:AvCostH_BinPhi}
\end{equation}
which may then be proved to coincide in the asymptotic $N$ limit 
with the corresponding CRB \eref{eq:FI&CRB_BinPhi}, as $\left\langle \mathcal{C}_{\textrm{H}}(\tilde{\varphi}_{N}^\t{\tiny H})\right\rangle \overset{N\rightarrow\infty}{=}1/N$.
Revisiting the discussions of \secref{sub:AvMSE},
as the average cost \eref{eq:AvCostH_BinPhi} is asymptotically equivalent to the $\overline{\textrm{MSE}}$ \eref{eq:AvMSE},
\eqnref{eq:MinAvMSE_Jensen} relating local and global results applies.
Furthermore, as the corresponding CRB \eref{eq:FI&CRB_BinPhi}, $1/N$, is parameter independent,
\eqnref{eq:MinAvMSE_Jensen} takes its simple form, so that for any
regular prior distribution the average cost must indeed converge to the CRB as $N\!\rightarrow\!\infty$.

For completeness, let us also consider the problem of $\mathsf{p}$-estimation,
which we have shown to allow for a global efficient estimator within the frequentist approach.
As the binary outcome probability $\mathsf{p}$ does not constitute a circular parameter, we may directly
minimise the $\overline{\textrm{MSE}}$ \eref{eq:AvMSE}, $\left\langle \Delta^{2}\tilde{\mathsf{p}}_{N}^\t{\tiny MMSE}\right\rangle$, 
for which the MMSE estimator \eref{eq:MMSE_Est}
is always optimal. Thus, assuming a uniform prior distribution $p(\mathsf{p})\!=\!1$ for $\mathsf{p}\!\in\![0,1]$,
we calculate the posterior PDF defined with help of \eqnref{eq:PostPDF}
after substituting for the binomial PDF \eref{eq:PDF_BinPhi}, $p^{N}_\mathsf{p}\!(k)\!\equiv\!p^{N}\!(k|\mathsf{p})$, to obtain
$p^{N}\!(\mathsf{p}|k)\!=\!(N+1)\,p^{N}\!(k|\mathsf{p})$.
As a result, we may compute the MMSE estimator defined as the mean of $p^{N}\!(\mathsf{p}|k)$:
\begin{equation}
\tilde{\mathsf{p}}_{N}^\t{\tiny MMSE}(k)=\int_{0}^{1}\!\! \t{d}\mathsf{p}\; \mathsf{p}\; p^N\!(\mathsf{p}|k)=\frac{k+1}{N+2},
\label{eq:MMSE_BinP}
\end{equation}
and the corresponding minimal $\overline{\textrm{MSE}}$ \eref{eq:MinAvMSE}:
\begin{equation}
\left\langle \Delta^{2}\tilde{\mathsf{p}}_{N}^\t{\tiny MMSE}\right\rangle \;=\;\int_{0}^{1}\!\! d\mathsf{p}\; p(\mathsf{p})\sum_{k=0}^{N}p^{N}\!(k|\mathsf{p})\,\left(\frac{k+1}{N+2}-\mathsf{p}\right)^{2}\;=\;\frac{1}{6\left(N+2\right)}.
\label{eq:AvMSE_BinP}
\end{equation}
Again, in accordance with discussions of \secref{sub:AvMSE}, the (optimal Bayesian) 
MMSE estimator \eref{eq:MMSE_BinP} converges with $N$ to the global unbiased (and thus the ML) estimator
$\tilde{\mathsf{p}}_{N}^\t{ML}(k)\!=\! k/N$ derived within 
the frequentist approach.
On the other hand, as the corresponding CRB \eref{eq:CRBp} 
is $\mathsf{p}$-dependent, \eqnref{eq:MinAvMSE_Jensen} relating local
and global precision measures assures the minimal $\overline{\textrm{MSE}}$ \eref{eq:AvMSE_BinP} 
to converge to the average version of the CRB \eref{eq:CRBp}, so that
\begin{equation}
\left\langle \Delta^{2}\tilde{\mathsf{p}}_{N}^\t{\tiny MMSE}\right\rangle\;\overset{N\rightarrow\infty}{=}\;\left\langle \left.\Delta^{2}\tilde{\mathsf{p}}_{N}^\t{\tiny ML}\right|_{\mathsf{p}}\right\rangle_{p(\mathsf{p})}\;=\;\int_{0}^{1}\!\! d\mathsf{p}\,\frac{\mathsf{p}(\mathsf{p}-1)}{N}=\frac{1}{6N},
\end{equation}
what is consistent with \eqnref{eq:AvMSE_BinP} 
also yielding 
$\left\langle \Delta^{2}\tilde{\mathsf{p}}_{N}^\t{\tiny MMSE}\right\rangle\!\overset{N\rightarrow\infty}{=}\!1/(6N)$.

Lastly, let us emphasise once more that due to $p_{\varphi/\mathsf{p}}(\mathbf{x})\!=\!\prod_{i=1}^{N}p_{\varphi/\mathsf{p}}(x_{i})$
the asymptotic $N$ limit corresponds to the regime of the infinitely sized, \emph{independently} distributed data.
From the frequentist perspective, this means that the locality of estimation is always guaranteed when $N\!\rightarrow\!\infty$,
whereas within the Bayesian framework one is thus able to invoke the asymptotic connection \eref{eq:MinAvMSE_Jensen} 
with the local results and relate the corresponding estimators and precision measures.
In the case of the Mach-Zehnder interferometer of \figref{fig:MZinter_cl}, 
such condition is fulfilled owing to the assumption that the photons employed are uncorrelated between 
one another and individually measured at the end. As a result, each of them constitutes an \emph{independent 
statistical object} and the infinite sampling regime may be attained just by increasing the photon number. 
Importantly, such an assumption is not valid when considering general quantum systems, 
for which we would like to investigate the positive impact of correlations in 
between the constituent particles. Thus, in the quantum setting we must not only account for the quantum nature 
of the process, but also slightly modify the overall estimation scheme, so that the regime of independently distributed 
data can be clearly identified.


\section{Quantum estimation theory} 
\label{sec:QEst}

\subsection{The parameter estimation problem in the quantum setting}
\label{sub:QEstProb}

\begin{figure}[!t]
\begin{center}
\includegraphics[width=0.9\columnwidth]{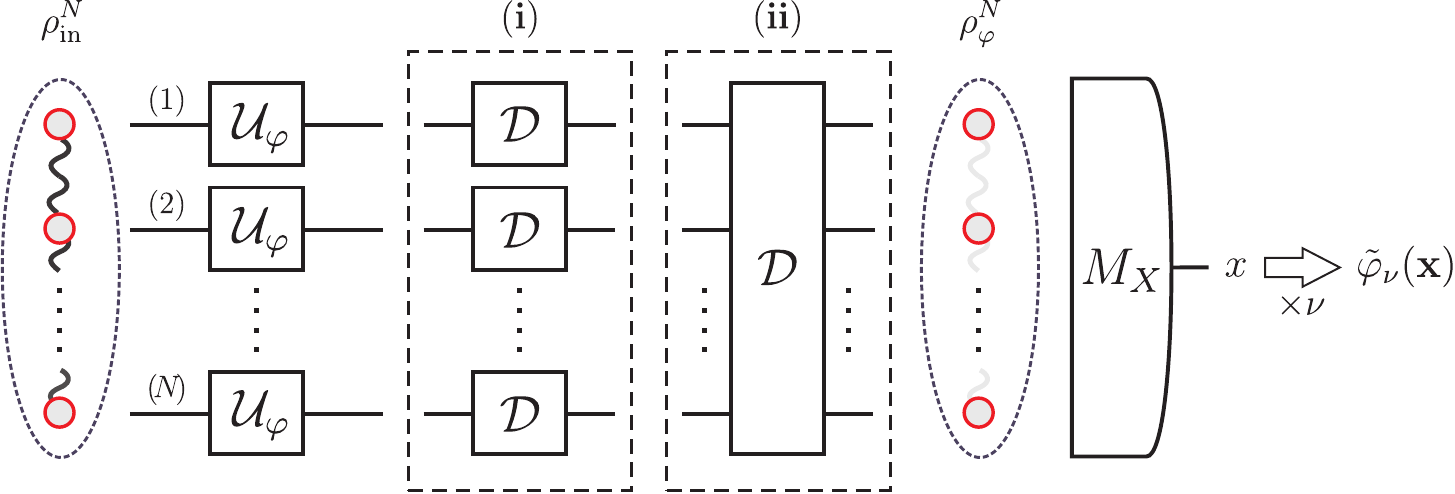}
\end{center}
\caption[Noisy phase estimation scheme in the quantum setting]{%
\textbf{Noisy phase estimation scheme in the quantum setting} \citep{Giovannetti2004,Giovannetti2006}.
The system is prepared in an \emph{input state} $\rho_{\t{in}}^N$ consisting of $N$ particles.
The parameter is then \emph{independently} encoded onto each 
of the constituent particles as a \emph{phase} dictated by the unitary transformation $\mathcal{U}_\varphi$. 
During the process, the system is affected by the \emph{noise} $\mathcal{D}$
that disturbs the particles in an uncorrelated, (\textbf{i}),  or correlated, (\textbf{ii}), fashion. 
Afterwards, a \emph{quantum measurement} $M_X$ is performed on the system \emph{output state} $\rho_{\varphi}^N$
yielding an outcome $x$. If necessary, the whole procedure is \emph{repeated} $\nu$ times,
in order to assure the independent character of the sample collected:
$\mathbf{x}\!=\!\left\{ x_{1},x_{2},\dots,x_{\nu}\right\}$. Finally, an \emph{estimator} $\tilde\varphi_\nu(\mathbf{x})$ is constructed
on the data, which performance may be determined and bounded with use of the classical estimation techniques of \secref{sec:ClEst}.
\label{fig:Ph_Est_Scheme}}
\end{figure}

In the quantum setting, we consider a general estimation scenario 
\citep{Giovannetti2004,Giovannetti2006} depicted in \figref{fig:Ph_Est_Scheme},
which for compatibility with the classical estimation problem of \secref{sub:ClEstProb} is only constrained so 
that the estimated parameter $\varphi$ is encoded \emph{independently} onto each of the $N$ particles 
contained in the initial, \emph{input state} of the system:~$\rho_{\t{in}}^N$ (see \secref{sub:QSboundtates} 
for introduction to the density-matrix description). As a result, $\varphi$ may in general represent 
any \emph{latent} variable discussed in \secref{sec:qmet_latpars} parametrising 
the \emph{single-particle} evolution. Yet,
before considering a general channel estimation scenario, 
we restrict ourselves in \figref{fig:Ph_Est_Scheme} to the 
\emph{phase estimation problem}, in which the parameter specifies the 
``angle" of a \emph{unitary} rotation%
\footnote{%
Following the notation of \chapref{chap:q_sys}, we do not alter the fount when labelling the \emph{operators} acting on quantum wave-functions --
e.g.~$U$, $\mathbb{I}$ and $K$  respectively representing unitary, identity and Kraus operators, 
whereas we utilise the calligraphic fount to denote \emph{super-operators} acting on density matrices --
e.g.~$\mathcal{U}[\bullet]\!=\!U\bullet U^\dagger$, $\mathcal{I}[\bullet]\!=\!\bullet$ and 
$\mathcal{D}[\bullet]\!=\!\sum_i K_i\bullet K_i^\dagger$
now corresponding respectively to unitary, identity super-operators and a general quantum channel.}:
$\mathcal{U}_\varphi[\bullet]\!=\!U_\varphi \bullet U_\varphi^\dagger$ with $U_\varphi\!=\!\e^{-\ii \hat{H}\varphi}$, 
generated by some single-particle Hamiltonian $\hat H$.
Such a choice allows us to clearly explain the mechanism behind the quantum enhancement 
of attainable precision, but the frameworks described in this section apply also
to non-unitary estimation tasks, e.g.~see the decoherence-strength estimation scenarios discussed in \chapref{sec:loc_dec_est}. 
Nevertheless, the scheme of \figref{fig:Ph_Est_Scheme}
importantly accounts for the noise present in the setup, which we
represent by a general quantum channel $\mathcal{D}$ (see \secref{sub:Evo_QCh}) modelling 
the decoherence that may affect the particles in an uncorrelated, (\textbf{i}),  or correlated, (\textbf{ii}), manner.
One should note that such a picture is an oversimplification for quantum 
systems in which the parameter-encoding (here unitary) part of the evolution cannot 
be separated from the noise, i.e.~when they do not
commute with one another \citep{Chaves2013}.
In the case of uncorrelated noise and the frequentist approach, 
however, we show later in \chapref{chap:local_noisy_est}
that such an issue becomes irrelevant after 
utilising the language of $\varphi$-parametrised quantum channels.
Generally for the scenario of \figref{fig:Ph_Est_Scheme}, we may write the final, \emph{output state} of the whole system 
both as $\rho_{\varphi}^N\!=\!\mathcal{D}\!\left[\mathcal{U}_\varphi^{\otimes N}[\rho_{\t{in}}^N]\right]\!=\!\mathcal{U}_\varphi^{\otimes N}\!\left[\mathcal{D}[\rho_{\t{in}}^N]\right]$,
where in the case of uncorrelated noise $\mathcal{D}$ also factorises to $\mathcal{D}^{\otimes N}$.
A quantum measurement $M_{X}$ that is performed on the \emph{whole} system yields in principle a \emph{single}
outcome described by a random variable $X$, which is then distributed according to the PDF 
$p_{\varphi}^N(x)\!=\!\textrm{Tr}\!\left\{ \rho_{\varphi}^N M_{x}\right\}$.
The quantum measurement operators $M_{x}$ correspond to the elements
of a \emph{Positive Operator Valued Measure} (POVM) introduced in \secref{sub:Qmeas}
that are positive semi-definite, $M_{x}\!\geq\!0$, and form a complete basis by satisfying either $\int\! \t{d}\! x\, M_{x}\!=\!\mathbb{I}$ 
or $\sum_x M_{x}\!=\!\mathbb{I}$ for respectively continuous or discrete (and hence more experimentally relevant) set of the outcomes%
\footnote{%
We will use the notation $\sumint\!\t{d}x\dots$ to represent both the integral and the sum over the outcomes,
in order to indicate that the analysis is valid regardless of their continuous or discrete nature.}.
$M_{X}$ is assumed to be \emph{collective} (acting on all the constituent particles) so 
that the description encompasses all the potential measurement scenarios, e.g.~the ones 
in which the particles are measured separately and $M_X\!=\!\bigotimes_{n=1}^N M_{X_n}^{(n)}$, 
but also the \emph{adaptive schemes} in which after performing measurements 
on some fraction of the constituent particles further measurements may be 
adjusted basing on the outcomes already gathered \citep{Wiseman1997,Wiseman1998,Wiseman2009}.

Notice that the quantum character of the process has impact only on
the outcome statistics and plays no role at the data inference stage, 
during which all the techniques developed previously in \secref{sec:ClEst} apply. 
However, if one requires the final data set obtained to be \emph{independently distributed} and thus 
equivalent to the one introduced in \secref{sub:ClEstProb}, the whole estimation procedure 
must be in principle \emph{repeated $\nu$ times}, so that $\mathbf{x}\!=\!\left\{ x_{1},x_{2},\dots,x_{\nu}\right\}$. 
In any case, the procedure is completed as before by constructing an optimal estimator $\tilde{\varphi}_\nu(\mathbf{x})$ 
most accurately predicting the parameter value.
Yet, in general, it is now the repetition number $\nu$ that describes the sample size 
and \emph{not} the particle number $N$. As a consequence,
the asymptotic regime of the frequentist approach corresponds to the $\nu\!\rightarrow\!\infty$ 
limit and it is in principle \emph{not} enough to consider sufficiently large $N$
in order to guarantee the locality of estimation, as in the previous case
of \secref{sub:LocConseq}. Moreover, only in the $\nu\!\rightarrow\!\infty$ limit 
the Bayesian results may be associated with the frequentist ones, as only then the corresponding estimators 
are guaranteed to converge and the minimal $\overline{\t{MSE}}$ \eref{eq:MinAvMSE} may be related to
the CRB \eref{eq:CRB} via \eqnref{eq:MinAvMSE_Jensen}.

However, for the special case of:~\emph{product input states}
$\rho_{\t{in}}^N\!=\!\rho_\t{in}^{\otimes N}$, \emph{uncorrelated noise} $\mathcal{D}^{\otimes N}$, and 
\emph{separable measurements} $M_X\!=\!\bigotimes_{n=1}^N M_{X_n}^{(n)}$; the PDF dictating the outcome statistics factorises with 
$p_{\varphi}^N(x)\!=\!\prod_{n=1}^N p_{\varphi}(x_n)$ 
and $p_{\varphi}(x_n)\!=\!\t{Tr}\!\left\{\mathcal{D}\!\left[\rho_\t{in}\right]M_{x_n}^{(n)}\right\}$, 
so that we recover in ``one shot'' the classical parameter estimation problem of \secref{sub:ClEstProb} with
an $N$-point, independently distributed data set $\mathbf{x}\!=\!\left\{ x_{1},x_{2},\dots,x_{N}\right\}$.
Thus, in such a \emph{classical scenario}%
\footnote{%
In general, the classical setting also allows for \emph{classical correlations} in between the particles and
their measurements, which are not accounted for in the above scenario. Yet, we show later in \secref{sub:QFIproperties}
that these are insufficient to surpass the asymptotic SQL-like scaling of precision with $N$ even for $\nu\!\rightarrow\!\infty$, 
so that for simplicity we assume their absence in any scenario that we name later in this work to be \emph{classical}.
}, 
we may set without loss of generality $\nu\!=\!1$, as it is indeed the particle number 
$N$ that plays the role of the sample size. 
In particular, such an observation explicitly proves that we correctly assumed the estimation problem 
to be classical in the Mach-Zehnder interferometer example of \secref{sub:ClEst_MZInter_SQL}, where
we considered:~the input to be in the Fock state $\ket{N}$ that (as shown in \noteref{note:ent_Fock}) is
a product state of photons, no noise to be present in the setup, and a photon-number--counting measurement
which does not explore any of the photonic correlations.

In conclusion, the \emph{quantum parameter estimation problem} may 
always be divided into its quantum and classical parts. In the quantum part, one must find
the \emph{optimal input states} leading to the highest parameter sensitivity, but also perform 
a non-trivial optimisation over the class of all POVMs to find the \emph{best measurement scheme}.
Meanwhile, the outcomes collected 
must be \emph{efficiently interpreted} via the classical estimation methods to yield the maximal precision,
which then may be compared with the corresponding classical scenario, in order
to quantify the quantum enhancement and, in particular, verify if one is able to surpass
the $1/N$ \emph{SQL-like scaling} with the particle number.  
However, the tools of \emph{quantum estimation theory} often allow to circumvent the
classical part of the estimation procedure. Within the frequentist approach, 
the \emph{Quantum Cram\'{e}r-Rao Bound} dictates fundamental local limits on precision 
that are already optimised over both the quantum measurement strategies and the estimators. 
The quantum Bayesian methods, on the other hand, allow to benefit from the symmetry
of a given estimation problem by utilizing the structure of the so-called
\emph{covariant POVMs}, which are already designed in a way to incorporate the best 
data inference strategy and simplify the procedure of the measurement-scheme optimisation. 
We describe these tools in detail below.

\subsection{Frequentist approach -- \emph{local} estimation of a \emph{deterministic} parameter}
\label{sub:QEstLocal}

\subsubsection{Quantum Cram\'{e}r-Rao Bound and Quantum Fisher Information}
\label{sub:QCRBandQFI}

As shown in the previous section, after specifying the system input state and 
a particular measurement scheme, any quantum estimation problem becomes 
fully classical with the only difference being the potential necessity of procedure 
repetitions to assure the independent character of the data gathered.
Hence, the CRB \eref{eq:CRB} of \secref{sub:ClEstLocal} naturally
applies at this stage and lower-bounds the MSE \eref{eq:MSE} of any locally
unbiased estimator constructed on the outcomes. However, one may always
determine a further, ultimate lower bound, i.e.~the \emph{Quantum Cram\'{e}r-Rao
Bound} (QCRB) \citep{Helstrom1976,Holevo1982}, which importantly is valid 
regardless of the measurement strategy chosen. Following the notation 
of \figref{fig:Ph_Est_Scheme}, it reads:
\begin{equation}
\Delta^{2}\tilde{\varphi}_{\nu}\ge\frac{1}{\nu\, F_{\textrm{Q}}\!\left[\rho_{\varphi}^{N}\right]}\,,\textrm{\qquad where\qquad}F_{\textrm{Q}}\!\left[\rho_{\varphi}^{N}\right]=\textrm{Tr}\!\left\{ \rho_{\varphi}^{N}\!\left.L_\t{\tiny S}[\rho_{\varphi}^{N}]\right.^{2}\right\}
\label{eq:QCRB}
\end{equation}
is the \emph{Quantum Fisher Information} (QFI) that is solely determined
by the dependence of $\rho_{\varphi}^{N}$ on the estimated parameter.
The Hermitian operator $L_\t{\tiny S}$ is the \emph{Symmetric Logarithmic Derivative} (SLD), which can
be unambiguously defined for any state $\varrho_{\varphi}$ via the
relation%
\footnote{%
One may show that \emph{non-symmetric} (or in other words non-Hermitian) logarithmic derivatives satisfying
$\dot{\varrho}_{\varphi}\!=\!\frac{1}{2}\left(\varrho_{\varphi}\tilde{L}_\t{\tiny S}[\varrho_{\varphi}]\!+\! \tilde{L}_\t{\tiny S}[\varrho_{\varphi}]^\dagger\varrho_{\varphi}\right)$
may only lead to weaker bounds, i.e.~$F_{\textrm{Q}}\!\left[\rho_{\varphi}^{N}\right]\!\le\!F_{\tilde{L}}\!\left[\rho_{\varphi}^{N}\right]$
\citep{Holevo1982,Hayashi2005a}.
}
$\dot{\varrho}_{\varphi}\!=\!\frac{1}{2}\left(\varrho_{\varphi}L_\t{\tiny S}[\varrho_{\varphi}]\!+\! L_\t{\tiny S}[\varrho_{\varphi}]\varrho_{\varphi}\right)$, 
so that in the eigenbasis of $\varrho_{\varphi}\!=\!\sum_{i}\lambda_{i}(\varphi)\left|e_{i}(\varphi)\right\rangle \!\left\langle e_{i}(\varphi)\right|$
with $\left\{ \left|e_{i}(\varphi)\right\rangle \right\} _{i}$ forming
a complete basis ($\forall_{i}\!:\,0\!\le\!\lambda_{i}\!\le\!1$):
\begin{equation}
L_\t{\tiny S}[\varrho_{\varphi}]=\sum_{\underset{\lambda_{i}+\lambda_{j}\ne0}{i,j}}\frac{2\left<e_{i}(\varphi)\right|\dot{\varrho}_{\varphi}\left|e_{j}(\varphi)\right>}{\lambda_{i}(\varphi)+\lambda_{j}(\varphi)}\left|e_{i}(\varphi)\right>\!\left<e_{j}(\varphi)\right|.
\label{eq:SLD}
\end{equation}
In other words, for any concrete POVM yielding for $\rho_{\varphi}^{N}$ a PDF $p_{\varphi}^{N}(x)$,
the QFI upper-limits the corresponding classical FI \eref{eq:FI},
so that%
\footnote{%
Notice that in contrast to the classical estimation problem of \secref{sub:ClEstProb}
we may not assume here $p_{\varphi}^{N}(x)$ to be factorisable.
}
$F_{\textrm{cl}}\!\left[p_{\varphi}^{N}\right]\!\le\! F_{\textrm{Q}}\!\left[\rho_{\varphi}^{N}\right]$. 
Most importantly, however, there always exists a measurement strategy 
\citep{Nagaoka1989,Braunstein1994}---a projective measurement in the eigenbasis
of the SLD \eref{eq:SLD}, $L_\t{\tiny S}[\rho_{\varphi}^N]\!=\!\sum_{i}\mu_{i}(\varphi)\left|f_{i}(\varphi)\right\rangle \!\left\langle f_{i}(\varphi)\right|$,
with POVM elements reading $M_{i}(\varphi)\!=\!\left|f_{i}(\varphi)\right\rangle \!\left\langle f_{i}(\varphi)\right|$---for
which the $F_{\textrm{cl}}$ attains the $F_{\textrm{Q}}$. One should
note that such an optimal POVM is in principle not only collective, i.e.~acting 
on \emph{all} the particles, but also, as emphasised by the notation, it
depends on the true value of the estimated parameter, what is consistent 
with the locality assumption of the frequentist approach.
On the other hand, one may thus always interpret the QCRB \eref{eq:QCRB} 
as the classical CRB \eref{eq:CRB} evaluated for such an
optimal measurement scheme. Hence, all the discussions
of \secref{sub:CRBSat} analysing the CRB-saturability issues apply,
meaning that the QCRB is assured to be tight 
only within the regime of local estimation discussed in \secref{sub:LocConseq}, 
which 
may always be guaranteed in the $\nu\!\rightarrow\!\infty$ limit of infinitely many procedure 
repetitions.

Moreover, the QFI may be interpreted similarly to the FI \eref{eq:FI}
as an \emph{information measure} \citep{Barndorff2000}, which now 
is generalised from the space of PDFs to the space of quantum states.
In fact, writing explicitly the form of the QFI according to \eqnref{eq:QCRB} 
for the above introduced $\varrho_{\varphi}$ as
\begin{equation}
F_{\textrm{Q}}\!\left[\varrho_{\varphi}\right]\;=\;\sum_{\underset{\lambda_{i}\ne0}{i}}\frac{\dot{\lambda}_{i}(\varphi)^{2}}{\lambda_{i}(\varphi)}\;+\;2\!\!\sum_{\underset{\lambda_{i}+\lambda_{j}\ne0}{i,j}}\!\!\frac{\left[\lambda_{i}(\varphi)-\lambda_{j}(\varphi)\right]^{2}}{\lambda_{i}(\varphi)+\lambda_{j}(\varphi)}\left|
\braket{\dot{e}_{i}(\varphi)}{e_{j}(\varphi)}\right|^{2},
\label{eq:QFI}
\end{equation}
one realizes that $F_{\textrm{Q}}$ splits into its classical and quantum parts.
The first, ``classical term'' in \eqnref{eq:QFI} represents the
classical FI, $F_{\textrm{cl}}\!\left[\lambda(\varphi)\right]$,
quantifying the information about the parameter encoded in the discrete PDF 
of the eigenvalues $\lambda_i(\varphi)$, whereas the second, 
non-negative term accounts for the quantum nature of the state adding 
a contribution from the rotation of the eigenvectors with the parameter change. 
Consistently, for any classical state with an invariant eigenbasis, $\varrho_{\varphi}^\t{\tiny cl}\!=\!\sum_{i}p_{\varphi,i}\left|e_{i}\right\rangle \!\left\langle e_{i}\right|$,
the QFI \eref{eq:QFI} reproduces the FI, i.e.~the discrete version of \eqnref{eq:FI},
as the ``quantum term'' in \eqnref{eq:QFI} then vanishes and
$F_{\textrm{Q}}\!\left[\varrho_{\varphi}^\t{\tiny cl}\right]\!=\! F_{\textrm{cl}}\!\left[p_{\varphi}\right]$.
On the other hand, in the case of pure quantum states, $\varrho_{\varphi}\!=\!\left|\psi_{\varphi}\right\rangle \!\left\langle \psi_{\varphi}\right|$,
for which the ``classical term'' in \eqnref{eq:QFI} is absent, 
the SLD \eref{eq:SLD} takes a more appealing form:
$L_\t{\tiny S}\!\left[\left|\psi_{\varphi}\right\rangle \right]\!=\!2\left(\left|\dot{\psi}_{\varphi}\right\rangle\!\left\langle\overset{}{\psi_{\varphi}}\right|\!+\!\left|\overset{}{\psi_{\varphi}}\right\rangle\!\left\langle \dot{\psi}_{\varphi}\right|\right)$,
so that the expression for the QFI simplifies dramatically to:%
\footnote{%
For simplicity, we shorten the notation of functions and super-operators
of pure states, so that e.g.~$F\!\left[\left|\psi\right\rangle \right]\equiv F\!\left[\left|\psi\right\rangle \!\left\langle \psi\right|\right]$,
$D(\ket{\psi},\!\ket{\phi})\equiv D(\ket{\psi}\!\bra{\psi},\!\ket{\phi}\!\bra{\phi})$
and $\mathcal{U}\!\left[\left|\psi\right\rangle \right]\equiv\mathcal{U}\!\left[\left|\psi\right\rangle \!\left\langle \psi\right|\right]$.%
}
\begin{equation}
F_{\textrm{Q}}\!\left[\left|\psi_{\varphi}\right\rangle \right]\!=\!4\left(\!\left\langle\!\left.\dot{\psi}_{\varphi}\right|\!\dot{\psi}_{\varphi}\right\rangle \!-\!\left|\left\langle\!\left.\dot{\psi}_{\varphi}\right|\!\psi_{\varphi}\right\rangle \right|^{2}\right).
\label{eq:QFI_pure}
\end{equation}

Similarly to the FI, the QFI is a \emph{local} quantity,
as for a given parameter true value $\varphi_{0}$ it is fully specified
by $\varrho_{\varphi_{0}}$ and $\dot{\varrho}_{\varphi_{0}}$, what
may be verified by inspecting \eqnsref{eq:QCRB}{eq:QFI}.
Thus, any two quantum states $\varrho_{\varphi,1}$ and $\varrho_{\varphi,2}$
are \emph{equivalent} at a given $\varphi_{0}$ from the QFI perspective, as long as they coincide there up to $O(\delta\varphi^{2})$,
so that $\varrho_{\varphi_{0},1}\!=\!\varrho_{\varphi_{0},2}$
and $\dot{\varrho}_{\varphi_{0},1}\!=\!\dot{\varrho}_{\varphi_{0},2}$
yielding $\left.F_{\textrm{Q}}\!\left[\varrho_{\varphi,1}\right]\right|_{\varphi_{0}}\!=\!\left.F_{\textrm{Q}}\!\left[\varrho_{\varphi,2}\right]\right|_{\varphi_{0}}$.
Consequently, the geometrical interpretation of the FI and \eqnref{eq:FIviaStDist} 
may also be naturally carried over into the
quantum picture \citep{Braunstein1994,Barndorff2000,Hayashi2005a}%
\footnote{%
For geometrical approach to quantum estimation and information theory also see \citep{Petz1996,Petz1999,Petz2002,Amari2007}.
}. 
The previously introduced angular distance $D_{\textrm{cl}}$
between PDFs may be directly generalised to quantum states \citep{Wootters1981}
to obtain the so-called \emph{quantum (Bures) angular distance} \citep{Bengtsson2006}:
$D_{\textrm{Q}}(\varrho_{1},\varrho_{2})\!=\!\arccos\left[\mathsf{Fid}_\t{Q}(\varrho_{1},\varrho_{2})\right]$
with $\mathsf{Fid}_\t{Q}(\varrho_{1},\varrho_{2})\!=\!\textrm{Tr}\!\left\{ \!\sqrt{\sqrt{\varrho_{1}}\varrho_{2}\sqrt{\varrho_{1}}}\right\} $
representing now the \emph{quantum fidelity} \citep{Nielsen2000} defined for any two states $\varrho_{1/2}$.
Notice that in the classical case of states sharing a common eigenbasis, i.e.~$\varrho_{1/2}^\t{\tiny cl}\!=\!\sum_{i}p_{1/2,i}\left|e_{i}\right\rangle \!\left\langle e_{i}\right|$,
$D_\t{Q}$ accordingly reproduces the classical angular distance as%
\footnote{%
For pure states, on the other hand, $D_{\textrm{Q}}(\ket{\psi},\!\ket{\phi})\!=\!\arccos\!\left|\braket{\psi}{\phi}\right|$
reproduces the so-called Fubini-Study metric, which is the natural metric used in the
geometric approaches to quantum mechanics \citep{Bengtsson2006}, for instance
widely utilised in study of the \emph{Berry phase} phenomenon \citep{Anandan1990}.
}
$D_\t{Q}(\varrho_{1}^\t{\tiny cl},\varrho_{2}^\t{\tiny cl})\!=\!D_\t{cl}(p_{1},p_{2})$. 
Importantly, expanding the quantum angular distance between 
the neighbouring states $\varrho_{\varphi_{0}}$
and $\varrho_{\varphi_{0}+\delta\varphi}$ for small $\delta\varphi\!\ge\!0$,
we naturally obtain the quantum equivalent of \eqnref{eq:FIviaStDist}:
\begin{equation}
D_{\textrm{Q}}\!\left(\varrho_{\varphi_{0}},\varrho_{\varphi_{0}+\delta\varphi}\right)=\frac{1}{2}\sqrt{\left.F_{\textrm{Q}}\!\left[\varrho_{\varphi}\right]\right|_{\varphi_{0}}}\,\delta\varphi+O\!\left(\delta\varphi^{2}\right),
\label{eq:QFIviaStDist}
\end{equation}
so that the QFI \eref{eq:QFI} may be similarly interpreted as the
the square of the speed, $F_{\textrm{Q}}\!\left[\varrho_{\varphi}\right]\!=\!4\left(\textrm{d}D_{\textrm{Q}}/\textrm{d}\varphi\right)^2$,
with which the quantum state $\varrho_{\varphi}$ is ``moving'' along the path of $\varphi$ for a given parameter value%
\footnote{%
For completeness, let us comment on the consequences if the \emph{quantum relative entropy} \citep{Nielsen2000, Bengtsson2006}
was used instead in \eqnsref{eq:FIviaStDist}{eq:QFIviaStDist} as an (asymmetric) distance measure 
between neighbouring PDFs and quantum states, which is the adequate quantity employed in
the, complementary to estimation, \emph{hypothesis testing} problems \citep{Helstrom1976,Bengtsson2006}.
Importantly, it would also locally reproduce the FI at the classical level, but \emph{not} the QFI in the quantum case \citep{Hayashi2005a}. 
Such a behaviour indicates that one may interpret any classical local-estimation problem as an (asymmetric) discrimination task between two 
infinitely close PDFs differing by $\delta\varphi$, but such an interpretation does \emph{not} 
generalise to quantum states \citep{Hayashi2005a}.
}.

In the case of the phase estimation scheme of \figref{fig:Ph_Est_Scheme}, in which
the parameter-encoding part of the evolution is assumed to be unitary and commuting
with the noise, the output state may be written as $\ensuremath{\rho_{\varphi}^{N}\!=\!\mathcal{U}_{\varphi}^{\otimes N}\!\left[\rho_{0}^{N}\right]}$
with $\rho_{0}^N\!=\!\mathcal{D}[\rho_{\textrm{in}}^{N}]$, $U_{\varphi}^{\otimes N}\!=\!\bigotimes_{n=1}^N\e^{-\ii{\hat H}^{(n)}\varphi}\!=\!\textrm{e}^{-\textrm{i}\hat{H}_{N}\varphi}$ and the $N$-particle Hamiltonian defined as $\hat H_N\!=\!\sum_{n=1}^N \hat H^{(n)}$.
Hence, the expression for the QFI simplifies,
as performing the eigendecomposition of $\rho_{0}^{N}\!=\!\sum_{i}\lambda_{i}\left|e_{i}\right\rangle \!\left\langle e_{i}\right|$
we only obtain the ``quantum term'' in \eqnref{eq:QFI}:
\begin{equation}
F_{\textrm{Q}}\!\left[\rho_{\varphi}^{N}\right]\;=\;2\!\!\sum_{\underset{\lambda_{i}+\lambda_{j}\ne0}{i,j}}\!\!\frac{\left(\lambda_{i}-\lambda_{j}\right)^{2}}{\lambda_{i}+\lambda_{j}}\left|\left<e_{i}\right|\hat{H}_N\left|e_{j}\right>\right|^{2},
\label{eq:QFIscheme}
\end{equation}
which due to the \emph{unitary} parameter-encoding is always parameter independent.
Importantly, in order to maximise the precision of the protocol, we must seek the optimal input states
that maximise the QFI \eref{eq:QFIscheme} and thus minimise the QCRB \eref{eq:QCRB}.
In general, due to the presence of noise, such an optimisation task is non-trivial. 
Yet, when the noise is absent and the input state is assumed
to be pure, so that $\rho_{0}^{N}\!=\!\left|\psi^{N}\right\rangle \!\left\langle \psi^{N}\right|$
and $\rho_{\varphi}^{N}\!=\!\left|\psi_{\varphi}^{N}\right\rangle \!\left\langle \psi_{\varphi}^{N}\right|$
with $\left|\psi_{\varphi}^{N}\right\rangle \!=\!\textrm{e}^{-\textrm{i}\hat{H}_N\varphi}\!\left|\psi^{N}\right\rangle $,
\eqnref{eq:QFIscheme} simplifies further and the QFI
becomes proportional to the \emph{variance of the Hamiltonian} considered:
\begin{equation}
F_{\textrm{Q}}\!\left[\left|\psi_{\varphi}^{N}\right\rangle \right]\;=\;4\,\Delta^{2}\hat{H}_N\;=\;4\left(\left\langle \psi^{N}\right|\hat{H}_N^{2}\left|\psi^{N}\right\rangle -\left\langle \psi^{N}\right|\hat{H}_N\left|\psi^{N}\right\rangle ^{2}\right).
\label{eq:QFIPureVarH}
\end{equation}
As a result, one may directly see that the optimal input state
maximising the variance of $\hat H_N$, and hence the QFI \eref{eq:QFIPureVarH}, is an equally weighted superposition
of the eigenvectors corresponding to the Hamiltonian minimal and maximal eigenvalues, 
i.e.~${\ket{\psi_\t{in}^N}}_\t{\tiny opt}\!=\!\frac{1}{\sqrt 2}\!\left(\ket{\mu_\t{min}^N}+\e^{\ii\phi}\ket{\mu_\t{max}^N}\right)$
with arbitrary $\phi$, where $\hat H_N\ket{\mu^N}\!=\!\mu^N\ket{\mu^N}$, for which the maximal QFI thus reads
$F_\t{Q}\!\left[\textrm{e}^{-\textrm{i}\hat{H}_N\varphi}\,{\ket{\psi_\t{in}^N}}_\t{\tiny opt}\!\right]\!=\!\left(\mu_\t{max}^N\!-\!\mu_\t{min}^N\right)^2$ \citep{Giovannetti2006}.

Furthermore, one should also note that for the QFI \eref{eq:QFIPureVarH} and a single repetition $\nu\!=\!1$, 
the QCRB \eref{eq:QCRB} takes an appealing form of a \emph{``time-energy''--like uncertainty
relation} for the latent parameter (see \noteref{note:time_energy}):
\begin{equation}
\Delta^{2}\hat{H}_N\;\Delta^{2}{\tilde\varphi}\;\ge\;\frac{1}{4},
\label{eq:Phase_Et_UR}
\end{equation}
which importantly is \emph{not} inferred from a quantum observable with 
help of the error-propagation formula \eref{eq:ErrPropForm}, but rather directly
determined by the parametrisation of the quantum state $\varrho_\varphi$ that dictates 
the local estimation capabilities,  i.e.~the speed at which the system is ``moving'' with 
variations in $\varphi$.

\subsubsection{Purification-based definitions of the QFI} 
\label{sub:QFIPurifDefs}

As indicated by \eqnref{eq:QFI}, for mixed states the computation
of the QFI in principle involves diagonalisation of the density matrix,
what in the case of a general output state $\rho_{\varphi}^{N}$ of
\figref{fig:Ph_Est_Scheme} may be already infeasible for moderate $N$ due to the
dimension of the system Hilbert space, $\mathcal{H}_\t{\tiny S}$, growing exponentially with the number of particles. However,
in order to potentially circumvent this problem, alternative definitions
of the QFI were proposed that do not require the eigen-decomposition,
but are specified at the level of state purifications, i.e.~any $\left|\tilde{\Psi}_{\varphi}\right\rangle$
such that $\varrho_{\varphi}\!=\!\textrm{Tr}_\t{\tiny E}\!\left\{ \left|\tilde{\Psi}_{\varphi}\right\rangle \!\left\langle \tilde{\Psi}_{\varphi}\right|\right\}$
with E denoting the ancillary part (environment) of the extended system Hilbert space, $\mathcal{H}_\t{\tiny S}\!\otimes\!\mathcal{H}_\t{\tiny E}$, required for the purification. 
Interestingly, in \citep{Escher2011} the QFI of any $\varrho_{\varphi}$ has been
proved to be equal to the smallest QFI of its purifications (see also \appref{chap:appEquivPurif} for an alternative proof):
\begin{equation}
F_{\textrm{Q}}\!\left[\varrho_{\varphi}\right]=\min_{\tilde{\Psi}_{\varphi}}F_{\textrm{Q}}\!\left[\left|\tilde{\Psi}_{\varphi}\right\rangle \right]=4\min_{\tilde{\Psi}_{\varphi}}\left\{ \left\langle\!\left.\dot{\tilde{\Psi}}_{\varphi}\,\right|\dot{\tilde{\Psi}}_{\varphi}\right\rangle -\left|\left\langle \tilde{\Psi}_{\varphi}\left|\,\dot{\tilde{\Psi}}_{\varphi}\right.\!\right\rangle \right|^{2}\right\},
\label{eq:QFIPurifEscher}
\end{equation}
whereas in \citep{Fujiwara2008} another purification-based
QFI definition has been constructed:
\begin{equation}
F_{\textrm{Q}}\!\left[\varrho_{\varphi}\right]=4\min_{\tilde{\Psi}_{\varphi}}\left\langle\!\left.\dot{\tilde{\Psi}}_{\varphi}\,\right|\dot{\tilde{\Psi}}_{\varphi}\right\rangle.
\label{eq:QFIPurifFuji}
\end{equation}
Despite the apparent difference, the above definitions
are indeed equivalent, as one may prove (see \appref{chap:appEquivPurif}) that 
any purification minimizing one of them is likewise optimal for the other 
and satisfies the condition 
$\left|\dot{\tilde{\Psi}}_{\varphi}^\t{\tiny opt}\right\rangle\!=\!\frac{1}{2}L_\t{\tiny S}\!\left[\varrho_{\varphi}\right]\otimes\mathbb{I}^\t{\tiny E}\left|\tilde{\Psi}_{\varphi}^\t{\tiny opt}\right\rangle$
causing the second term in \eqnref{eq:QFIPurifEscher} to vanish.
The minimisation over all purifications, $\min_{\tilde{\Psi}_{\varphi}}$, may at first sight incorrectly seem as an abstract
and thus a non-computable procedure. However, owing to the local character of the QFI, 
also the minimised terms on the r.h.s.~of \eqnsref{eq:QFIPurifEscher}{eq:QFIPurifFuji}
are sensitive only to small---up to $O(\delta\varphi^2)$---variations of $\varphi$ around a given $\varphi_0$.
Consequently, they locally depend only on:~$\ket{\tilde{\Psi}_{\varphi_0}}$ which is fixed due to  
$\varrho_{\varphi_0}\!=\!\textrm{Tr}_\t{\tiny E}\!\left\{ \left|\tilde{\Psi}_{\varphi_0}\right\rangle \!\left\langle \tilde{\Psi}_{\varphi_0}\right|\right\}$, and $\ket{\dot{\tilde{\Psi}}_{\varphi_0}}$
which is the only one that may vary between purifications. That is why, 
the optimisation can always be systematically performed starting from any purification valid at $\varphi_0$, 
say $\ket{\Psi_{\varphi}}$%
\footnote{%
In fact, the minimisation at $\varphi_0$ requires only the knowledge of $\ket{\Psi_{\varphi_0}}$ and $\ket{\dot\Psi_{\varphi_0}}$.
}, and searching through purifications that are generated 
by a unitary rotation of the environment subspace such that 
$\left|\tilde{\Psi}_{\varphi}\right\rangle=u_{\varphi}^{\textrm{\tiny E}}\!\left|\Psi_{\varphi}\right\rangle$
with $u_\varphi^\t{\tiny E}\!=\!\textrm{e}^{-\textrm{i}\hat{h}_\t{\tiny E}(\varphi-\varphi_{0})}$,
which cover all the necessary potential shifts of the first derivative at $\varphi_0$, as then:
$\left|\tilde{\Psi}_{\varphi_{0}}\right\rangle \!=\!\left|\Psi_{\varphi_{0}}\right\rangle$ and
$\left|\dot{\tilde{\Psi}}_{\varphi_{0}}\right\rangle \!=\!\left|\dot{\Psi}_{\varphi_{0}}\right\rangle-\ii\hat{h}_\t{\tiny E}\!\left|\Psi_{\varphi_{0}}\right\rangle$.
Crucially, the minimisation over purifications at a given $\varphi_0$ is thus equivalent 
to the optimisation over all Hermitian generators $\hat{h}_\t{\tiny E}$, 
which importantly are of dimension equal to the rank of the density matrix $\varrho_{\varphi_0}$ 
and not its size, what leaves room for potential numerical efficiency.
\clearpage
\begin{note}[Purification-based QFI of a pure state when estimating phase]
As an example, let us consider the simplest example of the phase estimation scheme
of \figref{fig:Ph_Est_Scheme} with the noise absent and a pure input state,
for which $\rho_{\varphi}^{N}\!=\!\left|\psi_{\varphi}^{N}\right\rangle \!\left\langle \psi_{\varphi}^{N}\right|$
with $\left|\psi_{\varphi}^{N}\right\rangle \!=\!\textrm{e}^{-\textrm{i}\hat{H}\varphi}\!\left|\psi^{N}\right\rangle $.
As the output state is pure---rank-one---all the relevant purifications
are generated by just varying the phase of $\left|\psi_{\varphi}^{N}\right\rangle$, which from
the point of view of the state---but \emph{not} its derivative---is
irrelevant, i.e.~$\left|\tilde{\psi}_{\varphi}^{N}\right\rangle \!=\!\textrm{e}^{-\textrm{i}h_{\textrm{\tiny E}}(\varphi-\varphi_{0})}\!\left|\psi_{\varphi}^{N}\right\rangle $
with $h_{\textrm{\tiny E}}$ now being an arbitrary scalar variable. \eqnsref{eq:QFIPurifEscher}{eq:QFIPurifFuji}
then respectively read 
\begin{equation}
4\min_{h_{\textrm{\tiny E}}}\left\{\!\left\langle\psi_{\varphi}^{N}\right|\!\left(\hat{H}+h_{\textrm{\tiny E}}\right)^{2}\!\left|\psi_{\varphi}^{N}\right\rangle - \left|\left\langle \psi_{\varphi}^{N}\right|\!\left(\hat{H}+h_{\textrm{\tiny E}}\right)\!\left|\psi_{\varphi}^{N}\right\rangle \right|^{2}\right\}\!\quad\textrm{and}\quad4\min_{h_{\textrm{\tiny E}}}\left\{\!\left\langle \psi_{\varphi}^{N}\right|\!\left(\hat{H}+h_{\textrm{\tiny E}}\right)^{2}\!\left|\psi_{\varphi}^{N}\right\rangle\!\right\},
\label{eq:PurifDefsUnitary}
\end{equation}
and consistently simplify to $F_{\textrm{Q}}\!\left[\left|\psi_{\varphi}^{N}\right\rangle \right]\!=\!4\Delta^{2}\hat{H}$
of \eqnref{eq:QFIPureVarH} for the optimal $h_{\textrm{\tiny E}}^\t{\tiny opt}\!=\!-\left\langle \psi_{\varphi}^{N}\right|\hat{H}\left|\psi_{\varphi}^{N}\right\rangle $.
Moreover, $h_{\textrm{\tiny E}}^\t{\tiny opt}$ can be equivalently determined 
by solving the adequate necessary condition $\left|\dot{\tilde{\psi}}_{\varphi}^{N}\right\rangle \!=\!\frac{1}{2}L_\t{\tiny S}\!\left[\left|\psi_{\varphi}^{N}\right\rangle \right]\!\left|\tilde{\psi}_{\varphi}^{N}\right\rangle $
with the SLD \eref{eq:SLD} taking a simpler form:~$L_\t{\tiny S}\!\left[\left|\psi_{\varphi}^{N}\right\rangle \right]\!=\!2\textrm{i}\!\left[\left|\psi_{\varphi}^{N}\right\rangle \!\left\langle \psi_{\varphi}^{N}\right|\!,\hat{H}\right]$, for the unitary encoding.
Although such an example may seem trivial as $h_{\textrm{\tiny E}}$ does
not constitute an operator, it gives the correct intuition about
the role of the optimal-purification generator. $h_{\textrm{\tiny E}}^\t{\tiny opt}$
produces a counter-rotation of the state phase, which may be interpreted as
an erasure operation that minimises the information about the parameter
encoded in the arbitrary phase---more generally the environment. 
We elaborate more on this issue in \secref{sec:ChEstLocal},
where we apply the purifications-based QFI definitions to quantum channels.
\end{note}

On the other hand, 
we may diminish the number of free variables one has to minimise over,
when utilising the purification-based QFI definitions,
by restricting to some subclass of e.g.~physically motivated generators
$\hat{h}_\t{\tiny E}$. As a result, \eqnsref{eq:QFIPurifEscher}{eq:QFIPurifFuji} 
may also serve as an effective tool for establishing upper bounds on $F_\t{Q}\!\left[\varrho_{\varphi}\right]$.
Although the definition \eref{eq:QFIPurifFuji} can never provide for some 
suboptimal $\left|\tilde{\Psi}_{\varphi}\right\rangle$ 
a tighter upper bound on the QFI than \eqnref{eq:QFIPurifEscher},
it allows for more agility when applied to channel-estimation scenarios discussed in
\chapref{chap:local_noisy_est} \citep{Fujiwara2008}. That is why,
we utilise \eqnref{eq:QFIPurifFuji} explicitly within this work, 
in particular, when reformulating later the relevant purification-minimisation procedures 
into \emph{semi-definite programs} (SDPs).

\subsubsection{Key properties of the QFI and their consequences}
\label{sub:QFIproperties}

We discuss the key properties of the QFI \eref{eq:QFI} that play an important
role when studying the ultimate bounds on precision dictated by the QCRB \eref{eq:QCRB}
and, in particular, when analysing the impact of
the correlations (see \secsref{sub:ent}{sub:sys_indist_part}) in between the constituent particles of the system 
on the potentially achievable precision-scaling with the particle number $N$ .

\paragraph{Non-negativity and additivity}~\\
As shown already in \secref{sub:QCRBandQFI}, the QFI may be
treated as not only an upper bound on the FI \eref{eq:FI}, but also
as the FI itself corresponding to the optimal measurement strategy.
Such fact naturally makes the QFI an information measure \citep{Barndorff2000}
that must be \emph{non-negative}, what is indeed assured by both $\varrho_{\varphi}$
and $\left.L_\t{\tiny S}[\varrho_{\varphi}]\right.^{2}$ in 
\eqnref{eq:QCRB} being positive semi-definite matrices. Most importantly, the
QFI also generalises the notion of \emph{additivity} of the FI from
PDFs to density matrices, what may be verified by considering a general
bipartite, parameter-dependent, product state $\varrho_{\varphi}^\t{\tiny AB}\!=\!\varrho_{\varphi}^\t{\tiny A}\otimes\varrho_{\varphi}^\t{\tiny B}$,
for which then $F_{\textrm{Q}}\!\left[\varrho_{\varphi}^\t{\tiny A}\otimes\varrho_{\varphi}^\t{\tiny B}\right]\!=\! F_{\textrm{Q}}\!\left[\varrho_{\varphi}^\t{\tiny A}\right]\!+\! F_{\textrm{Q}}\!\left[\varrho_{\varphi}^\t{\tiny B}\right]$.
Hence, in particular, for tensor product states $F_{\textrm{Q}}\!\left[\varrho_{\varphi}^{\otimes m}\right]\!=\! m\, F_{\textrm{Q}}\!\left[\varrho_{\varphi}\right]$,
indicating that by having access to identical copies of a given
state we may at most observe a linear precision improvement of the
QCRB leading to the \emph{SQL-like scaling}.
Furthermore, this shows that, at the level of the frequentist
bound, the scenario of possessing $m$ uncorrelated copies of a system
is fully equivalent to the protocol with $\nu\!=\!m$ estimation-procedure
repetitions accounted for in the scheme of \figref{fig:Ph_Est_Scheme}. 
The additivity property thus \emph{proves} that when estimating
locally or in the asymptotic $m$ limit, i.e.~when the QCRB is guaranteed
to be tight, one may \emph{not} benefit from collective measurements
that utilise all the available copies, as the same precision must also be
achievable performing single-copy measurements that mimic procedure
repetitions. 

As a result, when investigating the phase estimation scheme of \figref{fig:Ph_Est_Scheme}
with particles and noise assumed to be uncorrelated---for which most generally the output 
state reads $\rho_{\varphi}^{N}\!=\!\bigotimes_{n=1}^{N}\rho_{\varphi}^{(n)}$,
so that each particle $\rho_{\varphi}^{(n)}$ may be effectively treated as a separate ``copy''%
\footnote{%
For generality, we allow the particles to be in different states labelled by $(n)$, what encompasses
the most natural situation, when $\rho_{\varphi}^{N}\!=\!\rho_{\varphi}^{\otimes N}$ and we deal with
$N$ identical copies of a particle in a state $\rho_\varphi$.
}%
---the additivity of the QFI 
assures the QCRB to be saturated in the asymptotic $N$ limit without need
of procedure repetitions ($\nu\!=\!1$) and, most notably, without
use of collective measurements performed on all the particles.
However, as soon as we allow for \emph{correlations} in between
the particles of the input state such that the output is still \emph{separable}
but no longer a product state, i.e.~$\rho_{\varphi}^{N}\!=\!\sum_{i}p_{i}\,\bigotimes_{n=1}^{N}\rho_{\varphi,i}^{(n)}$,
we may not assume the optimal measurement POVMs to be particle-separable
unless we again let $\nu\!\rightarrow\!\infty$.
It is so, because only in the limit of infinitely many
repetitions we may rewrite $\left.\rho_{\varphi}^{N}\right.^{\otimes\nu}$ as a tensor product:
\begin{equation}
\left.\rho_{\varphi}^{N}\right.^{\otimes\nu}=\left(\sum_{i}p_{i}\;\bigotimes_{n=1}^{N}\rho_{\varphi,i}^{(n)}\right)^{\otimes\nu}\quad\overset{\nu\rightarrow\infty}{\equiv}\quad\bigotimes_{i}\left(\bigotimes_{n=1}^{N}\rho_{\varphi,i}^{(n)}\right)^{\otimes p_{i}\nu}\!\!,
\label{eq:SepInpInfRep}
\end{equation}
so that the probabilistic nature of $\rho_{\varphi}^{N}$ may be 
ignored by letting the mixing probabilities $p_{i}$ represent the fractions (frequencies) of 
various types (labelled by $i$) of the system states we possess \citep{Giovannetti2006}.
\eqnref{eq:SepInpInfRep} also shows that any \emph{non-entanglement--like correlations}%
\footnote{For a review of non-classical correlations without entanglement see for instance \citep{Streltsov2015}.} 
do \emph{not} allow to surpass the \emph{SQL-like scaling} when $\nu\!\rightarrow\!\infty$,
as due to additivity $F_{\textrm{Q}}\!\left[\!\left.\rho_{\varphi}^{N}\right.^{\otimes\nu}\right]\!\overset{\nu\rightarrow\infty}{=}\!F_{\textrm{Q}}\!\left[\bigotimes_{i}\!\left(\bigotimes_{n=1}^{N}\rho_{\varphi,i}^{(n)}\right)^{\otimes p_{i}\nu}\right]\!=\!\nu\sum_{n=1}^{N}\sum_{i}p_{i}F_{\textrm{Q}}\!\left[\rho_{\varphi,i}^{(n)}\right]\le\nu N\,const$ 
and the QFI must%
\footnote{%
Importantly, $const$ is independent of $N$, as it may always be upper-bounded by $\underset{n}{\max}\!\left\{\sum_{i}p_{i}F_{\textrm{Q}}\!\left[\rho_{\varphi,i}^{(n)}\right]\right\}$.
}
at most scale linearly in $N$.
In fact, such a restrictive conclusion is true irrespectively of the number of repetitions $\nu$,
owing to the convexity property of the QFI discussed in the last paragraph of this section. 

On the other hand, the argument of \eqnref{eq:SepInpInfRep} fails
when $\rho_{\varphi}^{N}$ is \emph{particle-entangled},
what may occur not only when one considers non-separable inputs,
but also when dealing with collective noises capable of entangling the system particles 
during the evolution (e.g.~effectively representing particle interactions).
In such a situation, in order to reach the QCRB, collective measurements
on all the particles are in principle required even in the asymptotic 
$\nu$ limit of many repetitions. Nevertheless, it may turn out that separable 
measurements are still sufficient in particular cases. For instance, it is so for
the optimal scenario of \figref{fig:Ph_Est_Scheme} under the idealistic assumption 
of no noise \citep{Giovannetti2006} (when maximally entangled input states
are optimal), which we discuss in \secref{sub:QEst_MZInter_HL}
with the example of the quantum-enhanced Mach-Zehnder interferometer.

\paragraph{Monotonicity under parameter-independent quantum maps}~\\
One should note that the QFI is invariant under any unitary, $\varphi$-independent
map, so that $\forall_{\mathcal{U}}\!:\, F_{\textrm{Q}}\!\left[\varrho_{\varphi}\right]\!=\! F_{\textrm{Q}}\!\left[\mathcal{U}\!\left[\varrho_{\varphi}\right]\right]$.
This may be explicitly verified by realizing that the adjoint transformation
$\mathcal{U}\!\left[\varrho_{\varphi}\right]\!=\! U\varrho_{\varphi}U^{\dagger}$
of the state results in an inverse mapping of the SLD \eref{eq:SLD}, i.e.~$L_\t{\tiny S}\!\left[U\varrho_{\varphi}U^{\dagger}\right]\!=\! U^{\dagger}L_\t{\tiny S}\!\left[\varrho_{\varphi}\right]U$,
so that the expression \eref{eq:QCRB} for the QFI is indeed unaffected
by any $\mathcal{U}$. On the other hand, such invariance may be intuitively explained
by exploring the fact that any unitary rotation is reversible. Hence, 
it can be always undone and thus also treated as a part of the
measurement stage of the protocol, which is known not to have any influence on
the QFI. Such notion, however, cannot be generalised to non-unitary
quantum maps, i.e.~the CPTP maps introduced in \secref{sub:Evo_QCh}, which generally
are irreversible and therefore should somehow affect the QFI. 
In fact, looking at the space of quantum 
states from the geometric perspective, any \emph{parameter-independent
CPTP map} $\Lambda$ can only diminish the relative distance between
any two states $\varrho_{1/2}$ \citep{Bengtsson2006,Petz1996,Petz1999,Petz2002}, so that $D_{\textrm{Q}}\!\left(\Lambda\!\left[\varrho_{1}\right],\Lambda\!\left[\varrho_{2}\right]\right)\!\le\! D_{\textrm{Q}}\!\left(\varrho_{1},\varrho_{2}\right)$ 
and thus according to \eqnref{eq:QFIviaStDist} the QFI may
also only decrease under the action of $\Lambda$:
\begin{equation}
F_{\textrm{Q}}\!\left[\Lambda\!\left[\varrho_{\varphi}\right]\right]\le F_{\textrm{Q}}\!\left[\varrho_{\varphi}\right].
\label{eq:QFImono}
\end{equation}
On the other hand, the above \emph{monotonicity} property of the
QFI can be straightforwardly proved by utilizing the purification-based
definition \eref{eq:QFIPurifEscher}, as follows 
\begin{eqnarray}
F_{\textrm{Q}}\!\left[\Lambda\left[\varrho_{\varphi}\right]\right] & \!\!= & \!\!F_{\textrm{Q}}\!\left[\textrm{Tr}_{\t{\tiny E}_{\Lambda}}\!\left\{ \mathcal{U}^{\t{\tiny S}\t{\tiny E}_{\Lambda}}\!\left[\varrho_{\varphi}\otimes\left|\xi\right\rangle _{\t{\tiny E}_{\Lambda}}\!\!\left\langle\xi\right|\right]\right\} \right]=F_{\textrm{Q}}\!\left[\textrm{Tr}_{\textrm{E}_{\varrho}\textrm{E}_{\Lambda}}\!\left\{ \mathcal{U}^{\t{\tiny S}\t{\tiny E}_{\Lambda}}\!\otimes\mathcal{I}^{\t{\tiny E}_{\varrho}}\!\left[\left|\Psi_{\varphi}\right\rangle _{\t{\tiny S}\t{\tiny E}_{\varrho}}\!\!\left\langle \Psi_{\varphi}\right|\otimes\left|\xi\right\rangle _{\t{\tiny E}_{\Lambda}}\!\!\left\langle \xi\right|\right]\right\} \right]\nonumber \\
 & \!\!= & \!\!\min_{\tilde{\Psi}^\t{\tiny ext}}\,F_{\textrm{Q}}\!\left[\left|\tilde{\Psi}_{\varphi}^\t{\tiny ext}\right\rangle _{\textrm{S}\textrm{E}_{\varrho}\textrm{E}_{\Lambda}}\right]\quad\le\quad\min_{\tilde{\Psi}}\,F_{\textrm{Q}}\!\left[\left|\tilde{\Psi}_{\varphi}\right\rangle _{\t{\tiny S}\t{\tiny E}_{\varrho}}\right]\,=\,F_{\textrm{Q}}\!\left[\rho_{\varphi}\right].
\label{eq:QFImonoPurif}
\end{eqnarray}
Following the prescription of \secref{sub:QFIPurifDefs}, we have 
firstly purified the map $\Lambda$ via the Stinespring dilation theorem
(see \thmref{thm:stinespring}) and then the input state, 
so that $F_{\textrm{Q}}\!\left[\Lambda\left[\varrho_{\varphi}\right]\right]$
eventually equals the minimal QFI of the `extended' purifications
$\left|\tilde{\Psi}_{\varphi}^\t{\tiny ext}\right\rangle \!=\!\! u_{\varphi}^{\textrm{\tiny E}_{\Lambda}\textrm{\tiny E}_{\varrho}}\!\left|\Psi_{\varphi}^\t{\tiny ext}\right\rangle $
with $\left|\Psi_{\varphi}^\t{\tiny ext}\right\rangle \!=\!\left(U^{\t{\tiny S}\t{\tiny E}_{\Lambda}}\otimes\mathbb{I}^{\t{\tiny E}_{\varrho}}\right)\!\left|\Psi_{\varphi}\right\rangle _{\t{\tiny S}\t{\tiny E}_{\varrho}}\!\left|\xi\right\rangle _{\t{\tiny E}_{\Lambda}}$,
which span the system subspace, $\mathcal{H}_\t{\tiny S}$, but also the ancillary ones
$\mathcal{H}_{\t{\tiny E}_{\Lambda}}$ and $\mathcal{H}_{\t{\tiny E}_{\varrho}}$ introduced due
to purifying the channel and the input respectively. Now, by considering
`extended' purifications that are generated \emph{only} by unitary rotations
in the subspace $\mathcal{H}_{\t{\tiny E}_{\varrho}}$, i.e.
$\left|\breve{\Psi}_{\varphi}^\t{\tiny ext}\right\rangle \!=\! u_{\varphi}^{\textrm{\tiny E}_{\rho}}\!\left|\Psi_{\varphi}^\t{\tiny ext}\right\rangle$,
and realizing that the unitary $U^{\t{\tiny S}\t{\tiny E}_{\Lambda}}$
may then be ignored by the argument from the beginning of the paragraph,
we conclude that such a subclass of the `extended' purifications represents
all the relevant purifications to be considered in the absence of the map
$\Lambda$, as $F_{\textrm{Q}}\!\left[\left|\breve{\Psi}_{\varphi}^\t{\tiny ext}\right\rangle _{\t{\tiny S}\t{\tiny E}_{\varrho}\t{\tiny E}_{\Lambda}}\right]\!=\! F_{\textrm{Q}}\!\left[\left|\tilde{\Psi}_{\varphi}\right\rangle _{\t{\tiny S}\t{\tiny E}_{\varrho}}\right]$
with%
\footnote{%
Notice that by additivity $F_{\textrm{Q}}\!\left[\left|\breve{\Psi}_{\varphi}^\t{\tiny ext}\right\rangle \right]\!=\! F_{\textrm{Q}}\!\left[\left|\tilde{\Psi}_{\varphi}\right\rangle \!\left|\xi\right\rangle \right]\!=\! F_{\textrm{Q}}\!\left[\left|\tilde{\Psi}_{\varphi}\right\rangle \right]\!+\! F_{\textrm{Q}}\!\left[\left|\xi\right\rangle \right]\!=\! F_{\textrm{Q}}\!\left[\left|\tilde{\Psi}_{\varphi}\right\rangle \right]$,
as trivially $F_{\textrm{Q}}\!\left[\left|\xi\right\rangle \right]\!=\!0$
being independent of the estimated parameter.
}
$\left|\tilde{\Psi}_{\varphi}\right\rangle \!=\! u_{\varphi}^{\textrm{\tiny E}_{\rho}}\left|\Psi_{\varphi}\right\rangle$. 
In other words, in order to reverse the action of $\Lambda$, we must impose further
constraints on the `extended' purifications when performing the minimisation
in \eqnref{eq:QFImonoPurif}. As this may only increase the minimum obtained,
the QFI could have been only diminished by the action of $\Lambda$ in the first place, and 
therefore it must correspond to quantity that monotonically decreases under parameter-independent CPTP maps.
\hfill {\color{myred} $\blacksquare$}

\begin{note}[Monotonicity of the QFI under partial trace]
As the \emph{partial trace} operation is an example of a quantum map (see
\noteref{note:partial_CPTP}), \eqnref{eq:QFImono} also proves that 
after tracing-out some parts of a quantum state, e.g.~the part supported by 
$\mathcal{H}_\t{\tiny A}$ of the state $\varrho_{\varphi}^{\t{\tiny AB}}\!\in\!\mathcal{L}\!\left(\mathcal{H}_\t{\tiny A}\!\otimes\!\mathcal{H}_\t{\tiny B}\right)$,
the QFI may only decrease, i.e.
\begin{equation}
F_{\textrm{Q}}\!\left[\textrm{Tr}_\t{\tiny A}\!\left\{ \varrho_{\varphi}^{\t{\tiny AB}}\right\} \right]\le F_{\textrm{Q}}\!\left[\varrho_{\varphi}^{\t{\tiny AB}}\right],\label{eq:QFImonoTrP}
\end{equation}
what is consistent with the natural intuition that by ignoring some part of a given 
system we may only lose information about the estimated parameter.
\end{note}

When investigating the estimation scheme of \figref{fig:Ph_Est_Scheme},
the monotonicity property explicitly proves an intuitive prediction that 
decoherence may only worsen the precision achieved. Owing to the commutativity of
noise with parameter encoding, the map $\mathcal{D}$ may always be assumed to act at the end
of the evolution, so that by monotonicity \eref{eq:QFImono} it may only decrease the QFI, as
$F_\t{Q}\!\left[\mathcal{D}\!\left[\mathcal{U}_\varphi^{\otimes N}[\rho_{\t{in}}^N]\right]\right]\!\le\!
F_\t{Q}\!\left[\mathcal{U}_\varphi^{\otimes N}[\rho_{\t{in}}^N]\right]$.

\paragraph{Convexity in quantum states}~\\
Lastly, let us note that the QFI \eref{eq:QFI} is a \emph{convex}
quantity with respect to density matrices, so that for a given
statistical ensemble, $\left\{ p_{i},\varrho_{\varphi,i}\right\} _{i}$,
of quantum states $\varrho_{\varphi,i}$ with $p_{i}$ representing
their corresponding probabilities, the QFI obeys 
\begin{equation}
F_{\textrm{Q}}\!\left[\sum_{i}p_{i}\,\varrho_{\varphi,i}\right]\le\sum_{i}p_{i}\, F_{\textrm{Q}}\!\left[\varrho_{\varphi,i}\right].
\label{eq:QFIconvex}
\end{equation}
Intuitively, one may interpret the l.h.s.~of \eqnref{eq:QFIconvex}
as describing the situation in which we are estimating $\varphi$
basing on an unknown state randomly chosen from the ensemble, whereas
the r.h.s.~corresponds to the case in which the state is also drawn
from the ensemble but is well known before performing the estimation.
Hence, in the first case the \emph{same} estimation scheme must be
used for all the states, while in the second case we
may utilise \emph{different} schemes depending on the state obtained.
Consistently, as in the second scenario we can perform only better and extract more
information about the parameter, the r.h.s.~of \eqnref{eq:QFIconvex} 
cannot be less than the l.h.s.%
\footnote{%
For completeness, let us also mention a very recent result proving that for
a general state $\varrho_\varphi\!=\!\sum_i\!p_i\varrho_{\varphi,i}$ with
$\varrho_{\varphi,i}\!=\!\ket{\psi_{\varphi,i}}\!\bra{\psi_{\varphi,i}}$ and the unitary encoding 
$\varrho_{\varphi}\!=\!U_\varphi\varrho_0U_\varphi^{\dagger}$,
there always exists a decomposition of $\varrho_0\!=\!\sum_i p_i\ket{\psi_{0,i}}\!\bra{\psi_{0,i}}$, 
i.e.~an ensemble $\{p_i , \ket{\psi_{0,i}}\}$ with non-orthogonal $\ket{\psi_{0,i}}$, for which
the \eqnref{eq:QFIconvex} is tight \citep{Toth2013,Yu2013}.
}. 

On the other hand, in order to prove \eqnref{eq:QFIconvex}
rigorously, it is enough to consider a binary ensemble with two states
$\varrho_{\varphi,1/2}\!\in\!\mathcal{H}$ distributed according to probabilities
$\mathsf{p}$ and $1-\mathsf{p}$ respectively. One may then construct a new state in $\mathcal{H}\!\oplus\!\mathcal{H}$
as a direct sum of $\varrho_{\varphi,1/2}$, i.e.~$\tilde{\varrho}_{\varphi}\!=\mathsf{p}\varrho_{\varphi,1}\oplus(1-\mathsf{p})\varrho_{\varphi,2}$, 
which by the direct-sum structure does not possess any coherences between the two constituent subspaces.
As a result, although the states $\varrho_{\varphi,1/2}$ are still randomly distributed, they may be
perfectly distinguished before performing the estimation procedure,
so that $F_{\textrm{Q}}\!\left[\tilde{\varrho}_{\varphi}\right]\!=\!\mathsf{p}F_{\textrm{Q}}\!\left[\varrho_{\varphi,1}\right]+(1-\mathsf{p})F_{\textrm{Q}}\!\left[\varrho_{\varphi,2}\right]$%
\footnote{%
In general, $\forall_{\varrho,\sigma}\!:\, F_{\textrm{Q}}\!\left[\varrho\oplus\sigma\right]\!=\! F_{\textrm{Q}}\!\left[\varrho\right]+F_{\textrm{Q}}\!\left[\sigma\right]$,
what may be most easily proved by utilizing the QFI-definition \eref{eq:QCRB}.}. 
Now, as there exists a CPTP map $\Lambda$ such that $\Lambda\!\left[\mathsf{p}\varrho_{\varphi,1}\oplus(1-\mathsf{p})\varrho_{\varphi,2}\right]\!=\!\mathsf{p}\varrho_{\varphi,1}+(1-\mathsf{p})\varrho_{\varphi,2}$ for any $\varrho_{\varphi,1/2}$---consider
a quantum channel, \emph{$\mathcal{H}\!\oplus\!\mathcal{H}\!\rightarrow\!\mathcal{H}$},
with two Kraus operators that respectively select the subspaces
$1/2$ of a given state---by monotonicity \eref{eq:QFImono} $F_{\textrm{Q}}\!\left[\mathsf{p}\varrho_{\varphi,1}+(1-\mathsf{p})\varrho_{\varphi,2}\right]\!\le\! F_{\textrm{Q}}\!\left[\tilde{\varrho}_{\varphi}\right]$,
what completes the proof \citep{Fujiwara2001}.\hfill {\color{myred} $\blacksquare$}

Importantly, the convexity \eref{eq:QFIconvex} of the QFI allows
to assure that, when considering the phase estimation scheme of \figref{fig:Ph_Est_Scheme}, the \emph{inter-particle
entanglement} is \emph{necessary} to surpass the SQL-like
scaling in $N$. Assuming as before, when discussing the additivity of the QFI, the most general \emph{separable} output state consisting of $N$ particles:~$\rho_{\varphi}^{N}\!=\!\sum_{i}p_{i}\,\bigotimes_{n=1}^{N}\rho_{\varphi,i}^{(n)}$,
we may always lower-bound its corresponding QCRB \eref{eq:QCRB} for
any $\nu$ by utilizing the convexity property \eref{eq:QFIconvex} as
\begin{equation}
\Delta^{2}\tilde{\varphi}_{\nu}\ge\frac{1}{\nu\, F_{\textrm{Q}}\!\left[\sum_{i}p_{i}\,\bigotimes_{n=1}^{N}\rho_{\varphi,i}^{(n)}\right]}\quad\ge\quad\frac{1}{\nu\,\sum_{i}p_{i}\, F_{\textrm{Q}}\!\left[\,\bigotimes_{n=1}^{N}\rho_{\varphi,i}^{(n)}\right]}=\frac{1}{\nu\,\sum_{n=1}^{N}\sum_{i}p_{i}\, F_{\textrm{Q}}\!\left[\rho_{\varphi,i}^{(n)}\right]}\ge\frac{const}{\nu N},
\label{eq:QFIconvexQCRB_SQL}
\end{equation}
where the constant factor is again not scalable with%
\footnote{%
Being lower-limited by $\left(\underset{n}{\max}\sum_{i}p_{i}F_{\textrm{Q}}\!\left[\rho_{\varphi,i}^{(n)}\right]\right)^{-1}$. 
}
$N$. As a consequence, the QFI may also serve as 
a \emph{witness} \citep{Bengtsson2006,Horodecki2009} of entanglement present in between the particles, as by \eqnref{eq:QFIconvexQCRB_SQL}
$F_\t{Q}\!\left[\rho^N_\varphi\right]\!>\!N$ can occur only for non-separable states $\rho^N_\varphi$ \citep{Pezze2009}, what opens doors for
theoretical applications of the QFI \eref{eq:QFI} as a quantity sensing and potentially quantifying the multi-particle entanglement \citep{Hyllus2012,Toth2012}.

Moreover, for the phase estimation scheme of \figref{fig:Ph_Est_Scheme},
the convexity property explicitly proves that it is optimal to use \emph{pure input states}, 
as \eqnref{eq:QFIconvex} implies that 
given a general $\rho_\t{in}^{N}\!=\!\sum_{i}p_{i}\left|\psi_{i}^N\right\rangle \!\left\langle \psi_{i}^N\right|$
and defining $\Lambda_\varphi^N\!=\!\mathcal{D}\circ\mathcal{U}_\varphi^{\otimes N}\!\!=\!\mathcal{U}_\varphi^{\otimes N}\!\circ\mathcal{D}$:
$F_\t{Q}\!\left[\Lambda_\varphi^N[\rho_{\t{in}}^N]\right]\!=\!
F_\t{Q}\!\left[\sum_{i}p_i\Lambda_\varphi^N[\ket{\psi_i^N}]\right]
\!\le\!\sum_{i}p_i F_\t{Q}\!\left[\Lambda_\varphi^N[\ket{\psi_i^N}]\right]\!\le\!
\underset{i}{\max}\, F_\t{Q}\!\left[\Lambda_\varphi^N[\ket{\psi_i^N}]\right]$.
This agrees with the natural intuition that when estimating a latent parameter encoded by the system evolution
(or equivalently by a 
quantum channel, here $\Lambda_\varphi^N$, as described later in \chapref{chap:local_noisy_est})
and inputting a random state belonging 
to an ensemble of pure states, i.e.~a mixed state, we may perform only worse than if
we had used a perfectly known optimal state from the ensemble.

\subsection{Bayesian approach -- \emph{global} estimation of a \emph{stochastic} parameter} 
\label{sub:QEstGlobal}

Surely, at the classical level of the estimation scheme of \figref{fig:Ph_Est_Scheme} 
with the input fixed and a particular measurement scheme chosen,
nothing prevents us to:~reinterpret the established outcome PDF as
the conditional distribution, i.e.~$p_{\varphi}^{N}\!(x)\!\equiv\! p^{N}\!(x|\varphi)$,
assume some prior distribution of the parameter, $p(\varphi)$, and also
apply the Bayesian techniques described in \secref{sub:ClEstGlobal}.
Yet, if we incorporate the quantum part of the problem into the optimisation
procedure and, in particular, try to minimise the \emph{average cost}
\eref{eq:AvCost} over the choice of input states and quantum measurements, the task
in general turns out to be highly non-trivial. Furthermore, as the
Bayesian approach is designed to accurately account for the progressive
improvement of the knowledge about the estimated parameter with growth
of the sample size (see \secref{sub:AvMSE}), in order to correctly apply it
to the estimation scheme of \figref{fig:Ph_Est_Scheme}, 
we should \emph{update the prior} PDF, $p(\varphi)$, after 
\emph{each} of $\nu$ procedure rounds and repetitively minimise the average cost to
determine the optimal form of both the input state and the POVM $M_{X}$,
which may vary from ``shot to shot'' for the protocol to be most efficient \citep{Demkowicz2011}. 
This contrasts the case of \eqnref{eq:PostPDFprogress}, which in the 
classical setting at the level of PDFs assured the progressive updating scenario 
to be equivalent to a single minimisation of the overall $\overline{\textrm{MSE}}$ \eref{eq:AvMSE} 
evaluated on the whole data.

On the other hand, as the independent character of the sample is \emph{not}
a necessary requirement within the global approach, we may also set $\nu\!=\!1$ and 
consider a \emph{single repetition} of the protocol, in which the measurement
capabilities are unrestricted and one must thus optimise over \emph{all} $M_{X}$ 
satisfying:~$M_{x}\!\ge\!0$, $\sumint\!\textrm{d}x\, M_{x}\!=\!\mathbb{I}$;
and acting potentially on all $N$ particles. Although (as remarked in \secref{sub:QEstProb}) such 
a general ``single-shot'' estimation scheme encompasses all 
classical scenarios of uncorrelated:~particles, noise and measurements; 
the outcomes in principle may not be assumed to be independently distributed, so that
the Bayesian solution cannot be always related to the frequentist 
results in the asymptotic $N$ limit via \eqnref{eq:MinAvMSE_Jensen}. 
Nevertheless, we may greatly benefit from the general form of the POVM, 
especially in the case of the phase estimation scheme of \figref{fig:Ph_Est_Scheme},
in which the parameter is encoded unitarily and
the estimation problem enjoys the \emph{circular symmetry} described in
\secref{sub:AvCostCircSymm}. As a result, if also a complete \emph{lack of
prior knowledge} about the parameter, i.e.~a uniform
prior distribution $p(\varphi)\!=\!1/(2\pi)$, is assumed,
such symmetry is maintained at the level of the average cost \eref{eq:AvCost} and 
the estimation task may always be solved after restricting to
\emph{covariant measurements} that explore the circular nature of
the parameter, but in principle require correlated operations 
on all the particles \citep{Holevo1982}.

Considering a single round of the phase estimation protocol we drop
for simplicity the repetition-subscript of the estimator, so that
$\tilde{\varphi}(x)\!\equiv\!\tilde{\varphi}_{\nu=1}(x)$, and adapt
the circularly symmetric average cost (\ref{eq:AvCost_H}) to the
quantum setting as
\begin{equation}
\left\langle \mathcal{C}_{\textrm{H}}(\tilde{\varphi})\right\rangle =\!\int\!\frac{\textrm{d}\varphi}{2\pi}\sumint\!\textrm{d}x\;\textrm{Tr}\!\left\{ \rho_{\varphi}^{N}\, M_{x}\right\}\, C_{\textrm{H}}\!\left(\tilde{\varphi}(x)-\varphi\right)
\label{eq:QAvCost}
\end{equation}
with the outcome probability reading $p^{N}\!(x|\varphi)\!=\!\textrm{Tr}\!\left\{ \rho_{\varphi}^{N}M_{x}\right\}$
for the output state $\rho_{\varphi}^{N}\!=\!\mathcal{U}_{\varphi}^{\otimes N}\!\left[\rho_{0}^{N}\right]$,
where $\rho_{0}^{N}\!=\!\mathcal{D}\!\left[\rho_{\textrm{in}}^{N}\right]$ as before. 
As shown in \appref{chap:appCovMeas} following \citep{Holevo1982},
any POVM utilised in \eqnref{eq:QAvCost} may always be replaced
without affecting $\left\langle \mathcal{C}_{\textrm{H}}(\tilde{\varphi})\right\rangle $
by a \emph{covariant POVM} of the form
\begin{equation}
M_{\tilde{\varphi}}=U_{\tilde{\varphi}}^{\otimes N}\,\Xi^{N}\,U_{\tilde{\varphi}}^{\dagger\otimes N},
\label{eq:CovPOVM}
\end{equation}
which is parametrised by the estimated values and satisfies $\int\!\!\frac{\textrm{d}\tilde{\varphi}}{2\pi}M_{\tilde{\varphi}}\!=\!\mathbb{I}$,
$M_{\tilde{\varphi}}\!\ge\!0$ with its \emph{seed element} reading
$\Xi^{N}\!=\!\sumint\!\textrm{d}x\, U_{\tilde{\varphi}(x)}^{\dagger\otimes N}M_{x}\,U_{\tilde{\varphi}(x)}^{\otimes N}$
for a particular $M_{X}$ and an estimator $\tilde{\varphi}(x)$.
Importantly, as covariant measurements constitute a subclass of \emph{all}
legal POVMs, the above fact proves that, in order to establish the
\emph{minimal} average cost, we may always make \eqnref{eq:QAvCost} independent of the estimator
chosen by rewriting it with use of $M_{\tilde{\varphi}}$ \eref{eq:CovPOVM} as
\begin{eqnarray}
\left\langle \mathcal{C}_{\textrm{H}}\right\rangle  & = & \iint\!\frac{\textrm{d}\varphi}{2\pi}\frac{\textrm{d}\tilde{\varphi}}{2\pi}\;\textrm{Tr}\!\left\{ \rho_{\varphi}^{N}\, M_{\tilde{\varphi}}\right\} \, C_{\textrm{H}}\!\left(\tilde{\varphi}-\varphi\right) \nonumber \\
 & = & \textrm{Tr}\!\left\{ \left\langle \rho_{\varphi}^{N}\right\rangle _{C_{\textrm{H}}}\Xi^{N}\right\}, 
\label{eq:QAvCostCov}
\end{eqnarray}
and minimise \eqnref{eq:QAvCostCov} over all covariant POVMs, i.e.~over all possible 
seed elements that are \emph{positive
semi-definite}: $\Xi^{N}\!\!\ge\!0$, and satisfy the \emph{completeness 
constraint}:~$\int\!\!\frac{\textrm{d}\varphi}{2\pi}U_\varphi^{\otimes N}\Xi^NU_\varphi^{\dagger \otimes N}\!=\!\mathbb{I}$. 
The second expression
for the average cost $\left\langle \mathcal{C}_{\textrm{H}}\right\rangle$
follows from the invariance of the Haar measure---see \appref{chap:appCovMeas}---with 
respect to which both $\varphi$ and $\tilde{\varphi}$ are integrated.
Crucially, it indicates that the average cost \eref{eq:QAvCostCov}, due to the covariant symmetry, 
may be interpreted as an outcome of a quantum measurement represented
by the seed element, $\Xi^N$, evaluated on an effective state, which is the output, $\rho_{\varphi}^{N}$,
averaged over the cost function $C_{\textrm{H}}$ \eref{eq:CostFun_H}:~%
$\left\langle \rho_{\varphi}^{N}\right\rangle_{C_{\textrm{H}}}\!\!=\!\int\!\frac{\textrm{d}\varphi}{2\pi}\,\rho_{\varphi}^{N}\,4\sin^{2}\!\left(\frac{\varphi}{2}\right)$.

Note that, if a given phase estimation problem allows for a \emph{consistent} estimator---see \eqnref{eq:EstCons}---to exist,
then, as \eqnref{eq:QAvCostCov} is optimised by construction over all estimators, $\tilde\varphi$ 
is bound to approach $\varphi$ with $N$ for the optimal $\Xi^N$. Hence,
the \emph{minimal} average cost \eref{eq:QAvCostCov} converges asymptotically to the $\overline{\t{MSE}}$,
i.e.~$\left\langle\mathcal{C}_{\textrm{H}}\right\rangle\!\!\overset{N\to\infty}{=}\!\!\left\langle\Delta^2\tilde\varphi\right\rangle$,
and the MMSE estimator \eref{eq:MMSE_Est} representing the mean of the posterior distribution---$p(\tilde\varphi|\varphi)$ 
now being continuously parametrised by the outputs of the covariant POVM---is 
always optimal in the $N\!\to\!\infty$ limit.
However, we may not assume the adequate MMSE estimator to converge with $N$ to the ML estimator and
directly relate the minimal $\left\langle\mathcal{C}_{\textrm{H}}\right\rangle$ to the CRB \eref{eq:QCRB} with $\nu\!=\!1$
via \eqnref{eq:MinAvMSE_Jensen}, unless we consider a classical scenario with uncorrelated noise 
for which $\rho_{\varphi}^{N}\!=\!\rho_\varphi^{\otimes N}$ and establish an equivalently-optimal 
measurement scheme that ignores correlations in between the particles.

Summarising, the phase estimation problem described in \figref{fig:Ph_Est_Scheme}
is solved for a \emph{single repetition} within the global approach by minimising \eqnref{eq:QAvCostCov}
over the input states $\rho_{\textrm{in}}^{N}$ and seed elements $\Xi^{N}$.
Let us emphasise that the input which corresponds to the minimum is optimal 
from the Bayesian perspective which assumes \emph{no prior
knowledge} about the parameter. Hence, it may differ dramatically 
from the state maximising the QFI \eref{eq:QFIscheme}, which is optimal within the complementary frequentist approach
designed to indicate the inputs leading to highest sensitivity to small 
parameter fluctuations from a \emph{known} value---see \secref{sub:LocConseq}.
On the other hand, the optimal covariant POVM \eref{eq:CovPOVM} may seem unrealistic 
being continuously parametrised and thus probably not realisable in
a real-life experiment. Whence, in order for the solution to be of practical importance,
one should attempt to find another  POVM with a \emph{finite} number of elements,
i.e.~$\sum_x \!M_{x}\!=\!\mathbb{I}$, which also
attains the minimal cost \eref{eq:QAvCostCov}. Such a construction, however,
has been shown to be typically feasible for finite dimensional systems with the discrete outcomes
of the POVM being then directly associated with particular values of the estimated parameter,
what still circumvents the problem of the estimator optimisation \citep{Derka1998}.

\subsection{\caps{Example:} Mach-Zehnder interferometry at the Heisenberg Limit}
\label{sub:QEst_MZInter_HL}

\begin{figure}[!t]
\begin{center}
\includegraphics[width=0.9\columnwidth]{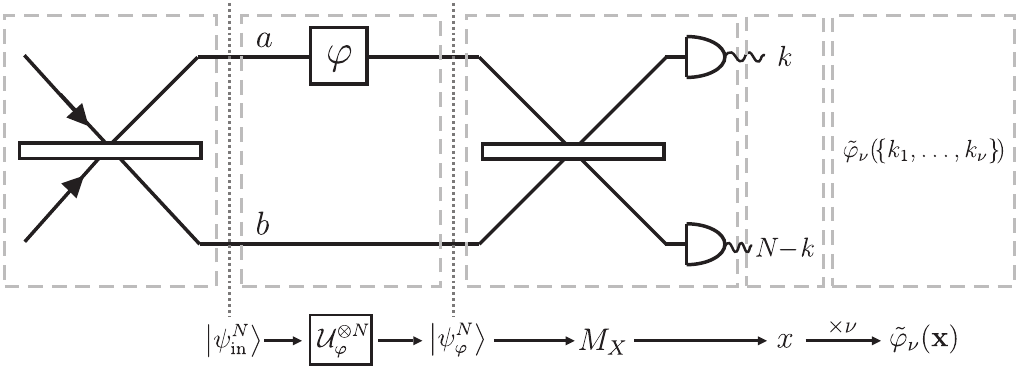}
\end{center}
\caption[Mach-Zehnder interferometry in the quantum setting]{%
\textbf{Mach-Zehnder interferometer} (\figref{fig:MZinter_cl}) \textbf{as an instance of a \emph{noiseless} phase estimation scheme} (\figref{fig:Ph_Est_Scheme}).
The input state $\ket{\psi_\t{in}^N}$ corresponds to the two-mode, $N$-photon state of light after the
first beam-splitter, whereas the output state $\ket{\psi_\varphi^N}$ differs only by the estimated
phase, $\varphi$, accumulated due to the path difference between the arms $a$ and $b$. The photon-counting measurement 
constitutes an example of a general POVM $M_X$, which outcome $x$ corresponds to the photon number recorded in one of the ports:~$k$ 
(other is then fixed to $N\!-\!k$). If necessary, the procedure is repeated $\nu$ times to build an independently 
distributed sample $\mathbf{x}\!=\!\{x_1\!\equiv\!k_1,\dots,x_\nu\!\equiv\!k_\nu\}$, from which the parameter value is finally inferred 
via the estimator $\tilde\varphi_\nu$.}
\label{fig:MZinter_Q}
\end{figure}

We revisit the \emph{Mach-Zehnder interferometer} example of \secref{sub:ClEst_MZInter_SQL}, in
order to generalise the previous discussion and fully account for the quantum aspect of the setup.
Regarding the interferometer of \figref{fig:MZinter_cl} with a definite number of photons, $N$, as a special case of the 
general \emph{phase estimation scheme} of \figref{fig:Ph_Est_Scheme}, we are able to 
directly apply the frequentist and Bayesian frameworks developed in \secsref{sub:QEstLocal}{sub:QEstGlobal}.
Yet, as shown in \figref{fig:MZinter_Q},  in order to exactly match their corresponding notation, one must identify the \emph{input state}
as a two-mode, $N$-photon state (see \eqnref{eq:Nph_state}) of light \emph{inside} the interferometer, i.e.
\begin{equation}
\ket{\psi_\t{in}^N}=\sum_{n=0}^N\, \alpha_n\; {\ket{n}}_a\,{\ket{N-n}}_b=\sum_{n=0}^N\, \alpha_n \; \ket{n,N-n},
\label{eq:Input_Q_MZ}
\end{equation}
where $a$ and $b$ label the interferometer arms---modes---and the general coefficients satisfy $\sum_{n=0}^N\left|\alpha_n\right|^2\!=\!1$.
Although the state \eref{eq:Input_Q_MZ} is written 
in the occupation number representation, i.e.~the second quantisation, in order to study its metrological properties, one should treat its
constituent photons as separate particles, what---see \secref{sub:sys_indist_part}---formally corresponds 
to returning to the first-quantisation picture and treating photons as distinguishable particles in a permutation invariant state \citep{Demkowicz2015}.
For example, the \emph{classical strategy} of  \secref{sub:ClEst_MZInter_SQL}, in which  $N$ uncorrelated photons, 
i.e.~a Fock $\ket{N}$ state of \noteref{note:ent_Fock}, are impinged on a balanced beam-splitter, results in binomially distributed $\alpha_n$-coefficients
in the ``modal picture'':
\begin{equation}
{\ket{\psi_\t{in}^N}}_\t{\tiny cl}=\sum_{n=0}^N\, \sqrt{\frac{1}{2^N}\binom{N}{n}}\; \ket{n,N-n}=\left[\frac{1}{\sqrt{2}}\left(\ket{a}+\ket{b}\right)\right]^{\otimes N}.
\label{eq:Cl_Input_Q_MZ}
\end{equation}
Whereas in the ``particle picture'', as shown in the second expression above, the state inside the interferometer corresponds then to 
a \emph{product state} of $N$ photons each being in an equally weighted superposition of states $\ket{a}$ and $\ket{b}$,
which form a qubit-like (spin-$1/2$--like) basis of each photon and represent it travelling in either of the arms/modes.
Crucially, by considering now a general $\ket{\psi_{\t{in}}^N}$ and varying $\alpha_n$ in 
the quantum setting, we can introduce the \emph{inter-particle entanglement}
that may allow to surpass the $1/N$ SQL-like scaling of precision accounted in \eqnsref{eq:FI&CRB_BinPhi}{eq:AvCostH_BinPhi},
which assessed the performance of the product input state \eref{eq:Cl_Input_Q_MZ} within local and global approaches respectively 
for the classical scenario of \figref{fig:MZinter_cl} with a photon-counting measurement.

As in this section we aim to determine the ultimate quantum-enhancement of precision theoretically attainable
in a Mach-Zehnder interferometer, similarly to the classical case, we assume a noiseless scenario in which
the photons during the evolution only accumulate the estimated phase that---see \figref{fig:MZinter_Q}---is encoded unitarily 
onto each of them via $U_\varphi^{\otimes N}\!=\!\e^{-\ii\hat H_N\varphi}$. $\hat H_N$ is the overall $N$-photon Hamiltonian 
that is decomposable in the ``particle picture'' as $\hat H_N\!=\!\sum_{n=1}^N \hat H^{(n)}$ with 
$\hat H^{(n)}\!=\!\frac{1}{2}\!\left(\ket{a}\!\bra{a}-\ket{b}\!\bra{b}\right)$ acting on the ``$n$-th photon''.
Equivalently, $\hat H_N$ may be represented in the ``modal picture'' as $\hat H_N\!=\!\frac{1}{2}\!\left(\hat n_a - \hat n_b\right)$
with $\hat n_a\!=\!{\hat a}^\dagger \hat a$ and $\hat n_b\!=\!{\hat b}^\dagger \hat b$ being the photon-number operators
of arms $a$ and $b$ respectively. As shown in \figref{fig:MZinter_Q}, the \emph{output state} should be associated
with $\ket{\psi_\varphi^N}\!=\!\e^{-\ii\hat H_N\varphi}\ket{\psi_\t{in}^N}$, as the
second beam-splitter can always be incorporated into the measurement part 
of the general phase estimation scheme of \figref{fig:Ph_Est_Scheme}.

\subsubsection{Frequentist approach}
\label{sub:QEst_MZInter_HL_Freq}

As the input state is assumed to be pure (what must be optimal due 
the convexity property of the QFI described in \secref{sub:QFIproperties}),
the scenario of \figref{fig:MZinter_Q} is noiseless, and the estimated phase is unitarily 
encoded onto photons, \eqnref{eq:QFIPureVarH} for the QFI directly applies,
so that the QFI is proportional to the variance of the $N$-photon Hamiltonian $\hat{H}_N$ and explicitly reads:
\begin{equation}
F_{\textrm{Q}}\!\left[\left|\psi_{\varphi}^{N}\right\rangle \right]\;=\;4\,\Delta^{2}\hat{H}_N\;=\;4\left[\sum_{n=0}^N |\alpha_n|^2\,n^2 - \left(\sum_{n=0}^N |\alpha_n|^2 \,n\right)^2\right].
\label{eq:QFI_MZ_Q}
\end{equation}

\paragraph{Classical input states}~\\
Let us briefly mention first the performance of inputs consisting of uncorrelated photons 
employed in the \emph{classical strategy} of \secref{sub:ClEst_MZInter_SQL},
i.e.~the states \eref{eq:Cl_Input_Q_MZ}, ${\ket{\psi_\t{in}^N}}_\t{\tiny cl}$, which in the ``modal picture''  lead to
binomially distributed coefficients $\alpha_n$. Evaluating \eqnref{eq:QFI_MZ_Q},
$F_{\textrm{Q}}\!\left[{\left|\psi_\varphi^{N}\right\rangle}_\t{\tiny cl} \right]\!=\!N$ and one arrives
at the QCRB $1/N$ coinciding the classical CRB \eref{eq:FI&CRB_BinPhi} obtained for the photon-counting measurement
strategy of \figref{fig:MZinter_cl}. Hence, the above fact \emph{proves} that indeed such a measurement scheme is optimal
for classical inputs \eref{eq:Cl_Input_Q_MZ} from the local perspective.

\paragraph{Optimal input states -- NOON states}~\\
On one hand, one may  perform explicitly the maximisation 
of \eqnref{eq:QFI_MZ_Q} over the coefficients $\alpha_n$, e.g.~with
use of the method of Lagrange multipliers, in order to show that it is 
optimal to choose only non-zero $|\alpha_0|\!=\!|\alpha_N|\!=\!1/\sqrt 2$.
However, without any calculation, it is straightforward to identify the maximal/minimal eigenvalues of 
the Hamiltonian $\hat H_N$ in the ``modal picture'' as $\pm N/2$ representing the maximal difference 
in the photon-number in between the interferometer arms and the corresponding eigenvectors $\ket{N,0}$ and $\ket{0,N}$. 
Thus, according to the discussion of \secref{sub:QCRBandQFI}, the \emph{optimal input state} must be the
equally weighted superposition of these eigenvectors%
\footnote{%
For simplicity, we set $\phi\!=\!0$ in $\frac{1}{\sqrt 2}\!\left(\ket{N,0}+\e^{\ii\phi}\ket{0,N}\right)$ and ignore the arbitrary relative phase in between the eigenvectors.
} 
\citep{Giovannetti2006}, i.e.
\begin{equation}
{\ket{\psi_\t{in}^N}}_\t{{\tiny NOON}}=\frac{1}{\sqrt 2}\!\left(\ket{N,0}+\ket{0,N}\right)=\frac{1}{\sqrt 2}\left( \ket{a}^{\otimes N}+\ket{b}^{\otimes N}\right),
\label{eq:NOON}
\end{equation}
what is consistent with the explicit maximisation of \eqnref{eq:QFI_MZ_Q}. The state \eref{eq:NOON} is 
commonly referred to as the \emph{NOON state} due to its form in the modal representation \citep{Lee2002,Bollinger1996},
what may be slightly misleading. That is why, we also write it explicitly in \eqnref{eq:NOON} in the ``particle picture'', in order to emphasise that 
it corresponds to a \emph{maximally entangled state}, i.e.~the \emph{Greenberger-Horne-Zeilinger} (GHZ) state \citep{GHZ}, 
of the particles. Importantly, it is the \emph{inter-particle entanglement} \citep{Demkowicz2015}, which
assures that if one photon travels in a particular arm of the interferometer then so must the others, that leads to an $N$-fold 
winding of the off-diagonal terms, e.g.~$\e^{\ii N \varphi}\ket{0,N}\!\bra{N,0}$, so that the effective density matrix
of the NOON state resembles the one of a single-photon state but with an $N$-times greater phase resolution. 
As a consequence, the sensitivity to $\varphi$-variations scales quadratically with the particle number $N$, 
what is explicitly manifested by the \emph{maximal} QFI \eref{eq:QFI_MZ_Q} reading
\begin{equation}
F_{\textrm{Q}}\!\left[{\ket{\psi_\t{in}^N}}_\t{{\tiny NOON}}\right]=N^2 \qquad\implies\qquad \Delta^2\tilde\varphi_\nu\;\ge\;\frac{1}{\nu\,N^2}
\label{eq:QFIandQCRB_NOON}
\end{equation}
and yielding the ultimate $1/N^2$ scaling of the QCRB \eref{eq:QCRB} dictated by the maximal inter-photon 
correlations allowed by quantum mechanics.

The QCRB \eref{eq:QFIandQCRB_NOON} defines the so-called \emph{Heisenberg Limit} (HL) \citep{Holland1993} imposed
on the precision scaling with the particle number (here \emph{definite} photon number):~$1/N^2$. 
However, when interferometric schemes of \figref{fig:MZinter_Q} with \emph{indefinite} photon number are considered, 
one must be more rigorous when quantifying the resources \citep{Zwierz2010,Zwierz2012}, as naively replacing $N$ in \eqnref{eq:QFIandQCRB_NOON} with
the average photon number $\bar N$ may lead to incorrect conclusions of ``surpassing the HL'',
which must then be properly redefined \citep{Hyllus2010,Berry2012,Hall2012a,Giovannetti2012a,Demkowicz2015,Hofmann2009}. 
Furthermore, one should bear in mind that \eqnref{eq:QFIandQCRB_NOON} represents a frequentist bound
and the HL is guaranteed to be attainable only in the local estimation regime, which is warranted only when $\nu\!\rightarrow\!\infty$. 
In fact, for moderate $\nu$, one may construct bounds on the estimator MSE \eref{eq:MSE}, $\Delta^2\tilde\varphi_\nu$,
of a different type%
\footnote{%
In particular, the Ziv-Zakai bounds \citep{Tsang2012}.
}
that turn out to be indeed tighter than the QCRB \eref{eq:QFIandQCRB_NOON} \citep{Giovannetti2012}.
On the other hand, the necessity of procedure repetitions in NOON-based scenarios
becomes evident when analysing the optimal measurement schemes discussed below that lead to precision
saturating the QCRB \eref{eq:QFIandQCRB_NOON}.

\paragraph{Optimal measurements}~\\
Firstly, let us consider the quantum measurement
assured to be optimal by construction, i.e.~the projective
measurement in the eigenbasis of the SLD \eref{eq:SLD}
\citep{Nagaoka1989,Braunstein1994}. Utilizing the expression below
\eqnref{eq:PurifDefsUnitary} we derive the SLD for the NOON-based strategy
with the output ${\ket{\psi_\varphi^N}}_\t{{\tiny NOON}}\!\!=\!\e^{-\ii\hat H_N \varphi}\,{\ket{\psi_\t{in}^N}}_\t{{\tiny NOON}}$ as
$L_\t{\tiny S}\!\left[{\ket{\psi_{\varphi_0}^N}}_\t{{\tiny NOON}}\right]\!=\!\ii N\e^{\ii N{\varphi_0}}\left|0,N\right\rangle\!\left\langle N,0\right|+h.c.$
for the parameter true value $\varphi_0$. In general, the eigenvectors of the SLD read
$\ket{E_\pm(\varphi_0)}\!=\!\frac{1}{\sqrt 2} \e^{-\ii\hat H_N\varphi_0}(\ket{N,0}\pm\ii\ket{0,N})$,
but as the QFI \eref{eq:QFIandQCRB_NOON} is parameter independent we may without loss of generality set
$\varphi_0\!=\!\pi/(2N)$ to simplify the form of the measurement elements obtained,
which then project the output onto states:~$\ket{E_\pm}\!=\!\frac{1}{\sqrt 2}(\ket{N,0}\pm\ket{0,N})$. 
Notice that these correspond to NOON-like states, which are maximally entangled in between the particles, making such a scheme 
extremely hard to implement in an experiment. Furthermore, for a single repetition of the protocol ($\nu\!=\!1$), 
the above quantum measurement yields---independently of the photon number $N$---only two outcomes that occur
with probabilities $p_{\varphi,\pm}\!=\!\left|\left<E_\pm\!\left|\psi_\varphi^N\right.\!\right>_\t{{\tiny NOON}}\right|^2$ 
reading $\cos^2(N \varphi/2)$ and $\sin^2(N \varphi/2)$ respectively. 
Such behaviour perfectly agrees with the previously mentioned intuition that we may treat the NOON state as 
a \emph{single statistical object} or in other words a single ``entanglement-enhanced photon''
which is $N$-times more sensitive to phase variations. Thus, it should not be 
surprising that we must require many repetitions $\nu$, in order to gather enough data
and attain the QCRB \eref{eq:QFIandQCRB_NOON} and hence the HL.
In fact, we may interpret the above scenario as a classical Mach-Zehnder interferometry 
scheme of \secref{sub:ClEst_MZInter_SQL} with $\nu$ such ``entanglement-enhanced photons''
\citep{Higgins2007,Higgins2009}, 
where each of them independently leads to detection probabilities at the output oscillating 
$N$-times quicker with $\varphi$, so that now $\mathsf{p}\!=\!\cos^2(N \varphi/2)$ in \figref{fig:MZinter_cl}.
Whence, basing on the binomial PDF \eref{eq:PDF_BinPhi}, we may directly write the distribution of 
registering $r$ ``+'' outcomes (and hence $\nu-r$ ``-'' ones) after carrying out $\nu$ repetitions of the protocol, as
\begin{equation}
p_{\varphi}^{\nu}(r)=\binom{\nu}{r}\left(\cos^{2}\!\frac{N\varphi}{2}\right)^{r}\left(\sin^{2}\!\frac{N\varphi}{2}\right)^{\nu-r},
\label{eq:PDF_BinNOON}
\end{equation}
which leads to $F_\t{cl}\!\left[p_{\varphi}^{\nu}\right]\!=\!\nu N^2$ and the CRB indeed coinciding with the QCRB \eref{eq:QFIandQCRB_NOON}. 
Moreover, we may straightforwardly establish the locally efficient and the ML estimators 
by modifying the corresponding classical-scenario expressions \eref{eq:LocEffEst_BinPhi} and \eref{eq:MLE_BinPhi}, as it is enough to 
substitute $k\!\rightarrow\!r$ and $N\!\rightarrow\!\nu$ (the repetitions now stand for the number of uncorrelated photons) and
rescale all parameter true values and estimators via $\varphi\!\rightarrow\!N\varphi$ to account for the $N$-fold resolution
improvement. For example, the ML estimator \eref{eq:MLE_BinPhi} now reads 
$\tilde\varphi_\nu^\t{ML}\!=\!\frac{2}{N}\t{arccot}\!\left(\sqrt{r/(\nu-r)}\right)$
and applies to the interval $[0,\pi/N]$.
Notice that previously in the classical scenario we suffered from the sign ambiguity of $\varphi$, which
forced us to consider only to a $\pi$-wide region of parameter values. Now, such 
an unambiguous sector is further shrank to $\pi/N$, because only then $\varphi$ may be conclusively inferred 
from $p_{\varphi,\pm}$. Such requirement is equivalent to the possession of prior knowledge about the parameter
with MSE $\approx\!\pi^2/N^2$, which is assumed to be available for free due to locality of the approach.
Importantly, it is this \emph{prior knowledge}---which quantification is beyond the capabilities of the frequentist 
framework---that leads to HL-like scaling in the above scenario!

On the other hand, it has been generally proved in \citep{Giovannetti2006} that in the \emph{noiseless} version
of the phase estimation scenario of \figref{fig:Ph_Est_Scheme}, in order to achieve the ultimate precision
in the local regime of $\nu\!\rightarrow\!\infty$ with optimal input states employed, it is sufficient to consider only
uncorrelated measurements acting separately on the constituent particles, i.e.~$M_X\!=\!\bigotimes_{n=1}^N M_{X_n}^{(n)}$.
In case of the quantum-enhanced Mach-Zehnder interferometer of \figref{fig:MZinter_Q} this 
have been shown in various ways \citep{Bollinger1996,Kok2002,Lee2002,Gerry2010}, but an instructive example of such 
an efficient uncorrelated scheme is again the photon-counting measurement previously employed in the classical scenario of \figref{fig:MZinter_cl}.
In fact, it allows to locally attain the corresponding QCRB not only for the optimal NOON-based strategy but also 
for any path(mode)-symmetric $\ket{\psi_\t{in}^N}$ \citep{Hofmann2009}, which possess all coefficients satisfying $|\alpha_n|\!=\!|\alpha_{N-n}|$.
After propagating the output state through the second beam-splitter%
\footnote{%
What corresponds to the transformations on the NOON state:
$\left|N,0\right\rangle\!\rightarrow\!\sum_{n=0}^{N}\sqrt{\frac{1}{2^{N}}\binom{N}{n}}\left|n,N-n\right\rangle$ and 
$\left|0,N\right\rangle\!\rightarrow\!\sum_{n=0}^{N}\sqrt{\frac{1}{2^{N}}\binom{N}{n}}(-1)^{n}\left|n,N-n\right\rangle$.
}, 
${\ket{\psi_\varphi^N}}_\t{{\tiny NOON}}\!\rightarrow\!{\ket{\tilde\psi_\varphi^N}}_\t{{\tiny NOON}}$, 
one may determine the quantum equivalent of \eqnref{eq:PDF_BinPhi}, i.e.~the probability of detecting $k$ and $N-k$ photons 
at the interferometer output ports in \figref{fig:MZinter_Q}, as 
\begin{equation}
p_{\varphi}^N(k)=\left|\left\langle k,N-k\left|\tilde{\psi}_{\varphi}^N\right.\!\right\rangle_\t{{\tiny NOON}} \right|^{2}=\frac{1}{2^{N}}\binom{N}{k}\left[1+(-1)^{k}\cos(N\varphi)\right],
\label{eq:PDF_PhCount_MZinterQ}
\end{equation}
which indeed yields classical FI, $F_\t{cl}\!\left[p_{\varphi}^N\right]\!=\!N^2$, saturating the QCRB \eref{eq:QFIandQCRB_NOON} for $\nu\!=\!1$.
However, apart from the combinatorial factor which does not carry any information
about the parameter, the distribution \eref{eq:PDF_PhCount_MZinterQ} varies only
depending on whether $k$ is even or odd.
Hence, complete information about $\varphi$ that may be retrieved from $p_\varphi^N(k)$
resides in the \emph{parity} of the photon-number registered \citep{Gerry2010,Seshadreesan2013,Chiruvelli2011,
Anisimov2010}. 
Furthermore, evaluating thus $p_{\varphi,+}\!=\!\sum_{k=0,2,..}^N \!p_\varphi^N(k)\!=\!\cos^2(N\varphi/2)$
and $p_{\varphi,-}\!=\!\sum_{k=1,3,..}^N\! p_\varphi^N(k)\!=\!\sin^2(N\varphi/2)$
we reproduce the outcomes of the SLD-based measurement strategy considered above. 
Therefore, although we have shown that a photon-counting measurement 
indeed suffices, all the discussions and results from the previous paragraph apply.
In particular, we suffer again from the necessity of locality (assured only in the asymptotic limit
of many repetitions) and, 
in particular, the notion of the HL, $1/N^2$, is again contained within the prior knowledge
which analysis lies beyond the scope of the frequentist approach.

\subsubsection{Bayesian approach}
\label{sub:QEst_MZInter_HL_Bay}

In order to study from the Bayesian perspective the performance of the quantum-enhanced Mach-Zehnder interferometer of \figref{fig:MZinter_Q} for a 
single repetition ($\nu\!=\!1$), we minimise the adequate average cost
\eref{eq:QAvCostCov}, $\left<\mathcal{C}_\t{H}\right>$, over all pure input states \eref{eq:Input_Q_MZ} inside 
the interferometer, $\ket{\psi_\t{in}^N}$, and all positive semi-definite seed elements, $\Xi^N\!\ge\!0$, from which the effective
measurement schemes based on covariant POVMs \eref{eq:CovPOVM} may be constructed \citep{Hradil1996, Berry2000}.
As $\int\!\frac{\t{d}\varphi}{2\pi}\,4 \sin^2(\varphi/2) \e^{-\ii\varphi(n-m)}\!=\!2\delta_{n,m} -(\delta_{n,m-1} + \delta_{n,m+1})$, 
the effective matrix $\left\langle \rho_{\varphi}^{N}\right\rangle _{C_{\textrm{H}}}$, which describes 
the output state $\rho_{\varphi}^{N}\!=\!\ket{\psi_\varphi^N}\!\bra{\psi_\varphi^N}$ averaged over the 
cost function \eref{eq:CostFun_H}, is tridiagonal, i.e.~possesses non-zero entries on its main and closest to the main diagonals.
Hence, we may most generally write the average cost \eref{eq:QAvCostCov} as
\begin{equation}
\left<\mathcal{C}_\t{H}\right>\;=\;2\left(\sum_{n=1}^N \left|\alpha_n\right|^2 \,\Xi_{n,n}^N\right) - 2\,\t{Re}\!\left\{\sum_{n=1}^N \alpha_n^*\alpha_{n-1} \,\Xi_{n,n-1}^N\right\} \;=\;2\left(1 - \t{Re}\!\left\{\sum_{n=1}^N \alpha_n^*\alpha_{n-1} \,\Xi_{n,n-1}^N\right\}\right),
\label{eq:QAvCostCovMZ}
\end{equation}
where we have acknowledged the fact that the completeness condition $\int\!\!\frac{\t{d}\varphi}{2\pi}U_\varphi^{\otimes N}\Xi^NU_\varphi^{\dagger \otimes N}\!\!=\!\mathbb{I}$ 
forces all the diagonal entries of any seed element to be equal to one, i.e.~$\Xi_{n,n}^N\!=\!1$ for all $n$.

\paragraph{Optimal covariant measurements}~\\
As within the Bayesian approach the precision limits are not measurement-strategy independent,
we firstly proceed with minimisation of $\left<\mathcal{C}_\t{H}\right>$ over the choice of seed elements.
Because by replacing the input-state coefficients and the seed-element entries by their absolute values one
may only decrease the average cost \eref{eq:QAvCostCovMZ}, it is optimal to choose $\Xi^N$
such that for all $n$:~$\alpha_n^*\alpha_{n-1}\Xi_{n,n-1}^N\!=\!\left|\alpha_n\right|\left|\alpha_{n-1}\right|\left|\Xi_{n,n-1}^N\right|$.
Moreover, due to the positive semi-definiteness of the seed element, $\Xi^{N}\!\!\ge\!0$, 
we may upper-bound then the absolute value of its entries as $\left|\Xi^{N}_{n,m}\right|\!\leq\!\sqrt{\Xi^{N}_{n,n}\,\Xi^{N}_{m,m}}\!=\!1$.
Hence, the subtracted term in \eqnref{eq:QAvCostCovMZ} can at most read $\sum_{n=1}^N |\alpha_n| |\alpha_{n-1}|$,
what may be always achieved by setting $\Xi^{N}_{n,m} \!=\!\e^{\ii(\phi_n - \phi_{m})}$, where $\phi_{n} \!=\! \arg(\alpha_n)$.
This corresponds to the choice of the \emph{optimal seed element} which may be expressed as:~$\Xi^{N}_\t{\tiny opt}\!\!=\!\ket{e^N}\!\bra{e^N}$ with
$\ket{e^N} = \sum_{n=0}^N \e^{\mathrm{i}\phi_n}\ket{n,N-n}$, what proves that it is positive semi-definite  
as required \citep{Holevo1982,Chiribella2005}.
As a consequence, we obtain the expression for the average cost \eref{eq:QAvCostCovMZ}
that is optimised over the choice of \emph{all} measurement strategies:
\begin{equation}
\left<\mathcal{C}_\t{H}\right>\;=\;2\left(1 - \sum_{n=1}^N \left|\alpha_n\right|\left|\alpha_{n-1}\right|\right).
\label{eq:QAvCostCovMZ_OptXi}
\end{equation}
and the form of the \emph{optimal covariant POVM} \eref{eq:CovPOVM} 
reading:~$M_{\tilde{\varphi}}\!=\!U_{\tilde{\varphi}}^{\otimes N}\ket{e^N}\!\bra{e^N}U_{\tilde{\varphi}}^{\dagger\otimes N}$.
Note that the elements of $M_{\tilde{\varphi}}$ are \emph{not} separable with respect to particles%
\footnote{%
Notice that $\ket{e^N}$ is a \emph{non-normalised} vector that is \emph{entangled} in between the particles,
as for it to be a product state of particles its coefficient would need to be binomially distributed as in \eqnref{eq:Cl_Input_Q_MZ}.
}, so that they define to a \emph{collective} measurement performed on all the particles.

\paragraph{Classical and locally optimal input states}~\\
Firstly, let us consider the performance of inputs employed in the \emph{classical strategy} of \secref{sub:ClEst_MZInter_SQL},
i.e.~the states ${\ket{\psi_\t{in}^N}}_\t{\tiny cl}$ defined in \eqnref{eq:Cl_Input_Q_MZ} that consist of uncorrelated photons.
Substituting their binomial distribution of coefficients $\alpha_n$ into \eqnref{eq:QAvCostCovMZ_OptXi},
we obtain the corresponding measurement-optimised average cost:
\begin{equation}
\left<\mathcal{C}_\t{H}\right>_\t{\tiny cl}= 2 \left(1-\frac{1}{2^N}\sum_{n=1}^{N}\sqrt{\binom{N}{n}\binom{N}{n-1}}\right) \overset{N\to\infty}{=} \frac{1}{N},
\label{eq:QAvCost_MZcl}
\end{equation}
which exhibits an SQL-like scaling for $N\!\to\!\infty$, previously observed within the local approach.
Moreover, as \eqnref{eq:QAvCost_MZcl} asymptotically coincides with the QCRB established for the classical states \eref{eq:Cl_Input_Q_MZ}
in the first paragraph of the previous section, one may \emph{incorrectly} assume that \eqnref{eq:MinAvMSE_Jensen} 
connecting two complementary approaches applies, and thus there must also exist 
an uncorrelated measurement acting separately on the particles, which asymptotically achieves the average cost \eref{eq:QAvCost_MZcl}.
Notice that it \emph{cannot} be the photon-counting measurement analysed in the classical scenario of \secref{sub:ClEst_MZInter_SQL},
as, although its corresponding Bayesian cost \eref{eq:AvCostH_BinPhi} also asymptotically achieves 
the $1/N$ CRB \eref{eq:FI&CRB_BinPhi} and hence the QCRB,
it requires the parameter to be \emph{a priori} restricted to the unambiguous $[0,\pi]$ region.
In contrast, the Bayesian strategy described by \eqnref{eq:QAvCost_MZcl} assumes a \emph{complete} ignorance of the parameter
and, as one may numerically verify that the ratio of \eqnsref{eq:AvCostH_BinPhi}{eq:QAvCost_MZcl} is in fact less than one,
i.e.~$\forall_N\!:\left\langle \mathcal{C}_{\textrm{H}}(\tilde{\varphi}_{N}^\t{\tiny H})\right\rangle/\left<\mathcal{C}_\t{H}\right>_\t{\tiny cl}\!\le\!1$,
the photon-counting scheme with $\varphi\!\in\![0,\pi]$ performs actually better for finite $N$ than
the optimal Bayesian strategy for $\varphi\!\in\![-\pi,\pi]$! The measurement scheme 
that achieves the average cost \eref{eq:QAvCost_MZcl} for the uniform prior distribution, $p(\varphi)\!=\!1/2\pi$, 
is the \emph{adaptive strategy} \citep{Wiseman1997,Wiseman1998,Wiseman2009}, 
in which the measurement of a subsequent photon is adjusted depending on the results previously gathered,
what progressively narrows down the region of confidence even when starting from a complete flat prior distribution. 
Note that for such a measurement strategy to be applicable, the photons must be distinguishable, e.g.~arriving 
at the detector in consecutive time bins, so that they may be targeted individually in a sequence. 
Although \eqnref{eq:QAvCostCovMZ_OptXi} has been derived assuming their indistinguishability,
$\left<\mathcal{C}_\t{H}\right>$ still quantifies the minimal cost for distinguishable-particles input states. 
We return to this issue in \chapref{chap:MZ_losses}, 
where we consider the lossy Mach-Zehnder interferometer scenario and 
prove (see \appref{chap:appMZinterAdaptM}) that the average cost 
cannot be decreased by benefiting from the distinguishability of the particles employed.

For completeness, before optimising \eqnref{eq:QAvCostCovMZ_OptXi} over all inputs, let us 
convince the reader that NOON states, suffering from the $\pi/N$ ambiguity explained in 
\secref{sub:QEst_MZInter_HL_Freq}, are useless when no prior knowledge 
about parameter is available. Evaluating the measurement-optimised 
average cost \eref{eq:QAvCostCovMZ_OptXi} for ${\ket{\psi_\t{in}^N}}_\t{{\tiny NOON}}$,
$\forall_{N>1}\!:\left<\mathcal{C}_\t{H}\right>_\t{\tiny NOON}\!=\!2$ and
the NOON-based estimation scheme does not provide any information about the parameter
\emph{regardless of} $N$, even though general measurements are taken into account.
Crucially, being designed for the local regime, in the global setting
NOON states are outperformed by the classical inputs \eref{eq:Cl_Input_Q_MZ}, 
which (as shown above) still lead to the vanishing cost with $N$ at the $1/N$ SQL-like rate.

\paragraph{Optimal input states -- Berry-Wiseman states}~\\
In order to minimise the average cost \eref{eq:QAvCostCovMZ_OptXi} over all input states,
we rewrite it in a more appealing matrix-form after denoting by $\boldsymbol{\alpha}\!=\!\{\alpha_0,\dots,\alpha_N\}$
the vector containing the coefficients $\alpha_n$, which 
optimally are taken to be real:
\begin{equation}
\left<\mathcal{C}_\t{H}\right>\;=\;2-\boldsymbol{\alpha}^T 
\left(
\begin{array}{ccccc}
0 & 1 & 0 & \dots\\
1 & 0 & 1 & \dots\\
0 & 1 & 0 & \dots\\
\vdots & \vdots & \vdots & \ddots
\end{array}\right)
\boldsymbol{\alpha}
\;=\;
2-\boldsymbol{\alpha}^T \!\bar{\mathbf{A}}\,\boldsymbol{\alpha}
\label{eq:AvCost_MZInter_matrix}
\,.
\end{equation}
Thus, it becomes apparent that the \emph{minimal} average cost equals $2-\bar\lambda_\t{max}$, where
$\bar\lambda_\t{max}$ is the maximal eigenvalue of the matrix $\bar{\mathbf{A}}$, whereas the optimal input 
state coefficients are determined by the eigenvector corresponding to $\bar\lambda_\t{max}$. The above problem
has been solved independently in \citep{Berry2000} and \citep{Buzek1999},
but, as the work of \citet{Berry2000} considered explicitly the interferometric
scenario here described, we will refer to the optimal input state as 
the \emph{Berry-Wiseman} (BW) state%
\footnote{%
Yet, both \eqnsref{eq:BW}{eq:AvCostBW} have been established for the first time 
before the Bayesian analysis has been applied to Mach-Zehnder interferometry
by \citet{Summy1990}, who have approached the phase estimation problem by
utilising the Hermitian phase operator method of \citep{Pegg1988}.
}:
\begin{equation}
{\ket{\psi_\t{in}^N}}_\t{\tiny BW}\; = \;\sum_{n=0}^{N}\,\sqrt{\frac{2}{N+2}}\;\sin\left(\frac{n+1}{N+2}\pi\right)\,\ket{n,N-n} ,
\label{eq:BW}
\end{equation}
which corresponds to $\bar\lambda_\t{max}\!=\!2\cos\left[\pi/(N+2)\right]$ and thus yields the 
\emph{minimal average cost} now optimised over both input states and measurements:
\begin{equation}
\left<\mathcal{C}_\t{H}\right>_\t{\tiny BW} \; = \; 2\,\left[1-\cos\left(\frac{\pi}{N+2}\right)\right]\;\overset{N\to\infty}{=} \;\frac{\pi^2}{N^2}\,.
\label{eq:AvCostBW}
\end{equation}
Importantly, \eqnref{eq:AvCostBW} proves that, despite the extra $\pi^2$ factor, the $1/N^2$ HL-like scaling 
is also exhibited when analysing the noiseless phase estimation scheme of \figref{fig:Ph_Est_Scheme} from the global perspective.
This is assured, because for $N\!\to\!\infty$ the average cost \eref{eq:AvCostBW} may be interpreted as the $\overline{\t{MSE}}$,
i.e.~$\left<\mathcal{C}_\t{H}\right>_\t{\tiny BW}\!\!\!\overset{N\to\infty}{=}\!\!\left<\Delta^2\tilde\varphi^\t{\tiny MMSE}\right>_\t{\tiny BW}
\!\!=\! \int\!\!\t{d}\varphi\, p(\varphi)\!\left.\Delta^2\tilde\varphi^\t{\tiny MMSE}\right|_{\varphi}\!=\!\pi^2/N^2$ with $p(\varphi)\!=\!1/2\pi$,
even though $\left<\mathcal{C}_\t{H}\right>_\t{\tiny BW}$ does not asymptotically coincide with the QCRB 
\eref{eq:QFIandQCRB_NOON} (\eqnref{eq:MinAvMSE_Jensen} does not hold due to the presence of inter-particle correlations).
Furthermore, \eqnref{eq:AvCostBW} has been recently generalised in \citep{Jarzyna2014} to asymptotically apply
even when any regular prior distribution, $p(\varphi)$, is considered. Hence, for this to be true, we conjecture that the ``local MSE''
of the MMSE estimator \eref{eq:MMSE_Est} reads $\left.\Delta^2\tilde\varphi^\t{\tiny MMSE}\right|_{\varphi_0}\!\!=\!\pi^2/N^2$ for any $\varphi_0$,
what indicates that $\tilde\varphi^\t{\tiny MMSE}$ does \emph{not} converge asymptotically to the ML estimator \eref{eq:MLE}.
In particular, its variance is always widened by the $\pi^2$ factor due to the global character of the approach,
even if a very narrow but regular prior PDF is assumed\textsuperscript{\ref{foot1}}.

Lastly, let us comment on the applicability of the Bayesian approach, when 
one considers the phase estimation problem of the noiseless 
Mach-Zehnder interferometer, but accounts for the possibility of 
\emph{many repetitions}, i.e.~$\nu\!>\!1$ in the general
protocol of \figref{fig:Ph_Est_Scheme}. 
In such a case, as mentioned in the first paragraph of \secref{sub:QEstGlobal},
the strategy that utilises BW-states \eref{eq:BW} and covariant POVMs \eref{eq:CovPOVM}
is optimal only in the first round, for which the above analysis applies. 
After each run of the protocol, the effective prior distribution
must be updated, and the derivation of the optimal input states and measurements
minimising the average cost \eref{eq:QAvCost} must be consecutively repeated.
In particular, as the prior is no longer flat after some information about the parameter
is gathered during the first round,
the circular symmetry of the problem is lost and 
one may not restrict any more only to covariant POVMs. However, a general procedure
has been recently proposed in \citep{Demkowicz2011} that allows
for a numerical minimisation of the general average cost \eref{eq:QAvCost}
regardless of the prior distribution assumed. Although
the method of \citet{Demkowicz2011} has been demonstrated to be generally efficient
only for relatively low numbers of particles,  it was sufficient to indicate for a given moderate $N$ 
a smooth transition of the optimal inputs from BW-states to NOON-states,
as the prior becomes gradually narrowed with increase of the knowledge about the parameter%
\footnote{\label{foot1}%
Let us remind the reader that eventually in the limit of an infinitely narrow prior, 
i.e.~for a Dirac delta distribution that is no longer \emph{regular}, 
the Bayesian approach ignores the optimisation, as it indicates then that it is optimal
just to output the a priori known parameter value, so that
$\left<\mathcal{C}\right>\!=\!\left<\Delta^2\tilde\varphi\right>\!=\!0$, and do not perform any estimation (see \noteref{note:deltaprior}).
}. Hence, we conjecture that such a transition of the optimal input states---which 
crucially may now be adaptively varied between the procedure repetitions---should 
also occur with an increase of the repetition number $\nu$. Yet,
when focusing only on any of the single protocol repetitions, we suspect the precision to still 
be limited by the ``global HL''  \eref{eq:AvCostBW}:~$\pi^2/N^2$, as such bound is 
independent of the prior considered \citep{Jarzyna2014} (which just smoothly
narrows down with an increase of $\nu$).

%% file: Chapters/local_noisy_est.tex
\chapter{Local estimation in the presence of uncorrelated noise} 
\label{chap:local_noisy_est} 
\lhead{Chapter 4. \emph{Local estimation in the presence of uncorrelated noise}}

\section{$N$-parallel--channels estimation scheme}
\label{sec:NChEstScheme}

\begin{figure}[!b]
\begin{center}
\includegraphics[width=0.7\columnwidth]{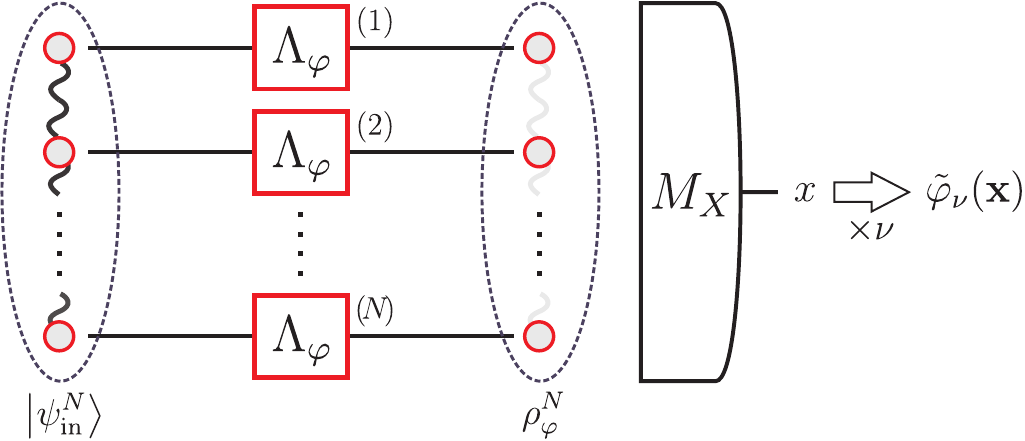}
\end{center}
\caption[$N\!$-parallel--channels estimation scheme]{%
\textbf{$N\!$-parallel--channels estimation scheme} as a generalisation of the phase estimation scheme
of \figref{fig:Ph_Est_Scheme} that includes the \emph{uncorrelated noise}. A pure input state $\ket{\psi_\t{in}^N}$
consists of $N$ particles, each of which evolves according to a general quantum map $\Lambda_\varphi$
that may in principle account for \emph{any} parameter encoding and uncorrelated noise. As before, 
a quantum measurement on the output state $\rho_\varphi^N\!=\!\Lambda_\varphi^{\otimes N}\left[\ket{\psi_\t{in}^N}\right]$ 
is represented by a POVM $M_X$ and the estimator $\tilde\varphi_\nu$ is constructed
on the sampled data after $\nu$ repetitions of the protocol.
\label{fig:NCh_Est_Scheme}}
\end{figure}

In this chapter we generalise the phase estimation scheme of \figref{fig:Ph_Est_Scheme}
with \emph{uncorrelated noise}, so that the quantum estimation problem
may be solved without 
imposing any constraints on the form of the single-particle evolution.
We depict the estimation scenario as shown in \figref{fig:NCh_Est_Scheme},
where now the parameter is encoded independently onto each particle
by some general \emph{$\varphi$-parametrised quantum channel} $\Lambda_\varphi$.
Naturally, such a description encapsulates the scheme of \figref{fig:Ph_Est_Scheme} with uncorrelated noise
and unitary encoding, for which then $\Lambda_\varphi\!=\!\mathcal{D}\circ\mathcal{U}_\varphi\!\!=\!\mathcal{U}_\varphi\!\circ\mathcal{D}$.
As in this chapter we analyse such an \emph{$N\!$-parallel--channels estimation scheme}
within the \emph{frequentist approach}, we may restrict (by convexity 
of the QFI discussed in \secref{sub:QFIproperties}) ourselves already
to \emph{pure input states}, $\ket{\psi_\t{in}^N}$, that 
can perform only better than mixed inputs.
Hence, the output state most generally reads:~$\rho_\varphi^N\!=\!\Lambda_\varphi^{\otimes N}\!\left[\ket{\psi_\t{in}^N}\right]$,
and a quantum measurement $M_X$ follows as in the scheme of \figref{fig:Ph_Est_Scheme}.
As before, after repeating the protocol $\nu$ times, an unbiased estimator, $\tilde\varphi_\nu$, is built
on the data gathered and its MSE \eref{eq:MSE} (or equivalently its variance \eref{eq:EstVar}) is 
lower-limited by the QCRB \eref{eq:QCRB} which is guaranteed to be attainable for $\nu\!\to\!\infty$.

\section{Local estimation of a \emph{single} quantum channel} 
\label{sec:ChEstLocal}

Before studying the attainable precision from the local perspective for
the general $N$-particle scheme of \figref{fig:NCh_Est_Scheme}, we
analyse the local properties of a \emph{single}, $\varphi$-parametrised quantum channel $\Lambda_\varphi$.
In particular, we study the possibilities of generalising the notion of the QFI \eref{eq:QFI} to
quantum maps---precisely, families of \emph{Completely Positive Trace Preserving}
(CPTP) maps%
\footnote{See \secsref{sub:Evo_QCh}{sec:qch_geom} for introduction to quantum channels and discussion of their geometric properties respectively.}
parametrised by $\varphi$, $\{\Lambda_\varphi\}_\varphi$---so that the attainable precision of estimation can be quantified 
by inspecting the form of a given channel, without need to explicitly study
its output states with help of the methods described previously in \secref{sub:QCRBandQFI}.

\subsection{Channel QFI}
\label{sub:ChQFI}

\begin{figure}[!t]
\begin{center}
\includegraphics[width=0.95\textwidth]{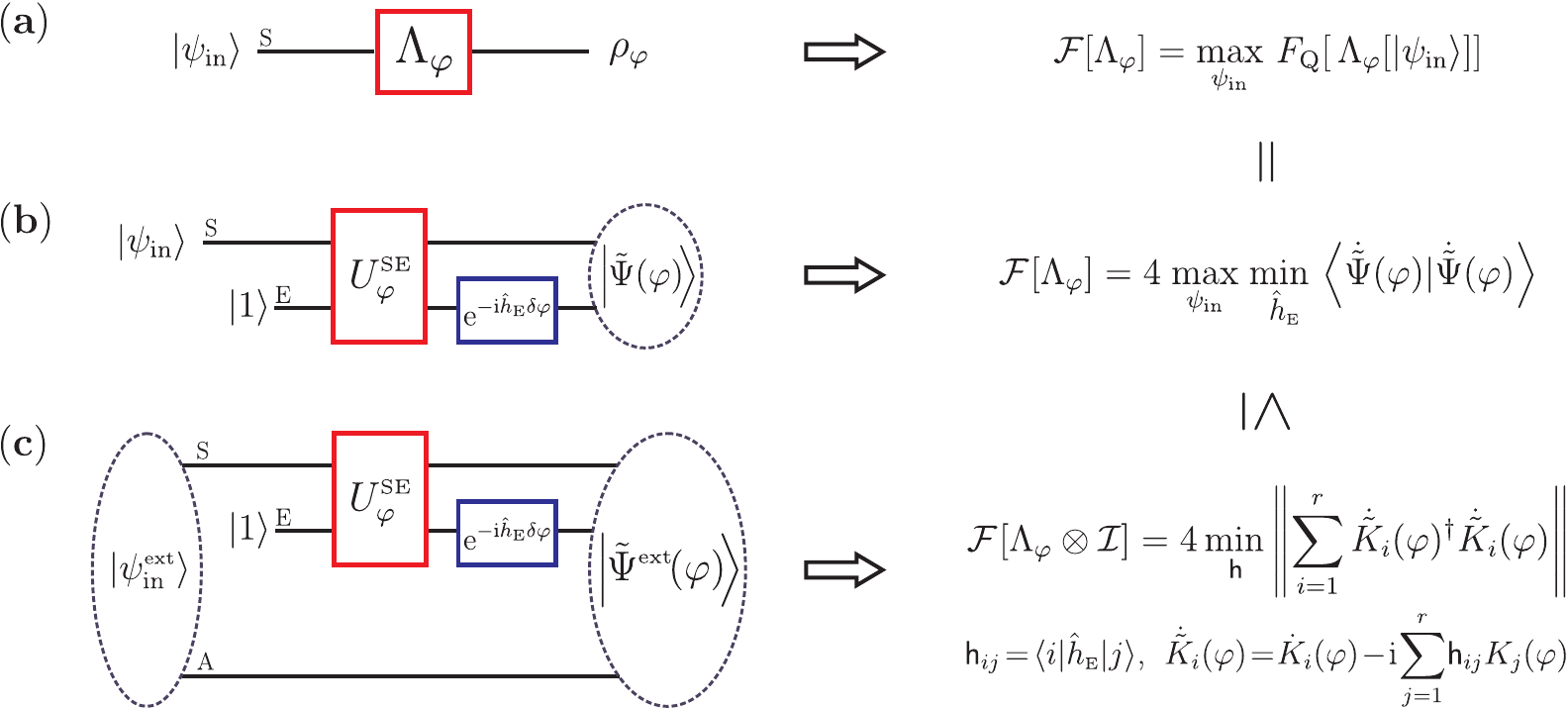}
\end{center}
\caption[Channel QFI evaluated basing on the output state purification]{
\textbf{Channel QFI evaluated basing on the output state purification}.\\
(\textbf{a}) \emph{Channel QFI} defined as the QFI of the output state maximised over all pure input states.\\
(\textbf{b}) \emph{Channel QFI} equivalently obtained by considering the output state purification
after a \emph{local, fictitious, parameter-dependent rotation of the environment},  $u_{\varphi}^\t{\tiny E}\!=\!\e^{-\textrm{i}{\hat h}_\t{\tiny E} \delta\varphi}$
with $\delta\varphi\!=\!\varphi\!-\!\varphi_{0}$ for a given $\varphi_0$, which hinders 
as much as possible information about the parameter.\\
(\textbf{c}) \emph{Extended-channel QFI} being independent of the maximisation
over the input states. In the purification picture, the environment rotation $u_{\varphi}^\t{\tiny E}$ 
can be just understood as a ``shift'' in the derivatives of the \emph{Kraus operators}, $K_i(\varphi)$, describing the action of the channel.}
\label{fig:ch_purif_ext}
\end{figure}

Given a single particle in the scheme of \figref{fig:NCh_Est_Scheme},
or more generally a quantum system in an initial pure state $\ket{\psi_\t{in}}$,
we identify its final (output) state after the evolution as $\rho_{\varphi}\!=\!\Lambda_{\varphi}[\ket{\psi_{\textrm{in}}}]$
with the parameter encoded by the action of the CPTP map $\Lambda_\varphi$.
The ultimate $\varphi$-estimation precision is then dictated
by the QCRB \eref{eq:QCRB} with the QFI reading:~$F_{\textrm{Q}}\!\left[\Lambda_{\varphi}[\ket{\psi_{\textrm{in}}}]\right]$,
and varies depending on the input state chosen.
Hence, as shown in \figref{fig:ch_purif_ext}(\textbf{a}), we define
the \emph{channel QFI }as the maximal QFI after performing the input
optimisation, so that it has a concrete operational and application-like
interpretation:
\begin{equation}
\mathcal{F}\!\left[\Lambda_{\varphi}\right]=\max_{\psi_{\textrm{in}}}\, F_{\textrm{Q}}\!\left[\Lambda_{\varphi}\!\left[\left|\psi_{\textrm{in}}\right\rangle \right]\right].\label{eq:ChQFI}
\end{equation}
For instance, in case of the noiseless phase estimation scenario previously describing
the Mach-Zehnder interferometer of \figref{fig:MZinter_Q},
$\Lambda_\varphi\!=\!\mathcal{U}_\varphi$ and
the output is a pure state:~$\e^{-\textrm{i}\hat H\varphi}\left|\psi_{\textrm{in}}\right\rangle$,
with $\hat H$ being the Hamiltonian generating the phase variation. Hence, as the QFI equals then
the variance of the Hamiltonian according to \eqnref{eq:QFIPureVarH}, the definition \eref{eq:ChQFI} just corresponds to 
\begin{equation}
\mathcal{F}\!\left[\mathcal{U}_{\varphi}\right]=\max_{\psi_{\textrm{in}}}\,F_{\textrm{Q}}\!\left[\e^{-\textrm{i}\hat H \varphi}\!\left|\psi_{\textrm{in}}\right\rangle \right]=4 \max_{\psi_{\textrm{in}}}\,\Delta^{2}\hat H=\left(\mu_\t{max}\!-\!\mu_\t{min}\right)^2,
\label{eq:ChQFIUnit}
\end{equation}
where $\mu_{\t{max}/\t{min}}$ are respectively the maximal/minimal eigenvalues of $\hat H$
and the maximum occurs for ${\ket{\psi_\t{in}}}$  being 
an equally weighted superposition of the eigenvectors corresponding to $\mu_{\t{max}/\t{min}}$.

\begin{note}[Time-energy uncertainty relation from the channel QFI perspective]
\label{note:time_energy_rel_channel}
An interesting result is obtained, if one applies \eqnref{eq:ChQFIUnit} in case of the natural latent
parameter of the evolution, i.e.~the \emph{time} $t$, for which $U_t\!=\!\e^{-\textrm{i}\hat H t}$
and $\mathcal{F}\!\left[\mathcal{U}_t\right]\!=\!4\,\Delta^2 \hat H\!=\!(E_\t{max}\!-\!E_\t{min})^2$, where
$E_{\t{max}/\t{min}}$ represent 
the maximal/minimal energies in the spectrum:~$\hat H \,\ket{E}\!=\!E\,\ket{E}$.
Note that the QCRB-equivalent \eref{eq:Phase_Et_UR} constitutes then exactly
the \emph{time-energy uncertainty relation} discussed in \noteref{note:time_energy}:
\begin{equation}
\Delta^2 \hat H\;\Delta^2 \tilde{t}\;\ge\;\frac{1}{4} \quad \implies \quad \Delta \tilde{t}\;\ge\;\frac{1}{E_\t{max}\!-\!E_\t{min}}\,,
\label{eq:TimeEnergyUR}
\end{equation}
stating that the maximal variance of the Hamiltonian---specified by the energy difference $E_\t{max}\!-\!E_\t{min}$---defines 
the ultimate resolution with which the duration estimator $\tilde t$ may be resolved
\citep{Aharonov2002}. Moreover, as \eqnref{eq:TimeEnergyUR}
is similarly to \eqnref{eq:Phase_Et_UR} optimised over all potential measurement and inference strategies, 
the time-energy uncertainty relation \eref{eq:TimeEnergyUR}
is more general than the Mandelstam-Tamm inequality \eref{eq:MTineq},
which (as shown in \noteref{note:time_energy}) is derived 
basing on a particular observable measurement that determines the 
ultimate sensitivity of the estimator $\tilde t$ to the variations of the actual 
elapsed time $t$ \citep{Braunstein1996}.
\end{note}

\paragraph{Purification-based definition of the channel QFI}~\\
In order to express the \emph{channel QFI} \eref{eq:ChQFI} with help of
the \emph{purification-based definition} \eref{eq:QFIPurifFuji} introduced 
previously in \secref{sub:QFIPurifDefs} for quantum states,
we utilise
the Stinespring dilation theorem (see \thmref{thm:stinespring} and \figref{fig:stinespring})
and rewrite a general channel
$\Lambda_{\varphi}$ as a unitary map $U_{\varphi}^\t{\tiny SE}$ acting
on the system, S, and the environment, E, disregarded after the evolution.
As a result, the combined state $\left|\Psi(\varphi)\right\rangle \!=\! U_{\varphi}^\t{\tiny SE}\left|\psi_{\textrm{in}}\right\rangle_\t{\tiny S}\!\left|1\right\rangle_\t{\tiny E}$
naturally constitutes the output purification such that $\Lambda_{\varphi}\!\left[\left|\psi_{\textrm{in}}\right\rangle \right]\!=\!\textrm{Tr}_\t{\tiny E}\!\left\{ \left|\Psi(\varphi)\right\rangle \!\left\langle \Psi(\varphi)\right|\right\}$,
where $\left|1\right\rangle$ is an arbitrary fixed state (i.e.~$\ket{\xi}$ in \figref{fig:stinespring}) 
chosen to be the first vector in the basis $\left\{ \left|i\right\rangle \right\} _{i=1}^{r}$
of the environment Hilbert space $\mathcal{H}_\t{\tiny E}^{r}$.
As depicted in \figref{fig:ch_purif_ext}(\textbf{b}), by specifying the dimension $r$ of $\mathcal{H}_\t{\tiny E}^{r}$
to be equal to the rank of $\Lambda_{\varphi}$, we may directly follow
the recipe of \secref{sub:QFIPurifDefs} and construct
all the \emph{locally relevant} output purifications, $\ket{\tilde{\Psi}(\varphi)}$, for the parameter 
true value $\varphi_0$ by applying
fictitious unitary rotations $u_{\varphi}^\t{\tiny E}\!=\!\e^{-\textrm{i}{\hat h}_\t{\tiny E}(\varphi-\varphi_{0})}$
to the environment that are generated by any Hermitian ${\hat h}_\t{\tiny E}$, so that 
$\left|\tilde{\Psi}(\varphi)\right\rangle \!=\!\tilde{U}_{\varphi}^\t{\tiny SE}\left|\psi_{\textrm{in}}\right\rangle_\t{\tiny S}\!\left|1\right\rangle_\t{\tiny E}$ with $\tilde{U}_{\varphi}^\t{\tiny SE}\!=\! u_{\varphi}^\t{\tiny E}\, U_{\varphi}^\t{\tiny SE}$.
Hence, as this corresponds again to a local shift of the purification first derivative,
we may express the channel QFI \eref{eq:ChQFI} as the ``channel version'' of \eqnref{eq:QFIPurifFuji}:
\begin{equation}
\mathcal{F}\!\left[\Lambda_{\varphi}\right]  =
4\max_{\psi_{\textrm{in}}}\,\min_{{\hat h}_\t{\tiny E}}\;\braket{\dot{\tilde{\Psi}}(\varphi)}{\dot{\tilde{\Psi}}(\varphi)}\,,
\label{eq:ChQFIPurif}
\end{equation}
where for a given $\varphi_0$ we must search only through
purifications such that $\ket{\tilde{\Psi}(\varphi_0)}\!=\!\ket{\Psi(\varphi_0)}$
and $\ket{\dot{\tilde{\Psi}}(\varphi_0)}\!=\!(\dot{U}^\t{\tiny SE}_{\varphi_0} - \textrm{i} {\hat h}_\t{\tiny E} U^\t{\tiny SE}_{\varphi_0})\left|\psi_{\textrm{in}}\right\rangle_\t{\tiny S}\!\left|1\right\rangle_\t{\tiny E}$. On the other hand, writing the action of the channel $\Lambda_\varphi$ in its Kraus representation
form introduced in \secref{sub:Evo_QCh}, i.e.~$\Lambda_{\varphi}\!\left[\left|\psi_{\textrm{in}}\right\rangle \right]\!=\!\sum_{i=1}^{r}K_i(\varphi)\left|\psi_{\textrm{in}}\right\rangle\!\left\langle \psi_{\textrm{in}}\right|K_i(\varphi)^{\dagger}$,
we can identify the Kraus operators corresponding to $\ket{\tilde{\Psi}(\varphi)}$ as
\begin{equation}
\tilde{K}_{i}(\varphi)=\left\langle i\right|\tilde{U}_{\varphi}^\t{\tiny SE}\left|1\right\rangle =
\sum_{j=1}^{r}\mathsf{u}(\varphi)_{ij}\, K_{j}(\varphi)\,,
\label{eq:KrausReps}
\end{equation}
where $\mathsf{u}(\varphi)_{ij}\!=\!\left\langle i\right|u_{\varphi}^\t{\tiny E}\left|j\right\rangle$
and $K_{j}(\varphi)\!=\!\left\langle j\right|U_{\varphi}^\t{\tiny SE}\left|1\right\rangle $
are the Kraus operators of the original purification $\ket{\Psi(\varphi)}$.
Thus, we may further rewrite \eqnref{eq:ChQFIPurif} as:
\begin{equation}
\mathcal{F}\!\left[\Lambda_{\varphi}\right]  = 4\max_{\psi_{\textrm{in}}}\,\min_\mathsf{h}\; \bra{\psi_{\textrm{in}}} \sum_{i=1}^{r}\dot{\tilde{K}}_{i}(\varphi)^{\dagger}\dot{\tilde{K}}_{i}(\varphi)\, \ket{\psi_{\textrm{in}}},
\label{eq:ChQFIPurifK}
\end{equation}
where now a particular choice of ${\hat h}_\t{\tiny E}$ corresponds
to a local shift of first derivatives of the Kraus operators. Whence,
at a given $\varphi_0$ we must search through Kraus representations
generated from the starting one via:~$\tilde{K}_{i}(\varphi_0)\!=\!K_{i}(\varphi_0)$ and 
$\dot{\tilde{K}}_{i}(\varphi_0)\!=\!\dot{K}_{i}(\varphi_0)-\textrm{i}\sum_{j=1}^{r}\mathsf{h}_{ij}K_{j}(\varphi_0)$
with $\mathsf{h}_{ij}\!=\!\left\langle i\right|\hat h_\t{\tiny E} \left|j\right\rangle$, what 
corresponds to the minimisation in \eqnref{eq:ChQFIPurifK} over all Hermitian matrices $\mathsf{h}$ of size%
\footnote{%
Following \secref{sub:Evo_QCh}, we denote the matrices $\mathsf{u}(\varphi)$ and $\mathsf{h}$ 
with a different fount to indicate that these are \emph{not} operators, and 
should be just treated as matrices with complex entries specifying linear transformations of Kraus operators. 
}
$r\!\times\!r$.

Importantly, by inspecting \figref{fig:ch_purif_ext}(\textbf{b})
and \eqnref{eq:ChQFIPurif}/\eref{eq:ChQFIPurifK},
it becomes evident that the optimal purification/Kraus representation---which by the previous argumentation of \secref{sub:QFIPurifDefs} would be identified
for the output state $\rho_{\varphi}$ as the one satisfying either $\left|\dot{\tilde{\Psi}}_{\varphi}^\t{\tiny opt}\right\rangle\!=\!\frac{1}{2}L_\t{\tiny S}\!\left[\rho_{\varphi}\right]\otimes\mathbb{I}^\t{\tiny E}\left|\tilde{\Psi}_{\varphi}^\t{\tiny opt}\right\rangle$ 
or 
$\dot{\tilde K}_i^\t{\tiny opt}(\varphi)\left|\psi_{\textrm{in}}\right\rangle \!=\!\frac{1}{2}L_\t{\tiny S}\!\left[\rho_\varphi\right]{\tilde K}_{i}^\t{\tiny opt}(\varphi)\left|\psi_{\textrm{in}}\right\rangle$
respectively---can now be 
neatly interpreted as the one for which an artificial environment is chosen that 
hinders as much as possible information about the estimated parameter due to its extra
local rotation $u_{\varphi}^\t{\tiny E}$.
As a result, the environment part of the optimal purification, $\left|\tilde{\Psi}_{\varphi}^\t{\tiny opt}\right\rangle$, 
does not carry any locally extractable information about $\varphi$, so that consistently with the 
purification-based definition \eref{eq:QFIPurifEscher}:~$F_\t{Q}\!\left[\left|\tilde{\Psi}_{\varphi}^\t{\tiny opt}\right\rangle\right]\!=\!F_\t{Q}\!\left[\rho_\varphi\right]$.

\subsection{Extended-channel QFI}
\label{sub:ExtChQFI}

From the geometrical perspective, similarly to the case of quantum states 
for which the QFI describes the ``speed'' of variations
in $\varphi$ (see \eqnref{eq:QFIviaStDist}), one may like
the channel QFI \eref{eq:ChQFI} also to quantify 
the local statistical distance but this time between the maps $\Lambda_{\varphi}$
and $\Lambda_{\varphi+\delta\varphi}$ for small $\delta\varphi$. 
However, at the level of quantum channels, one should bear in mind
that for any variation of $\varphi$ to be noticeable,
there must exist input states which lead to a measurable
change of the channel output that is at best in some ``orthogonal
direction''. As a consequence, the quantity $\min_\mathsf{\,h} \!\left\{ \dots\right\}$
in \eqnref{eq:ChQFIPurifK} non-trivially varies with the input $\ket{\psi_{\textrm{in}}}$,
as the minimum occurs for the optimal Kraus operators $\left\{ K_{i}^\t{\tiny opt}(\varphi)\right\} _{i=1}^{r}$
defined via condition $\dot{K}_{i}^\t{\tiny opt}(\varphi)\left|\psi_{\textrm{in}}\right\rangle \!=\!\frac{1}{2}L_\t{\tiny S}\!\left[\Lambda_{\varphi}\left[\left|\psi_{\textrm{in}}\right\rangle \right]\right] K_{i}^\t{\tiny opt}(\varphi)\left|\psi_{\textrm{in}}\right\rangle$
that explicitly depends on $\ket{\psi_\t{in}}$. 
As a result, one cannot in principle express the channel QFI \eref{eq:ChQFI} using
only the form of $\Lambda_\varphi$ and circumvent the problem of
input state maximisation, i.e.~$\max_{\ket{\psi_{\textrm{in}}}}$ in \eqnref{eq:ChQFIPurifK}, 
that is difficult in general, as for 
a given input chosen one must always establish the optimal purification/Kraus representation and thus \emph{cannot} naively
interchange the order of $\max$ and $\min$ in \eqnref{eq:ChQFIPurif}/\eref{eq:ChQFIPurifK} \citep{Fujiwara2008}.

Yet, as depicted in \figref{fig:ch_purif_ext}(\textbf{c}), one may
construct a natural upper bound on the channel QFI \eref{eq:ChQFI}
by \emph{extending} the input space, $\mathcal{H}_\t{\tiny S}$,
by an equally-large ancillary space, $\mathcal{H}_\t{\tiny A}$,
which is unaffected by the map but measured along with the channel
output. In this way, by employing
extended input states entangled between these two spaces,
$\left|\psi_{\textrm{in}}^\t{\tiny ext}\right\rangle \!\in\!\mathcal{H}_\t{\tiny S}\!\otimes\!\mathcal{H}_\t{\tiny A}$,
one may extract more information about $\varphi$ by inspecting also the ancilla, A,
which---despite not being affected by $\Lambda_\varphi$---due
to entanglement with S improves the capabilities of quantum measurements
performed on the whole, extended output state 
$\rho_\varphi^\t{\tiny ext}\!=\!\Lambda_\varphi\otimes\mathcal{I}\!\left[\ket{\psi_\t{in}^\t{\tiny ext}}\right]$ 
\citep{Ziman2008,Sedlak2009}. Such a \emph{channel extension} defines then the \emph{extended-channel QFI}:
\begin{equation}
\mathcal{F}\!\left[\Lambda_{\varphi}\otimes\mathcal{I}\right]=\max_{\psi_\textrm{in}^\t{\tiny ext}}\, F_{\textrm{Q}}\!\left[\Lambda_{\varphi}\otimes\mathcal{I}\left[\left|\psi_\textrm{in}^\t{\tiny ext}\right\rangle \right]\right]
\label{eq:ExtChQFI}
\end{equation}
that is naturally greater than the channel QFI \eref{eq:ChQFI} and $\mathcal{F}\!\left[\Lambda_{\varphi}\otimes\mathcal{I}\right]\!\ge\!\mathcal{F}\!\left[\Lambda_{\varphi}\right]$. Note that
\eqnsref{eq:ChQFI}{eq:ExtChQFI} coincide when the extension turns out \emph{not} to be beneficial,
what is manifested by the optimal extended input state, $\ket{\psi_\t{in}^\t{\tiny ext}}$, being separable,
i.e.~$\ket{\psi_\t{in}^\t{\tiny ext}}\!=\!{\ket{\psi_\t{in}}}_\t{\tiny S}{\ket{\xi}}_\t{\tiny A}
\!\Rightarrow\!\rho_\varphi^\t{\tiny ext}\!=\!\rho_\varphi\otimes{\ket{\xi}}_\t{\!\tiny A}\!\bra{\xi}$
for any $\ket{\xi}$, and one may freely trace out the ancillary subspace, $\mathcal{H}_\t{\tiny A}$,
without affecting the QFI of the extended output state.

\paragraph{Purification-based definition of the extended-channel QFI}~\\
The analogue of \eqnref{eq:ChQFIPurifK} that specifies the \emph{purification-based definition of the extended-channel QFI}
by utilising equivalent Kraus representations \eref{eq:KrausReps} of the channel reads \citep{Fujiwara2008}:
\begin{equation}
\mathcal{F}\!\left[\Lambda_{\varphi}\otimes\mathcal{I}\right]=4\max_{\rho_{\textrm{in}}^\t{\tiny S}}\min_\mathsf{h}\,
\textrm{Tr}_\t{\tiny S}\!\left\{ \rho_{\textrm{in}}^\t{\tiny S}\sum_{i=1}^{r}\!\dot{\tilde{K}}_{i}(\varphi)^{\dagger}
\dot{\tilde{K}}_{i}(\varphi)\right\} \!=4\min_{\mathsf{h}}\left\Vert \sum_{i=1}^{r}\!\dot{\tilde{K}}_{i}(\varphi)^{\dagger}
\dot{\tilde{K}}_{i}(\varphi)\right\Vert \!,
\label{eq:ExtChQFIPurif}
\end{equation}
where $\left\Vert \dots\right\Vert $ represents the operator norm.
The first expression above is obtained by tracing over the ancillary space
$\mathcal{H}_\t{\tiny A}$, what leads to the maximisation over all
mixed states
$\rho_{\textrm{in}}^\t{\tiny S}\!=\!\textrm{Tr}_\t{\tiny A}\!\left\{ \left|\psi_{\textrm{in}}^\t{\tiny ext}\right\rangle \!\left\langle
\psi_{\textrm{in}}^\t{\tiny ext}\right|\right\} $.
Thus, \eqnref{eq:ExtChQFIPurif} is exactly the purification-based expression for the (unextended) channel QFI \eref{eq:ChQFIPurifK} 
with pure states $\ket{\psi_\t{in}}$ replaced by mixed ones $\rho_{\textrm{in}}^\t{\tiny S}$, 
what crucially allows to interchange the order of $\max$ and $\min$ above
and obtain the second expression \citep{Fujiwara2008}.
However, one shall \emph{not} be mistaken that the extended-channel QFI \eref{eq:ExtChQFI} 
can be interpreted as the generalisation of the (unextended) channel QFI \eref{eq:ChQFI} 
to mixed-state inputs! Although one may replace without loss of generality pure input states with mixed ones in the 
primary channel QFI definition \eref{eq:ChQFI} due to convexity of the QFI,
the purification-based definitions \eref{eq:ChQFIPurif} and \eref{eq:ChQFIPurifK} after such an interchange become invalid.
It is so, because the purifications employed in \eqnsref{eq:ChQFIPurif}{eq:ChQFIPurifK} should then
also account for the fact of \emph{purifying the mixed input state} and 
otherwise lead to an overestimate of the actual output state QFI. 
As unintentionally shown by \eqnref{eq:ExtChQFIPurif}, the quantity obtained then is the extended-channel QFI \eref{eq:ExtChQFI}, 
which consistently constitutes an upper bound:~$\mathcal{F}\!\left[\Lambda_{\varphi}\otimes\mathcal{I}\right]\!\ge\!\mathcal{F}\!\left[\Lambda_{\varphi}\right]$.
On the other hand, by inspecting the optimal $\rho_{\textrm{in}}^\t{\tiny S}$ in \eqnref{eq:ExtChQFIPurif}
one may verify if the extension leads to a precision improvement.
If there does \emph{not} exist an optimal $\rho_{\textrm{in}}^\t{\tiny S}$ which is \emph{mixed}%
\footnote{%
What occurs only if the maximal eigenvalue of the operator $\sum_{i=1}^{r}\!\dot{\tilde{K}}_{i}(\varphi)^{\dagger}
\dot{\tilde{K}}_{i}(\varphi)$ in \eqnref{eq:ExtChQFIPurif} is \emph{non-degenerate} forcing
the optimal $\rho_{\textrm{in}}^\t{\tiny S}$ to be pure.
}, the optimal extended input $\left|\psi_{\textrm{in}}^\t{\tiny ext}\right\rangle$ must be separable, 
what (as discussed in the previous paragraph) assures
$\mathcal{F}\!\left[\Lambda_{\varphi}\otimes\mathcal{I}\right]\!=\!\mathcal{F}\!\left[\Lambda_{\varphi}\right]$.

Importantly, the expression \eref{eq:ExtChQFIPurif} for the extended-channel QFI 
involves \emph{only} the Kraus representation optimisation and may be reformulated 
as a \emph{semi-definite program} (SDP) that crucially is \emph{always} efficiently evaluable numerically. 
In \appref{chap:appFinNCEasSDP}, we demonstrate how to construct 
the relevant SDP for a more general task of upper-bounding the QFI of
$N$-parallel channels $\Lambda_\varphi^{\otimes N}$, i.e.~for the scheme of \figref{fig:NCh_Est_Scheme} explicitly
analysed from the local perspective later in \secref{sec:EstNChannels}.
Yet, because \eqnref{eq:ExtChQFIPurif} is just a special case of such a more 
general procedure with $N\!=\!1$, its SDP-reformulation directly follows (see \appref{chap:appFinNCEasSDP} for details).

\subsection{RLD-based upper bound on the extended-channel QFI}
\label{sub:RLD}

Nevertheless, as the evaluation of the extended-channel QFI via \eqnref{eq:ExtChQFIPurif} still
involves minimisation, which due to many free 
parameters ($\mathsf{h}$ is only constrained to be a Hermitian matrix) is
in general not easily solvable analytically, one may want to seek for further upper bounds on 
$\mathcal{F}\!\left[\Lambda_{\varphi}\otimes\mathcal{I}\right]$, and hence also on
$\mathcal{F}\!\left[\Lambda_{\varphi}\right]$, that do not involve any optimisation at all.
One possibility for such a construction is to relax the
QCRB itself by replacing the SLD in \eqnref{eq:QCRB} with
other logarithmic derivatives of the output state $\rho_\varphi$,
which are non-Hermitian but still satisfy $\dot\rho_{\varphi}\!=\!\frac{1}{2}\left(\rho_{\varphi}L[\rho_{\varphi}]\!+\! L[\rho_{\varphi}]^{\dagger}\rho_{\varphi}\right)$.
As proved in \citep{Holevo1982,Hayashi2005a}, for any such $L[\rho_{\varphi}]$
an upper limit on the QFI \eref{eq:QFI} is obtained:~$F_{\textrm{Q}}\!\left[\rho_{\varphi}\right]\!\le\!\textrm{Tr}\!\left\{ \rho_{\varphi}L[\rho_{\varphi}]L[\rho_{\varphi}]^{\dagger}\right\}$, 
that consistently is guaranteed to be tight for the SLD, i.e.~$L[\rho_{\varphi}]\!=\!L[\rho_{\varphi}]^\dagger\!=\!L_\t{\tiny S}[\rho_{\varphi}]$ \citep{Nagaoka2005}.
A commonly utilised example of $L[\rho_{\varphi}]$ is the
\emph{Right Logarithmic Derivative} (RLD):~$L_{\textrm{\tiny R}}[\rho_{\varphi}]\!=\!\rho_{\varphi}^{-1}\,\dot\rho_{\varphi}$,
which exists \emph{if and only if} $\dot\rho_{\varphi}$ \emph{is contained 
within the support of} $\rho_{\varphi}$, but simplifies often the calculations,
as its corresponding upper bound on the QFI,
$F_{\textrm{Q}}\!\left[\rho_{\varphi}\right]\!\le\!\textrm{Tr}\!\left\{\rho_{\varphi}^{-1}\!\left.\dot\rho_{\varphi}\!\right.^{2}\right\}$,
requires inversion of the output state $\rho_{\varphi}$ and not its full 
eigendecomposition%
\footnote{%
For completeness, let us note as aside that when considering
\emph{multi-parameter} estimation schemes, in which the SLD 
is no more the unique logarithmic derivative defining
the multi-parameter QCRB, the RLD may sometimes lead to
tighter bounds on the overall achievable precision of simultaneous estimation
of multiple parameters \citep{Fujiwara1994,Hayashi2005a,Genoni2012}.
}.

In \citep{Hayashi2011} the applicability of such an RLD-based bound has 
been addressed in the context of quantum channels. In particular,
by defining the Choi-Jamio\l{}kowski (CJ)
matrix%
\footnote{%
See \secref{sub:CJiso} for the description of Choi-Jamio\l{}kowski representation of a quantum channel.
}
representing the map $\Lambda_{\varphi}$,
i.e.~$\Omega_{\Lambda_{\varphi}}\!=\!\Lambda_{\varphi}\otimes\mathcal{I}\left[\left|\mathbb{I}\right\rangle \right]$
with $\left|\mathbb{I}\right\rangle \!=\!\sum_{i=1}^{\dim\mathcal{H}_\t{\tiny S}}\left|i\right\rangle_\t{\tiny S}\!\otimes\!\left|i\right\rangle_\t{\tiny A}$,
it has been proved that the extended-channel QFI \eref{eq:ExtChQFI} 
can be further upper-limited after replacing the SLD with the RLD
via:
\begin{equation}
\mathcal{F}\!\left[\Lambda_{\varphi}\otimes\mathcal{I}\right]\quad\le\quad\mathcal{F}^\t{\tiny RLD}\!\left[\Lambda_{\varphi}\otimes\mathcal{I}\right]=\left\Vert \textrm{Tr}_\t{\tiny A}\!\left\{ \dot{\Omega}_{\Lambda_{\varphi}}\Omega_{\Lambda_{\varphi}}^{-1}\dot{\Omega}_{\Lambda_{\varphi}}\right\} \right\Vert ,
\label{eq:RLDbound}
\end{equation}
where $\left\Vert \dots\right\Vert $ is again the operator norm and
$\Omega_{\Lambda_{\varphi}}^{-1}$ is the inverse of $\Omega_{\Lambda_{\varphi}}$.
Crucially, in contrast to \eqnsref{eq:ChQFI}{eq:ExtChQFI},
the above \emph{RLD-based upper bound on the extended-channel QFI}
is determined \emph{solely} by the form of
$\Lambda_{\varphi}$ (its CJ representation)---and does not require
any optimisation neither over the input states accepted by the map nor 
over its Kraus representations.
However, the condition for existence of the RLD defined on quantum states 
has a direct generalisation to the case of quantum channels, as the bound \eref{eq:RLDbound}
is non-divergent and thus applicable for a given map $\Lambda_\varphi$ specified by its CJ matrix $\Omega_{\Lambda_{\varphi}}$
\emph{if and only if} $\dot{\Omega}_{\Lambda_{\varphi}}$ 
\emph{is contained within the support of} $\Omega_{\Lambda_{\varphi}}$.
In \appref{chap:appRLDboundCond}, we explicitly prove this requirement 
stemming from the work of \citet{Hayashi2011} and stress
that this is \emph{exactly} the criterion for a given CPTP map 
to be $\varphi$-non-extremal (see \critref{crit:ch_phi_extrem} and \appref{chap:appPhiExtremCond}).
Thus, \emph{the RLD-based bound \eref{eq:RLDbound} applies to and only to
$\varphi$-non-extremal channels}, what importantly provides a clear
geometric explanation for what kinds of quantum maps, and with what parametrisations,
is the RLD-based approach valid.

One may wonder whether the bound \eref{eq:RLDbound} could be straightforwardly 
improved by considering the QFI calculated w.r.t.~the CJ matrix of the quantum channel considered.
However, let us explicitly note that 
$F_\t{Q}\!\left[\Omega_{\Lambda_\varphi}\right]\!=\!F_\t{Q}\!\left[\Lambda_{\varphi}\otimes\mathcal{I}\left[\left|\mathbb{I}\right\rangle \right]\right]$
and thus it \emph{lower}-limits the channel 
QFI:~$F_\t{Q}\!\left[\Omega_{\Lambda_\varphi}\right]\!\le\!\mathcal{F}\!\left[\Lambda_{\varphi}\otimes\mathcal{I}\right]$,
which is maximised over all input states. Hence, $F_\t{Q}\!\left[\Omega_{\Lambda_\varphi}\right]$ possesses 
only an operational meaning when the above lower bound is tight, what
occurs \emph{only if} the maximally entangled states, $\ket{\mathbb{I}}$, 
turn out to be optimal. This, however, happens very rarely 
especially as $\Lambda_\varphi$ is assumed to be noisy and thus \emph{not} unitary.

Last but not least, the bound \eref{eq:RLDbound} has been proved
in \citep{Hayashi2011} to be \emph{additive} on channels, so that
for any two, $\varphi$-non-extremal maps $\Lambda_{\varphi}^{(1)}$ and $\Lambda_{\varphi}^{(2)}$:
\begin{eqnarray}
&& \mathcal{F}^\t{\tiny RLD}\!\left[\left(\Lambda_{\varphi}^{(1)}\!\otimes\!\mathcal{I}\right)\!\otimes\!\left(\Lambda_{\varphi}^{(2)}\otimes\mathcal{I}\right)\right] = \mathcal{F}^\t{\tiny RLD}\!\left[\Lambda_{\varphi}^{(1)}\!\otimes\!\mathcal{I}\right]+\mathcal{F}^\t{\tiny RLD}\!\left[\Lambda_{\varphi}^{(2)}\!\otimes\!\mathcal{I}\right]\nonumber \\
&& \quad\implies\quad\mathcal{F}\!\left[\left(\Lambda_{\varphi}\!\otimes\!\mathcal{I}\right)^{\otimes N}\right]\;\;\le\;\;\mathcal{F}^\t{\tiny RLD}\!\left[\left(\Lambda_{\varphi}\!\otimes\!\mathcal{I}\right)^{\otimes N}\right]  = N\,\mathcal{F}^\t{\tiny RLD}\!\left[\Lambda_{\varphi}\!\otimes\!\mathcal{I}\right].
\label{eq:RLDboundN}
\end{eqnarray}
As remarked in the second expression, the RLD-based bound thus constrains not only the QFI of a single extended channel,
but also restricts the QFI of $N$ extended channels used in \emph{parallel}
to scale at most linearly with $N$. Crucially, as the extension can
only improve the precision, \eqnref{eq:RLDboundN} is also a valid
upper-bound on the QFI of $N$ parallel uses of an unextended channel,
i.e.~on the QFI of the output state in the scheme of \figref{fig:NCh_Est_Scheme}:~$F_\t{Q}\!\left[\rho_\varphi^N\right]\!\le\!N\mathcal{F}^\t{\tiny RLD}\!\left[\Lambda_{\varphi}\!\otimes\!\mathcal{I}\right]$. 
Strikingly, this proves that the estimation precision 
attained in the scheme of \figref{fig:NCh_Est_Scheme},
when considering \emph{any single-particle channel $\Lambda_\varphi$
that is $\varphi$-non-extremal}, must \emph{at most}
asymptotically follow the SQL-like scaling. 
We give an alternative, but maybe more intuitive, explanation 
to such a conclusion in \secref{sub:CSbound}, where we show 
that any locally $\varphi$-non-extremal channel can in fact be 
\emph{classically simulated}, what indeed assures the asymptotic 
scaling to follow $const/N$.
Nevertheless, as $\mathcal{F}^\t{\tiny RLD}\!\left[\Lambda_{\varphi}\!\otimes\!\mathcal{I}\right]$
provides a quantitative measure upper-limiting the maximum achievable quantum enhancement of precision,
we utilise it explicitly in \secref{sub:SQLAsBounds}, where we term it for short as
the \emph{RLD bound} and the above RLD-based precision-bounding procedure 
as the \emph{RLD method}, which we then compare with other approaches
also allowing to derive asymptotic SQL-like bounds on attainable precision.

\subsection{\caps{Example:} Noisy-phase--estimation channels}

In order to apply the notions of \emph{channel QFI} \eref{eq:ChQFI},
\emph{extended-channel QFI} \eref{eq:ExtChQFI} and the \emph{RLD bound} \eref{eq:RLDbound}
in a concrete setting, we consider the channels introduced in \secref{sub:noise_models} and previously depicted 
in \figref{fig:noise_models}, which model the evolution of a \emph{qubit} with 
the parameter being encoded as the angle, $\varphi$, of rotation around the $z$ axis and 
the decoherence corresponding to one of the noise-types:~\emph{dephasing},
\emph{depolarisation}, \emph{loss} and \emph{spontaneous
emission}; each of strength $\eta$. Notice that these constitute 
an example of the phase estimation scheme of \figref{fig:Ph_Est_Scheme} with $N\!=\!1$
and $\mathcal{D}$ representing one of the above-listed noise models.
For each of the channels, we determine the relevant quantities 
and present them in \tabref{tab:ChQFISingle} in an increasing order, as accordingly
$\mathcal{F}\!\left[\Lambda_{\varphi}\right]\!\le\!\mathcal{F}\!\left[\Lambda_{\varphi}\!\otimes\!\mathcal{I}\right]\!\le\!\mathcal{F}^\t{\tiny RLD}\!\left[\Lambda_{\varphi}\!\otimes\!\mathcal{I}\right]$.

Due to the low dimension of the system, we 
are able to explicitly calculate the \emph{channel QFI} \eref{eq:ChQFI}
for all of the cases and confirm that, as intuitively expected 
from \figref{fig:noise_models}, it is always optimal to
prepare the qubit in any state lying on the \emph{equator} of the Bloch ball,
as during the action of any of the noise-types considered it still remains represented by 
the furthest point from the $z$ axis and thus most sensitive to the 
parameter variations (rotations around the $z$ axis).

\begin{table}[!t]
\begin{center}
\begin{tabular}{|M{3.5cm}||M{2cm}|M{3cm}|M{1.5cm}|M{3.5cm}|N}
\hline 
\textbf{Noise model:} & \emph{Dephasing} & \emph{Depolarization} & \emph{Loss} & \emph{Spontaneous emission}  
&\\[12pt]
\hline
$\mathcal{F}\!\left[\Lambda_{\varphi}\right]$ \eref{eq:ChQFI}:~& $\eta^{2}$ & $\eta^{2}$ & $\eta$  & $\eta$ 
&\\[12pt]
\hline 
$\mathcal{F}\!\left[\Lambda_{\varphi}\!\otimes\!\mathcal{I}\right]$
\eref{eq:ExtChQFI}:~& $\eta^{2}$ & $\frac{2\eta^{2}}{1+\eta}$ & $\eta$  & $\frac{4\eta}{\left(1+\sqrt{\eta}\right)^{2}}$
&\\[12pt]
\hline 
$\mathcal{F}^\t{\tiny RLD}\!\left[\Lambda_{\varphi}\!\otimes\!\mathcal{I}\right]$
\eref{eq:RLDbound}:~& $\frac{\eta^{2}}{1-\eta^{2}}$ & $\frac{2\eta^{2}(1+\eta)}{\left(1-\eta\right)\left(1+3\eta\right)}$ & n.a. & n.a.
&\\[12pt]
\hline
\end{tabular}
\end{center}
\caption[Single channel QFI measures for the noisy-phase--estimation models]{%
\textbf{Channel QFIs}, $\mathcal{F}\!\left[\Lambda_{\varphi}\right]$, \textbf{extended-channel QFIs}, $\mathcal{F}\!\left[\Lambda_{\varphi}\!\otimes\!\mathcal{I}\right]$, and \textbf{RLD bounds}, $\mathcal{F}^\t{\tiny RLD}\!\left[\Lambda_{\varphi}\!\otimes\!\mathcal{I}\right]$, for the noisy-phase--estimation channels
introduced in \secref{sub:noise_models} and depicted in \figref{fig:noise_models}. RLD bounds cannot be constructed for the
\emph{loss} and \emph{spontaneous emission} noise-types, as these correspond to $\varphi$-extremal channels
(see \secref{sub:noise_models_geom}). \hfill [n.a.---not~applicable]
}
\label{tab:ChQFISingle}
\end{table}

In case of the \emph{extended-channel QFI} \eref{eq:ExtChQFI}---due to the
presence of the ancilla---the output generally corresponds to a mixed state of two qubits, 
for which the analytic computation of the QFI \eref{eq:QFI} with a generic input
is not straightforward any more. That is why, we utilise
\eqnref{eq:ExtChQFIPurif} and explicitly perform the purification-minimisation 
for each of the channels of \figref{fig:noise_models}, what may be always simplified by
decreasing the number of free parameters after inspecting
the numerical form of the optimal-purification generator $\mathsf{h}$, which
accuracy may be quantified and assured due to the SDP reformulation of
\eqnref{eq:ExtChQFIPurif} presented in \appref{chap:appFinNCEasSDP}.
We list in \appref{chap:appOptHExtCh} the analytic forms of the optimal generators $\mathsf{h}$
that determine the adequate shift of the first derivatives of Kraus operators, when starting 
from the canonical Kraus representations stated in \tabref{tab:noise_models} 
for each of the channels considered. Importantly, the results of
\tabref{tab:ChQFISingle} justify that \emph{extension enhances} the precision only for
\emph{depolarisation} and \emph{spontaneous emission} channels, for which the optimal 
inputs correspond to maximally entangled (Bell-like) 
states:~$\frac{1}{\sqrt 2}\left(\ket{0}_\t{\tiny S}\ket{\psi}_\t{\tiny A}+\e^{\ii\phi}\ket{1}_\t{\tiny S}\ket{\psi_\perp}_\t{\tiny A}\right)$.
These may be intuitively interpreted to consist of the
system qubit prepared again on equator for highest parameter sensitivity
and an entangled to it ancillary state with $\braket{\psi}{\psi_\perp}\!=\!0$,
which leads to improved precision of the measurements performed on the 
overall extended output state%
\footnote{The ambiguity of choosing any state 
of the ancilla may be neatly explained realising the freedom of local unitary rotations
that may always be performed on it at the measurement stage.
}. 
Let us also note that in the absence of noise:~$\mathcal{F}\!\left[\Lambda_{\varphi}\!\otimes\!\mathcal{I}\right]\!\overset{\eta\to1}{=}\!
\mathcal{F}\!\left[\Lambda_{\varphi}\right]$, as the extension 
may not be beneficial for a unitary channel \citep{DAriano2001a}.

Lastly, we compute the corresponding \emph{RLD bounds} \eref{eq:RLDbound} for
the \emph{dephasing} and \emph{depolarisation} noise models, which are the only ones---as 
shown in \secref{sub:noise_models_geom}---that
lead to $\varphi-$\emph{non-extremal} channels.
As $\mathcal{F}^\t{\tiny RLD}\!\left[\Lambda_{\varphi}\!\otimes\!\mathcal{I}\right]$
constitutes also an SQL-like bound on the asymptotic scaling (see \eqnref{eq:RLDboundN}), RLD 
bounds in \tabref{tab:ChQFISingle} correctly diverge when $\eta\!=\!1$ for the 
noiseless estimation scenario, in which the HL must be attainable.
However, as a result, the RLD bounds, when interpreted as upper limits on the extended-channel QFI,  
become useless in the regime of $\eta\!\to\!1$, in which $(\mathcal{F}^\t{\tiny RLD}\!\left[\Lambda_{\varphi}\!\otimes\!\mathcal{I}\right]/\mathcal{F}\!\left[\Lambda_{\varphi}\!\otimes\!\mathcal{I}\right])\!\to\!\infty$.

\section{Local estimation of $N$ quantum channels in \emph{parallel}}
\label{sec:EstNChannels}

We now consider explicitly the $N\!$-parallel--channels estimation
scheme of \figref{fig:NCh_Est_Scheme},
for which the $N$-particle output state of the system reads
$\rho_{\varphi}^{N}\!=\!\Lambda_{\varphi}^{\otimes N}\!\left[\left|\psi_{\textrm{in}}^{N}\right\rangle \right]$.
Thus, analogously to the single channel measures, we define the $N$-\emph{channel QFI} as:
\begin{equation}
\mathcal{F}\!\left[\Lambda_{\varphi}^{\otimes N}\right]=\max_{\psi_{\textrm{in}}^{N}}\, F_{\textrm{Q}}\!\left[\,\Lambda_{\varphi}^{\otimes N}\!\left[\left|\psi_{\textrm{in}}^{N}\right\rangle \right]\right],
\label{eq:NChQFI}
\end{equation}
which linear or quadratic dependence on $N$ dictates 
respectively the SQL- or HL-like scaling of precision. 
In case of a classical strategy, for which a further constraint on \eqnref{eq:NChQFI} must be imposed
restricting the inputs to product states: $\left|\psi_{\textrm{in}}^{N}\right\rangle \!=\!\left|\psi_{\textrm{in}}\right\rangle^{\otimes N}$,
consistently the $N$-channel QFI becomes $N$ times the single channel QFI \eref{eq:ChQFI}, 
i.e.~$\!\left.\mathcal{F}\!\left[\Lambda_{\varphi}^{\otimes N}\right]\right|_\t{cl}\!=\! N\,\mathcal{F}\!\left[\Lambda_{\varphi}\right]$.

\subsection{SQL-like bounds on the asymptotic precision}
\label{sub:SQLAsBounds}

As we would like to investigate various methods that allow 
to prove and quantify the asymptotic SQL-like precision scaling emergent
in the scheme of 
\figref{fig:NCh_Est_Scheme} due to the impact of uncorrelated noise, we define
the \emph{asymptotic channel QFI} as:
\begin{equation}
\mathcal{F}_{\textrm{as}}\!\left[\Lambda_{\varphi}\right]=\lim_{N\rightarrow\infty}\frac{\mathcal{F}\!\left[\Lambda_{\varphi}^{\otimes N}\right]}{N},
\label{eq:QFIAs}
\end{equation}
which is always bounded from above, unless
a given channel $\Lambda_{\varphi}$ allows for an asymptotic super-classical precision scaling.
In general, $\mathcal{F}_{\textrm{as}}\!\left[\Lambda_{\varphi}\right]\!\ge\!\mathcal{F}\!\left[\Lambda_{\varphi}\right]$
with equality indicating the asymptotic optimality of classical estimation scenarios and no room for any
quantum enhancement of precision.

On the other hand,  we may thus quantify with
help of \eqnref{eq:QFIAs} the \emph{maximal quantum enhancement of precision}, $\chi\!\left[\Lambda_{\varphi}\right]$,
as the ratio of the asymptotic estimation errors between the classical
and optimal-quantum strategies dictated by their corresponding QCRBs \eref{eq:QCRB}, i.e.
\begin{equation}
\chi\!\left[\Lambda_{\varphi}\right]=\lim_{N\rightarrow\infty}\sqrt{\frac{\left.\Delta^2\tilde{\varphi}_\nu\right|_\t{cl}}{\left.\Delta^2\tilde{\varphi}_\nu\right|_\t{Q}}}=\sqrt{\frac{\nu\,\mathcal{F}_{\textrm{as}}\!\left[\Lambda_{\varphi}\right]}{\nu\,\mathcal{F}\!\left[\Lambda_{\varphi}\right]}}\ge1,
\label{eq:qEnh}
\end{equation}
where we have left on purpose the repetition number in the last expression 
to stress that, although \eqnref{eq:qEnh}
is $\nu$-independent, it is guaranteed to be saturable only
in the $\nu\!\to\!\infty$ limit due to the locality of the frequentist approach.
\eqnref{eq:qEnh} has rather only a formal meaning, as
it involves the computation of $\mathcal{F}\!\left[\Lambda_{\varphi}^{\otimes N}\right]$
for arbitrary large $N$, what is generally infeasible due to the
complexity of the QFI \eref{eq:QFI} rising exponentially with $N$,
not to mention the impossibility of performing the maximisation
over all input states in \eqnref{eq:NChQFI}.

In what follows we present methods that allow to upper-bound the
asymptotic channel QFI \eref{eq:QFIAs}, and hence
the maximal quantum precision enhancement \eref{eq:qEnh},
basing \emph{purely} on the form of a \emph{single} channel $\Lambda_\varphi$---without 
need of neither considering its tensor-product structure nor
performing any optimisation over the input states.
Example of such a procedure is the already-mentioned \emph{RLD method}
that, due to the additivity property of the RLD bound \eref{eq:RLDbound} 
introduced in \secref{sub:RLD}, leads to
$\mathcal{F}\!\left[\Lambda_{\varphi}^{\otimes N}\right]\le N\,
\mathcal{F}^\t{\tiny RLD}\!\left[\Lambda_{\varphi}\otimes\mathcal{I}\right]$
for any $\varphi$-non-extremal channel.

\begin{figure}[!t]
\begin{center}
\includegraphics[width=0.95\columnwidth]{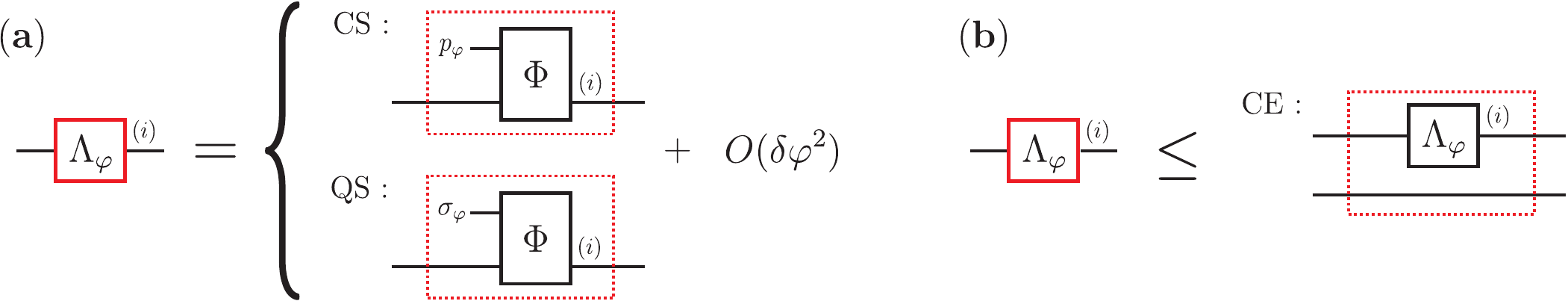}
\end{center}
\caption[SQL-bounding methods on the asymptotic precision scaling]{%
\textbf{SQL-bounding methods on the asymptotic precision scaling}
that stem only from the form of a \emph{single} channel, $\Lambda_\varphi$, 
representing the single-particle evolution in the scheme of \figref{fig:NCh_Est_Scheme}.
(\textbf{a}): In the \emph{CS} and \emph{QS methods}, the channel must be \emph{locally}
equivalent to a parameter-independent map $\Phi$ that is also fed either
a classical diagonal state $p_{\varphi}$ in CS, or a general, quantum state
$\sigma_{\varphi}$ in QS.
(\textbf{b}): Nevertheless, the tightest bound that is applicable
to the widest class of quantum channels is obtained via the \emph{CE method}, 
in which $\Lambda_{\varphi}$ is replaced by its extension $\Lambda_{\varphi}\otimes\mathcal{I}$,
what trivially can only improve the estimation capabilities.}
\label{fig:SQLmethods}
\end{figure}

In general, we can write
\begin{equation}
\mathcal{F}_{\textrm{as}}\!\left[\Lambda_{\varphi}\right] \le \mathcal{F}_{\textrm{as}}^\t{\tiny bound}
,\qquad\qquad\chi\!\left[\Lambda_{\varphi}\right]\le\sqrt{\frac{\mathcal{F}_{\textrm{as}}^\t{\tiny bound}}
{\mathcal{F}\!\left[\Lambda_{\varphi}\right]}},
\label{eq:QFIAsBound}
\end{equation}
where for $\mathcal{F}_{\textrm{as}}^\t{\tiny bound}$
we may substitute not only the RLD-based $\mathcal{F}^\t{\tiny RLD}\!\left[\Lambda_{\varphi}\otimes\mathcal{I}\right]$,
but also:~$\mathcal{F}_{\textrm{as}}^\t{\tiny CS}\ge\mathcal{F}_{\textrm{as}}^\t{\tiny QS}\ge\mathcal{F}_{\textrm{as}}^\t{\tiny CE}$;
corresponding to the bounds derived via respectively the
\emph{Classical Simulation} (CS), \emph{Quantum Simulation} (QS)
 and \emph{Channel Extension} (CE) methods 
schematically explained in \figref{fig:SQLmethods},
but described in detail consecutively below.

\subsubsection{Classical Simulation bound}
\label{sub:CSbound}

\begin{figure}[!t]
\begin{center}
\includegraphics[width=0.60\columnwidth]{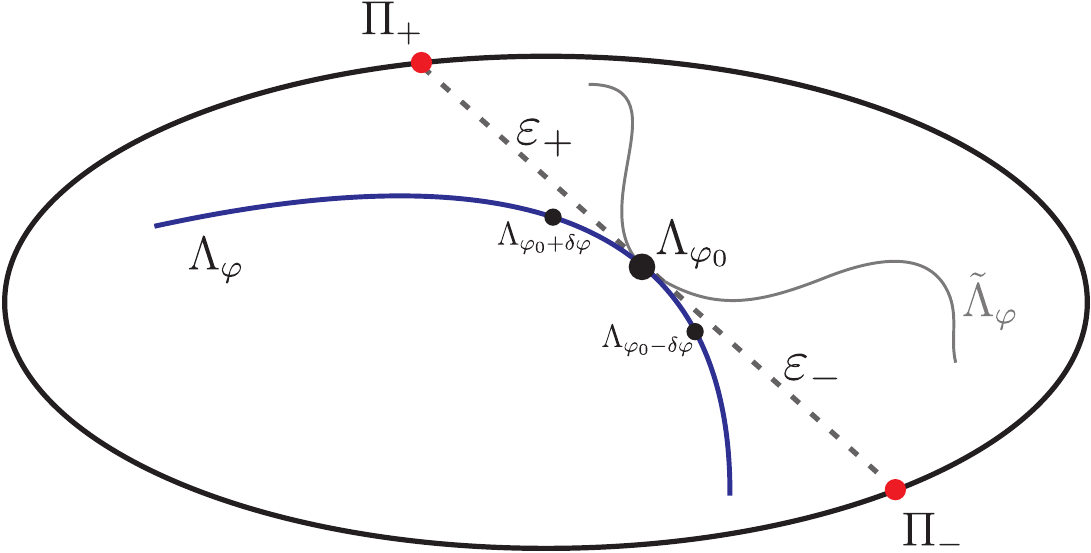}
\end{center}
\caption[Local CS of a $\varphi$-non-extremal quantum channel]{
\textbf{Local CS of a $\varphi$-non-extremal quantum channel}.
The family of $\varphi$-parametrised CPTP maps, $\{\Lambda_\varphi\}_\varphi$,
is depicted, similarly to \figref{fig:ExtremCh}(\textbf{b}), as a trajectory (\emph{blue curve}) in the convex set
representing the space of all relevant quantum channels sharing the same
input and output Hilbert spaces. 
As the map $\Lambda_\varphi$ is $\varphi$\emph{-non-extremal} at $\varphi_0$
it is possible to construct its local \emph{Classical Simulation} (CS) there. In case
of the \emph{unitary} parameter-encoding \citep{Demkowicz2012}, the optimal CS corresponds
to a probabilistic mixture of two channels lying at the intersection of the tangent
and the boundary of the set:~$\Pi_{\pm}$.
Note that from the point of view of the QFI, any two channel
trajectories, e.g.~$\Lambda_{\varphi}$ and $\tilde{\Lambda}_{\varphi}$
(\emph{grey curve}), are equivalent at a given $\varphi_{0}$ as
long as the form of the maps and their first derivatives w.r.t.~$\varphi$ coincide there.}
\label{fig:CSpic}
\end{figure}

Crucially, the notion of $\varphi$\emph{-non-extremality} of
a parametrised map $\Lambda_\varphi$ (introduced in 
\secref{sub:ExtremCh}) carries a \emph{natural geometric explanation} 
of why a given channel 
is bound to asymptotically follow the SQL-like scaling of precision.
Returning in \figref{fig:CSpic} to the picture representing the family of CPTP maps
parametrised by $\varphi$ as a trajectory in the convex space of
quantum channels, we explain---following \citep{Matsumoto2010}---that 
the channel $\varphi$-non-extremality at a given $\varphi_0$
assures a local \emph{Classical Simulation} (CS) of 
the map $\Lambda_\varphi$ to be feasible there, 
so that when considering the parallel action of channels, $\Lambda_\varphi^{\otimes N}$,
in \figref{fig:NCh_Est_Scheme} and the $N\!\to\!\infty$ limit, the precision scaling in $N$
is forced to behave as if the estimation problem was classical.

We term a channel $\Lambda_\varphi$
to be \emph{classically simulable} at the parameter value $\varphi_0$, 
if it can be written as a classical mixture of $\varphi$-independent
channels $\left\{ \Pi_{x}\right\} _{x}$ such that for $\varphi\!=\!\varphi_{0}\!+\!\delta\varphi$ and 
any $\varrho$:
\begin{equation}
\Lambda_{\varphi}\!\left[\varrho\right]=\sum_{x}\, p_{\varphi}(x)\,\Pi_{x}\!\left[\varrho\right]+O\!\left(\delta\varphi^{2}\right)=\Phi\!\left[\varrho\otimes p_{\varphi}\right]+O\!\left(\delta\varphi^{2}\right),
\label{eq:CS}
\end{equation}
so that the action of $\Lambda_\varphi$ is mimicked
on average by probabilistically choosing a \emph{fixed} channel
from the set $\{\Pi_x\}_x$ according to a random variable $X$
distributed with $p_\varphi(X)$.
Equivalently, as indicated in the second expression above and \figref{fig:SQLmethods}(\textbf{a}), 
one may write the decomposition \eref{eq:CS} in a form, in which the channel
$\Lambda_\varphi$ is expressed as a $\varphi$-independent CPTP map, $\Phi$, 
which apart from the input acts also on a diagonal density 
matrix, $p_{\varphi}\!=\!\sum_{x}p_{\varphi}(x)\left|e_{x}\right\rangle \!\left\langle e_{x}\right|$,
and obeys:~$\forall_{\varrho,x}\!:\Phi\!\left[\varrho\otimes\!\left|e_{x}\right\rangle \!\left\langle e_{x}\right|\right]\!=\!\Pi_{x}\!\left[\varrho\right]$.
Crucially, for the CS to be valid, \eqnref{eq:CS} must be satisfied only \emph{locally}, 
as the QFI \eref{eq:QFI} just quantifies the sensitivity to
parameter variations (see \eqnref{eq:QFIviaStDist}) what
at the level of quantum maps means that, from the point of view 
of the QFI, all channels
are equivalent if they output density matrices that 
are identical up to the first order in $\delta\varphi$.
On the other hand, as the action of any channel
may be written with use of its CJ representation (see \eqnref{eq:ch_action_via_CJ}) 
a given $\Lambda_\varphi$ may thus be defined
to be tantamount to another $\tilde\Lambda_\varphi$ at $\varphi_0$
from the local-estimation perspective, 
as long as their CJ matrices coincide there up to $O\!\left(\delta\varphi^{2}\right)$,
i.e.~$\Omega_{\Lambda_{\varphi_0}}\!\!=\!\Omega_{\tilde\Lambda_{\varphi_0}}$
and $\dot\Omega_{\Lambda_{\varphi_0}}\!\!=\!\dot\Omega_{\tilde\Lambda_{\varphi_0}}$,
as geometrically shown in \figref{fig:CSpic}. Hence, for the CS \eref{eq:CS} to be feasible, we must find 
an ensemble $\{p_{\varphi}(x),\Pi_x\}_x$ that satisfies
$\sum_x p_{\varphi_0}(x)\Omega_{\Pi_x}\!=\!\Omega_{\Lambda_{\varphi_0}}$
and $\sum_x{\dot p}_{\varphi_0}(x)\Omega_{\Pi_x}\!=\!\dot\Omega_{\Lambda_{\varphi_0}}$.
This corresponds exactly to the requirement of $\varphi$\emph{-non-extremality} of $\Lambda_\varphi$ at $\varphi_0$
(see \defref{def:ch_phi_extrem}), as
these conditions can be matched \emph{if and only if} there exist
physical maps lying in both directions along the tangent to the trajectory at $\varphi_0$, 
what---as formally specified in \secref{sub:ExtremCh}---is true if one may find
 $\epsilon\!>\!0$ such that both matrices $\Omega_{\Pi_{\pm}}\!=\!\Omega_{\Lambda_{\varphi_0}}\!\!\pm\!\epsilon\,\dot{\Omega}_{\Lambda_{\varphi_0}}$
are positive semi-definite%
\footnote{Or equivalently,  if $\dot{\Omega}_{\Lambda_{\varphi}}$ is contained 
within the support of $\Omega_{\Lambda_{\varphi}}$ at $\varphi_0$---see \critref{crit:ch_phi_extrem}.},
i.e.~their corresponding $\Pi_\pm$ lie within the convex set depicted in \figref{fig:CSpic}.

Now, as the maps $\Lambda_\varphi$ in the $N\!$-parallel--channels estimation scheme of \figref{fig:NCh_Est_Scheme}
act \emph{independently} on each of the particles, we can simulate the overall action of $\Lambda_\varphi^{\otimes N}$
with help of \eqnref{eq:CS} by just associating $N$ \emph{independent} variables, $X^N$, with each of the channels.
The estimation procedure may thus be described as follows
\begin{equation}
\varphi\rightarrow X^{N}\rightarrow\Lambda_{\varphi}^{\otimes N}\rightarrow\Lambda_{\varphi}^{\otimes N}\!\left[\left|\psi_{\textrm{in}}^{N}\right\rangle \right]\rightarrow\tilde{\varphi},
\label{eq:CSMarkovChain}
\end{equation}
so that the parameter is firstly encoded
by infinite sampling of $X^N$  onto the form of 
quantum channels acting 
on the input, $\ket{\psi_\t{in}^N}$, and then decoded by
performing measurements on the output state, $\rho_\varphi^N$.
Clearly, such a protocol can only be less efficient than
the strategy in which we could infer the parameter directly
from $X^{N}$, i.e.~$\varphi\!\rightarrow\! X^{N}\!\!\rightarrow\!\tilde{\varphi}$,
what corresponds to a \emph{classical} estimation problem
with maximal attainable precision dictated by the (classical) FI 
\eref{eq:FI}:~$F_{\textrm{cl}}\!\left[p_{\varphi}^{N}\right]\!=\!N F_{\textrm{cl}}\!\left[p_{\varphi}\right]$.
As result, we obtain a linearly scaling upper bound on the QFI 
of the output $\rho_{\varphi}^{N}$
that yields the desired $\mathcal{F}_{\textrm{as}}^\t{\tiny bound}$
of \eref{eq:QFIAsBound}, i.e.
\begin{equation}
F_{\textrm{Q}}\!\left[\rho_{\varphi}^{N}\right]\le N F_{\textrm{cl}}\!\left[p_{\varphi}\right] \qquad\implies\qquad
\mathcal{F}_{\textrm{as}}\!\left[\Lambda_{\varphi}\right] \le F_{\textrm{cl}}\!\left[p_{\varphi}\right],
\label{eq:CSbound}
\end{equation}
being fully determined by the single-channel--mixing PDF $p_\varphi(X)$ of \eqnref{eq:CS}.

Importantly, the notion of $\varphi$\emph{-non-extremality} of a given channel,
as introduced in \figref{fig:ExtremCh}, naturally provides a valid CS \eref{eq:CS}
of $\Lambda_\varphi$ that corresponds to the mixture of channels $\Pi_\pm$,
which lie along the tangent to the trajectory at $\varphi_0$ in opposite directions and
thus possess CJ matrices:~$\Omega_{\Pi_{\pm}}\!=\!\Omega_{\Lambda_{\varphi_0}}\!\pm\epsilon_\pm\dot{\Omega}_{\Lambda_{\varphi_0}}$ with $\epsilon_\pm\!>\!0$.
Consequently, a locally equivalent channel $\tilde{\Lambda}_{\varphi}\!=\! p_{\varphi,+}\Pi_{+}\!+\!p_{\varphi,-}\Pi_{-}$
may always be constructed at $\varphi_0$, as one can always 
adequately choose $p_{\varphi,\pm}$, so that 
$\Omega_{\tilde{\Lambda}_{\varphi_0}}\!\!=\!\Omega_{\Lambda_{\varphi_0}}$
and $\dot{\Omega}_{\tilde{\Lambda}_{\varphi_0}}\!\!=\!\dot{\Omega}_{\Lambda_{\varphi_0}}$.
Moreover, one may easily verify that the correct binary PDF then yields
$F_{\textrm{cl}}\!\left[p_{\varphi,\pm}\right]\!=\!1/(\epsilon_{+}\epsilon_{-})$,
which constitutes a valid example of the bound \eref{eq:CSbound}.
Importantly, as proved in \citep{Demkowicz2012}, for channel estimation problems 
in which the parameter is \emph{unitarily} encoded, it is always \emph{optimal} 
to employ such a two-point CS that---as depicted in \figref{fig:CSpic}---mixes 
channels $\Pi_\pm$ that lie at the intersection of the 
tangent line with the boundary of the quantum maps set.
Such a choice leads to the tightest bound 
\eref{eq:CSbound}:~$\mathcal{F}_\t{as}^\t{\tiny CS}\!=\!1/(\varepsilon_{+}\varepsilon_{-})$,
which we refer to as the \emph{CS bound}.
Notably, the CS bound is thus fully dictated by the ``distances'' $\varepsilon_\pm$
between $\Lambda_{\varphi_0}$ and the boundary along the 
tangent (see \figref{fig:CSpic}) that mathematically correspond to 
the largest possible values of $\epsilon_\pm$ for which the
CJ matrices of $\Pi_\pm$ are still positive semi-definite.

\subsubsection{Quantum Simulation bound}
\label{sub:QSbound}

In \citep{Matsumoto2010}, a natural generalisation of the channel
CS  has been proposed, which---as schematically presented in \figref{fig:SQLmethods}(\textbf{a})---corresponds 
to replacing the classical diagonal state $p_\varphi$ in the second expression of \eqnref{eq:CS}
with a general quantum state $\sigma_\varphi$.
Consequently, the so-called \emph{Quantum Simulation} (QS) of channel $\Lambda_\varphi$ 
is obtained%
\footnote{%
Note that the notion of \emph{quantum simulability} of a quantum map
is equivalent to the \emph{channel} \emph{programmability} concept
introduced in \citep{Ji2008}.
}:
\begin{equation}
\Lambda_{\varphi}\!\left[\varrho\right]=\Phi\!\left[\varrho\otimes\sigma_{\varphi}\right]+O(\delta\varphi^2)=
\textrm{Tr}_{\t{\tiny E}_\Phi \t{\tiny E}_\sigma}\!\!\left\{ U\left(\varrho\otimes
\left|\xi_{\varphi}\right\rangle\!\left\langle \xi_{\varphi}\right|\right)U^{\dagger}\right\} +O(\delta\varphi^2),
\label{eq:QS}
\end{equation}
which again must hold only locally for a given $\varphi_0$.
In order to be able to later utilise the purification-based definitions \eref{eq:QFIPurifEscher}/\eref{eq:QFIPurifFuji} of the QFI,
we have written the purified version of the QS in the second expression above,
in which we have purified both the channel $\Phi$ (by utilising the Stinespring 
dilation theorem \refcol{thm:stinespring} represented pictorially in \figref{fig:stinespring}) and
the auxiliary state $\sigma_\varphi$ containing the complete information about the parameter,
so that $\sigma_{\varphi}\!=\!\textrm{Tr}_{\t{\tiny E}_\sigma}\!\!\left\{ \left|\xi_{\varphi}\right\rangle \!\left\langle \xi_{\varphi}\right|\right\}$.
$\t{E}_\Phi$ and $\t{E}_\sigma$ thus represent the adequate extra Hilbert spaces required
for the purifications to be performed.

By analogous reasoning to the CS case, when \eqnref{eq:QS} holds, we may upper-bound the QFI
of the $N$-particle output state in the scheme of \figref{fig:NCh_Est_Scheme}
as follows:
\begin{eqnarray}
F_{\textrm{Q}}\!\left[\rho_{\varphi}^{N}\right]=F_{\textrm{Q}}\!\left[\Lambda_{\varphi}^{\otimes N}\!\left[\left|\psi_{\textrm{in}}^{N}\right\rangle \right]\right] 
& = &  
F_{\textrm{Q}}\!\left[\Phi^{\otimes N}\!\left[\left|\psi_{\textrm{in}}^{N}\right\rangle\!\left\langle\psi_{\textrm{in}}^{N}\right| \otimes \sigma_\varphi^{\otimes N}\right]\right]
= 
F_{\textrm{Q}}\!\left[\bar\Pi\!\left[\sigma_\varphi^{\otimes N}\right]\right] \nonumber \\
 & \le & 
F_{\textrm{Q}}\!\left[\sigma_\varphi^{\otimes N}\right]\;=\; N\,F_{\textrm{Q}}\!\left[\sigma_\varphi\right],
\label{eq:QFIQSbound}
\end{eqnarray}
where we have introduced the $\varphi$\emph{-independent} map 
$\bar\Pi\!\left[\bullet\right]\!=\!\Phi^{\otimes N}\!\left[\left|\psi_{\textrm{in}}^{N}\right\rangle\!\left\langle\psi_{\textrm{in}}^{N}\right|\otimes\bullet\right]$
to make it clear that it may only decrease the overall QFI (see \secref{sub:QFIproperties}), what 
leads then to the linearly-scaling upper limit \eref{eq:QFIQSbound},
and thus the desired $\mathcal{F}_{\textrm{as}}^\t{\tiny bound}\!=\!F_\t{Q}\!\left[\sigma_\varphi\right]$.
Notice that consistently, by replacing $\sigma_\varphi$ with a diagonal state $p_\varphi$,
we would recover \eqnref{eq:CSbound} of the CS method.
Hence, the CS may indeed be treated as a special type of the QS and
\eqnref{eq:QFIQSbound} actually serves as an alternative proof
of the asymptotic SQL-like precision scaling for classically simulable maps, which we have previously derived
basing on the concept of $N$ independent random variables associated with each channel use. 

Similarly as in the case of the CS method, a \emph{quantum simulable} channel may admit many decompositions \eref{eq:QS}
and the optimal one must yield the lowest $F_{\textrm{Q}}\!\left[\sigma_{\varphi}\right]$.
Therefore, without loss of generality, in the search for the optimal QS,
we may take $U$ in \eqnref{eq:QS} to act on the full purified system, i.e.~also on the $\t{E}_{\sigma}$ 
space. This enlarges the set of all possible
QSs beyond the original ones, for which $U\!=\! U^{\t{\tiny S}\t{\tiny E}_{\Phi}}\!\otimes\!\mathbb{I}^{\t{\tiny E}_\sigma}$,
and yields $\mathcal{F}_{\textrm{as}}^\t{\tiny bound}\!=\! F_{\textrm{Q}}\!\left[\left|\xi_{\varphi}\right\rangle \right]$,
which, due to the purification-based definition \eref{eq:QFIPurifEscher} of the QFI,
cannot be smaller than $F_{\textrm{Q}}\!\left[\sigma_{\varphi}\right]$.
As a matter of fact, \eqnref{eq:QFIPurifEscher} ensures that for any QS employing
$\sigma_{\varphi}$, there exists an ``enlarged'' decomposition
\eref{eq:QS} leading to the same $\mathcal{F}_{\textrm{as}}^\t{\tiny bound}\!=\!F_{\textrm{Q}}\!\left[\sigma_{\varphi}\right]\!=\!F_{\textrm{Q}}\!\left[\left|\xi_{\varphi}\right\rangle \right]$
with $\ket{\xi_{\varphi}}$ being then the minimal purification in \eqnref{eq:QFIPurifEscher}.

Importantly, we prove in \appref{chap:appQSasCE} that, in order for the QS \eref{eq:QS}
to be locally feasible at $\varphi_0$ and lead to a finite asymptotic bound, $\Lambda_{\varphi}$ of rank
$r$ must admit Kraus operators $\left\{ K_{i}(\varphi)\right\} _{i=1}^{r}$
that satisfy at $\varphi_0$ conditions:
\begin{equation}
\textrm{i}\sum_{i=1}^{r}\dot{K}_{i}(\varphi_0)^{\dagger}K_{i}(\varphi_0)=0\qquad\textrm{and}\qquad\sum_{i=1}^{r}\dot{K}_{i}(\varphi_0)^{\dagger}\dot{K}_{i}(\varphi_0)=\frac{1}{4}\!\left.F_{\textrm{Q}}\!\left[\left|\xi_{\varphi}\right\rangle\right]\right|_{\varphi_0}\mathbb{I}.
\label{eq:QSconds}
\end{equation}
Hence, by optimising over all locally equivalent Kraus representations
of $\Lambda_{\varphi}$---the ones related to one another by rotations
\eref{eq:KrausReps} generated by any Hermitian $\mathsf{h}$---that
satisfy constraints \eref{eq:QSconds}, we may determine the
asymptotic bound given by the optimal local QS, which we refer to
as the \emph{QS bound} -- $\mathcal{F}_{\textrm{as}}^\t{\tiny QS}$,
as follows:
\begin{equation}
\mathcal{F}_{\textrm{as}}^\t{\tiny QS}=\min_{\mathsf{h}}\,\lambda\quad\textrm{ s.t. \;}
\;\alpha_{\tilde{K}}=\frac{\lambda}{4}\,\mathbb{I},\;
\beta_{\tilde{K}}=0,
\label{eq:QSbound}
\end{equation}
where $\alpha_{\tilde{K}}\!=\!\sum_{i=1}^{r}\dot{\tilde{K}}_{i}(\varphi)^{\dagger}\dot{\tilde{K}}_{i}(\varphi)$,
$\beta_{\tilde{K}}\!=\!\textrm{i}\sum_{i=1}^{r}\dot{\tilde{K}}_{i}(\varphi)^{\dagger}\tilde{K}_{i}(\varphi)$,
$\mathsf{h}$ represents as before Hermitian generators locally shifting
the first derivatives of Kraus operators,
and $\lambda$ has the interpretation of $\mathcal{F}_{\textrm{as}}^\t{\tiny bound}\!=\!F_{\textrm{Q}}\!\left[\left|\xi_{\varphi}\right\rangle \right]$.

Yet, one should note that by generalising 
the CS to QS, we have paid the price of losing the intuitive geometrical
interpretation, as the set of quantum simulable channels now
contains all maps that locally admit a Kraus representation
satisfying conditions \eqref{eq:QSconds}, which, however,
cannot be rewritten neatly at the level of the channel CJ
representation. On the other hand, as the CS
method may be interpreted as the more general
QS method with an extra constraint forcing $\sigma_\varphi$ in \eqnref{eq:QS}
to be diagonal, not only \emph{all $\varphi$-non-extremal channels
must be quantum simulable}, but also the QS bound must be
at least as tight for them as the CS one.
In other words, whenever the CS bound \eref{eq:CSbound} is 
constructable:~$\mathcal{F}_\t{as}^\t{\tiny QS}\!\le\!\mathcal{F}_\t{as}^\t{\tiny CS}$.

\subsubsection{Channel Extension bound}
\label{sub:CEbound}

In the \emph{Channel Extension} (CE) method---as shown in 
\figref{fig:SQLmethods}(\textbf{b})---each of the $N$ channels
in the scheme of \figref{fig:NCh_Est_Scheme} is \emph{extended},
i.e.~appended an ancillary particle unaffected by the
action of $\Lambda_\varphi$ exactly in the same fashion as 
discussed in \secref{sub:ExtChQFI} while introducing 
the extended-channel QFI \eref{eq:ExtChQFI}. 
As such an extension can only improve the overall 
attainable precision,
the $N$-channel QFI \eref{eq:NChQFI}
can then be trivially upper-limited by the
corresponding
$N$\emph{-extended-channel QFI},
which importantly can always be upper-bounded as follows
\citep{Fujiwara2008}:
\begin{equation}
\mathcal{F}\!\left[\Lambda_{\varphi}^{\otimes N}\right]\le\mathcal{F}\!\left[\left(\Lambda_{\varphi}\otimes\mathcal{I}\right)^{\otimes N}\right]\quad\le\quad4\,\min_{\mathsf{h}}\left\{ N\,\left\Vert \alpha_{\tilde{K}}\right\Vert +N(N-1)\left\Vert \beta_{\tilde{K}}\right\Vert ^{2}\right\} ,
\label{eq:FujiBound}
\end{equation}
where as before:~$\left\Vert \dots\right\Vert $ represents the operator norm, $\alpha_{\tilde{K}}\!=\!\sum_{i=1}^{r}\dot{\tilde{K}}_{i}(\varphi)^{\dagger}\dot{\tilde{K}}_{i}(\varphi)$,
$\beta_{\tilde{K}}\!=\!\textrm{i}\sum_{i=1}^{r}\dot{\tilde{K}}_{i}(\varphi)^{\dagger}\tilde{K}_{i}(\varphi)$,
and $\mathsf{h}$ is the generator of local Kraus-representation rotations
\eref{eq:KrausReps}. Crucially, if there exists a Kraus representation
for which $\beta_{\tilde{K}}\!=\!0$ so that the second term in \eqnref{eq:FujiBound} vanishes, 
$\mathcal{F}\!\left[\Lambda_{\varphi}^{\otimes N}\right]$
must asymptotically scale at most linearly in $N$. Hence \citep{Fujiwara2008}:

\begin{mydef}[$\beta_{\tilde{K}}\!=\!0$ condition]
\label{def:BetaKcond}~\\
\emph{The asymptotic SQL-like scaling of precision is assured in the scheme of \figref{fig:NCh_Est_Scheme} 
for a given channel $\Lambda_\varphi$ of rank $r$ and the parameter value $\varphi_0$,
if one may find, for a particular set Kraus operators
$\{K_{i}(\varphi)\}_{i=1}^{r}$ of $\Lambda_\varphi$, a Hermitian matrix $\mathsf{h}$
that satisfies at $\varphi_0$:}
\begin{equation}
\sum_{i,j=1}^{r}\mathsf{h}_{ij}\, K_{i}(\varphi_0)^{\dagger}K_{j}(\varphi_0)\;=\;\textrm{i}\sum_{i=1}^{r}\dot{K}_{i}(\varphi_0)^{\dagger}K_{i}(\varphi_0).
\label{eq:BetaKCond}
\end{equation}
\end{mydef}

Notice that the above requirement,
which we term as the \emph{$\beta_{\tilde{K}}\!=\!0$ condition},
turns out to be very effective, as for it to be applicable one needs only
to know a particular Kraus representation%
\footnote{In fact, only the form of all $K_i$ and $\dot K_i$ for a given, fixed $\varphi_0$.}
of a single channel without any further details of the single-particle evolution \citep{Chaves2013}.
Furthermore, according to our best knowledge, 
no example of a parametrised quantum channel has been found that 
does not fulfil \eqnref{eq:BetaKCond}, but still is 
asymptotically constrained to follow the SQL-like precision scaling%
\footnote{%
Moreover, \eqnref{eq:BetaKCond} has been proved to ensure the
asymptotic SQL-like scaling of precision even when one allows for \emph{feedback}
in the $N$-parallel--channels scheme of \figref{fig:NCh_Est_Scheme} \citep{Escher2011}.
}.

What is more, \eqnref{eq:FujiBound} allows to construct for any channel, that 
admits a generator $\mathsf{h}$ fulfilling the condition \eref{eq:BetaKCond}, 
an upper bound on the \emph{asymptotic
extended-channel QFI} as follows%
\footnote{Yet, we conjecture that the bound \eref{eq:CEbound} is actually tight, so that the CE bound coincides with the asymptotic
extended-channel QFI, i.e.~$\mathcal{F}_{\textrm{as}}\!\left[\Lambda_{\varphi}\otimes\mathcal{I}\right]=\mathcal{F}_{\textrm{as}}^\t{\tiny CE}$.}:
\begin{equation}
\mathcal{F}_{\textrm{as}}\!\left[\Lambda_{\varphi}\otimes\mathcal{I}\right]=
\lim_{N\rightarrow\infty}\!\frac{\mathcal{F}\!\left[\left(\Lambda_{\varphi}\otimes\mathcal{I}\right)^{\otimes N}\right]}{N}
\quad \le \quad
\mathcal{F}_{\textrm{as}}^\t{\tiny CE}=4\,\min_{\underset{\beta_{\tilde{K}}=0}{\mathsf{h}}}\;\left\Vert \sum_{i=1}^{r}\dot{\tilde{K}}_{i}(\varphi)^{\dagger}\dot{\tilde{K}}_{i}(\varphi)\right\Vert,
\label{eq:CEbound}
\end{equation}
which also naturally  consitutes the required asymptotic bound
$\mathcal{F}_{\textrm{as}}\!\left[\Lambda_{\varphi}\right]\!\le\!\mathcal{F}_{\textrm{as}}^\t{\tiny bound}$ in \eqnref{eq:QFIAsBound}
that we refer to as the \emph{CE bound} -- $\mathcal{F}_{\textrm{as}}^\t{\tiny CE}$.
We have explicitly written the form of $\alpha_{\tilde K}$ in \eqnref{eq:CEbound} to emphasise
the similarity between the CE bound and the extended-channel QFI \eref{eq:ExtChQFIPurif}. The
essential difference between \eqnsref{eq:ExtChQFIPurif}{eq:CEbound} is the 
$\beta_{\tilde{K}}\!=\!0$ condition \eref{eq:BetaKCond}
yielding consistently $\mathcal{F}_{\textrm{as}}^\t{\tiny CE}\!\ge\!\mathcal{F}\!\left[\Lambda_{\varphi}\!\otimes\!\mathcal{I}\right]\!\ge\!\mathcal{F}\!\left[\Lambda_{\varphi}\right]$
and leaving room for potential asymptotic quantum enhancement of precision.
Despite the extra constraint \eref{eq:BetaKCond} imposed
in \eqnref{eq:CEbound}, $\mathcal{F}_{\textrm{as}}^\t{\tiny CE}$
may always be computed similarly to \eqnref{eq:ExtChQFIPurif} by
reformulating the minimisation in \eqnref{eq:CEbound} 
into an SDP---see \appref{chap:appFinNCEasSDP}.
Although the corresponding SDP is always efficiently solvable only
numerically, it may be utilised to identify the non-zero entries of the optimal generator $\mathsf{h}$ in \eqnref{eq:CEbound}
and their complex structure, as the numerical accuracy of the SDP solutions
may always be quantified. 
Hence, with help of the SDP we may then construct an ansatz for $\mathsf{h}$ in \eqnref{eq:CEbound}, 
with help of which the minimisation over $\mathsf{h}$ may be eventually performed analytically.

Importantly, one should note that the CE bound \eref{eq:CEbound}
resembles the QS bound \eref{eq:QSbound}, but \emph{without}
the additional constraint in \eqnref{eq:QSconds} forcing the operator $\alpha_{\tilde{K}}$ to be
proportional to identity. Hence, such an observation \emph{proves} that 
not only the CE method applies to a wider class of parametrised 
quantum channels than the QS method---and hence
also than the CS and RLD methods being further restricted only to the $\varphi$-non-extremal channels---but 
also the QS bound \eref{eq:QSbound} can never 
outperform its CE equivalent, so that 
most generally:~$\mathcal{F}_{\textrm{as}}^\t{\tiny CE}\le\mathcal{F}_{\textrm{as}}^\t{\tiny QS}\le\mathcal{F}_{\textrm{as}}^\t{\tiny CS}$.
On the other hand, we prove in \appref{chap:appCEvsRLD} that
also the RLD bound \eref{eq:RLDbound} applied
to any $\varphi$-non-extremal map can never lead to a tighter bound
on the asympotic QFI, so that also $\mathcal{F}_{\textrm{as}}^\t{\tiny CE}\le\mathcal{F}^\t{\tiny RLD}\!\left[\Lambda_{\varphi}\otimes\mathcal{I}\right]$
and the CE method is indeed most effective out of the ones presented in this work.

Lastly, let us note that by appending \emph{more than one ancillary particle per channel},
while performing the extension in \eqnref{eq:FujiBound}, we
could only improve the estimation precision (as trivially for any $k\!>\!1$:
$\mathcal{F}\!\left[\Lambda_{\varphi}\otimes\mathcal{I}\right]\!\le\!\mathcal{F}\!\left[\Lambda_{\varphi}\otimes\mathcal{I}^{\otimes k}\right]$
implying 
$\mathcal{F}_{\textrm{as}}\!\left[\Lambda_{\varphi}\otimes\mathcal{I}\right]\!\le\!\mathcal{F}_{\textrm{as}}\!\left[\Lambda_{\varphi}\otimes\mathcal{I}^{\otimes k}\right]$)
and obtain a larger $\mathcal{F}_{\textrm{as}}^\t{\tiny bound}$,
and thus a weaker upper bound on $\mathcal{F}_{\textrm{as}}\!\left[\Lambda_{\varphi}\right]$,
what worsens the method. However, for a given  
$\Lambda_\varphi$ which satisfies \eqnref{eq:BetaKCond}, 
one may wonder whether 
the asymptotic SQL-like scaling can be beaten
by just increasing $k$ at some sufficiently large rate with $N$.
Unfortunately, this can never be the case, as
by adequately extending an effective 
\emph{$k$-ancilla--channel}:~$\Lambda_{\varphi}\otimes\mathcal{I}^{\otimes k}$,
and applying to it the bound \eref{eq:FujiBound},
one may easily verify that the operator norms in \eqnsref{eq:FujiBound}{eq:CEbound}
remain unaltered for any $k\!\ge\!1$. Hence, not only
\eqnref{eq:BetaKCond} still constitutes a sufficient condition for the asymptotic SQL, but 
also the CE bound takes the form of \eref{eq:CEbound} regardless of $k$.
Such a behaviour, might have been expected at least for the $\varphi$-non-extremal channels,
as a single ancilla is enough to provide a sufficient extension for the CJ representation of the map $\Lambda_\varphi$
to be constructable, what is manifested by the RLD bound \eref{eq:RLDbound} also taking the same form 
independently of the ancilla number $k$.

Hence, in general, the \emph{only} way to surpass the CE bound \eref{eq:CEbound}
is to find a way to alter the form of the channel considered, so that its Kraus representation 
\emph{ceases} to fulfil the $\beta_{\tilde{K}}\!=\!0$ condition \eref{eq:BetaKCond}.
For instance, this has been achieved by considering channels that 
due to decoherence satisfy \eqnref{eq:BetaKCond} at finite times, but not in the limit
of infinitely short evolution. Such a behaviour
has been shown to emerge when accounting for the
Non-Markovianity effects in the single-particle evolution 
\citep{Matsuzaki2011,Chin2012}, but also when considering
dephasing noise perpendicular to the phase-encoding evolution part \citep{Chaves2013}.
In fact, for the second case it has been also shown that, 
due to the $\beta_{\tilde{K}}\!=\!0$ condition \eref{eq:BetaKCond}
being violated at small times, one may actually benefit in this regime from
increasing the number of ancillary particles, which may be then
utilised to perform the \emph{quantum error-correction}
and fully retain the HL \citep{Duer2014}.

\subsubsection{\caps{Example:} Noisy-phase--estimation channels}
\label{sub:SQLboundsExamples}

Similarly to the discussion of the single channel QFI measures in \secref{sec:ChEstLocal}, 
we consider the noisy-phase--estimation models introduced in \secref{sub:noise_models},
in which the phase is encoded onto the particle represented by a qubit via the $\e^{-\ii\hat\sigma_3\varphi}$ rotation and 
the decoherence is specified by one of the \emph{dephasing, depolarisation, loss}
and \emph{spontaneous emission} maps (see \figref{fig:noise_models} and \tabref{tab:noise_models}).
In case of the $N$-parallel--channel estimation scheme of \figref{fig:NCh_Est_Scheme},
we model the evolution of each of the $N$ particles by one of the above models, 
which thus determines the single-particle channel $\Lambda_\varphi$.
In \tabref{tab:SQLboundsQEnh}, we present in an increasing order 
the corresponding bounds on the asymptotic QFI \eref{eq:QFIAs}
obtained via the methods analysed in this section, where in the last row we
list the upper limits on the maximal quantum enhancement of precision \eref{eq:qEnh}
specified by the ratios of the adequate CE bounds and the single-channel QFIs presented in \tabref{tab:ChQFISingle}, 
i.e.~$\chi\!\left[\Lambda_\varphi\right]\!\le\!\sqrt{\mathcal{F}_\t{as}^\t{\tiny CE}\!\left[\Lambda_\varphi\right]/\mathcal{F}\!\left[\Lambda_\varphi\right]}$.

\begin{table}[t!]
\begin{center}
\begin{tabular}{|M{3cm}||M{2cm}|M{3cm}|M{1.5cm}|M{3.5cm}|N}
\hline 
\textbf{Noise models:} & \emph{Dephasing} & \emph{Depolarization} & \emph{Loss} & \emph{Spontaneous emission}
&\\[12pt]
\hline 
$\mathcal{F}_{\textrm{as}}^\t{\tiny CE}$ {\footnotesize \eref{eq:CEbound}} & $\frac{\eta^{2}}{1-\eta^{2}}$ & $\frac{2\eta^{2}}{\left(1-\eta\right)\left(1+2\eta\right)}$ & $\frac{\eta}{1-\eta}$ & $\frac{4\eta}{1-\eta}$
&\\[12pt]
\hline 
$\mathcal{F}_{\textrm{as}}^\t{\tiny QS}$ {\footnotesize\eref{eq:QSbound}} & $\frac{\eta^{2}}{1-\eta^{2}}$ & $\frac{2\eta^{2}}{\left(1-\eta\right)\left(1+2\eta\right)}$ & $\frac{\eta}{1-\eta}$ & n.a.
&\\[12pt]
\hline 
$\mathcal{F}^\t{\tiny RLD}\!\left[\Lambda_{\varphi}\!\otimes\!\mathcal{I}\right]$  {\footnotesize\eref{eq:RLDbound}} & $\frac{\eta^{2}}{1-\eta^{2}}$ & $\frac{2\eta^{2}(1+\eta)}{\left(1-\eta\right)\left(1+3\eta\right)}$ & n.a. & n.a.
&\\[12pt]
\hline 
$\mathcal{F}_{\textrm{as}}^\t{\tiny CS}$ {\footnotesize\eref{eq:CSbound}} & $\frac{\eta^{2}}{1-\eta^{2}}$ & $\frac{4\eta^{2}}{\left(1-\eta\right)\left(1+3\eta\right)}$ & n.a. & n.a.
&\\[12pt]
\hline 
\hline 
$\chi\!\left[\Lambda_{\varphi}\right]$ {\footnotesize\eref{eq:qEnh}} & $=\sqrt{\frac{1}{1-\eta^{2}}}$ & $\le\sqrt{\frac{2}{(1-\eta)(1+2\eta)}}$ & $=\sqrt{\frac{1}{1-\eta}}$ & $\le\sqrt{\frac{4}{1-\eta}}$
&\\[12pt]
\hline
\end{tabular}
\end{center}
\caption[CE, QS, RLD and CS bounds on the asymptotic channel QFI]{%
\textbf{CE, QS, RLD and CS bounds on the asymptotic channel QFI \eref{eq:QFIAs}}
for the noisy-phase--estimation channels introduced in \secref{sub:noise_models} and depicted in \figref{fig:noise_models}.
For each noise model, we also present an upper bound on the \emph{maximal quantum enhancement of precision} \eref{eq:qEnh}, 
which is obtained by utilising the corresponding CE bound and the channel QFI presented in \tabref{tab:ChQFISingle}, 
so that $\chi\!\left[\Lambda_\varphi\right]\!\le\!\sqrt{\mathcal{F}_\t{as}^\t{\tiny CE}\!\left[\Lambda_\varphi\right]/\mathcal{F}\!\left[\Lambda_\varphi\right]}$.
Yet, in the case of dephasing and loss channels the corresponding values of the limits on $\chi\!\left[\Lambda_{\varphi}\right]$
have been shown to be attainable \citep{Ulam2001,Caves1981}.\\$\t{~}$\hfill[n.a.---not~applicable]}
\label{tab:SQLboundsQEnh}
\end{table}

Again, as only the phase estimation scenarios with \emph{dephasing} and \emph{depolarisation} noise-types 
lead to the effective $\varphi$\emph{-non-extremal} channels, only for these we may 
limit the asymptotic precision by utilising the CS and RLD methods. 
In case of the \emph{dephasing} noise, which yields the simplest, rank-2 
$\Lambda_\varphi$ channel (see \secsref{sub:noise_models}{sub:noise_models_geom}) all
the asymptotic SQL-like bounds in \tabref{tab:SQLboundsQEnh} take
the same form, proving that the geometric CS method is sufficient
to determine the maximal quantum enhancement of precision.
It is so, as the CS bound \eref{eq:CSbound} has indeed been shown to 
coincide with the asymptotic channel QFI \eref{eq:QFIAs} \citep{Ulam2001,Demkowicz2015}.
As the \emph{depolarisation} noise yields a full-rank channel
that does not lie on any of the boundaries of the CPTP-maps space---see 
\secref{sub:noise_models_geom}---the RLD bound \eref{eq:CSbound}
proves for it to be tighter than the CS one. Nevertheless, the more general 
QS method provides an even better bound, but most importantly 
also applies to the $\varphi$\emph{-extremal} channel
obtained accounting for the \emph{loss} noise-type. 
Furthermore, for all these three channels
the QS bounds \eref{eq:QSbound} turn out to be as accurate as the CE ones,
what may be also verified by inspecting the relevant optimal 
Kraus-representation generators $\mathsf{h}$ of the CE method 
listed in \appref{chap:appOptHExtCh}, which indeed
yield $\alpha_{\tilde{K}}$ to be proportional to identity---satisfying 
the extra QS constraint \eref{eq:QSconds}.
However, in the case of \emph{spontaneous-emission} noise, the QS 
method ceases to work, as the $\beta_{\tilde{K}}\!=\!0$
condition \eref{eq:BetaKCond} fixes $\alpha_{\tilde{K}}$ to be disproportional to identity.
It is so, because---as explained in \secref{sub:noise_models_geom}---the 
spontaneous-emission channel is \emph{extremal} and
the CE method is the only one able to deal with it.

One should also note that the CE bounds, $\mathcal{F}_{\textrm{as}}^\t{\tiny CE}$,
indeed provide the best limits on the asymptotic QFI for all the cases and, that is why, we 
employ them to upper-bound the maximal quantum enhancement of precision, $\chi\!\left[\Lambda_{\varphi}\right]$, 
presented in the last row of \tabref{tab:SQLboundsQEnh}.
Let us remark that, for \emph{dephasing} and \emph{loss} noise models,
phase estimation strategies have been found that asymptotically attain the corresponding CE bounds
\citep{Caves1981,Ulam2001,Demkowicz2015}. Hence in these two cases, not only 
$\mathcal{F}_\t{as}\!\left[\Lambda_\varphi\right]\!=\!\mathcal{F}_\t{as}^\t{\tiny CE}$,
but also one may not asymptotically benefit from the channel extension and
$\mathcal{F}_{\textrm{as}}\!\left[\Lambda_{\varphi}\right]\!=\!\mathcal{F}_{\textrm{as}}\!
\left[\Lambda_{\varphi}\otimes\mathcal{I}\right]$ (see \eqnref{eq:CEbound}).
On the other hand, as shown in \tabref{tab:ChQFISingle}, these two channels 
are also examples of ones for which the extension does not improve the precision at
the single-channel level, i.e.~$\mathcal{F}\!\left[\Lambda_{\varphi}\right]\!=\!\mathcal{F}\!
\left[\Lambda_{\varphi}\otimes\mathcal{I}\right]$, but the problem of relating the two regimes 
we leave open for future research.

Lastly, let us comment that all the bounds presented in \tabref{tab:SQLboundsQEnh} 
correctly diverge as $\eta\!\to\!1$, when we return to the noiseless
unitary phase estimation problem, in which the $1/N^2$ HL limit must be attainable
and any SQL-bounding methods of \secref{sub:SQLAsBounds} must fail.
Geometrically, for $\varphi$-non-extremal channels (here dephasing 
and depolarisation noises) such 
limit corresponds to decreasing the 
distances $\varepsilon_\pm$ to the boundary of the quantum channels set 
in \figref{fig:CSpic}, so that the $\varphi$-non-extremality is eventually lost
and both CS and RLD methods cease to apply (i.e.~the 
CS \eref{eq:CSbound} and RLD \eref{eq:RLDbound} bounds diverge,
as respectively $\mathcal{F}_\t{as}^\t{\tiny CS}\!=\!1/(\varepsilon_-\varepsilon_+)\!\overset{\!\varepsilon_\pm\to0}{\,=}\!\infty$
and $\dot\Omega_{\mathcal{U}_\varphi}$ is no longer contained within the support of $\Omega_{\mathcal{U}_\varphi}$).
On the other hand, in the case of the QS and CE methods, the necessary 
$\beta_{\tilde{K}}\!=\!0$ condition \eref{eq:BetaKCond}
cannot be satisfied for the noiseless unitary evolution,
so that (in contrast to extended-channel QFI \eref{eq:ExtChQFI}
which definition \eref{eq:ExtChQFIPurif} crucially lacks 
the $\beta_{\tilde{K}}\!=\!0$ constraint) $\left\Vert\alpha_{\tilde K}\right\Vert$ diverges in both
\eqnsref{eq:QSbound}{eq:CEbound} as $\eta\!\to\!1$, and
the QS and CE bounds become unbounded.

\subsection{Finite-$N$ CE bound}
\label{sub:FinN_CEbound}

In \secref{sub:SQLAsBounds}, we have presented the \emph{CE method} as 
the most effective one that applies to the broadest class of parametrised 
channels (i.e.~ones satisfying the $\beta_{\tilde K}\!=\!0$ condition \eref{eq:BetaKCond})
and provides the tightest upper limits on the asymptotic QFI \eref{eq:QFIAs} and the
maximal quantum enhancement of precision \eref{eq:qEnh}.
On the other hand, when moderate numbers of particles, $N$, are considered
in the scheme of \figref{fig:NCh_Est_Scheme}, the CE bound \eref{eq:CEbound}
derived for the asymptotic regime, despite still being valid, is far too weak to be useful.
Although for very low values of $N$ the precision can be quantified
numerically, for instance, by brute-force--type methods computing explicitly
the QFI, in the intermediate-$N$ regime---being beyond
the reach of computational power, yet with $N$ too low for the asymptotic
methods to be effective---more accurate bounds should play an 
important role.

We propose the \emph{finite-$N$ CE method} which, despite
still being based \emph{only} on the properties of a \emph{single} channel, provides
bounds on precision that are relevant in the intermediate-$N$ regime.
We utilise the upper limit \eref{eq:FujiBound} on the $N$-extended-channel
QFI and construct the \emph{finite-$N$ CE bound}, $\mathcal{F}_{N}^\t{\tiny CE}$,
as follows
\begin{equation}
\frac{\mathcal{F}\!\left[\left(\Lambda_{\varphi}\otimes\mathcal{I}\right)^{\otimes N}\right]}{N}\quad
\le\quad\mathcal{F}_{N}^\t{\tiny CE}=4\,\min_{\mathsf{h}_N}\left\{ \left\Vert \alpha_{\tilde{K}}\right\Vert +(N-1)\left\Vert \beta_{\tilde{K}}\right\Vert ^{2}\right\},
\label{eq:CEboundN}
\end{equation}
where in contrast to the CE bound \eref{eq:CEbound} we do not 
impose the SQL-bounding condition $\beta_{\tilde{K}}\!=\!0$ \eref{eq:BetaKCond}. 
We rather search for the minimal Kraus representation at each $N$,
so that the optimal generator, $\mathsf{h}_N$, now varies depending on 
the particle number. Importantly, we show in \appref{chap:appFinNCEasSDP} that
$\mathcal{F}_{N}^\t{\tiny CE}$ can similarly to $\mathcal{F}_{\textrm{as}}^\t{\tiny CE}$ 
be efficiently evaluated numerically by recasting the minimisation over $\mathsf{h}_N$ in \eqnref{eq:CEboundN}
into an SDP. Note that for $N\!=\!1$, the finite-$N$ CE bound \eref{eq:CEboundN} coincides with
the extended-channel QFI \eref{eq:ExtChQFIPurif}, i.e.~$\mathcal{F}_{N=1}^\t{\tiny CE}\!=\!\mathcal{F}\!\left[\Lambda_{\varphi}\otimes\mathcal{I}\right]$,
whereas for $N\!\to\!\infty$, (if a given channel allows the
$\beta_{\tilde{K}}\!=\!0$ condition to be fulfilled) $\mathcal{F}_{N}^\t{\tiny CE}$
attains the CE bound \eref{eq:CEbound}, i.e.~$\mathcal{F}_{N\rightarrow\infty}^\t{\tiny CE}\!=\!\mathcal{F}_{\textrm{as}}^\t{\tiny CE}$,
and the optimal $\mathsf{h}_N$ in \eqnref{eq:CEboundN} converges to
$\mathsf{h}$ minimising \eqnref{eq:CEbound}.
As a consequence, $\mathcal{F}_{N}^\t{\tiny CE}$ varies smoothly
between these two regimes and it may provide much more accurate bounds 
on precision than its asymptotic version%
\footnote{%
Nevertheless, we conjecture that,
in order to actually construct a tight bound in \eqnref{eq:CEboundN}
that when optimised coincides with $\mathcal{F}\!\left[\!\left(\Lambda_\varphi\otimes\mathcal{I}\right)^{\otimes N}\right]$
(or even $\mathcal{F}\!\left[\Lambda_\varphi^{\otimes N}\right]$) for finite $N$,
one may not restrict to \emph{single}-channel generators $\mathsf{h}_N$ as in 
the case of \eqnref{eq:CEboundN}, but 
must also account for purifications that are generated by operations performed
collectively on the output of many channels acting in parallel.
}.

On the other hand, let us emphasise that the finite-$N$ CE bound \eref{eq:CEboundN}
also applies to channels for which it is impossible to
fulfil the $\beta_{\tilde{K}}\!=\!0$ condition \eref{eq:BetaKCond}
and thus the CE method fails.
Moreover, as \eqnref{eq:CEboundN} is solved independently for each $N$, 
the finite-$N$ CE method may also be utilised in scenarios
in which the single-particle evolution depends on the number of particles employed,
or in other words, in which the overall scheme may be viewed as the one of \figref{fig:NCh_Est_Scheme} 
but with each channel now denoted as $\Lambda_{\varphi,N}$ to indicate that
its form may change with $N$. 
Physically, for instance, such a description is valid 
when analysing schemes in which the strength of decoherence varies 
depending on the number of particles involved in an experiment \citep{Wasilewski2010},
or in frequency estimation scenarios discussed in \secref{sec:loc_freq_est},
in which one may vary the time duration of each experimental ``shot''
given a particular $N$ \citep{Huelga1997}.
As a matter of fact, $\mathcal{F}_{N}^\t{\tiny CE}$ has been explicitly used
in \citep{Chaves2013}, where a frequency estimation scheme 
was considered and it has been numerically
demonstrated that the finite-$N$ CE bound \eref{eq:CEboundN} allows to prove 
the correct asymptotic super-classical precision scaling, $1/N^{5/3}$,
reaching \emph{beyond} the SQL.

Returning again to the noisy-phase--estimation channels of \figref{fig:noise_models}
with uncorrelated noise modelled by respectively:~\emph{dephasing, depolarisation, loss} and \emph{spontaneous emission} maps;
we study numerically via SDPs the form of their corresponding finite-$N$ CE bounds \eref{eq:CEboundN}.
Surprisingly, we observe that in all four cases $\mathcal{F}_N^\t{\tiny CE}$ may be simply related to its asymptotic form%
\footnote{In case of the spontanous-emission--noise the formula \eref{eq:F_CE_anal} is valid only for $N\ge2$, what we suspect to be a consequence
of the spontaneous emission channel being an extremal map (see \secref{sub:noise_models_geom} and \tabref{tab:noise_models_extrem}).} 
as
\begin{equation}
\mathcal{F}_{N}^\t{\tiny CE}=\frac{N\,\mathcal{F}_{\textrm{as}}^\t{\tiny CE}}{N+\mathcal{F}_{\textrm{as}}^\t{\tiny CE}},
\label{eq:F_CE_anal}
\end{equation}
where one should substitute for $\mathcal{F}_{\textrm{as}}^\t{\tiny CE}$
the corresponding CE bounds presented in \tabref{tab:SQLboundsQEnh}.
Notice, that for all but the spontaneous emission noise-models the form of the finite-$N$
CE bound \eref{eq:F_CE_anal} allows us to establish a connection between the extended-channel QFI \eref{eq:ExtChQFI} and
the CE bound \eref{eq:CEbound}, as by writing \eqnref{eq:F_CE_anal} for $N\!=\!1$ we obtain
the relation $\mathcal{F}\!\left[\Lambda_\varphi\otimes\mathcal{I}\right]\!=\!\mathcal{F}_{\textrm{as}}^\t{\tiny CE}/(1+\mathcal{F}_{\textrm{as}}^\t{\tiny CE})$
that may be verified for dephasing, depolarisation and loss noise-types by substituting their 
$\mathcal{F}\!\left[\Lambda_\varphi\otimes\mathcal{I}\right]$ and $\mathcal{F}_{\textrm{as}}^\t{\tiny CE}$
listed in \tabsref{tab:ChQFISingle}{tab:SQLboundsQEnh}.

\subsection{\caps{Example:} $N$-qubit phase estimation in the presence of loss and dephasing}
\label{sub:Example_PhaseEst_Loss_Deph}

\begin{figure}[!t]
\includegraphics[width=1\columnwidth]{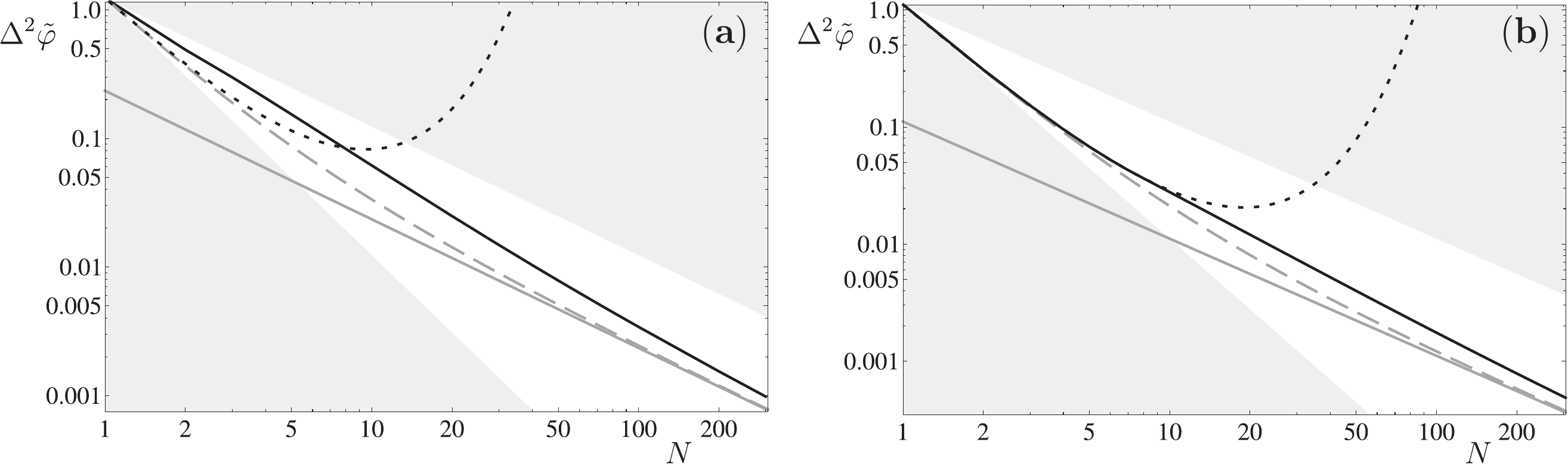}
\caption[Applicability of the CE bounds in noisy-phase--estimation scenarios]{%
\textbf{Applicability of the CE bounds in noisy-phase--estimation scenarios.}\\
Asymptotic (\emph{solid grey line}, \ref{eq:CEbound}) and finite-$N$ (\emph{dashed grey line}, \ref{eq:CEboundN})
CE bounds compared with $\Delta^2\tilde\varphi$ achieved by various strategies
in noisy phase estimation. Shaded areas represent regions in which the estimator MSE
is either worse than SQL, $1/N$, or surpasses the HL, $1/N^2$.\\
(\textbf{a}) \emph{Dephasing noise} of strength $\eta=0.9$: (\emph{solid black line}) -- 
spin-squeezed states and Ramsey-type measurements \citep{Ma2011}, 
(\emph{dotted black line}) -- QCRB \eref{eq:QCRB} evaluated for GHZ/NOON states \eref{eq:NOON}.\\
(\textbf{b}) \emph{Loss model} with $\eta=0.9$ particle survival probability: (\emph{solid black line}) -- QCRB \eref{eq:QCRB} 
minimised over input states, (\emph{dotted black line}) -- QCRB \eref{eq:QCRB} evaluated for GHZ/NOON states \eref{eq:NOON}.
}
\label{fig:plots_CEbounds}
\end{figure}

For \emph{dephasing} and \emph{loss} decoherence models of \figref{fig:noise_models}, we show explicitly
in \figref{fig:plots_CEbounds} both the asymptotic \eref{eq:CEbound} and the finite-$N$ \eref{eq:CEboundN} CE bounds
accompanied by the plots of the actual estimator variances and QCRBs \eref{eq:QCRB} evaluated for
particular phase estimation strategies that are optimal either
in the small or large particle-number regime.

In the case of \emph{dephasing noise}, we consider a Ramsey 
spectroscopy setup of \citep{Wineland1992,Wineland1994,Bollinger1996} in which the particles
(atoms) are prepared in a spin-squeezed state \citep{Ulam2001,Ma2011}.
The parameter is then encoded in the phase of a unitary rotation generated
by the total angular momentum of the atoms that independently experience
dephasing \citep{Huelga1997}, what constitutes an example of
the phase estimation scheme of \figref{fig:Ph_Est_Scheme} with uncorrelated noise.
By measuring the total angular momentum perpendicular to the one 
generating the estimated phase change, the parameter may be reconstructed
by registering the atomic population distribution. Such a scenario 
may be interpreted exactly as the Mach-Zehder interferometry 
setup of \figref{fig:MZinter_cl}
with a photon-number difference measurement, where 
the atoms are represented by a finite number of photons involved
and the uncorrelated dephasing noise accounts for the modal mismatch leading
to imperfect visibility of the interferometer \citep{Demkowicz2015}.
We plot the MSE achieved in such a protocol as the \emph{solid black line}
in \figref{fig:plots_CEbounds}(\textbf{a}),
which importantly saturates the dephasing CE bound of \tabref{tab:SQLboundsQEnh} (\emph{solid grey line})
proving the scheme to be asymptotically optimal.
For comparison, the maximal precision theoretically achievable
by the GHZ input states (or equivalently the NOON states when viewed in the modal picture---see \eqnref{eq:NOON}) is 
also shown (\emph{dotted black line}), i.e.~the 
QCRB \eref{eq:QCRB} with QFI:~$\mathcal{F}_{N}^{\textrm{\tiny NOON}}\!\!=\!\eta^{2N}N^2$,
which for low $N$ attains the finite-$N$ CE bound (\emph{dashed grey line}).
Such an observation proves that in experiments with only few particles involved, such as
\citep{Leibfried2004}, it has been correctly assumed to use the GHZ states and completely ignore the
uncorrelated dephasing present.

In \figref{fig:plots_CEbounds}(\textbf{b}),
we depict the precision achieved in the phase estimation
scenario in the presence of losses. The
\emph{loss noise-model} of \figref{fig:noise_models}
acting independently on each particle simulates a process in which
each particle may survive and sense the estimated parameter
with probability $\eta$. In case of the Mach-Zehnder interferometer
of \figref{fig:MZinter_cl},
this effectively corresponds to setting the power transmittance
of both interferometer arms to $\eta$, what we explicitly show in \chapref{chap:MZ_losses}
while discussing in detail the lossy interferometer scenario.
Here, we plot the QCRB \eref{eq:QCRB} for the numerically optimised
$N$-particle (or equivalently $N$-photon) input states
(\emph{solid black line}) and for GHZ/NOON states \eref{eq:NOON} (\emph{dotted black line}),
which for losses lead to the QFI:~$\mathcal{F}_{N}^{\textrm{\tiny NOON}}\!=\!\eta^{N}N^2$.
Crucially, for the loss noise-model one may prove that 
it is optimal to consider only inputs consisting of indistinguishable, bosonic 
particles (see \chapref{chap:MZ_losses} and \citep{Demkowicz2009}), so 
that it is possible to numerically compute and minimise the QCRB,
and thus explicitly perform the optimisation over the $N$-photonic input states.
As a result, we are able to assess the tightness of the
finite-$N$ CE bound, i.e.~$\mathcal{F}\left[\Lambda_\varphi^{\otimes N}\right]\!\le\!\mathcal{F}_N^\t{\tiny CE}$, 
by inspecting the gap between the \emph{solid black} and \emph{dashed grey} lines in \figref{fig:plots_CEbounds}(\textbf{b}).
Although evidently being a good approximation to $\mathcal{F}\left[\Lambda_\varphi^{\otimes N}\right]$
for low and very large particle numbers, 
$\mathcal{F}_N^\t{\tiny CE}$ ceases to be tight in the moderate-$N$ regime,
in which more accurate bounds would require more information about the estimation
capabilities than is contained in the form of a single channel.
As in the previous case, for very low $N$ the effects of decoherence may be ignored
and the GHZ/NOON-state--based strategy (\emph{dotted black line}) 
saturates the finite-$N$ CE bound (\emph{dashed grey line}).

\section{Local frequency estimation in atomic models}
\label{sec:loc_freq_est}

In this section, we study the precision measures as well as the bounds
developed in \secsref{sec:ChEstLocal}{sec:EstNChannels}
in the case of \emph{frequency estimation} problems of \emph{atomic spectroscopy}.
The general Ramsey spectroscopy
setup---as introduced in \citep{Wineland1992,Wineland1994,Bollinger1996}---describes
$N$ identical two-level atoms, i.e.~spin-$1/2$ particles or qubits,
and the estimated parameter is typically
the \emph{detuning}, $\omega$, of an external oscillator frequency
from the atoms transition frequency dictated by their energy spacing.
In general, the evolution of such an $N$-qubit system is modelled by a 
master equation \eref{eq:MEPic} (in the LGKS form \eref{eq:LGKS})  introduced in \secref{sub:Evo_MEq}:
\begin{equation}
\frac{\t{d}}{\t{d} t}\rho_{\omega}^{N}(t)=\sum_{n=1}^{N}\;-\textrm{i}\,\frac{\omega}{2}\left[\hat\sigma_{z}^{(n)},\rho_{\omega}^{N}(t)\right]+\mathcal{L}^{(n)}\!\left[\rho_{\omega}^{N}(t)\right],
\label{eq:freqMasterEq}
\end{equation}
where $\hat\sigma_{z}^{(n)}$ is the Pauli operator generating a unitary
rotation of the $n$-th atom around the $z$ axis in its Bloch-ball
representation%
\footnote{In case of channels specified in \tabref{tab:noise_models}
this just corresponds to setting $\varphi\!=\!\omega t$.}. 
The \emph{uncorrelated noise} 
is represented by the Liouvillian
part $\mathcal{L}^{(n)}$ acting independently on each 
(here the $n$-th) particle, so 
that the integral of \eqnref{eq:freqMasterEq} 
for a given input state $\rho_{\omega}^{N}(0)\!=\!\left|\psi_{\textrm{in}}^N\right\rangle\!\left\langle\psi_{\textrm{in}}^N\right|$
may be interpretted as the $t$-dependent output state of the 
\emph{$N\!$-parallel--channels estimation scheme} 
depicted in \figref{fig:NCh_Est_Scheme}:
$\rho_{\omega}^{N}(t)\!=\!\Lambda_{\omega,t}^{\otimes N}\!\left[\left|\psi_{\textrm{in}}^N\right\rangle\right]$.
As generally the master equation \eref{eq:freqMasterEq} 
accounts for the fact that the decoherence process 
occurs simultaneously and is non-separable from the Hamiltonian part of the evolution,
the form of the single-particle channel $\Lambda_{\omega,t}$ may
be highly non-trivial. Nevertheless, it is in priciple always computable,
as both the CJ and the Kraus representations of $\Lambda_{\omega,t}$ may be
established basing on \eqnref{eq:freqMasterEq} either by rewriting the action of the channel 
with use of the CJ matrix introduced in \secref{sub:CJiso} \citep{Bengtsson2006}, or directly from 
the LGKS form by a more general construction \citep{Andersson2007}. 
As a consequence, this means that in principle all the techniques described
in \secsref{sec:ChEstLocal}{sec:EstNChannels} that assess
the estimation capabilities stemming from the geometry of maps $\Lambda_{\omega,t}$ parametrised by $\omega$
or from the Kraus-representation--optimisation procedures should be applicable.

However, in order to correctly quantify the resources \citep{Huelga1997},
we must assume the complete spectroscopy experiment to take
a \emph{fixed} overall time%
\footnote{For an alternative approach, where rather than the overall time the rate of particle production is fixed, see \citep{Shaji2007}.
} $T$, 
during which the estimation procedure is repeated
$\nu\!=\!T/t$ times, so that by varying the single-shot duration, $t$,
the repetition number, $\nu$, in the scheme of \figref{fig:NCh_Est_Scheme} is altered.
Consequently, the QCRB \eref{eq:QCRB}
on the MSE \eref{eq:MSE} of an estimator $\tilde{\omega}$ of the detuning frequency can be conveniently
rewritten as
\begin{equation}
\Delta^2\tilde{\omega}\,\quad\ge\quad\frac{1}{T\, f_t \!\left[\rho_{\omega}^{N}(t)\right]}\quad\ge\quad\frac{1}{T}\,\frac{1}{\underset{0\le t\le T}{\max} \;f_t \!\left[\Lambda_{\omega,t}^{\otimes N}\!\left[\left|\psi_{\textrm{in}}^N\right\rangle\right]\right]},
\label{eq:freqQCRB}
\end{equation}
where $f_t \!\left[\rho_{\omega}^{N}(t)\right]\!=\! F_{\textrm{Q}}\!\left[\rho_{\omega}^{N}(t)\right]/t$
is now the effective `\emph{QFI per shot duration}'. 
As remarked  in the second expression above,
in order to establish the \emph{minimal} QCRB for a given input, one must also optimise
the single-shot duration, which importantly may change depending on the 
number of the particles employed.
As a result, the form of the single-particle evolution map $\Lambda_{\omega,t}$
in the picture of \figref{fig:NCh_Est_Scheme} varies with $N$ via $t\!=\!t_\t{opt}(N)$, so that
one must be careful when applying the precision-bounding techniques of
\secref{sec:EstNChannels}, which have been derived assuming 
a fixed form of the single-particle channel.
On the other hand, notice that the total time $T$ in \eqnref{eq:freqQCRB} 
plays now the role of $\nu$ in \eqnref{eq:QCRB}, so that the local estimation 
regime (see \secref{sub:LocConseq}) may always be 
guaranteed for a given $t$ by letting $T\!\gg\!t$.
Furthermore, this means that for protocols in which $t_\t{opt}(N)\!\to\!0$ as $N\!\to\!\infty$, the 
QCRB \eref{eq:freqQCRB} is always assured to be saturable in the
asymptotic $N$ limit, which then additionally warrants 
$\nu\!\to\!\infty$.

As now $f_t \!\left[\varrho_\omega(t)\right]$ in \eqnref{eq:freqQCRB}
plays the role of the QFI in \eqnref{eq:QCRB},
we may define with its help the frequency estimation equivalents of all 
the single-channel precision measures discussed in \secref{sec:ChEstLocal},
as well as the $N$-parallel--channel quantities and the bounds derived in \secref{sec:EstNChannels}.
For instance, the channel QFI  \eref{eq:ChQFI} may be reformulated as:
\begin{equation}
\mathfrak{f}\!\left[\Lambda_{\omega}\right]=\max_{0\le t\le T}\max_{\psi_{\textrm{in}}}\, f_{t}\!\left[\Lambda_{\omega,t}\!\left[\left|\psi_{\textrm{in}}\right\rangle \right]\right]
\label{eq:ChQFIfreq}
\end{equation}
that represents now the \emph{maximal channel QFI per shot duration},
which is optimised over not only the single-particle inputs $\left|\psi_{\textrm{in}}\right\rangle$ but 
also the shot duration $t$. In a similar manner, we may define:
$\mathfrak{f}\!\left[\Lambda_{\omega}\otimes\mathcal{I}\right]$ for \eref{eq:ExtChQFI}, 
$\mathfrak{f}^\t{\tiny RLD}\!\left[\Lambda_{\omega}\otimes\mathcal{I}\right]$ for \eref{eq:RLDbound},
$\mathfrak{f}_\t{as}\!\left[\Lambda_{\omega}\right]\!\le\!\mathfrak{f}_\t{as}^\t{\tiny bound}$ for \eref{eq:QFIAsBound}
with CE, QS and CS bounds respectively then reading: 
$\mathfrak{f}_{\textrm{as}}^\t{\tiny CS}\ge\mathfrak{f}_{\textrm{as}}^\t{\tiny QS}\ge\mathfrak{f}_{\textrm{as}}^\t{\tiny CE}$,
and the finite-$N$ CE bound \eref{eq:CEboundN} equivalent $\mathfrak{f}_{N}^\t{\tiny CE}$, 
which now needs to be maximised over $t$
independently for each $N$. Consequently, a $t$-optimised upper bound
on the maximal quantum enhancement of precision \eref{eq:qEnh} may also
be constructed as $\chi\!\left[\Lambda_\varphi\right]\!\le\!\sqrt{\mathfrak{f}_\t{as}^\t{\tiny CE}\!\left[\Lambda_\varphi\right]/\mathfrak{f}\!\left[\Lambda_\varphi\right]}$.

\begin{table}[!t]
\begin{center}
\begin{tabular}{|M{2.5cm}||M{2.3cm}|M{4.3cm}|M{2.2cm}|M{2.2cm}|N}
\hline 
\textbf{Noise models:} & \emph{Dephasing} & \emph{Depolarisation} & \emph{Loss} & \emph{Spontaneous emission}
&\\[16pt]
\hline 
$\mathfrak{f}\!\left[\Lambda_{\omega}\right]$ & $\frac{1}{2\,\textrm{e}\gamma}$ & $\frac{3}{4\,\textrm{e}\gamma}$ & $\frac{1}{\textrm{e}\gamma}$ & $\frac{1}{\textrm{e}\gamma}$ &\\[12pt]
\hline 
$\mathfrak{f}\!\left[\Lambda_{\omega}\!\otimes\!\mathcal{I}\right]$ & $\frac{1}{2\,\textrm{e}\gamma}$  & {\scriptsize $\approx\!1.27\times$}$\frac{3}{4\,\textrm{e}\gamma}$  & $\frac{1}{\textrm{e}\gamma}$ & $\frac{4\tilde{w}}{\gamma\left(1+\textrm{e}^{\tilde{w}/2}\right)^{2}}$
&\\[12pt]
\hline 
$\mathfrak{f}_{N}^\t{\tiny CE}$ \;{\small ($N\ge2$)} & $\!\frac{N}{2\gamma}\frac{w_{1}\!\left[N\right]}{1+\left(\textrm{e}^{w_{1}\!\left[N\right]}-1\right)N}$ & $\!\frac{3N}{4\gamma}\frac{\alpha\, w_{\beta}\left[N\right]}{2+\left(\textrm{e}^{\frac{\alpha}{4}w_{\beta}\left[N\right]}-1\right)\left(\textrm{e}^{\frac{\alpha}{4}w_{\beta}\left[N\right]}+2\right)N}$  & $\!\frac{N}{\gamma}\frac{w_{1}\!\left[N\right]}{1+\left(\textrm{e}^{w_{1}\!\left[N\right]}-1\right)N}$ & $\!\frac{N}{\gamma}\frac{4\, w_{4}\left[N\right]}{4+\left(\textrm{e}^{w_{4}\left[N\right]}-1\right)N}$
&\\[18pt]
\hline 
$\mathfrak{f}_{\textrm{as}}^\t{\tiny CE}$ & $\frac{1}{2\gamma}$ & $\frac{1}{\gamma}$  & $\frac{1}{\gamma}$ & $\frac{4}{\gamma}$
&\\[12pt]
\hline 
\hline 
$\chi\!\left[\Lambda_{\omega}\right]$ & {\scriptsize $=$}$\sqrt{\textrm{e}}$  & {\scriptsize $\le$}$\sqrt{\frac{4\textrm{e}}{3}}$  & {\scriptsize $=$}$\sqrt{\textrm{e}}$  & {\scriptsize $\le$}$\sqrt{2\textrm{e}}$
&\\[12pt]
\hline 
\end{tabular}
\end{center}
\caption[Channel QFIs, CE bounds and quantum enhancements in frequency estimation]{%
\textbf{Channel QFIs, CE bounds and quantum enhancements in frequency estimation.} In frequency estimation tasks, the precision is maximised by adjusting
the single experimental shot duration $t$. The $t$-optimised (extended)
channel QFIs as well as their finite-$N$ and asymptotic CE bounds
are presented, where $w_{x}\!\left[N\right]\!=\!1\!+\! W\!\left[\frac{x-N}{\textrm{e}N}\right]$,
$\tilde{w}\!=\!1\!+\!2\, W\!\left[\frac{1}{2\sqrt{\textrm{e}}}\right]$
and $W\!\left[x\right]$ is the Lambert $W$ function. As in the case
of depolarizing channel not all the solutions possess an analytic form, 
only their numerical approximations are shown with $\alpha\!\approx\!2.20$ and
$\beta\!\approx\!1.32$. 
In the last row, upper bounds on the 
maximal quantum enhancements of precision,  $\chi\!\left[\Lambda_\varphi\right]\!\le\!\sqrt{\mathfrak{f}_\t{as}^\t{\tiny CE}\!\left[\Lambda_\varphi\right]/\mathfrak{f}\!\left[\Lambda_\varphi\right]}$,
are listed that now account for the $t$-optimisation of the most efficient 
classical and quantum strategies. 
As before in \tabref{tab:SQLboundsQEnh} when estimating phase, for depolarisation and spontaneous emission maps
these limits may possibly be not achievable due
to the channel extension assumed by the CE bound.}
\label{tab:freqQEnh}
\end{table}

As before, we consider the decoherence models
represented by the \emph{dephasing}, \emph{depolarisation}, \emph{loss}
and \emph{spontaneous emission} qubit maps,
which corresponding Liouvillians of \eqnref{eq:freqMasterEq}
have been specified in \secref{sub:noise_models}.
However, as these noise-types \emph{commute}
with the unitary evolution part, the overall process may still be described
with use of the noisy-phase--estimation channels depicted in \figref{fig:noise_models}
after just setting $\varphi\!=\!\omega t$ and%
\footnote{%
For an example of the application of the precision-bounding
methods to a frequency estimation model described by \eqnref{eq:freqMasterEq}
with a non-commutative dissipative part, see \citep{Chaves2013}.
} $\eta\!\to\!\eta(t)$. The adequate substitutions
for the time-dependent strength of decoherence,
which importantly agree with the corresponding Liouvillians, can also be found 
in \secref{sub:noise_models}. 
Crucially, the commutativity property allows
us to directly establish the $t$-optimised frequency estimation
equivalents of all quantities, $\mathcal{F}$,
listed in \tabsref{tab:ChQFISingle}{tab:SQLboundsQEnh}
by just substituting in each of them for $\eta(t)$ according to \eqnref{eq:freqMasterEq}
and computing $\mathfrak{f}\!=\!\underset{0\le t\le T}{\max}\,\mathcal{F}\, t$,
which form becomes evident after realising that the `QFI per shot duration' in \eqnref{eq:ChQFIfreq}
may always be rewritten 
in terms of the QFI computed w.r.t.~$\varphi$, as for $\varphi\!=\!\omega t$
$F_{\textrm{Q}}\!\left[\varrho_{\omega}\right]\!
=\! F_{\textrm{Q}}\!\left[\varrho_{\varphi}\right]t^2$.

In \tabref{tab:freqQEnh}, we present the channel QFIs relevant for frequency
estimation, their asymptotic and finite-$N$ CE bounds, as well as
the upper bounds on the maximal quantum enhancements of precision for each noise-model considered.
In the case of dephasing, we correctly recover the results of \citet{Huelga1997,Escher2011}.
Let us emphasize that all the quantities listed are independent of the total 
experimental time $T$, what is really a consequence of the noise present,
which forces the optimal single-shot duration, $t_\t{opt}$, to be finite, as by letting
the system evolve for too long the decoherence effects dominate making
the $\omega$-estimation impossible. Moreover, by increasing 
the number of particles involved the noise has even more severe impact
on the attainable precision, so that one must conduct shorter 
experimental shots as $N$ grows and $t_\t{opt}(N)\!\!\overset{N\to\infty}{\longrightarrow}\!\!0$.
As a result, one may always set large enough $T$, so that 
$\forall_N\!:t_\t{opt}(N)\!\ll\!T$, and the time characteristics 
become indeed fully determined by the decoherence process.
This dramatically contrasts the noiseless case, in which it is always 
optimal to estimate as long as possible to just gather the maximal information
about $\omega$, and thus set $t\!=\!T$ independently of%
\footnote{%
Note that due to the presence of noise also the locality 
of estimation is assured in the asymptotic $N$ limit, whereas
in the noiseless scenario we run again into the problem of correctly
quantifying the resources, as we must still repeat the whole
experiment enough times for the QCRB \eref{eq:freqQCRB} to be meaningful.
} $N$.
Such a sudden change of the characteristics causes 
the channel and extended-channel measures:~$\mathfrak{f}\!\left[\Lambda_{\omega}\right]$ and
$\mathfrak{f}\!\left[\Lambda_{\omega}\otimes\mathcal{I}\right]$ in \tabref{tab:freqQEnh}, to diverge
in the limit of vanishing decoherence $\gamma\!\to\!0$, what contrasts the
case of their phase-estimation equivalents listed in \tabref{tab:ChQFISingle}
that remain finite when evaluated for $\eta\!=\!1$ (\emph{not} to be confused with the
asymptotic bounds of \tabref{tab:SQLboundsQEnh} that also diverge when $\eta\!\to\!1$).

\section{Local estimation of the decoherence-strength parameter}
\label{sec:loc_dec_est}

Lastly, we would like to explicitly demonstrate that the SQL-bounding methods
introduced in \secref{sub:SQLAsBounds}, i.e.~the CS, RLD, QS and CE methods, may 
also be applied to estimation tasks
in which the estimated parameter is \emph{not} unitarily encoded,
so that the estimation problem may \emph{not} be any more described by the noisy phase estimation
scenario of \figref{fig:Ph_Est_Scheme}, but rather \emph{only} by the 
\emph{$N$-parallel--channels estimation scheme} of \figref{fig:NCh_Est_Scheme}.
In order to do so, we study the qubit decoherence models
previously employed in the noisy-phase--estimation channels of \figref{fig:noise_models},
but this time with the parameter to be estimated being the \emph{decoherence strength}
$\eta$, so that $\Lambda_\eta\!=\!\mathcal{D}_\eta$ (see \tabref{tab:noise_models}). 
This kind of a problem has been widely considered not only
in the estimation theory \citep{Monras2007,Adesso2009,Hotta2005,Frey2011},
but also when examining issues of channel discrimination \citep{Sacchi2005,DAriano2005a}
with particular application in quantum reading \citep{Pirandola2011,Pirandola2011a,Nair2011a}.

As compared to unitary rotations, the nature of the estimated parameter
is then dramatically different. In case of a phase-like parameter
estimation (see \secref{sub:QEst_MZInter_HL_Freq}) the use of maximally
entangled input state of $N$ particles \eref{eq:NOON} results in an 
effective $N$-times greater ``angular speed'' 
of rotation leading to the HL-like scaling in the absence of noise.
In decoherence-strength estimation
tasks, a change in the parameter value can be geometrically interpreted
in the picture of \figref{fig:ExtremCh} as  a``movement'' in the direction away from the boundary,
i.e.~\emph{inside} the space
of all relevant quantum channels, so that 
unless such a trajectory coincides with another non-flat boundary face
(as in the case of spontaneous emission map that is \emph{extremal})
the channel must be trivially $\eta$\emph{-non-extremal} for 
any $\eta\!<\!1$, as depicted in \figref{fig:ExtremCh}(\textbf{b}).
Yet, the ``speed''  along such a trajectory cannot be naively
amplified $N$-times by just employing $N$ channels in parallel%
\footnote{%
From another point of view, in case of the noiseless \emph{unitary} parameter encoding, one may translate the parallel-channel scenario
of \figref{fig:NCh_Est_Scheme} into a sequential one, what naturally explains the  $N$-fold ``speed'' in parameter 
variations \citep{Maccone2013}. Such a picture, however, fails when considering \emph{decoherence-strength} estimation.
}, so that in contrast to the noiseless phase estimation scenario
the optimal entangled inputs must \emph{not} lead to a scaling but 
at most to a \emph{constant-factor} quantum enhancement,
which, however, can be quantified by the methods of \secref{sub:SQLAsBounds}. 

Moreover, due to the above interpretation it should not be surprising
that for the noise-models:~\emph{dephasing}, \emph{depolarisation}
and \emph{loss}
the purely geometrical notion of
\emph{classical simulability} is enough to bound most tightly the asymptotic precision of
estimation, so that 
$\mathcal{F}_\t{as}^\t{\tiny CS}\!=\!\mathcal{F}_\t{as}^\t{\tiny QS}\!=\!\mathcal{F}^\t{\tiny RLD}\!\left[\Lambda_{\varphi}\!\otimes\!\mathcal{I}\right]\!=\!\mathcal{F}_\t{as}^\t{\tiny CE}$. Remarkably, in all three cases the distances from the boundary of the channels set
are generally not symmetric, i.e.~$\varepsilon_+\!\ne\!\varepsilon_-$, and
vary with the decoherence-strength parameter. In the CS picture of \figref{fig:CSpic},
we obtain for \emph{dephasing}:~$\varepsilon_\pm\!=\!1\pm\eta$;~for \emph{depolarisation}:~$\varepsilon_+\!=\!\eta+\frac{1}{3}$, 
$\varepsilon_-\!=\!1-\eta$;~and 
for \emph{loss}:~$\varepsilon_+\!=\!\eta$, $\varepsilon_-\!=\!1-\eta$;
so that the corresponding CS bounds \eref{eq:CSbound}, $\mathcal{F}_\t{as}^\t{\tiny CS}\!=\!1/(\varepsilon_+ \varepsilon_-)$,
explicitly depend on $\eta$.
On the other hand, the \emph{spontaneous emission} map
is special as it is extremal (and thus $\eta$\emph{-extremal} for all $\eta$), so 
that both CS and RLD methods fail. 
Hence, we approach it by means of the CE method, for which it turns out that 
the necessary $\beta_{\tilde K}\!=\!0$ condition \eref{eq:BetaKCond}
is only satisfied trivially by setting $\mathsf{h}\!=\!\mathbf{0}$.
Therefore, the CE bound \eref{eq:CEbound} may be constructed, but
with no room for $\mathsf{h}$-optimisation in \eqnref{eq:CEbound},
what also constrains $\alpha_{\tilde K}\!\propto\!\!\!\!\!\!/\;\mathbb{I}$
disallowing the QS method to be applicable.
The results are summarised in \tabref{tab:dec_est} together
with (unextended-) \eref{eq:ChQFI} and extended- \eref{eq:ExtChQFI} channel QFIs 
for each of the noise-models (evaluated now w.r.t.~the decoherence-strength parameter
in contrast to phase estimation analysis of \tabref{tab:ChQFISingle}).

\begin{table}[!t]
\begin{center}
\begin{tabular}{|M{2.5cm}||M{2cm}|M{3cm}|M{2.2cm}|M{3.4cm}|N}
\hline 
\textbf{Noise models:} & \emph{Dephasing} & \emph{Depolarisation} & \emph{Loss} & \emph{Spontaneous emission}
&\\[12pt]
\hline 
$\mathcal{F}\!\left[\Lambda_{\eta}\right]$ {\footnotesize\eref{eq:ChQFI}} & $\frac{1}{1-\eta^{2}}$ & $\frac{1}{1-\eta^{2}}$ 
& $\frac{1}{\eta\left(1-\eta\right)}$ & $\frac{1}{\eta\left(1-\eta\right)}$ 
&\\[12pt]
\hline
$\mathcal{F}\!\left[\Lambda_{\eta}\!\otimes\!\mathcal{I}\right]$ {\footnotesize\eref{eq:ExtChQFI}} & 
$\frac{1}{1-\eta^{2}}$ & $\frac{3}{\left(1-\eta\right)\left(1+3\eta\right)}$  
& $\frac{1}{\eta\left(1-\eta\right)}$ & $\frac{1}{\eta\left(1-\eta\right)}$ 
&\\[12pt]
\hline 
$\mathcal{F}_{\textrm{as}}^{\textrm{\tiny bound}}$ {\footnotesize \eref{eq:QFIAsBound}} & 
$\mathcal{F}_{\textrm{as}}^{\textrm{\tiny CS}}=\frac{1}{1-\eta^{2}}$ & 
$\mathcal{F}_{\textrm{as}}^{\textrm{\tiny CS}}=\frac{3}{\left(1-\eta\right)\left(1+3\eta\right)}$ &
$\mathcal{F}_{\textrm{as}}^{\textrm{\tiny CS}}=\frac{1}{\eta\left(1-\eta\right)}$ & 
$\mathcal{F}_{\textrm{as}}^{\textrm{\tiny CE}}=\frac{1}{\eta\left(1-\eta\right)}$
&\\[12pt]
\hline 
\end{tabular}
\end{center}
\caption[Channel QFIs and asymptotic bounds in decoherence-strength estimation scenarios]{%
\textbf{Channel QFIs and asymptotic bounds in decoherence-strength estimation scenarios}
for the noise-models introduced in \secref{sub:noise_models}---the ones of \tabref{tab:noise_models} 
with $\Lambda_\eta\!=\!\mathcal{D}_\eta$ after setting $\varphi\!=\!0$---.
Owing to the different nature of the estimated parameter, the geometrical
CS method is enough to provide the tightest asymptotic SQL-like bounds on precision
for all but the \emph{spontaneous emission} map, which being \emph{extremal}---and thus $\eta$-extremal---allows
only for the CE bound \eref{eq:CEbound} to be applicable.
In \emph{all} the cases, as $\mathcal{F}_\t{as}^\t{\tiny bound}\!=\!\mathcal{F}\!\left[\Lambda_{\eta}\otimes\mathcal{I}\right]$,
the asymptotic bounds \eref{eq:QFIAsBound} are saturable \emph{classically} by employing \emph{extended} channels.
Moreover, only when considering the \emph{depolarisation} map there is room for quantum enhancement in the scheme of \figref{fig:NCh_Est_Scheme},
as in all other cases also $\mathcal{F}\!\left[\Lambda_{\eta}\right]\!=\!\mathcal{F}\!\left[\Lambda_{\eta}\otimes\mathcal{I}\right]$,
so that $\mathcal{F}_\t{as}^\t{\tiny bound}$ may be attained classically without need of extending each channel.}
\label{tab:dec_est}
\end{table}

Importantly, for \emph{all} the noise-models considered the asymptotic bounds \eref{eq:QFIAsBound} 
coincide with the corresponding extended-channel QFIs \eref{eq:ExtChQFI},
i.e.~$\mathcal{F}_\t{as}^\t{\tiny bound}\!=\!\mathcal{F}\!\left[\Lambda_{\eta}\otimes\mathcal{I}\right]$,
so that the maximal quantum enhancement of precision \eref{eq:qEnh} in each case must be attainable by a
\emph{classical} estimation strategy that employs \emph{extended} channels.
The fact that the product input states---uncorrelated in between the extended $N$ channels but possibly requiring
entanglement between each single particle and its ancilla---are optimal for noise-strength estimation
with extended channels, has already been noticed for the low-noise channels \citep{Hotta2006} and
for generalised Pauli channels \citep{Fujiwara2003}, of which the latter contain the
dephasing and depolarisation maps studied here.

Furthermore, in case of \emph{dephasing}, \emph{loss} and \emph{spontaneous emission} maps, 
we realise that the corresponding CS bounds are actually attainable \emph{classically} without need of the ancillae,
as also the extension at the single-channel level is unnecessary, i.e.~$\mathcal{F}\!\left[\Lambda_{\eta}\right]\!=\!\mathcal{F}\!\left[\Lambda_{\eta}\otimes\mathcal{I}\right]$.
For \emph{dephasing} noise, each single qubit must be optimally prepared in any pure state 
lying on the equator of the Bloch sphere, whereas while estimating the strength of \emph{losses}
the form of the input state (even mixed) is completely irrelevant.
The latter fact explicitly proves that in the optical interferometry experiments
sensing the power-transmission parameter of the interferometer arms
the entanglement between the photons entering the interferometer
is unnecessary and the total photon-number fluctuations are really 
the ones that limit the precision \citep{Monras2007,Adesso2009}.
These may be reduced by employing Gaussian states \citep{Monras2007} or in principle
fully eliminated with use of any definite photon-number
states that indeed attain the CS bound of \tabref{tab:dec_est} \citep{Adesso2009}.
In order for the scheme to be most sensitive
to variations of the strength---`decay rate'---of the \emph{spontaneous emission}, 
one should indeed intuitively prepare all qubits in the `excited', $\ket{1}$,
state represented by the south pole in \figref{fig:noise_models}(\textbf{d}).
As a result, one then achieves the coinciding unextended- and extended- channel QFIs of
\tabref{tab:dec_est}, which have been derived for the first time in \citep{Fujiwara2004}.

\begin{figure}[!t]
\begin{center}
\includegraphics[width=1\columnwidth]{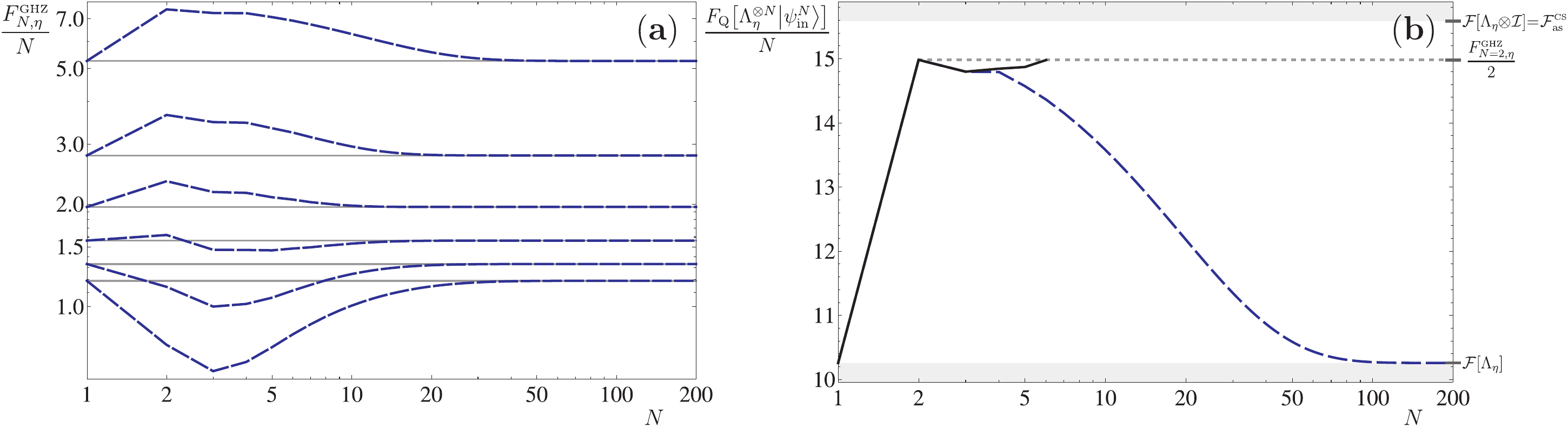}
\end{center}
\caption[$N$-parallel--channels depolarisation-strength estimation with GHZ inputs]{%
\textbf{$N$-parallel--channels depolarisation-strength estimation with GHZ inputs.}\\
(\textbf{a}) \emph{QFI per particle} attained by the 
GHZ inputs (\emph{blue dashed lines}) for depolarisation strengths:~$\eta\!=\!\{0.4,0.5,\dots,0.9\}$,
from bottom to top.
(\emph{Gray horizontal lines}) -- maximal precision
achieved by classical strategies, i.e.~$\mathcal{F}\!\left[\Lambda_\eta\right]$ for each $\eta$.
GHZ-based inputs outperform classical strategies provably only for moderate $N$ 
and $\eta>1/\sqrt{3}\!\approx\!0.58$.\\
(\textbf{b}) \emph{QFI per particle} attained by the
GHZ inputs (\emph{blue dashed line}) and optimal inputs (\emph{solid black line}) 
for $\eta\!=\!0.95$. Results indicate that when considering GHZ-based strategies,
it is optimal to employ uncorrelated \emph{pairs} of particles as inputs with
each doublet in ${\ket{\psi_\t{in}^{N=2}}}_\t{{\tiny GHZ}}$ (\emph{horizontal grey dotted line}).
\emph{Shaded regions} represent `worse than classical' (\emph{bottom}) and `better than the CS bound'
(\emph{top}) regimes.
}
\label{fig:dep_strength_est}
\end{figure}

The example of \emph{depolarisation} map is different, as it is known that
for qubits \citep{Fujiwara2001,Frey2011} (see also \tabref{tab:dec_est}), the
precision of estimation may be improved by extending the channel, 
i.e.~$\mathcal{F}\!\left[\Lambda_{\eta}\right]\!<\!\mathcal{F}\!\left[\Lambda_{\eta}\otimes\mathcal{I}\right]\!=\!\mathcal{F}_\t{as}^\t{\tiny CS}$
for $\eta\!<\!1$. This leaves space for potential quantum 
enhancement, so that if the CS bound of \tabref{tab:dec_est} is tight
then it may be asymptotically attained in the $N$-parallel--channels
scheme of \figref{fig:NCh_Est_Scheme}
only with use of entangled input particles.
Consistently, this notion has been confirmed when considering
two depolarisation channels acting in parallel, for which 
it is optimal to input maximally 
entangled two-qubit (Bell or equivalently ${\ket{\psi_\t{in}^{N=2}}}_\t{{\tiny GHZ}}$) states
for $1/\sqrt{3}\!<\!\eta\!<\!1$ and separable pure states otherwise \citep{Fujiwara2001}.
Thus, generalising such an observation, we consider the GHZ input states%
\footnote{%
As the depolarisation noise does
not preserve \emph{bosonicity} of the particles, we must
assume them to be distinguishable from the beginning, 
so we avoid naming the state \eref{eq:NOON} as
NOON, reserving such name to the modal description.
}
\eref{eq:NOON} of arbitrary particle number,
${\ket{\psi_\t{in}^N}}_\t{{\tiny GHZ}}$, and
utilise the results of \citep{Simon2002}, in which the evolution of GHZ states 
undergoing uncorrelated depolarisation has been studied,
in order to obtain the symbolic expression for the QFI \eref{eq:QFI}:
\begin{equation}
F_{N,\eta}^\t{\tiny GHZ}=F_\t{Q}\!\left[\Lambda_\eta^{\otimes N}{\ket{\psi_\t{in}^N}}_\t{\tiny GHZ}\right]=
\frac{N^{2}}{4}\,\frac{\eta^{2(\!N-1)}\,\alpha_{N-1}^{2}}{\alpha_{N}\!\left(\alpha_{N}^{2}-\eta^{2N}\right)}+\frac{2}{\left(1-\eta^{2}\right)^{2}}\,\sum_{k=0}^{N}\,\binom{N}{k}\,\frac{\left(N\beta_{N-k,k+1}^{-}-k\,\beta_{N-k,k}^{-}\right)^{2}}{\text{ }\beta_{N-k,k}^{+}},
\label{eq:QFI_dep_GHZ}
\end{equation}
where $\alpha_{n}\!=\!\left(\frac{1+\eta}{2}\right)^{n}\!\!+\!\left(\frac{1-\eta}{2}\right)^{n}$
and $\beta_{m,n}^{\pm}\!=\!\left(\frac{1+\eta}{2}\right)^{m}\!\!\left(\frac{1-\eta}{2}\right)^{n}\!\!\pm\left(\frac{1-\eta}{2}\right)^{m}\!\!\left(\frac{1+\eta}{2}\right)^{n}$.
Unfortunately, \eqnref{eq:QFI_dep_GHZ} proves that $F_{N,\eta}^\t{\tiny GHZ}/N$
is not monotonically rising with $N$ and, in fact,
$F_{N,\eta}^\t{\tiny GHZ}/N\!\overset{N\to\infty}{=}\!\!\mathcal{F}\!\left[\Lambda_\varphi\right]$, so
in the asymptotic $N$ limit the GHZ input states do \emph{not} lead to any quantum enhancement. 
Yet, for moderate particle numbers  and $\eta\!>\!1/\sqrt{3}$, as depicted in \figref{fig:dep_strength_est}(\textbf{a}),
the GHZ inputs outperform the classical strategies
attaining their best precision per particle-number always for $N\!=\!2$.
Hence, in order to reach the maximal quantum enhancement \eref{eq:qEnh} allowed
by the GHZ-based strategies, it is optimal to group particles into $N/2$ pairs of Bell 
states, ${\ket{\psi_\t{in}^{N=2}}}_\t{\tiny GHZ}$, what then allows to attain
$\chi_\t{\tiny GHZ}\!\left[\Lambda_\eta\right]$ that 
is indeed greater than one if $\eta\!>\!1/\sqrt3$ \citep{Fujiwara2001}, and reads:
\begin{equation}
\chi_\t{\tiny GHZ}\!\left[\Lambda_\eta\right]=\sqrt{\frac{F_{N=2,\eta}^\t{\tiny GHZ}/2}{\mathcal{F}\!\left[\Lambda_\eta\right]}}=
\frac{6\eta^2}{1+3\eta ^2}
\qquad\le\qquad
\sqrt{\frac{\mathcal{F}_\t{as}^\t{\tiny CS}}{\mathcal{F}\!\left[\Lambda_\eta\right]}}=
\sqrt{1+\frac{2}{1+3\eta }},
\label{eq:qEnh_dec_est_GHZ}
\end{equation}
where we have adequately upper-limited $\chi_\t{\tiny GHZ}\!\left[\Lambda_\eta\right]$
with help of the the CS bound of \tabref{tab:dec_est}.
The gap between the two is clearly shown in \figref{fig:dep_strength_est}(\textbf{b})
corresponding to the spacing between the horizontal
\emph{dotted grey line}  and the \emph{upper shaded region}.
Notice that despite very low depolarisation-strength, $\eta\!=\!0.95$,
the GHZ input states (\emph{blue dashed line}) rapidly
lose their quantum advantage attaining the classically achievable regime 
(\emph{lower shaded region}) already for $N\!\approx\!100$.
In order to benchmark their performance, we perform brute-force 
numerical optimisation over the input states,
what we are only able to do for $N\!\le\!6$ (\emph{solid black line}).
Yet, we already observe significant improvement over the GHZ states,
what indicates their sub-optimality and suggests the potential of
attaining the CS-based bound on quantum enhancement  
of \eqnref{eq:qEnh_dec_est_GHZ}.
Nevertheless, being beyond the scope of this work, we leave 
the problem of finding the optimal entangled input states 
attaining the maximal precision for future research.

%% file: Chapters/noisy_phase_est.tex
\chapter{Mach-Zehnder interferometry with photonic losses} 
\label{chap:MZ_losses} 
\lhead{Chapter 5. \emph{Mach-Zehnder interferometry with photonic losses}}

\section{Lossy Mach-Zehnder interferometer}
\label{sec:LossyInter}

\begin{figure}[!t]
\begin{center}
\includegraphics[width=0.8\columnwidth]{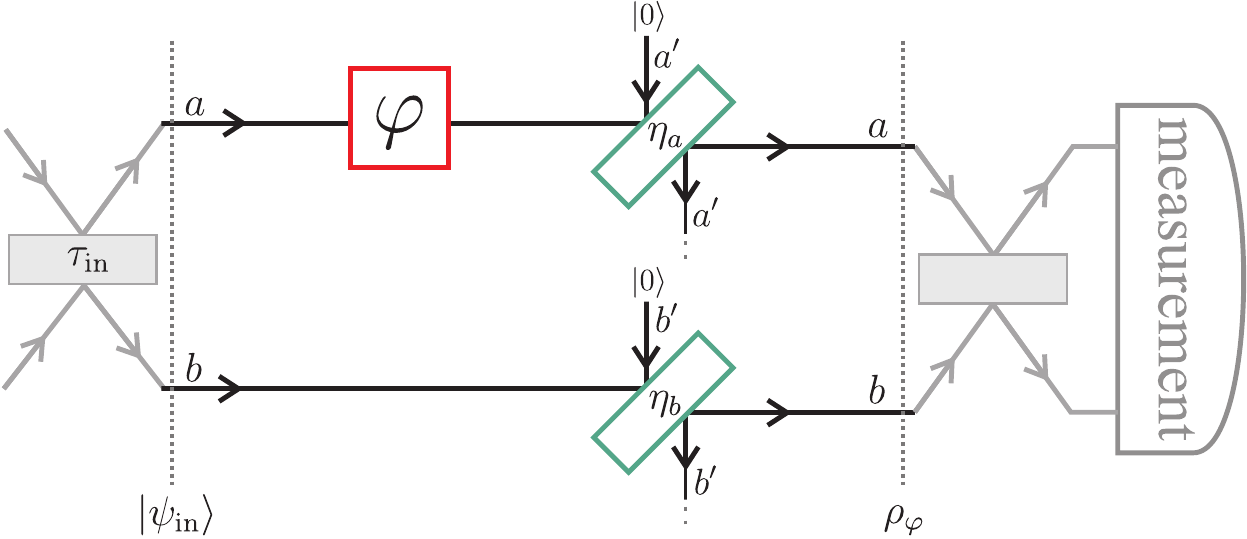}
\end{center}
\caption[Lossy interferometer]{\textbf{Lossy interferometer} corresponding to 
the generalisation of the Mach-Zehnder of \figref{fig:MZinter_Q} that 
accounts for photonic losses leading to non-perfect power transmittances $\eta_{a/b}$
of the interferometer arms $a/b$. As in \secref{sub:QEst_MZInter_HL},
an optimal $N$-photon input state \eref{eq:Cl_Input_Q_MZ}, $\ket{\psi_\t{in}}\!=\!\ket{\psi_\t{in}^N}$, 
is sought that despite losses leads to the
maximal precision of phase estimation. Its performance is compared against 
a \emph{classical strategy},
in which a coherent state $\ket{\alpha}$ of light with mean photon-number $|\alpha|^2\!=\!N$
is impinged on the 
input beam-splitter, so that then $\ket{\psi_\t{in}}\!=\!\left|\sqrt{\tau_{\textrm{in}}}\,\alpha\right\rangle _{a}\left|\sqrt{1-\tau_{\textrm{in}}}\,\alpha\right\rangle _{b}$.\\
$\t{ }$\hfill{\scriptsize The extra labelling of the environmental modes $a'$ and $b'$ is introduced to match the later notation of \figref{fig:MZinter_losses_JS}.}
}
\label{fig:MZinter_losses}
\end{figure}

In the last chapter of this work, we revisit again the Mach-Zehder interferometer 
setup previously discussed at the classical and quantum levels in 
\secsref{sub:ClEst_MZInter_SQL}{sub:QEst_MZInter_HL} respectively.
Yet, this time, we thoroughly analyse the corresponding phase estimation problem
but accounting for the \emph{photonic losses} that lead to potentially 
different power-transmission coefficients, $\eta_{a/b}$, of the interferometer arms labelled by
$a/b$. As depicted in \figref{fig:MZinter_losses}, such a process is 
generally modelled
by introducing fictitious beam-splitters of transmittances
$\eta_{a/b}$ with vacuum states,
$\ket{0}$, impinged on their auxiliary input ports.
Such a loss model is relatively general, as due to the commutativity
of the noise with the phase accumulation \citep{Demkowicz2009},
it accounts for photonic losses occurring at any time during the phase-sensing stage.
Moreover, any losses taking place during the preparation
and detection stages, provided they are equal in both arms,
are also included, as these may be commuted through the input 
and output beam-splitters pictured in \figref{fig:MZinter_losses},
and thus accommodated into the transmittances of the fictitious ones.
As a result, the above description finds its applicability 
in typical quantum-enhanced interferometry
experiments \citep{Kacprowicz2010,Vitelli2010,Spagnolo2012},
and most notably (as shown by \citet{Demkowicz2013}), when describing the
gravitational-wave detectors \citep{LIGO2011,LIGO2013}.
Notice that when following the notation of \secref{sub:sys_indist_part} 
and treating each photon as a distinguishable
particle in an overall permutation invariant state,
then in the special case of equal losses, $\eta_a\!=\!\eta_b$,
the channel representing the single-photon evolution
through the interferometer of \figref{fig:MZinter_losses}
is  exactly the noisy-phase--estimation \emph{loss map} illustrated in \figref{fig:noise_models}(\textbf{c}).

As before when analysing the Mach-Zehnder interferometer in the quantum setting
in \secref{sub:QEst_MZInter_HL},
we consider as the input a general pure, $N$-photon, two-mode (arm) state 
\eref{eq:Input_Q_MZ},
which we write again for convenience:
\begin{equation}
\ket{\psi_\t{in}^N}=\sum_{n=0}^N\, \alpha_n\; {\ket{n}}_a\,{\ket{N-n}}_b=\sum_{n=0}^N\, \alpha_n \; \ket{n,N-n}.
\label{eq:Input_Nph}
\end{equation}
Then, the output state of \figref{fig:MZinter_losses}
most generally reads
\begin{equation}
\rho_{\varphi}^{N}=\sum_{l_a=0}^{N}\sum_{l_b=0}^{N-l_a} p_{l_a,l_b}\,\left|\xi_{l_a,l_b}(\varphi)\right\rangle \!\left\langle \xi_{l_a,l_b}(\varphi)\right|=\bigoplus_{N^{\prime}=0}^{N}\sum_{l_a=0}^{N-N^{\prime}} p_{l_a,N-N'-l_a}\,\left|\xi_{l_a,N-N'-l_a}(\varphi)\right\rangle \!\left\langle \xi_{l_a,N-N'-l_a}(\varphi)\right|,
\label{eq:Output_Nph}
\end{equation}
where $p_{l_a,l_b}$ is the $\varphi$-independent,
binomially distributed probability of losing $l_a$ and
$l_b$ photons in arms $a$ and $b$ respectively, 
so that $p_{l_a,l_b}\!=\!\sum_{n=0}^{N}\left|\alpha_{n}\right|^{2} b_{n}^{(l_a,l_b)}$
with 
\begin{equation}
b_{n}^{(l_a,l_b)}=
\begin{cases}
\binom{n}{l_a}\,\eta_{a}^{n-l_a}\left(1-\eta_a\right)^{l_a}\;\binom{N-n}{l_b}\,\eta_{b}^{N-n-l_b}\left(1-\eta_{b}\right)^{l_b}&\!\! ,\, l_a\le n\le N-l_b\\
0 &\!\! ,\,\textrm{otherwise}
\end{cases}.
\label{eq:b^(la,lb)_n}
\end{equation}
The direct sum in the second expression of \eqnref{eq:Output_Nph} indicates that
the output states of different total number of surviving photons,
$N^{\prime}$, belong to orthogonal subspaces, which in principle
could be distinguished by a non-demolition, photon-number--counting
measurement. The pure states which are outputted according to $p_{l_a,l_b}$ read%
\footnote{%
The phase delay factor $\textrm{e}^{-\textrm{i}n\varphi}$ in \eqnref{eq:Xi_la_lb}
naturally arises if the losses occur \emph{after} the phase accumulation,
as drawn in \figref{fig:MZinter_losses}. However, as only the relative phase delay 
in between the modes $a$ and $b$ possesses physical significance, all
the later calculations in this chapter can be equivalently performed having the
order of the phase delay and losses reversed.
}:
\begin{equation}
\left|\xi_{l_a,l_b}\!(\varphi)\right\rangle =\frac{1}{\sqrt{p_{l_a,l_b}}}\sum_{n=0}^{N}\, \alpha_{n}\,\textrm{e}^{-\textrm{i}n\varphi}\,\sqrt{b_{n}^{(l_a,l_b)}}\left|n-l_a,N-n-l_b\right\rangle 
\label{eq:Xi_la_lb}
\end{equation}
and obey $\left\langle \!\left.\xi_{l_a,l_b}\right|\xi_{l'_{a},l'_{b}}\right\rangle \!=\!\delta_{l_a+l_b,l'_{a}+l'_{b}}$,
so that indeed no coherences exist between their versions for various total numbers
of surviving photons: $N^{\prime}\!=\! N-l_a-l_b$.

Motivated by typical optical implementations 
\citep{Hariharan2003},
in order to benchmark the quantum enhancement of precision
attained by the $N$-photon input states \eref{eq:Input_Nph}, 
we do \emph{not} compare them directly with the \emph{classical input states}
\eref{eq:Cl_Input_Q_MZ} of uncorrelated photons (previously 
considered in \secref{sub:ClEst_MZInter_SQL}), but rather with the 
natural optical strategy employing a \emph{coherent state} of light $\ket{\alpha}$ 
that despite not possessing a definite number of photons still
has them all uncorrelated with one another.
Although the state $\ket{\alpha}$ in principle leads to the input
${\left|\psi_{\textrm{in}}^{\ket{\alpha}}\right\rangle}\!=\!\left|\sqrt{\tau_{\textrm{in}}}\,\alpha\right\rangle _{a}\left|\sqrt{1-\tau_{\textrm{in}}}\,\alpha\right\rangle _{b}$,
when split on the input beam-splitter of transmittance $\tau_{\textrm{in}}$,
one should not forget that due to the lack of
global phase-reference ${\left|\psi_{\textrm{in}}^{\ket{\alpha}}\right\rangle}$ 
should also be phase-averaged \citep{Molmer1997,Bartlett2007,Jarzyna2012}.
Notice that the presence the global phase-reference would need 
to be accounted for by introducing a third mode of light%
\footnote{One may interpret then such a setup as an ``interferometer inside another
interferometer'', what clearly overcomplicates and unnecessarily modifies the phase estimation scenario.}
containing photons that would have to be also counted then as a part of the 
total resources $N$. As such an an extra beam cannot be assumed,
the actual input state must be really written as:
\begin{equation}
\rho_{\textrm{in}}^{\ket{\alpha}}=\bigoplus_{N=0}^{\infty}\; p_{N}\,\left|\xi_{\ket{\alpha}}^{N}\right\rangle\!\left\langle \xi_{\ket{\alpha}}^{N}\right|,
\label{eq:Input_coh}
\end{equation}
representing a mixture of $N$-photonic states that 
are Poissonian distributed according to $p_{N}\!=\!\textrm{e}^{-{\bar N}}\frac{{\bar N}^{N}}{N!}$
and read:
\begin{equation}
\left|\xi_{\ket{\alpha}}^{N}\right\rangle = \sum_{n=0}^{N}\sqrt{\binom{N}{n}\,\kappa_{a}^{n}\,\kappa_{b}^{N-n}}\;\left|n,N-n\right\rangle,
\label{eq:Xi_alpha^N}
\end{equation}
where ${\bar N}\!=\!\left|\alpha\right|^{2}$ is the mean number of
photons present in the interferometer that are split between the arms
$a$ and $b$ in fractions $\kappa_a\!=\!\tau_{\textrm{in}}$
and $\kappa_b\!=\!1-\tau_{\textrm{in}}$ respectively.
Importantly, notice that each state \eref{eq:Xi_alpha^N} is exactly the generalisation of the 
previously considered
classical $N$-photon input \eref{eq:Cl_Input_Q_MZ},
which would be obtained after setting $\tau_\t{in}\!=\!\kappa_a\!=\!\kappa_b\!=\!\frac{1}{2}$.
Hence, the photons are indeed uncorrelated with one another
with each of them being in the state
$\sqrt{\kappa_{a}}\left|{a}\right\rangle +\sqrt{\kappa_{b}}\left|{b}\right\rangle$
that in comparison to \eqnref{eq:Cl_Input_Q_MZ} differs only due to
the presence of the weights $\kappa_{a/b}$, which are additionally introduced 
to compensate for the potential unequal losses in the interferometer
arms, i.e.~$\eta_a\!\ne\!\eta_b$ in \figref{fig:MZinter_losses}. 
However, in order to adequately compare the precision attained by $\left|\psi_{\textrm{in}}^{N}\right\rangle $
and $\rho_{\textrm{in}}^{\ket{\alpha}}$,
we must take the mean number of photons in the state \eref{eq:Input_coh},
${\bar N}$, to equal the exact photon number of \eref{eq:Input_Nph},
$N$, so that $|\alpha|^2\!=\!N$.
On the other hand, as the global-phase--averaging may be 
equivalently performed after propagating the coherent state through the interferometer
of \figref{fig:MZinter_losses}, the form of the output state 
for the input \eref{eq:Input_coh} may be easily obtained:
\begin{equation}
\rho_{\varphi}^{\ket{\alpha}}=\bigoplus_{N=0}^{\infty}\; p_{N}\,\left|\xi_{\ket{\alpha}}^{N}(\varphi)\right\rangle \!\left\langle \xi_{\ket{\alpha}}^{N}(\varphi)\right|,\label{eq:Output_coh}
\end{equation}
where $\left|\xi_{\ket{\alpha}}^{N}(\varphi)\right\rangle $ is the straightforward
generalization of \eqnref{eq:Xi_alpha^N} with:~extra phase-factor
$\textrm{e}^{-\textrm{i}n\varphi}$ introduced, modified coefficients $\kappa_{a/b}$
that now due to losses read $\eta_{a}\tau_{\textrm{in}}$ and $\eta_{b}\!\left(1-\tau_{\textrm{in}}\right)$
respectively, and the overall average photon number diminished to
$\left|\alpha\right|^{2}\left[\eta_{a}\tau_{\textrm{in}}+\eta_{b}\left(1-\tau_{\textrm{in}}\right)\right]$.

\section{Frequentist approach -- local phase estimation}
\label{sec:MZ_losses_local}

The lossy interferometric setup of \figref{fig:MZinter_losses}
has been explicitly studied within the \emph{frequentist} approach in \citep{Dorner2009,Demkowicz2009},
where the corresponding form of the QFI \eref{eq:QFI} with respect to the estimated
phase $\varphi$ for the output state \eref{eq:Output_Nph}, $\rho_{\varphi}^{N}$,
has been investigated. In the most general case of unequal losses ($\eta_a\!\ne\!\eta_b$)
an upper bound on the QFI has been proposed which directly follows from the
convexity of the QFI (see \secref{sub:QFIproperties}):
\begin{equation}
F_{\textrm{Q}}\!\left[\rho_{\varphi}^{N}\right]\le\bar{F}_{\textrm{Q}}\!\left[\rho_{\varphi}^{N}\right]=\sum_{l_{a}=0}^{N}\sum_{l_{b}=0}^{N-l_{a}}p_{l_{a},l_{b}}F_{\textrm{Q}}\!\left[\left|\xi_{l_{a},l_{b}}(\varphi)\right\rangle \right],
\label{eq:FQ_le_FQub}
\end{equation}
where
\begin{equation}
\bar{F}_{\textrm{Q}}\!\left[\rho_{\varphi}^{N}\right]=2\sum_{l_{a}=0}^{N}\sum_{l_{b}=0}^{N-l_{a}}\frac{\mathbf{x}^{T}\mathbf{R}^{(l_{a},l_{b})}\mathbf{x}}{\mathbf{x}^{T}\mathbf{b}^{(l_{a},l_{b})}}
\label{eq:FQub}
\end{equation}
with $\mathbf{x}$ and $\mathbf{b}^{(l_{a},l_{b})}$
being vectors containing variables $x_{n}\!=\!\left|\alpha_{n}\right|^{2}$
and $b_{n}^{(l_{a},l_{b})}$ respectively. The elements
of the matrix $\mathbf{R}^{(l_{a},l_{b})}$
read $R_{nm}^{(l_{a},l_{b})}\!=\! b_{n}^{(l_{a},l_{b})}\!\left(n-m\right)^{2}\*b_{m}^{(l_{a},l_{b})}$.
Intuitively, the coefficients $b_{n}^{(l_{a},l_{b})}$
arise from the binomial distributions dictating the probability of
$l_{a}$ and $l_{b}$ photons being lost in either
of the interferometer arms for every $\left|n,N-n\right\rangle $
constituent of the $N$-photon input state \eref{eq:Input_Nph}.
Although inequality \eref{eq:FQ_le_FQub} is not generally tight,
for the case of equal losses ($\eta\!=\!\eta_a\!=\!\eta_b$) in both arms $\bar{F}_{\textrm{Q}}$
approximates the actual QFI within a negligible error margin \citep{Demkowicz2009}.
Yet, when losses occur in a single arm  ($\eta\!=\!\eta_a$, $\eta_b\!=\!1$), only the direct sum in
\eqnref{eq:Output_Nph} is present, and hence $F_{\textrm{Q}}\!\left[\rho_{\varphi}^{N}\right]\!=\!\bar{F}_{\textrm{Q}}\!\left[\rho_{\varphi}^{N}\right]$,
so that \eqnref{eq:FQub} becomes the expression for the QFI.

On the other hand, one may in principle construct the optimal POVM saturating the
real QFI \eref{eq:FQ_le_FQub} by the general recipe of \secref{sub:QCRBandQFI}, 
which for $\varrho_\varphi^N$ \eref{eq:Output_Nph} leads to a non-demolition
measurement---allowing to learn the number of surviving photons%
\footnote{What in real-life experiments is typically achieved by post-selection.}---followed 
by a projection onto the eigenvectors of the corresponding SLD \eref{eq:SLD}.
Yet, such a measurement is not very practical,
as due to locality of the approach its form generally depends on the particular value of the 
estimated phase $\varphi$ \citep{Demkowicz2009}.
It has been also proved in \citep{Demkowicz2009} that $\bar{F}_{\textrm{Q}}$ 
\eref{eq:FQub} cannot increase by allowing the
$N$ photons in the input state \eref{eq:Input_Nph}
to be \emph{distinguishable} (see \secref{sub:sys_indist_part} and \eqnref{eq:Nqubit_state}),
e.g.~by sending them in non-overlapping time bins. 
In fact, the knowledge of which photon was lost additionally harms
the quantum superposition, whereas the distinguishability 
does not provide any advantage in terms of phase
sensitivity.  On the other hand, the ability to target 
each photon individually allows for the adaptive measurement 
schemes \citep{Wiseman1997,Wiseman1998,Wiseman2009},
which being $\varphi$-independent turn out to be
much more experimentally approachable. Nevertheless, from the point of view of the QFI it is optimal 
to restrict only to the bosonic input states of the form \eref{eq:Input_Nph},
as there always exist a measurement scheme, e.g.~the one described above, 
that attains the QCRB without utilising the adaptivity.

\paragraph{Classical input states}~\\
In case of the coherent-state--based \emph{classical strategy}, one may directly calculate 
the QFI of the output state $\rho_{\varphi}^{\ket{\alpha}}$ due the direct-sum
structure in \eqnref{eq:Output_coh}:~$F_\t{Q}[\rho_{\varphi}^{\ket{\alpha}}]\!=\!\sum_N p_N F_\t{Q}\!\left[\left|\xi_{\ket{\alpha}}^{N}(\varphi)\right\rangle\right]$,
and after maximising it over $\tau_\t{in}$ construct the minimal QCRB \eref{eq:QCRB} on the achievable precision, which
correctly exhibits the SQL-like--scaling behaviour \citep{Demkowicz2009}:
\begin{equation}
\Delta^2\tilde{\varphi}_\t{cl}\;\ge\;\frac{1}{4}\,\left(\frac{1}{\sqrt{\eta_{a}}}+\frac{1}{\sqrt{\eta_{b}}}\right)^2\;\frac{1}{N},
\label{eq:Prec_loss_cl}
\end{equation}
where we have adequately fixed $\left|\alpha\right|^{2}\!=\!N$,
so that  \eqnref{eq:Prec_loss_cl} may be later compared with $\Delta^2\tilde{\varphi}_\t{Q}$ 
attained by the optimal $N$-photon input states \eref{eq:Input_Nph}.
Notably, \eqnref{eq:Prec_loss_cl} is obtained after setting the 
input beam-splitter transmittance to $\tau_{\textrm{in}}\!=\!1/\left(1\!+\!\sqrt{\eta_{a}/\eta_{b}}\right)$,
what may be intuitively explained, as such a choice yields the intensities of light-beams in 
arms $a$ and $b$ described by the output state
\eref{eq:Output_coh} to be equal, and thus leads to the highest \emph{visibility}
of the interferometer.

\subsection{Numerical solution for moderate $N$}

\begin{figure}[!t]
\begin{center}
\includegraphics[width=0.8\columnwidth]{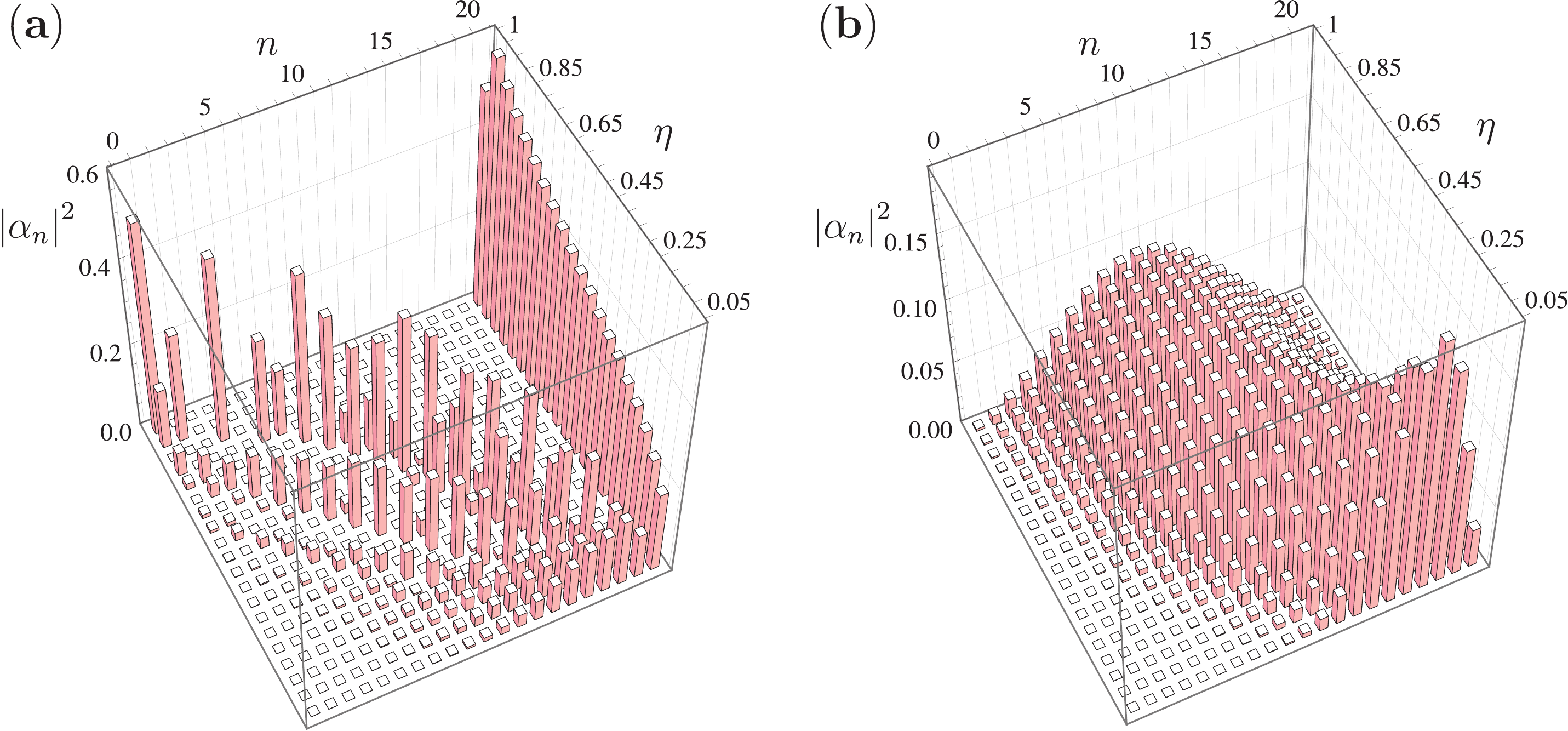}
\end{center}
\caption[Coefficients of the optimal $N$-photon input states]{%
\textbf{Coefficients}, $\left|\alpha_{n}\right|^{2}$,
\textbf{of the optimal $N$-photon input states} \eref{eq:Input_Nph}, $\left|\psi_{\textrm{in}}^{N}\right\rangle$,
plotted as a function of the transmittance coefficient
$\eta$ of the arm $a$, which is the only one subjected to photonic
losses ($N\!=\!20$). As $\eta$ decreases, more weight must be appointed
to coefficients in the mid-range of $n$-s to increase
the robustness of $\left|\psi_{\textrm{in}}^{N}\right\rangle $ against
losses. Due to the asymmetry of noise a bias towards higher values of
$n$ appears, which does not occur for the equal-losses scenario.
Such overall, qualitative behaviour is observed both when the \emph{frequentist}
(\textbf{a}) and \emph{Bayesian }(\textbf{b}) approaches are considered,
despite the contrasting structures of $\left|\alpha_{n}\right|$-distributions.
In the Bayesian case (\textbf{b}), the lack of prior knowledge effectively ``smooths
out'' the optimal distribution.
At $\eta\!=\!1$, the numerical results correctly reproduce the 
noiseless solutions of \secref{sub:QEst_MZInter_HL}:~for $(\textbf{a})$ the NOON states \eref{eq:NOON},
for $(\textbf{b})$ the BW states \eref{eq:BW}.}
\label{fig:coeffs}
\end{figure}

As $\bar{F}_{\textrm{Q}}$ in \eqnref{eq:FQub} is a
concave function with respect to the input-state coefficients 
$x_{n}\!=\!\left|\alpha_{n}\right|^{2}$ \citep{Demkowicz2009},
one may efficiently seek numerically for the optimal input states maximizing $\bar{F}_{\textrm{Q}}$ for
moderate values of $N$ ($\le\!100$). 
An example of an optimal $\left|\alpha_{n}\right|^{2}$-distribution
as a function of $\eta$ is illustrated in \figref{fig:coeffs}(\textbf{a}), 
where we have assumed single-arm losses  ($\eta\!=\!\eta_a$, $\eta_b\!=\!1$)
and set the photon number to $N\!=\!20$.
The numerical results of \figref{fig:coeffs}(\textbf{a})
indicate that although in accordance with \secref{sub:QEst_MZInter_HL_Freq}
the \emph{NOON input states} \eref{eq:NOON} are optimal in the noiseless scenario ($\eta\!=\!1$),
they rapidly cease to be efficient in the presence of noise,
as other coefficients than $\left|\alpha_{0}\right|$ and $\left|\alpha_{N}\right|$ 
must be gradually introduced with increase of losses,
in order to improve the robustness of the input.
The fact that NOON states are extremely fragile
should not be surprising, bearing in mind that
a loss of a single photon is enough for them
to erase all the information about
the estimated phase. Mathematically, such
behaviour is manifested by their QFI 
being determined solely by the subspace 
$N'\!=\!N$ in \eqnref{eq:Output_Nph},
so that when utilising \eqnref{eq:FQ_le_FQub} 
only the non-zero term with $p_{l_a=0,l_b=0}$
contributes and%
\footnote{%
For unequal-losses ($\eta_a\!\ne\!\eta_b$) scenarios, one may further
improve \eqnref{eq:QFI_NOON_losses} by
considering unbalanced NOON input states: 
$\alpha_N\ket{N,0}+\alpha_0\ket{0,N}$,
for which it is optimal to set $\alpha_0\!=\!\sqrt{\eta_a^{N/2}/(\eta_a^{N/2}+\eta_b^{N/2})}$
what yields
$p_{l_a=0,l_b=0}\!=\!\!\left(\eta_a \eta_b\right)^{N/2}$
and
$F_\t{Q}\!=\!4\frac{\!\left(\eta_a\eta_b\right)^N}{\left(\eta_a^{N/2}+\eta_b^{N/2}\right)^2} N^2$ \citep{Demkowicz2009}.
}:
\begin{equation}
F_\t{Q}\!\left[\rho_{\varphi}^\t{\tiny NOON}\right]=p_{l_a=0,l_b=0}^\t{\tiny NOON}\;F_{\textrm{Q}}\!\left[\left|\xi_{l_a=0,l_b=0}^\t{\tiny NOON}(\varphi)\right\rangle \right]
= 2\,\frac{\left(\eta_a\eta_b\right)^N}{\eta_a^N+\eta_b^N} \;N^2,
\label{eq:QFI_NOON_losses}
\end{equation}
what is consistent with the expression already noted in \secref{sub:Example_PhaseEst_Loss_Deph}
for equal-losses ($\eta_a\!=\!\eta_b\!=\!\eta$):~$\mathcal{F}_{N}^{\textrm{\tiny NOON}}\!=\!\eta^{N}N^2$.
Importantly, \eqnref{eq:QFI_NOON_losses} shows that,
although the $N^2$ term (which previously lead to 
the HL-like scaling in \eqnref{eq:QFIandQCRB_NOON}) is 
preserved, the losses introduce an exponentially decaying factor that yields
vanishing QFI and thus divergent precision in the asymptotic $N$ limit.
In other words, the losses degrade exponentially in $N$ the probability of \emph{not} losing any photon
in a NOON state, i.e.~$p_{l_a=0,l_b=0}^\t{\tiny NOON}\!=\!(\eta_a^N\!+\!\eta_b^N)/2$,
so that no information about the phase may be retrieved for sufficiently large $N$.

Crucially, as the numerical optimisation over the input states 
in \eqnref{eq:FQ_le_FQub} may be performed
only for moderate $N$, one may not be sure about answering the question
whether there exist inputs that despite losses lead
to asymptotic precision scaling beyond the SQL.
Yet, we have shown explicitly in \secref{sub:SQLboundsExamples} that 
it cannot be so in the equal-losses scenario, for which
the QS bound \eref{eq:QSbound} is sufficient to prove the asymptotic
SQL-like scaling of precision. Such an observation actually proves that
for any $\eta_a\!<\!\eta_b\!<\!1$ in \figref{fig:MZinter_losses} 
the asymptotic precision scaling
must also be SQL-like, as by increasing $\eta_a\!\to\!\eta_b$
and thus decreasing the overall noise, we may always construct an equal-losses 
scenario which is known to be SQL-bounded and may perform
only better. Such an argument, however, does not hold when considering
single-arm losses, but we show explicitly in the following section 
that the techniques of \secref{sub:SQLAsBounds} are then also
applicable, and in fact yield asymptotic SQL-like bounds for any $\eta_{a/b}$
in \figref{fig:MZinter_losses} (with the only exception 
being naturally the noiseless scenario ($\eta_{a}\!=\!\eta_{b}\!=\!1$)
discussed in \secref{sub:QEst_MZInter_HL_Freq} that attains the HL).

\subsection{Asymptotic SQL-like bound on precision}
\label{sub:asbounds_mzloss}

In the following paragraphs we show that 
the \emph{QS method} of \secref{sub:QSbound} is sufficient to determine
an asymptotic SQL-like bound on precision not only for equal losses (as 
already discussed in \secref{sub:SQLboundsExamples}), but also in 
the general lossy interferometry setting 
of \figref{fig:MZinter_losses}. Furthermore, we demonstrate
that such a bound may be independently derived by utilising the concept 
of \emph{minimisation over purifications} introduced in \secref{sub:QFIPurifDefs},
but returning to the \emph{modal}- rather than \emph{particle}-picture of the photonic
state employed as an input. Nevertheless, both approaches yield
the same upper bound on precision that, however, has been just 
recently proved to be tight \citep{Knysh2014}, i.e.~to be exactly the 
QCRB \eref{eq:QCRB} calculated for the 
optimal $N$-photon inputs \eref{eq:Input_Nph} in the asymptotic $N$ limit:
\begin{equation}
\Delta^2\tilde{\varphi}_\t{Q}\;\ge\;\frac{1}{4}\,\left(\sqrt{\frac{1-\eta_a}{\eta_a}}+\sqrt{\frac{1-\eta_b}{\eta_b}}\right)^2\;\frac{1}{N},
\label{eq:Prec_loss_Q}
\end{equation}
which thus also quantifies the \emph{maximal quantum enhancement of precision} \eref{eq:qEnh}
corresponding to the ratio of \eqnsref{eq:Prec_loss_cl}{eq:Prec_loss_Q}: 
\begin{equation}
\chi\!\left[\Lambda_{\varphi}^{(\eta_a,\eta_b)}\right]=\lim_{N\rightarrow\infty}\sqrt{\frac{\Delta^2\tilde{\varphi}_\t{cl}}{\Delta^2\tilde{\varphi}_\t{Q}}}=\frac{\sqrt{\eta_a}+\sqrt{\eta_b}}{\sqrt{\left(1-\eta_a\right)\eta_b}+\sqrt{\left(1-\eta_b\right)\eta_a}},
\label{eq:qEnh_loss}
\end{equation}
where by $\Lambda_\varphi^{(\eta_a,\eta_b)}$ we label the \emph{lossy interferometer channel}---the 
effective quantum map describing the evolution of each photon while propagating
through the interferometer of \figref{fig:MZinter_losses}.
Notice that, for \emph{equal losses} ($\eta\!=\!\eta_a\!=\!\eta_b$), \eqnref{eq:qEnh_loss}
correctly coincides with the expression:~$\sqrt{1/(1-\eta)}$, previously stated in \tabref{tab:SQLboundsQEnh},
whereas for \emph{single-arm losses} it simplifies to $\sqrt{\left(1+\sqrt{\eta}\right)/\left(1-\sqrt{\eta}\right)}$
originally derived in \citep{Knysh2011}.

\begin{figure}[t!]
\includegraphics[width=1\columnwidth]{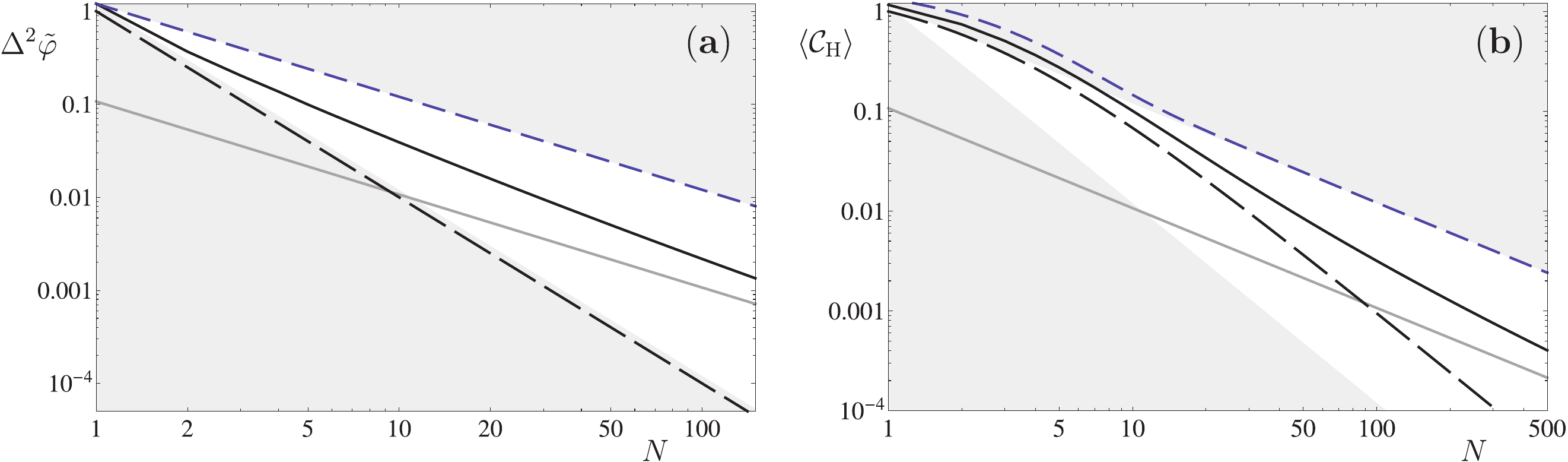}
\caption[Precision of phase estimation for a lossy interferometer]{%
\textbf{Precision of phase estimation for a lossy interferometer} 
when analysed from the \emph{frequentist} (\textbf{a}) and \emph{Bayesian} (\textbf{b})
perspectives. For a single-arm losses scenario, i.e.~$\eta_a\!=\!0.7$, $\eta_b\!=\!1$,
the precision attained by the optimal $N$-photon input states
of \figref{fig:coeffs} (\emph{solid black lines}) and the coherent-state--based \emph{classical strategy} 
(\emph{short-dashed blue lines}) is depicted. Within both approaches the ultimate precision 
is dictated by the bound \eref{eq:Prec_loss_Q}
(\emph{solid grey lines}). For comparison, the performance of the optimal $N$-photon 
strategy in the \emph{noiseless} scenario ($\eta_a\!=\!\eta_b\!=\!1$),
i.e.~NOON states \eref{eq:NOON} in (\textbf{a}) and BW states \eref{eq:BW}
in (\textbf{b}), is plotted (\emph{long-dashed black lines}) 
that in both cases attains the HL.
}
\label{fig:PrecPlotLossInter}
\end{figure}

In \figref{fig:PrecPlotLossInter}(\textbf{a}), we present a plot of 
the maximal achievable precision in the 
lossy interferometer with \emph{single-arm} losses ($\eta_a\!=\!\eta$, $\eta_b\!=\!1$)
after choosing $\eta\!=\!0.7$. The HL attained in the \emph{noiseless} setting
by the NOON states \eref{eq:NOON} (\emph{long-dashed black line})
is shown for comparison. The presence of losses forces
the asymptotic scaling to be SQL-like and follow the 
ultimate precision dictated by the bound \eref{eq:Prec_loss_Q} 
(\emph{solid grey line}). The optimal $N$-photon input states \eref{eq:Input_Nph}
with coefficients distributed according to \figref{fig:coeffs}(\textbf{a}) 
approach the bound, yet we are restricted by the numerical capabilities
of maximising the QFI \eref{eq:FQ_le_FQub},
$F_{\textrm{Q}}\!\left[\rho_{\varphi}^{N}\right]\!=\!\bar{F}_{\textrm{Q}}\!\left[\rho_{\varphi}^{N}\right]$, 
to relatively low photon numbers ($N\!\lesssim\!150$).
Nevertheless, apart from the explicit calculation of the 
asymptotic form of the $F_{\textrm{Q}}\!\left[\rho_{\varphi}^{N}\right]$ in \citep{Knysh2014},
the bound \eref{eq:Prec_loss_Q} has been shown to be attainable
by utilising particular indefinite-photon--number input states and measurements,
e.g.~inputs employing light-squeezing accompanied by the photon-counting detection 
\citep{Caves1981,Demkowicz2015}.
For completeness, the performance of coherent-state--based \emph{classical strategy} is also 
plotted (\emph{short-dashed blue line}) that is fully determined
by \eqnref{eq:Prec_loss_cl}.

Lastly, let us emphasise that due to the lack of 
global-phase reference any SQL-like bound, e.g.~the 
one of \eqnref{eq:Prec_loss_Q}, must also apply 
to \emph{indefinite-photon--number input states} of light with
the mean number of photons, $\bar N$, fixed.
Any such bound is derived basing on the previously discussed 
upper limit \eref{eq:QFIAsBound} on the QFI that scales linearly with the number of particles 
involved, i.e.~$F_{\textrm{Q}}\!\left[\rho_{\varphi}^N\right]\!\le\!N\,\mathcal{F}_{\textrm{as}}^\t{\tiny bound}$.
On the other hand, due to global-phase averaging,
any indefinite-particle(photon)--number input
takes the form:~$\sum_N\! p_N \rho_\t{in}^{N}$ 
with $\sum_N p_N\!=\!\bar N$. Thus, the output state
may always be written as $\sum_N p_N \rho_\varphi^{N}$, 
and its QFI may always be upper-bounded 
stemming from the convexity property of the QFI (see \secref{sub:QFIproperties}) as:
\begin{equation}
F_\t{Q}\!\left[\sum_N p_N \rho_\varphi^{N}\right] \;\le\; \sum_N p_N F_\t{Q}\!\left[\rho_\varphi^{N}\right] 
\;\le\; 
\sum_N p_N N\, \mathcal{F}_{\textrm{as}}^\t{\tiny bound}=\bar N\, \mathcal{F}_{\textrm{as}}^\t{\tiny bound}.
\label{eq:bounds_indefN_inputs}
\end{equation}
Hence, most generally, for any input that we may regard as an incoherent 
mixture of states occupying different particle-number sectors,
we may apply all the SQL-like bounding techniques discussed 
in \secref{sub:SQLAsBounds} after
accordingly replacing the definite particle number, $N$, with
the input mean particle number, $\bar N$, in all the 
eventually obtained bounds on precision%
\footnote{%
Yet, notice that the argumentation of \eqnref{eq:bounds_indefN_inputs}
may not be applied when considering the \emph{finite-$N$ CE bound} of
\secref{sub:FinN_CEbound} that does \emph{not} yield a linear, i.e.~concave in $N$,
upper limit on the QFI. As a result, the finite-$N$ CE bound \eref{eq:CEboundN} 
may only be confidently applied when dealing 
with systems consisting of a definite number of particles.
}.

\paragraph{Quantum Simulation of the lossy interferometer channel}~\\
Let us show that the precision bound \eref{eq:Prec_loss_Q}
may be derived by means of the \emph{QS method} discussed in \secref{sub:QSbound}, 
which previously turned to be already sufficient when considering the scenario of
equal losses (see \tabref{tab:SQLboundsQEnh}), i.e.~for the loss noise model
depicted in \figref{fig:noise_models}(\textbf{c}). In order to apply the QS bound \eref{eq:QSbound}
to the more general setting of the interferometer in \figref{fig:MZinter_losses},
we establish the form of the corresponding channel $\Lambda_\varphi^{(\eta_a,\eta_b)}$ acting
on each photon within the input state \eref{eq:Input_Nph},
so that the output state \eref{eq:Output_Nph} may be then written as
$\rho_\varphi^N\!=\!\Lambda_\varphi^{(\eta_a,\eta_b)\otimes N}\!\left[\ket{\psi_\t{in}^N}\right]$
allowing us to regard the interferometer of \figref{fig:MZinter_losses}
as an instance of the \emph{$N$-parallel--channels estimation scheme} of \figref{fig:NCh_Est_Scheme}.
Fortunately, this may be easily achieved by just trivially generalising
the canonical Kraus operators specified for the equal-loss model described in \tabref{tab:noise_models},
so that they now read:
\begin{equation}
K_{1}=\left(\!\!
\begin{array}{cc}
\sqrt{\eta_a} & 0\\
0 & \sqrt{\eta_b}\\
0 & 0
\end{array}\!\right)\!,\; 
K_{2}=\left(\!\!
\begin{array}{cc}
0 & 0\\
0 & 0\\
\sqrt{1-\eta_a} & 0
\end{array}\right)\!,\;
K_{3}=\left(
\begin{array}{cc}
0 & 0\\
0 & 0\\
0 & \sqrt{1-\eta_b}
\end{array}\!\right)\!,
\label{eq:LossKrauses}
\end{equation}
and thus account for different transmittances of the arms
$a$ and $b$. As before, the overall action
of the single-photon map may be written as $\Lambda_\varphi^{(\eta_a,\eta_b)}\!\left[\bullet\right]\!=\!\sum_{i=1}^{3} K_{i}(\varphi) \bullet K_{i}(\varphi)^{\dagger}$,
after setting all $K_{i}(\varphi)\!=\! K_{i}U_{\varphi}$
in accordance with \figref{fig:MZinter_losses},
but---as the losses and phase-accumulation commute with one another---we 
may have equivalently chosen $K_{i}(\varphi)\!=\!\tilde{U}_{\varphi}K_{i}$
with $\tilde{U}_{\varphi}$ representing an enlarged 
$U_{\varphi}\!=\!\textrm{e}^{-\textrm{i}\frac{\varphi}{2}\hat\sigma_{z}}$
that also trivially acts on the vacuum state $\left|0\right\rangle$.
Notice that, in order to emphasise the symmetry of the problem
and match the previous notation of noisy-phase--estimation models
of \figref{fig:noise_models}, we have without loss of generality assumed again
the phase to accumulate as%
\footnote{%
As aside, let us note that such a choice not only simplifies the calculations, but also, 
in case the output state $\rho_\varphi$ is path(mode)-symmetric, assures the redundancy
of a global-phase reference \citep{Jarzyna2012}.}
($\varphi/2$,$-\varphi/2$) in the arms ($a$,$b$) of the interferometer, and \emph{not} 
as ($\varphi$,$0$),
as originally drawn in \figref{fig:MZinter_losses}.

Having established the Kraus representation of 
the lossy interferometer channel $\Lambda_\varphi^{(\eta_a,\eta_b)}$,
we may directly compute%
\footnote{%
Actually, it is more efficient to compute the
\emph{CE bound} \eref{eq:CEbound}, what can be
achieved with help of the SDP discussed in \appref{chap:appFinNCEasSDP},
and then realise that the extra condition, $\alpha_{\tilde K_\t{opt}}\!\!\propto\!\mathbb{I}$, 
of the QS method is also naturally fulfilled.} 
the \emph{QS bound} \eref{eq:QSbound} following the recipe of 
\secref{sub:QSbound}:
\begin{equation}
\mathcal{F}_\t{as}^\t{\tiny QS} = \frac{4}{\left(\sqrt{\frac{1-\eta_a}{\eta_a}}+\sqrt{\frac{1-\eta_b}{\eta_b}}\right)^{2}}
\label{eq:QSbound_lossy_inter}
\end{equation}
corresponding to the minimum in \eqnref{eq:QSbound} that occurs 
after optimally choosing 
\begin{equation}
\mathsf{h}_\t{\tiny opt}=-\frac{1}{8}\,\textrm{diag}\left\{\chi,\frac{\eta_a}{1-\eta_a}\left(\frac{4}{\eta_a}-\chi\right),-\frac{\eta_b}{1-\eta_b}\left(\frac{4}{\eta_a}+\chi\right)\right\}
\quad\t{with}\quad\chi= \mathcal{F}_\t{as}^\t{\tiny QS}\,\frac{\eta_a-\eta_b}{\eta_a\eta_b}.
\label{eq:hOpt_QS_lossy_inter}
\end{equation}
Hence, we indeed recover the ultimate quantum limit 
\eref{eq:Prec_loss_Q} as $\Delta^2\tilde\varphi_\t{Q}\!\ge\!1/(\mathcal{F}_\t{as}^\t{\tiny QS}\,N)$,
which leads to the correct maximal quantum enhancement of precision \eref{eq:qEnh_loss}.
Consistently, for equal losses ($\eta\!=\!\eta_a\!=\!\eta_b$), the QS bound \eref{eq:QSbound_lossy_inter} coincides 
with the one quoted in \tabref{tab:SQLboundsQEnh} for the loss noise-model
and the optimal generator \eref{eq:hOpt_QS_lossy_inter} becomes then, as necessary, 
the adequate $\mathsf{h}_\t{\tiny opt}$ \eref{eq:hOpt_CE_loss} noted in \appref{chap:appOptHExtCh}.

\paragraph{Optimal purification of the output state}~\\
Interestingly, we show that in the case of the lossy interferometer
of \figref{fig:MZinter_losses} the ultimate bound on precision \eref{eq:Prec_loss_Q}
may also be independently established after moving away from the $N$-parallel--channel 
picture of \figref{fig:NCh_Est_Scheme}, in which the properties of the
single-particle map are investigated to determine the SQL-like bounds on precision.
We demonstrate that \eqnref{eq:Prec_loss_Q} may be equivalently derived by considering 
the overall $N$-particle output state, $\rho_\varphi^N$, and by directly utilising the \emph{purification-based
definition of its QFI} \eref{eq:QFIPurifEscher} introduced in \secref{sub:QFIPurifDefs}.
We follow the analysis of \citep{Escher2011} after adopting the \emph{modal
description} of light propagating through the lossy interferometer of \figref{fig:MZinter_losses},
but in addition we explicitly make use of the so-called \emph{Jordan-Schwinger} (JS) \emph{map} \citep{Jordan1935,Schwinger1965},
what greatly facilitates the intuition behind the search for the optimal purification of
the output. It  allows us to benefit from the notion previously introduced for finite-dimensional spaces
(purification-based methods introduced for states in \secref{sub:QFIPurifDefs} and for channels in \secref{sub:ChQFI}) 
stating that, in order to establish the tightest upper bound on the output state QFI, one must 
choose a local rotation of the environmental subspace that erases most information about the estimated parameter
potentially encoded in the environment. As shown below, such picture naturally generalises 
to the modal light-description, in case of which one must optimise over local rotations of the 
environmental \emph{modes} in order to establish the best possible output purification, 
$\left|\tilde\Psi_{\varphi}\right\rangle$  satisfying $\rho_{\varphi}^N\!=\!\textrm{Tr}_\t{\tiny E}\!\left\{ \left|\tilde\Psi_{\varphi}\right\rangle \!\left\langle \tilde\Psi_{\varphi}\right|\right\}$, 
that yields the tightest $F_\t{Q}\!\left[\rho_{\varphi}^N\right]\!\le\!F_\t{Q}\!\left[\left|\tilde\Psi_{\varphi}\right\rangle\right]$ 
in accordance with the purification-based definition \eref{eq:QFIPurifEscher}.

\begin{figure}[!t]
\begin{center}
\includegraphics[width=0.8\columnwidth]{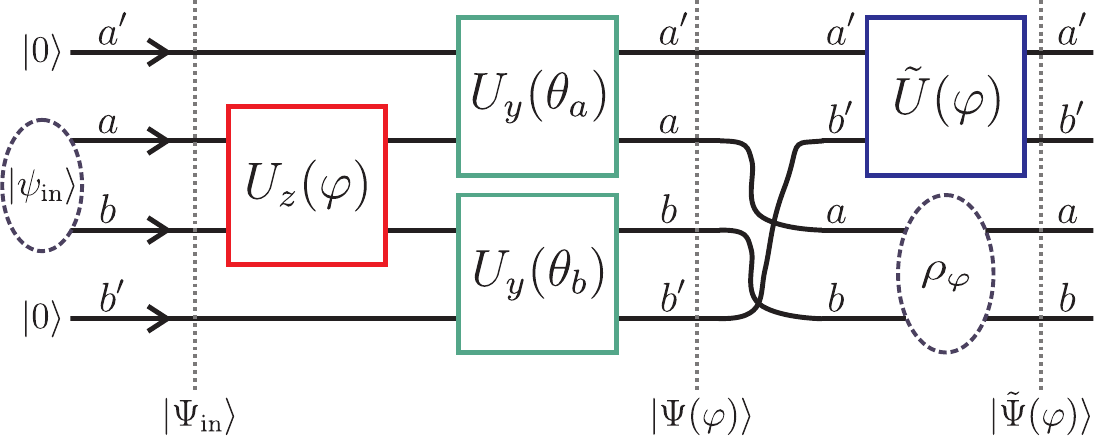}
\end{center}
\caption[Jordan-Schwinger representation of a lossy interferometer]{
\textbf{Jordan-Schwinger representation of the lossy interferometer}
depicted in \figref{fig:MZinter_losses}.
From left to right, the unitary transformations in the figure
correspond to: $U_{z}^{ab}(\varphi)$ -- phase
accumulation; $U_{y}^{aa'}(\theta_a)$
and $U_{y}^{bb'}(\theta_b)$ -- fictitious
beam-splitters introduced in arms $a$ and $b$ to mimic losses;
$\tilde{U}^{a'b'}\!(\varphi)\!=\!\textrm{e}^{-\textrm{i}\hat h_{a'b'}\delta\varphi}$ 
-- local ($\varphi\!=\!\varphi_0\!+\!\delta\varphi$) unitary rotation of the excluded environmental modes that 
must be optimised over all $\hat h_{a'b'}$ to establish the optimal purification, $\left|\tilde{\Psi}(\varphi)\right\rangle$,
of the output state.
$\left|\psi_{\textrm{in}}\right\rangle $
and $\rho_{\varphi}$ represent respectively the interferometer
input and output states of \figref{fig:MZinter_losses}, whereas $\ket{\Psi_\t{in}}$ and $\left|\Psi(\varphi)\right>$ 
denote their corresponding purifications that also include the environmental modes $a'$ and $b'$.\\
$\t{ }$\hfill{\scriptsize Note that the colours of above transformations match the adequate optical elements in \figref{fig:MZinter_losses}.}
}
\label{fig:MZinter_losses_JS}
\end{figure}

Importantly, the JS map allows us to express
the action of the interferometer of \figref{fig:MZinter_losses} in terms of the
algebra of the angular momentum operators \citep{Yurke1986}.
Denoting by $\hat a^\dagger$ and $\hat a$ respectively the creation 
and annihilation operators of the mode describing the light 
travelling in arm $a$, and similarly for arm $b$, we
define the angular momentum operators:
\begin{equation}
\hat{J}_x^{ab} = \frac{1}{2}(\hat{a}^\dagger \hat{b} + \hat{b}^\dagger \hat{a}), \quad
\hat{J}_y^{ab} = \frac{\ii}{2}(\hat{b}^\dagger \hat{a} - \hat{a}^\dagger \hat{b}), \quad
\hat{J}_z^{ab} = \frac{1}{2}(\hat{a}^\dagger \hat{a} - \hat{b}^\dagger \hat{b} ),
\label{eq:ang_mom_ops}
\end{equation}
which fulfil the commutation relations\footnote{$i\!=\!\{1\!\equiv\!x,2\!\equiv\!y,3\!\equiv\!z\}$.}:~$[\hat{J}_i^{ab},\hat{J}_j^{ab}]\!=\!\mathrm{i}\,\epsilon_{ijk} \hat{J}_k^{ab}$.
As later we consider such operators acting on various pairs of modes,
we explicitly label by a superscript the modes, e.g.~$m_1$ and $m_2$,
on which a given $\hat J_i^{m_1m_2}$ is defined.
Moreover, we may construct unitary operators generated by
the angular momentum operators \eref{eq:ang_mom_ops} that thus
describe the rotations in the abstract spin space:~$U_{i}^{m_1m_2}(\theta)\!=\!\textrm{e}^{-\textrm{i}\theta\hat J_{i}^{m_1m_2}}$.
Notably, a beam-splitter of transmittance $\eta$, acting on light modes $m_1$ and $m_2$ 
and transforming a given state $\ket{\t{in}}$  into $\ket{\t{out}}$, 
may then be represented as such a spin rotation around the $y$ axis \citep{Yurke1986,Campos1989}:
\begin{equation}
\left(\!\!\begin{array}{c}
\hat{m}_1\\
\hat{m}_2
\end{array}\!\right)_{\textrm{out}}=\left(\!\!\!\begin{array}{cc}
\sqrt{\eta} & -\sqrt{1-\eta}\\
\sqrt{1-\eta} & \sqrt{\eta}
\end{array}\!\!\right)\left(\!\!\begin{array}{c}
\hat{m}_1\\
\hat{m}_2
\end{array}\!\right)_{\textrm{in}}\quad\;\Longleftrightarrow\quad\;\left|\textrm{out}\right\rangle=U_{y}^{m_1m_2}(\theta)\left|\textrm{in}\right\rangle
\quad\t{with}\quad\theta=2\,\arccos\!\sqrt{\eta}\,.
\end{equation}
Similarly, the accumulation of phase $\varphi$ in between two modes corresponds 
just to a rotation around the $z$ axis:~$U_z^{m_1m_2}(\varphi)$.
As a consequence, we are able to fully re-express the action of the 
lossy interferometer of \figref{fig:MZinter_losses} employing the JS representation,
which we schematically depict in \figref{fig:MZinter_losses_JS} after already ignoring 
the input and output beam-splitters of \figref{fig:MZinter_losses} that do not contribute 
to the phase-sensing process.
On the other hand, in order to verify such description,
one may inspect the overall
input-output relation between $\ket{\psi_\t{in}^N}$ and $\rho_\varphi^N$ (previously 
decomposed into independent channels acting on the constituent 
photons to match the $N$-parallel--channel scheme of \figref{fig:NCh_Est_Scheme}), but 
this time working in the \emph{second-quantisation} picture, so that:
\begin{equation}
\rho_{\varphi}^{N}=
\sum_{l_a,l_b=0}^{\infty}\hat{K}_{l_a,l_b}(\varphi)\left|\psi_{\textrm{in}}^{N}\right\rangle \!\left\langle \psi_{\textrm{in}}^{N}\right|\hat{K}_{l_a,l_b}^{\dagger}(\varphi)\,,
\end{equation}
where we have employed the natural Kraus operators 
expressible with help of the modal bosonic operators
and parametrised by the numbers of photons lost in each arm \citep{Dorner2009}, i.e.
\begin{equation}
\hat{K}_{l_a,l_b}(\varphi)=\sqrt{\frac{\left(1-\eta_a\right)^{l_a}}{l_a!}}\sqrt{\eta_a}^{\hat{a}^{\dagger}\hat{a}}\,\hat{a}^{l_a}\;\sqrt{\frac{\left(1-\eta_b\right)^{l_b}}{l_b!}}\sqrt{\eta_b}^{\hat{b}^{\dagger}\hat{b}}\,\hat{b}^{l_b}\;\textrm{e}^{-\textrm{i}\frac{\varphi}{2}\left(\hat{a}^{\dagger}\hat{a}-\hat{b}^{\dagger}\hat{b}\right)}\,.
\label{eq:KrausOps_bos}
\end{equation}
In accordance with \figref{fig:MZinter_losses}, each $\hat{K}_{l_a,l_b}(\varphi)$ accounts 
first for the phase accumulation and afterwards for the losing $l_a$ and $l_b$ photons in arms 
$a$ and $b$ respectively. Furthermore, after simple algebra one may also show that each 
Kraus operator \eref{eq:KrausOps_bos} may be importantly written 
utilising the spin rotations in the JS picture, so that
\begin{equation}
\hat{K}_{l_a,l_b}(\varphi)=\prescript{}{a'}{\left\langle l_a\right|}\prescript{}{b'}{\left\langle l_b\right|}\,U_y^{aa'}(\theta_a)\,U_y^{bb'}(\theta_b)\,U_{z}^{ab}(\varphi)\left|0\right\rangle _{a'}\left|\textrm{0}\right\rangle _{b'},
\label{eq:KrausOps_JS}
\end{equation}
where $\theta_{a/b}\!=\!2\arccos\!\sqrt{\eta_{a/b}}$ and ${\ket{n}}_{m}$ generally represents
an $n$-photon Fock state in mode $m$. Notice that \eqnref{eq:KrausOps_JS} represents 
exactly the ``circuit'' of \figref{fig:MZinter_losses_JS} with  the environmental modes
$a'$ and $b'$ being accordingly prepared in the vacuum state $\ket{0}$ and eventually projected
onto the $l_a$- and $l_b$-photon Fock states.

Crucially, the JS representation of \figref{fig:MZinter_losses_JS}
allows us to straightforwardly vary the output state purification, i.e.~$\left|\Psi(\varphi)\right>$
containing the environmental modes, by introducing a general unitary 
transformation $\tilde{U}^{a'b'}\!(\varphi)$
acting only on modes $a'$ and $b'$, so that 
$\left|\tilde\Psi(\varphi)\right>\!=\!\tilde{U}^{a'b'}\!(\varphi)\!\left|\Psi(\varphi)\right>$.
Furthermore, due to the locality of the QFI, when estimating around a given $\varphi_0$ 
with $\varphi\!=\!\varphi_0\!+\!\delta\varphi$,
all the adequate ``environment-rotations'' may be parametrised as
$\tilde{U}^{a'b'}\!(\varphi)\!=\!\textrm{e}^{-\textrm{i}\hat h_{a'b'}\delta\varphi}$,
where the Hermitian generator, $\hat h_{a'b'}$, plays exactly
the role of $\hat h_\t{\tiny E}$ introduced in \secref{sub:QFIPurifDefs}, yet
being now generalised to an infinite-dimensional Hilbert space.
Previously, $\hat h_\t{\tiny E}$ yielded a Hermitian matrix $\mathsf{h}$
defined in some basis of the environmental subspace and the optimisation over 
purifications (see \secref{sub:QFIPurifDefs}) corresponded 
to a search through all such matrices. Now, as unfortunately we may build
\emph{any} Hermitian generator, $\hat h_{a'b'}$,
from the modal bosonic operators:~$\hat{a}'$, 
$\hat{a}'^{\dagger}$, $\hat{b}'$, $\hat{b}'^{\dagger}$;
that may be chosen to be of arbitrary order,
the minimisation of the QFI over all purifications cannot
be in principle explicitly performed.

Nevertheless, as we are only interested in establishing a 
bound $F_\t{Q}\!\left[\rho_{\varphi}^N\right]\!\le\!F_\t{Q}\!\left[\left|\tilde\Psi_{\varphi}\right\rangle\right]$
in the asymptotic $N$ limit, motivated by the work of \citep{Escher2011}
and the symmetry of the problem,
we constrain the generator
to have the form $\hat h_{a'b'}\!=\!-\frac{1}{2}\left(\gamma_a\hat{n}_a-\gamma_b\hat{n}_b\right)$,
where by $\hat{n}_{m}\!=\!\hat{m}^{\dagger}\hat{m}$ 
we denote the photon-number operator of a given mode $m$. 
Intuitively (similarly to $e^{-\textrm{i}{\hat h}_\t{\tiny E}\delta\varphi}$ in \figref{fig:ch_purif_ext}), we 
may thus interpret $\tilde{U}^{a'b'}\!(\varphi)$
as an \emph{erasure operation} which ``unwinds'' the phase encoded in 
the environmental modes. Such an interpretation becomes even more 
evident in the equal-losses scenario,
when we may assume the optimal input $\left|\psi_{\textrm{in}}^{N}\right\rangle $
to possess the modal ($a$/$b$) interchange symmetry and set $\gamma\!=\!\gamma_a\!=\!\gamma_b$,
so that $\tilde{U}^{a'b'}\!(\varphi)\!=\! U_{z}^{a'b'}(-\gamma\varphi)$
is exactly the reverse phase accumulation. Then, because of the
relation $U_{z}^{a'b'}(-\varphi)U_{y}^{aa'}(\theta_a)U_{y}^{bb'}(\theta_b)U_{z}^{ab}(\varphi)\!=\! U_{z}^{ab}(\varphi)U_{y}^{aa'}(\theta_a)U_{y}^{bb'}(\theta_b)$,
we may also interpret $\tilde{U}^{a'b'}\!(\varphi)$
as an operation that---by varying $\gamma$ from $0$ to $1$---commutes
the losses with the phase delay, so that when $\gamma\!=\!1$
they occur before the phase delay not allowing the environmental modes
to possess any information about%
\footnote{%
Such an interpretation may be slightly misleading, as
it suggests to always naively set 
$\gamma\!=\!1$, in order not to let any ``information'' about the parameter
leak out into the environmental modes. Such choice, however, is actually \emph{not} 
optimal even in the $N\!\to\!\infty$ limit \citep{Escher2011}.
} $\varphi$.
Although in the most general scenario the optimal purification has also 
to account for the asymmetry of the loss model and $\gamma_a\!\ne\!\gamma_b$,
we are always able to write the QFI of the generated purification, and thus the required bound, as:
\begin{eqnarray}
F_{\textrm{Q}}\!\left[\rho_{\varphi}^{N}\right] 
& \le &  
\min_{\gamma_{a},\gamma_{b}}F_{\textrm{Q}}\!\left[\left|\tilde{\Psi}(\varphi)\right\rangle \right]
=
\min_{\gamma_{a},\gamma_{b}}\!\left\{ \left.\Delta^{2}{\hat h}_{a'b'}^\t{in}\right|_{\textrm{in}}+\left.\Delta^{2}{\hat J}_{z}^{ab}\right|_{\textrm{in}}-2\left.\textrm{cov}\!\left({\hat h}_{a'b'}^\t{in},{\hat J}_{z}^{ab}\right)\right|_{\textrm{in}}\right\}=
\label{eq:QFI_PurifBound_JS}\\
& &= 
\min_{\gamma_{a},\gamma_{b}}\!\left\{ 
\begin{array}{c}
\left(1-\tilde{\gamma}_{a}\right)^{2}\!\left.\Delta^{2}\hat{n}_{a}\right|_{\textrm{in}}+\eta_{a}\tilde{\gamma}_{a}\!\left\langle \hat{n}_{a}\right\rangle _{\textrm{in}}
+
\left(1-\tilde{\gamma}_{b}\right)^{2}\!\left.\Delta^{2}\hat{n}_{b}\right|_{\textrm{in}}+\eta_{b}\tilde{\gamma}_{b}\!\left\langle \hat{n}_{b}\right\rangle _{\textrm{in}}+\\
-\,2\left(1-\tilde{\gamma}_{a}\right)\left(1-\tilde{\gamma}_{b}\right)\left.\textrm{cov}\!\left(\hat{n}_{a},\hat{n}_{b}\right)\right|_{\textrm{in}}
\end{array}\right\}.\nonumber 
\end{eqnarray}
The first expression in \eqnref{eq:QFI_PurifBound_JS} follows by
writing explicitly the QFI \eref{eq:QFI} of the purification $\left|\tilde{\Psi}(\varphi)\right\rangle$ after conveniently defining%
\footnote{Due to locality, we may without loss of generality set $\varphi_0\!=\!0$ to simplify calculations, what implies $U_z^{ab}(\varphi_0)\!=\!{\tilde U}^{a'b'}\!(\varphi_0)\!=\!\mathbb{I}$.}
${\hat h}_{a'b'}^\t{in}\!=\! U_{z}^{ab}(-\varphi_0)U_{y}^{aa'}(-\theta_a)U_{y}^{bb'}(-\theta_b){\hat h}_{a'b'} U_{y}^{aa'}(\theta_a)U_{y}^{bb'}(\theta_b)U_{z}^{ab}(\varphi_0)$
as the generator of $\tilde{U}^{a'b'}\!(\varphi)$ evolved back through the ``circuit'' of \figref{fig:MZinter_losses_JS} in the Heisenberg picture,
so that it now acts on the purified input state $\left|\Psi_{\textrm{in}}\right\rangle\!=\!{\ket{\psi_\t{in}^N}}_{\!ab}\,{\ket{0}}_{a'}{\ket{0}}_{b'}$, with respect to which all
the quantities in \eqnref{eq:QFI_PurifBound_JS} are evaluated, i.e.~$\left.\bullet\right|_{\textrm{in}}\!\equiv\!\left\langle \Psi_{\textrm{in}}\right|\bullet\textrm{\ensuremath{\left|\Psi_{\textrm{in}}\right\rangle }}$.
Notice that such a formula may be intuitively explained, as  
the parameter may be treated at the level of $\left|\Psi_{\textrm{in}}\right\rangle$
as if it was encoded by two non-commuting generators
($\hat J_z^{ab}$ and $\hat h_{a'b'}$ evaluated for the initial stage of the ``circuit''), 
so that the QFI then corresponds to the \emph{sum of their variances minus twice their 
covariance}%
\footnote{%
As defined in \eqnref{eq:un_rel} before,
the \emph{covariance} of any two observables $\hat A$ and $\hat B$
reads:~$\t{cov}(\hat A,\hat B)\!=\!\frac{1}{2}\left<\left\{\hat A,\hat B\right\}\right>-\left<\hat A\right>\!\left<\hat B\right>$.
} \citep{Demkowicz2015}.
In order to derive the second expression in \eqnref{eq:QFI_PurifBound_JS}
with $\tilde{\gamma}_{a/b}\!=\!\gamma_{a/b}(1-\eta_{a/b})$,
one must commute ${\hat h}_{a'b'}$ through
the ``circuit'' of \figref{fig:MZinter_losses_JS},
what is possible by utilising the Baker-Campbell-Hausdorff formula
and the adequate bosonic operators commutation relations%
\footnote{%
We do not include the derivation here, as it is equivalent to
the one presented in the Sup.~Mat.~of \citep{Escher2011}.}.
Finally, \eqnref{eq:QFI_PurifBound_JS}
may be explicitly minimised w.r.t.~$\gamma_{a/b}$,
and for the special case of the input state \eref{eq:Input_Nph}
consisting of a \emph{definite} number of photons, $N$,
it may be also upper-limited by \citep{Escher2011}: 
\begin{equation}
F_{\textrm{Q}}\!\left[\rho_{\varphi}^{N}\right]\le\left(\frac{2N}{\sqrt{1+\frac{1-\eta_a}{\eta_a}N}+\sqrt{1+\frac{1-\eta_b}{\eta_b}N}}\right)^{2}\le\frac{4\, N}{\left(\sqrt{\frac{1-\eta_a}{\eta_a}}+\sqrt{\frac{1-\eta_b}{\eta_b}}\right)^{2}},
\end{equation}
so that the required SQL-like bound $\mathcal{F}_{as}^{\tiny bound}$ of 
\eqnref{eq:QFIAsBound} is recovered, which indeed reproduces 
the ultimate quantum limit on precision \eref{eq:Prec_loss_Q}.

\section{Bayesian approach -- global phase estimation}

Finally, we analyse the lossy interferometer of \figref{fig:MZinter_losses}
within the complementary \emph{Bayesian approach} to phase estimation.
Following the recipe of \secref{sub:QEstGlobal}, we assume 
\emph{no prior knowledge} about the estimated phase
and utilise the \emph{average cost} \eref{eq:QAvCostCov} that reads:
\begin{equation}
\left\langle \mathcal{C}_{\textrm{H}}\right\rangle = \textrm{Tr}\!\left\{ \left\langle \rho_{\varphi}^{N}\right\rangle _{C_{\textrm{H}}}\Xi^{N}\right\}
=
2-\boldsymbol{\alpha}^T \!\mathbf{A}\,\boldsymbol{\alpha}
\label{eq:AvCost_losses}
\end{equation}
with $\rho_\varphi^N$ now representing the
general output state of the interferometer \eref{eq:Output_Nph}.
The second expression in \eref{eq:AvCost_losses}
constitutes an analogue of \eqnref{eq:AvCost_MZInter_matrix} obtained
previously for the noiseless MZ interferometer, where again 
the matrix $\mathbf{A}$ possesses only non-zero terms on its 
first off-diagonals that this time read:
\begin{equation}
A_{n,n-1}=A_{n-1,n}=\sum_{l_a,l_b=0}^{n,N-n} \sqrt{b_{n}^{(l_a,l_b)}\,b_{n-1}^{(l_a,l_b)}}
\label{eq:OffDiagEls}
\end{equation}
with $b_{n}^{(l_a,l_b)}$ being the binomial
coefficients defined in \eqnref{eq:b^(la,lb)_n}.
Before analysing the problem of optimisation
of the average cost over the input state \eref{eq:Input_Nph} 
real coefficients:~$\boldsymbol{\alpha}\!=\!\{\alpha_0,\dots,\alpha_N\}$, 
we discuss the derivation of \eqnref{eq:AvCost_losses}
in more detail.

As in the case of the noiseless interferometer analysis of \secref{sub:QEst_MZInter_HL_Bay},
in order to obtain the second expression in \eref{eq:AvCost_losses},
we must firstly minimise the average cost over all \emph{covariant POVMs}, i.e.~their 
seed elements:~$\Xi^N\!\!\ge\!0$. Yet, the direct-sum structure of 
the lossy interferometer output state \eref{eq:Output_Nph}
implies that without loss of optimality we may also assume:~$\Xi_\t{\tiny opt}^N\!=\!\bigoplus_{N^{\prime}=0}^{N}\Xi_\t{\tiny opt}^{N'}$.
Physically, the block-diagonal form of $\Xi_{\textrm{\tiny opt}}^N$
indicates that the optimal covariant measurement requires a non-demolition
photon-number measurement to be performed before carrying out any
phase measurements, so that the orthogonal subspaces, labelled by the
number of surviving photons $N^{\prime}$, may be firstly distinguished.
One may correctly expect that, subsequently after learning $N'$, it is optimal
to just perform the measurement derived in \secref{sub:QEst_MZInter_HL_Bay}
for the noiseless scenario,
i.e.~$\Xi^{N'}_\t{\tiny opt}\!\!=\!\ket{e^{N'}}\!\bra{e^{N'}}$ with
$\ket{e^{N'}} = \sum_{n=0}^{N'} \ket{n,N'-n}$. However, due to the losses 
potentially present in \emph{both} arms, one cannot compensate any more
at the measurement stage for the complex phases of the input state coefficients, 
i.e.~$\phi_n$ in each $\alpha_n\!=\!|\alpha_n|\e^{\ii\phi_n}$, as this 
would also require the exact knowledge of how many photons were lost 
in each of the arms. 
Note that evaluating the ``lossy'' equivalent of \eqnref{eq:QAvCostCovMZ}
for the output state \eref{eq:Output_Nph}, $\rho_\varphi^N$, we most generally obtain:
\begin{eqnarray}
\left<\mathcal{C}_{\textrm{H}}\right> 
& = & 
2\left(1-\textrm{Re}\left\{ \sum_{N'=0}^{N}\;\sum_{l_{a}=0}^{N-N'}\sqrt{b_{n+l_{a}}^{(l_{a},N-N^{\prime}-l_{a})}b_{n-1+l_{a}}^{(l_{a},N-N^{\prime}-l_{a})}}\sum_{n=0}^{N'}\alpha_{n+l_{a}}^{*}\alpha_{n-1+l_{a}}\Xi_{n,n-1}^{N'}\right\} \right)\nonumber \\
& \ge & 
2\left(1-\sum_{N'=0}^{N}\sum_{l_{a}=0}^{N-N'}\sqrt{b_{n+l_{a}}^{(l_{a},N-N^{\prime}-l_{a})}b_{n-1+l_{a}}^{(l_{a},N-N^{\prime}-l_{a})}}\sum_{n=0}^{N'}\left|\alpha_{n+l_{a}}\right|\left|\alpha_{n-1+l_{a}}\right|\left|\Xi_{n,n-1}^{N'}\right|\right)\nonumber \\
 & \ge & 
2\left(1-\sum_{n,m=0}^{N}A_{n,n-1}\left|\alpha_{n}\right|\left|\alpha_{n-1}\right|\right).
\label{eq:AvCost_losses_expl}
\end{eqnarray}
For the first inequality above to be saturated---as $l_a$ is in principle not measurable---we 
must just impose $\forall_{n=0}^N\!:\alpha_n\!=\!|\alpha_n|$,
so that after indeed choosing $\Xi^{N}\!=\!\Xi^{N}_\t{\tiny opt}$
we obtain the third expression coinciding with \eqnref{eq:AvCost_losses}.
On the other hand, for the case of \emph{single-arm} losses such requirement is unnecessary,
as by measuring $N'$ we also then learn $l_a\!=\!N-N'$ and may thus always set 
$\ket{e^{N'}} = \sum_{n=0}^{N'} \e^{\ii\phi_{n+l_a}}\ket{n,N'-n}$ in $\Xi^{N}_\t{\tiny opt}$,
in order to saturate the first inequality independently of the input state complex phases.

Nevertheless, after taking without loss of optimality 
all $\alpha_n$ in \eqnref{eq:Input_Nph} to be real, 
we may identify similarly to \secref{sub:QEst_MZInter_HL_Bay} the 
\emph{minimal} average cost \eref{eq:AvCost_losses} for the lossy
interferometer as $2-\lambda_{\textrm{max}}$,
where $\lambda_{\textrm{max}}$ is again the maximal eigenvalue of the matrix
$\mathbf{A}$ and the corresponding eigenvector $\boldsymbol{\alpha}$
provides the \emph{optimal} input state coefficients. 
Recall (see \secref{sub:AvCost}) that the minimal average cost
quantifies the maximal achievable precision and in the $N\!\rightarrow\!\infty$ 
limit may be interpreted as the $\overline{\t{MSE}}$ \eref{eq:AvMSE}, $\left\langle \Delta^{2}\tilde{\varphi}\right\rangle$,
due to the convergence of the cost function
\eref{eq:CostFun_H} to the squared distance, $(\tilde\varphi-\varphi)^2$, 
as $\tilde{\varphi}\!\rightarrow\!\varphi$ with $N$. 

On the other hand, similarly to the case of the frequentist approach of
\secref{sec:MZ_losses_local} and, in particular, the investigation of the 
QFI bound \eref{eq:FQub} \citep{Demkowicz2009}, one should also verify if
the bosonic input states \eref{eq:Input_Nph} are really optimal from the Bayesian perspective,
as more general inputs consisting of \emph{distinguishable photons}
could in principle yield better precision. Yet, we explicitly demonstrate 
in \appref{chap:appMZinterAdaptM}
that by generalising the above analysis to input states which allow for
individual targeting of each of the constituent photons, the 
average cost \eref{eq:AvCost_losses} may only increase, so that
it is indeed sufficient to restrict only to the inputs \eref{eq:Input_Nph}.
Furthermore, we show that the corresponding covariant
POVMs that apply to such distinguishable-photons--strategies 
encompass the adaptive measurement strategies \citep{Wiseman1997,Wiseman1998,Wiseman2009},
and hence---similarly to the local approach of \secref{sec:MZ_losses_local}---the measurement 
adaptivity is
theoretically not necessary to attain the ultimate precision
determined within the Bayesian approach, although helpful from practical perspective.

\paragraph{Classical input states}~\\
As in the case of the local approach, we derive the minimal average cost \eref{eq:QAvCostCov} attained
by the \emph{classical strategy} employing a coherent state, $\ket{\alpha}$, 
split on the input beam-splitter of with transmittance $\tau_\t{in}$
shown in \figref{fig:MZinter_losses}.
This corresponds to evaluating \eref{eq:AvCost_losses} for
the state $\rho_{\varphi}^{\ket{\alpha}}$ \eref{eq:Output_coh},
what, however, requires a straightforward generalisation of the optimal covariant POVM 
to $\Xi_\t{\tiny opt}\!=\!\bigoplus_{N^\prime=0}^\infty \Xi^{N^\prime}_\t{\tiny opt}$,
so that its block-diagonal form matches \eqnref{eq:Output_coh}
for all the $N'$-photon sectors up to $N'\!\to\!\infty$.
After setting $|\alpha|^2\!=\!N$ and assuming a general $\tau_\t{in}$, we obtain:
\begin{equation}
\left<\mathcal{C}_{\textrm{H}}\right>_\t{cl}
\;=\;
2 -\frac{2\,\mathcal{B}\!\left[N \eta_{a}\tau_{\t{in}}\right]
\mathcal{B}\!\left[N \eta_{b}\!\left(1-\tau_{\t{in}}\right)\right]}{N\sqrt{\eta_{a}\tau_{\t{in}}\,\eta_{b}\left(1-\tau_{\t{in}}\right)}}
\;\overset{N\to\infty}{=}\;
\left(\frac{1}{\tau_{\t{in}} \eta_a} + \frac{1}{(1-\tau_{\t{in}}) \eta_b}  \right)\frac{1}{N}.
\label{eq:AvCost_losses_cl}
\end{equation}
where $\mathcal{B}(x)\!=\!\e^{-x}\sum_{n=0}^{\infty}\frac{x^{n}}{n!}\sqrt{n}$
is the Bell polynomial of order $1/2$.
Although for finite $N$ \eqnref{eq:AvCost_losses_cl} may only be numerically
minimised over $\tau_\t{in}$, in the strong-beam regime of 
$N\!\to\!\infty$ it takes a more appealing form shown in the second expression above, 
which similarly to the local approach indicates that 
it is optimal  to intuitively set $\tau_{\t{in}} = 1/(1+\sqrt{\eta_a/\eta_b})$
yielding the highest \emph{visibility} of the interferometer. Moreover, 
as indicated in \figref{fig:coeffs}(\textbf{b}) (\emph{short-dashed blue lines}),
asymptotically the same precision is then achieved as within the frequentist approach,
so that the minimal average cost \eref{eq:AvCost_losses_cl} attains the 
local classical bound \eref{eq:Prec_loss_cl}, i.e.
 \begin{equation}
\left<\mathcal{C}_{\textrm{H}}\right>_\t{cl}\;\overset{N\to\infty}{=}\; 
\frac{1}{4} \left( \frac{1}{\sqrt{\eta_a}} + \frac{1}{\sqrt{\eta_b}} \right)^2 \frac{1}{N}.
\label{eq:Prec_loss_cl_B}
\end{equation}
Let us note that the convergence of \eqnsref{eq:Prec_loss_cl}{eq:Prec_loss_cl_B} should have been expected, as
we are dealing with uncorrelated particles within the classical strategy, so that
the relation \eref{eq:MinAvMSE_Jensen}, which connects frequentist and Bayesian precision measures
at the classical level, directly applies.

\subsection{Numerical solution for moderate $N$}

In the noiseless case discussed in \secref{sub:QEst_MZInter_HL_Bay}, despite the deteriorating
assumption of the lack of prior knowledge, the minimal average cost \eref{eq:AvCostBW} has allowed
to prove the asymptotic HL-like scaling of precision, which is attained by 
the BW states \eref{eq:BW} employed as inputs.
In \figref{fig:coeffs}(\textbf{b}), we present the
change of the coefficient values defined by $\boldsymbol{\alpha}$
minimising \eqnref{eq:AvCost_losses}, as the losses are gradually introduced in the interferometer. 
In contrast to the frequentist approach depicted in \figref{fig:coeffs}(\textbf{a})---in 
which the optimal $|\alpha_n|^2$-distribution rapidly changes 
abandoning NOON states \eref{eq:NOON} for $\eta\!<\!1$, so that
other $\alpha_n$ coefficients must be discontinuously introduced as $\eta$
further decreases \citep{Knysh2011}---the optimal solution in the 
Bayesian case is ``smoothly`` modified with the growth of losses. 
Nevertheless, qualitatively a similar effect is observed in both cases, that  
as the amount of losses increases higher weights must be associated 
with intermediate $\alpha_{n}$-coefficients at the expense of the marginal ones,
in order to increase robustness of the input and preserve the
quantum superposition even after some photons are lost. On the other hand, a growing bias 
towards coefficients with higher $n$ must be introduced due to the asymmetry 
of the single-arm--losses model depicted in \figref{fig:coeffs}. In the regime of very 
high losses, $\eta\!\rightarrow\!0$, the $|\alpha_n|^2$-distributions derived within 
the complementary frequentist and Bayesian approaches attain a common one, 
suggesting that the role of the prior knowledge
is severely reduced when nearly all the photons are lost in the setup.

\subsection{Asymptotic SQL-like bound on precision}

Unfortunately, we can only numerically determine the minimal average cost 
\eref{eq:AvCost_losses} (corresponding to $2-\lambda_\t{max}$ with $\lambda_\t{max}$ being 
the maximal eigenvalue of the matrix $\mathbf{A}$), so that the asymptotic precision scaling 
with $N$ cannot be again verified analytically. However, similarly to the case of the local approach in which we
have utilised the QFI-upper--bounding techniques, we prove the 
asymptotic SQL-like scaling of precision within the Bayesian approach
by constructing a lower bound on the minimal average cost 
that scales classically, as $1/N$, with the number of photons, whenever any losses are present.

We achieve this by constructing a valid upper bound on $\lambda_\t{max}$
that applies if either $\eta_{a}\!<\!1$ or $\eta_{b}\!<\!1$.
Recall that $\mathbf{A}$ is a symmetric real matrix, and thus 
for any arbitrary normalised real vector $\boldsymbol{v}$, $\boldsymbol{v}^T \!\mathbf{A} \boldsymbol{v}\leq
 \lambda_{\t{max}}$. Let $\boldsymbol{\alpha}$ be
the eigenvector corresponding to $\lambda_{\t{max}}$, so that
$\boldsymbol{\alpha}^T \!\mathbf{A} \boldsymbol{\alpha}\!=\!\lambda_{\t{max}}$.
The fact that all matrix elements of $\mathbf{A}$ are non-negative
implies that also all $\alpha_n \geq 0$.
Let us now define a matrix $\mathbf{A}'$
such that all non-zero entries of $\mathbf{A}$ are replaced by
its largest element $A_{N}^\uparrow\!=\!\max_{n}
\left\{ A_{n,n-1}\right\}$, which is contained within
the only non-zero, off-diagonal entries \eref{eq:OffDiagEls}.
In case of the \emph{single-arm losses}, $A_{N}^\uparrow$ corresponds
either to the first or last term, i.e.~$A_{N,N-1}$ for ($\eta_a\!<\!1$, $\eta_b\!=\!1$) 
and $A_{1,0}$ for ($\eta_a\!=\!1$, $\eta_b\!<\!1$),
whereas for \emph{equal losses} ($\eta_a\!=\!\eta_b\!<\!1$) to the middle term
$A_{\left\lceil\frac{N}{2}\right\rceil,\left\lceil\frac{N}{2}\right\rceil-1}$.
For \emph{unequal-losses} scenario $A_{N}^\uparrow$ is one of the
entries in between, which unfortunately may only be determined numerically.
We have labelled $A_{N}^\uparrow$ with the subscript $N$, in order to emphasise that the 
maximal entry of $\mathbf{A}$ defined by \eqnref{eq:OffDiagEls} explicitly depends 
on the photon number, with the \emph{only} exception being the noiseless scenario
in which $\forall_n\!:A_{n,n-1}\!=\!1$ in accordance with \eqnref{eq:AvCost_MZInter_matrix}.

Since $\alpha_n \geq 0$ and $A^\prime_{n,m} \geq A_{n,m} \geq 0$,
we can write:
\begin{equation}
\lambda_{\t{max}} \;=\; \boldsymbol{\alpha}^T\! \mathbf{A} \boldsymbol{\alpha} \;\leq\;
\boldsymbol{\alpha}^T\! \mathbf{A}^\prime \boldsymbol{\alpha} \;\leq\; \lambda^\prime_{\t{max}},
\end{equation}
where $\lambda^\prime_{\t{max}}$ is now the maximal eigenvalue of $\mathbf{A}^\prime$. 
Crucially, $\lambda^\prime_{\t{max}}$ can be easily found analytically 
after noticing that $\mathbf{A}^\prime\!=\!A_{N}^\uparrow\bar{\mathbf{A}}$,
where $\bar{\mathbf{A}}$ is the matrix occurring in the noiseless scenario and
defined in \eqnref{eq:AvCost_MZInter_matrix}. As multiplication by a constant
just rescales equally all the eigenvalues, so that in particular
$\lambda^\prime_{\t{max}}\!=\!A_{N}^\uparrow\bar\lambda_\t{max}\!=\!2 A_{N}^\uparrow \cos\left[\pi/(N+2)\right]$
with $\bar\lambda_\t{max}$ corresponding to the maximal eigenvalue in the noiseless 
setting, we obtain the desired lower bound on the minimal average cost \eref{eq:AvCost_losses} that reads:
\begin{equation}
\left\langle \mathcal{C}_{\textrm{H}}\right\rangle \;\ge\; 2\,\left[1-A_{N}^\uparrow\cos\left(\frac{\pi}{N+2}\right)\right].
\label{eq:AvCost_losses_LB}
\end{equation}
As $A_{N}^\uparrow$ is analytically defined \emph{only}
for single-arm and equal losses, we can explicitly determine 
the analytical form of \eqnref{eq:AvCost_losses_LB} for $N\!\to\!\infty$
only in these two cases.
Nevertheless, we numerically verify that \eqnref{eq:AvCost_losses_LB} 
actually coincides in the asymptotic $N$ limit 
with the bound \eref{eq:Prec_loss_Q} derived within the local approach
independently of the $\eta_{a/b}$ chosen, so that most generally:
\begin{equation}
\left\langle \mathcal{C}_{\textrm{H}}\right\rangle\;\ge\;\frac{1}{4}\,\left(\sqrt{\frac{1-\eta_a}{\eta_a}}+\sqrt{\frac{1-\eta_b}{\eta_b}}\right)^2\;\frac{1}{N}.
\label{eq:Prec_loss_Q_B}
\end{equation}
In \figref{fig:PrecPlotLossInter}(\textbf{b}), we study 
the minimal average cost \eref{eq:AvCost_losses} attainable
in the \emph{single-arm} losses scenario ($\eta_a\!=\!\eta$, $\eta_b\!=\!1$)
after setting $\eta\!=\!0.7$.
The results show that the above asymptotic bound \eref{eq:Prec_loss_Q_B} (\emph{solid grey line}) is indeed 
saturated within the global approach by the optimal input states
(\emph{solid black line}) with $\alpha_n$-coefficients
distributed according to \figref{fig:coeffs}(\textbf{b}).
Although the results are numerical, one should note that within
the Bayesian approach---as the optimisation corresponds just
to determining the largest eigenvalue of the $\mathbf{A}$ matrix---much 
higher photon numbers ($N\!\lesssim\!1000$) are reachable allowing to numerically verify
the convergence of the minimal average cost to the limit dictated by \eqnref{eq:Prec_loss_Q_B} 
for the whole range of $\eta_{a/b}$ up to a convincing accuracy.
For comparison, as in case of the local approach analysed in \figref{fig:PrecPlotLossInter}(\textbf{a}),
we show also the cost, $\left<\mathcal{C}_{\textrm{H}}\right>_\t{cl}$,
attained by the coherent-state--based \emph{classical strategy} 
(\emph{short-dashed blue line}), which is determined for large $N$ by \eqnref{eq:Prec_loss_cl_B}.
Moreover, we plot $\left<\mathcal{C}_\t{H}\right>_\t{\tiny BW}$ \eref{eq:AvCostBW}
(\emph{long-dashed black line})
achieved by the BW states \eref{eq:BW} in the \emph{noiseless} setting, which should be really 
treated as a benchmark defining the HL within the Bayesian framework.

Let us emphasise that due to the presence of losses the
frequentist and Bayesian asymptotic precision bounds \eref{eq:Prec_loss_Q} and \eref{eq:Prec_loss_Q_B}
coincide, what importantly contrasts the noiseless case in which a discrepancy-factor of
$\pi^2$ has been observed in between \eqnsref{eq:QFIandQCRB_NOON}{eq:AvCostBW}.
In the case of classical inputs, i.e.~the coherent input states \eref{eq:Input_coh},
such behaviour could have been expected (see \eqnsref{eq:Prec_loss_cl}{eq:Prec_loss_cl_B}), 
as similarly to \secref{sub:ClEst_MZInter_SQL} the interfering photons are then uncorrelated,
so that the relation \eref{eq:MinAvMSE_Jensen} may be again invoked to prove
the convergence of the CRB \eref{eq:CRB} and the average Bayesian cost \eref{eq:AvCost}
at the classical level of probabilities. On the other hand, such explanation fails
for the quantum-enhanced strategies, as the optimal input states must then benefit 
from the inter-particle entanglement even in the presence of losses and
the constituent photons cannot be thus treated as independent ``statistical objects'' 
also in the asymptotic $N$ limit.
Yet, the convergence of the asymptotic bounds, combined with the
fact that they are known to be saturable for $N\!\to\!\infty$ without need of conducting
procedure repetitions ($\nu\!=\!1$ in \figref{fig:Ph_Est_Scheme}) \citep{Knysh2014,Demkowicz2015},
suggests that the optimal input states may be well-approximated 
for sufficiently large $N$ by inputs exhibiting only short-range correlations
of the particles. As a result the input states may then be effectively thought of as 
if they consisted of entangled groups of particles (photons) that do not possess any 
correlations in between the groupings, so that by setting $N\!\to\!\infty$ the regime 
of infinite sampling (infinite number of uncorrelated groups) is naturally attained in a single shot. 
Such an intuition has recently been shown to be indeed correct by efficiently approximating 
the optimal input states in the presence of losses with use of the so-called matrix-product states 
\citep{Jarzyna2013} that do not exhibit long range inter-particle correlations.

%% file: Chapters/conc.tex
\chapter{Conclusions and outlook} 
\label{chap:conc} 
\lhead{Chapter 6. \emph{Conclusions and outlook}}

In this thesis, we have analysed in detail the effects
of uncorrelated noise in quantum metrological scenarios, and
most importantly presented tools that allow to bound
the maximal attainable precision in the presence of decoherence.
We have shown that generic types of such noise, even infinitesimally 
small in magnitude, force the precision 
to scale at the Standard Quantum Limit (i.e.~limit the Mean Squared Error
to follow the $1/N$ scaling) when the number of particles employed, $N$,
becomes infinitely large, and thus constrain the ultimate quantum 
enhancement to a constant factor.
Furthermore, we have proposed a generalisation
of these asymptotic methods to the finite-$N$ regime, which
importantly is always efficiently computable due to its
semi-definite program form.

Yet, as the bounds obtained are typically saturated
in the asymptotic $N$ limit by employing non-complex 
quantum states and measurements, our results
crucially prove that, in order to improve large-scale experiments involving
many particles, the primary priority of an experimentalist should be 
the reduction of the noise present, as investing in more exotic quantum states and
detection strategies does not then lead to any significant precision improvement.
A spectacular application of our precision bounds has been demonstrated
in the analysis of gravitational wave-detectors \citep{LIGO2011,LIGO2013},
in which \citet{Demkowicz2013} have explicitly shown that the currently
conducted detection schemes are working already at the theoretical 
limits predicted by our techniques.
On the other hand, our finite-$N$ semi-definite programming
methods allow to verify the size of the small-$N$ regime, in which
the presence of decoherence may be effectively ignored.
As a consequence, they prove for the first time the optimality of the protocols 
utilised in the pioneering quantum metrology experiments (e.g.~\citep{Leibfried2003,Mitchell2004}), 
which assumed the absence of noise.

Moreover, as our results also apply to metrological scenarios
in which the single-particle evolution varies with the total number
of particles involved (e.g.~the optical depth of an atomic ensemble
changes with the number of atoms employed in an experiment \citep{Wasilewski2010}),
they may be equivalently utilised to quantify the attainable precision 
in frequency estimation models---see \secref{sec:loc_freq_est}---in 
which the time duration of each experimental trial
serves as an extra parameter that may in principle be freely adjusted. 
In \citep{Chaves2013}, our methods have been applied to a
similar model as the one presented in \secref{sec:loc_freq_est},
yet with non-commuting Liouvillian and Hamiltonian parts of the master equation 
(see \secref{sub:Evo_MEq}). 
In particular,
the tools proposed in this thesis have allowed to identify for the first time 
(according to our best knowledge) an example of a Markovian uncorrelated noise 
for which the SQL-like scaling may still be surpassed---corresponding to a dephasing-type 
noise directed in a transversal direction to the Hamiltonian encoding the frequency. 
As the attainability of super-classical precision scalings has also been demonstrated in
Non-Markovian noise models \citep{Matsuzaki2011,Chin2012}, an interesting question
arises, whether our bounding-techniques (which are then also applicable) could suggest
further room for improvement in such frequency estimation protocols with memory.

We would also like to point that our SQL-like bounds allow one to avoid 
some of the controversies characteristic for the idealised noiseless scenarios.
When decoherence is \emph{not} present and the probe states with indefinite number 
of particles are considered (e.g.~squeezed states of light),
the exact form of the Heisenberg Limit needs to be reconsidered 
\citep{Hyllus2010,Berry2012,Hall2012a,Giovannetti2012a,Hofmann2009},
since the direct replacement of $N$ with mean number of particles $\bar{N}$ may make the HL invalid.
Moreover, the final claims on the achievable precision scaling may strongly depend on the form
of the prior knowledge about the parameter assumed, and lead to apparent 
contradictions \citep{Anisimov2010, Giovannetti2012}.
These difficulties do not arise in the presence of uncorrelated noise, as
the asymptotic SQL-like bounds are then valid also when $N$ is replaced by 
$\bar{N}$---as explicitly shown in \secref{sub:asbounds_mzloss}. In particular,
the bounds obtained are saturated in a single-shot scenario, unlike the decoherence-free case 
when only after some number of independently repeated experiments one may expect to
approach the theoretical limits \citep{Pezze2009,Giovannetti2012a}.
This is due to the fact that by employing input states of
grouped particles, which possess no correlations in between the groupings,
and by letting the groups to be of finite but sufficiently large size,
one can attain the ultimate asymptotic SQL-like bound up to any precision
\citep{Jarzyna2014}. 
It is thus the uncorrelated noise that makes the asymptotic 
$N$ limit naturally account for the infinite sampling regime, 
what---by the arguments of \secref{sub:AvMSE}---also explains why the 
frequentist and Bayesian approaches yield same asymptotic measures of accuracy. 
We have demonstrated such
a behaviour explicitly for the lossy interferometer model in \chapref{chap:MZ_losses}, 
yet a general explanation has been recently provided in \citep{Jarzyna2014}.
As a result, we also expect the \emph{information theoretic} results
\citep{Hall2012,Nair2012}, complementary to frequentist and Bayesian methods, 
to recover bounds in the presence of uncorrelated noise that
are compatible with our techniques determined for the 
local (Fisher-Information--based) measures.
Notice that the above argument also suggests the asymptotically optimal form of 
the input states, which should include ones that do not possess 
long-range correlations in between the particles.
This observation has already been made in \citep{Sorensen2001} and indicates that in
methods designed to search for the optimal inputs in scenarios with uncorrelated noise
one may restrict himself to states with short-range correlations such as for example
the matrix product states of low bond dimensions \citep{Jarzyna2013}.

For completeness, in order to also be critical, let us note that 
the best-performing method we have proposed stems from the Channel Extension
presumption---see \secref{sub:CEbound}---assuming presence of ancillary non-evolving 
particles, what allows then to carry out the semi-definite program reformulation. 
Thus, although the Channel Extension method turns out to be most agile and effective, 
it may still provide asymptotic and finite-$N$ bounds that are not saturable in 
real-life protocols, in which such auxiliary particles are not available.
Hence, it is important always to verify the tightness of the bounds predicted
by our results, what, however, has very recently been shown to be possible 
by utilising a more involved approach based on the calculus of variations
\citep{Knysh2014}. What is more, throughout this thesis we have analysed 
models in which one deals with a \emph{single} latent parameter of interest 
to be estimated in a metrological scenario, so that a natural 
future work on our methods would be to generalise them and 
study their applicability in the \emph{multi-parameter} estimation schemes 
\citep{Genoni2013,Humphreys2013,Vidrighin2014,Crowley2014}.
However,  let us note that in such a case there does not
exist a clear equivalent of the Fisher Information (matrix) 
in the quantum setting \citep{Hayashi2005a}, what questions 
whether saturable non-trivial precision bounds can even be constructed.
However, as in the multi-parameter estimation scenario 
the non-commutativity of observables---in 
particular, the generators 
of the estimated parameters---starts to play a crucial role, 
many unexpected and interesting issues may arise 
when following these lines of research.
On the other hand, we have also entirely focussed
on the uncorrelated noise models, so that one may wonder
what are the effects of collective decoherence, which coherently affects
all the constituent particles of a given quantum system. 
As physically relevant types of such noise 
may be treated as fluctuations of the estimated parameter itself
\citep{Knysh2014,Macieszczak2014a,Genoni2011,Dorner2012},
correlated noise generally yields an extra constant offset-term in the error of an estimator, which
importantly does not diminish with an increase in the number of particles $N$.
Thus, one may intuitively expect the uncorrelated noise 
to vary the scaling of the degradable term to become SQL-like in accordance 
with our methods, while the other correlated-noise term persists unaffected \citep{Knysh2014}.
An interesting question arises, whether it is possible to
develop techniques similar to ours that would predict
bounds when considering noise-types of a fixed correlation size
that may be varied smoothly from a single particle (uncorrelated noise
studied in this thesis) to all particles (correlated noise) \citep{Macieszczak2014,Jeske2013}.

Lastly, let us remark that, as noted in various parts of this work 
(see Notes \refcol{note:time_energy} and \refcol{note:time_energy_rel_channel}),
the parameter estimation techniques can be naturally employed
to quantify precision with which the time duration of
system evolution may be determined. This is a very special metrology-like problem
that has been analysed in various ways not directly related to 
statistical inference techniques of estimation theory  \citep{Aharonov2002},
and most notably also with use of the so-called ``\emph{speed limits}'' that bound the 
speed with which a quantum system may evolve \citep{Margolus1998,Deffner2013a,Giovannetti2003}.
Furthermore, it has been demonstrated similarly to
the metrological scenario that, given a system consisting 
of $N$ particles, their inter-particle entanglement substantially 
speeds up the evolution process \citep{Frowis2012,Giovannetti2003a,Zander2007},
as  also predicted by the Quantum Cram\'{e}r-Rao Bound \eref{eq:TimeEnergyUR} 
yielding the HL-like 
scaling, $1/N^2$, of the time-duration estimator for a 
maximally entangled state.
Thus, it is a natural question to ask whether our methods 
could be adapted to such a particular setting, in order to 
account for the impact of uncorrelated noise; 
and how are they related to the recent results
which allow to incorporate decoherence-effects  into 
the `speed-limit' formalism \citep{Taddei2013,Campo2013}, or 
even the Non-Markovian behaviour of the evolution \citep{Deffner2013}.

%% file: Appendices/AppendixA.tex
\chapter{Choi criterion for extremality of a quantum channel} 
\label{chap:appChoiCrit}

\lhead{Appendix A. \emph{Choi criterion for extremality of a quantum channel}} 

Let us consider a quantum channel---a CPTP map 
(see \secref{sub:Evo_QCh})---$\Lambda\!:\mathcal{B}(\mathcal{H}_\t{in})\!\to\!\mathcal{B}(\mathcal{H}_\t{out})$
of rank $r$, so that it admits a set of linearly independent Kraus operators $\{K_i\}_{i=1}^r$.
For simplicity, let us also denote the space of all CPTP maps as $\mathcal{CPTP}$ and its 
subset containing the \emph{extremal} quantum channels as $\partial\mathcal{CPTP}$.
We explicitly prove, following \citep{Choi1975}, the \emph{Choi criterion} \refcol{crit:Choi} stating that:

\begin{mytheorem}
$\Lambda$ is an extremal map if and only if $\left\{ K_{i}^{\dagger}K_{j}\right\} _{ij}$
is a linearly independent set of $r^{2}$ matrices, i.e.~the only 
$r\!\times\!r$ matrix $\boldsymbol{\mu}$ that satisfies $\sum_{ij}\boldsymbol{\mu}_{ij}K_{i}^{\dagger}K_{j}=0$ 
is the trivial one $\boldsymbol{\mu}\!=\!0$.
\end{mytheorem} 

\textsf{\color{myred} Proof:}~\\
$\Lambda\in\partial\mathcal{CPTP}\;\implies\;\t{the set of matrices} \left\{ K_{i}^{\dagger}K_{j}\right\} _{ij}$
is linearly independent.\\
Suppose there exists $\boldsymbol{\mu}\!\ne\!0$ such that $\epsilon\sum_{ij}\boldsymbol{\mu}_{ij}K_{i}^{\dagger}K_{j}\!=\!0$
for any $\epsilon\!>\!0$. $\boldsymbol{\mu}$ can be assumed to be Hermitian, as by taking the Hermitian 
conjugate:~$\left(\sum_{ij}\boldsymbol{\mu}_{ij}K_{i}^{\dagger}K_{j}\right)^\dagger\!\!=\!\sum_{ij}\boldsymbol{\mu}_{ij}^{*}K_{j}^{\dagger}K_{i}
\!=\!\sum_{ij}\boldsymbol{\mu}_{ij}^\dagger K_{i}^{\dagger}K_{j}$, so that equivalently
$\epsilon\sum_{ij} \!\left(\boldsymbol{\mu}\!\pm\!\boldsymbol{\mu}^\dagger\right)_{ij}\! K_i^\dagger K_j\!=\!0$.
Let us define two maps $\Lambda_{\pm}\!:\mathcal{B}(\mathcal{H}_\t{in})\!\to\!\mathcal{B}(\mathcal{H}_\t{out})$
such that
$\forall_{\varrho\in\mathcal{B}(\mathcal{H}_\t{in})}\!:
\Lambda_{\pm}\!\left[\varrho\right]\!=\!\sum_{i,j=1}^{r}\left(\mathbb{I}\pm\epsilon\boldsymbol{\mu}\right)_{ij}K_{i}\varrho K_{j}^{\dagger}$.
Firstly, $\mathrm{Tr}\!\left\{ \Lambda_{\pm}\!\left[\varrho\right]\right\}\!=\!\mathrm{Tr}\!\left\{\varrho\right\}\!\pm\!\epsilon\,\mathrm{Tr}\!\left\{ \sum_{i,j=1}^{r}\boldsymbol{\mu}_{ij}K_{j}^{\dagger}K_{i}\,\varrho\right\}\!=\!1$,
what proves that both $\Lambda_{\pm}$ are TP. 
On the other hand, let us choose $\epsilon$ small enough,
so that $\mathbb{I}\!\pm\!\epsilon\boldsymbol{\mu}\!\ge\!0$, and construct
$\boldsymbol{\nu}\!=\!\sqrt{\mathbb{I}\!\pm\!\epsilon\boldsymbol{\mu}}$.
Then, $\Lambda_{\pm}\!\left[\varrho\right]\!=\!\sum_{i,j=1}^{r}\left(\boldsymbol{\nu}_\pm^2\right)_{ij}K_{i}\varrho K_{j}^{\dagger}
\!=\!\sum_{i=1}^{r}K_{i}^\pm\varrho K_{i}^{\pm\dagger}$
with effective Kraus operators $K_{i}^\pm\!=\!\sum_{j=1}^{r}\!\boldsymbol{\nu}_{\pm ji}K_{j}$,
what proves that $\Lambda_\pm\!\in\!\mathcal{CPTP}$.
Now, as the original channel is obtained by composing 
$\Lambda\!=\!\frac{1}{2}\left(\Lambda_{+}\!+\!\Lambda_{-}\right)$,
it \emph{cannot} be extremal according to the original \defref{def:ch_extrem}. \hfill {\color{myred} $\blacksquare$}

The set of matrices $\left\{ K_{i}^{\dagger}K_{j}\right\} _{ij}$ is linearly independent $\;\implies\;\Lambda\in\partial\mathcal{CPTP}$.\\
Suppose $\Lambda\!\notin\!\partial\mathcal{CPTP}$, so that
there exist two maps $\Lambda_\pm\!\in\!\mathcal{CPTP}$ that yield
$\Lambda\!=\!\frac{1}{2}\!\left(\Lambda_{+}\!+\!\Lambda_{-}\right)$
and possess Kraus representations such that
$\forall_{\varrho\in\mathcal{B}(\mathcal{H}_\t{in})}\!\!:\Lambda_\pm\!\left[\varrho\right]\!=\!\sum_{i=1}^{r_\pm}\!K_{i}^\pm\varrho K_{i}^{\pm\dagger}$ 
with $\sum_{i}^{r_\pm}\!K_{i}^{\pm\dagger}K_{i}^\pm\!=\!\mathbb{I}$.
On the other hand, in the CJ matrix \eref{eq:CJmatrix} representation $\Omega_\Lambda\!=\!\frac{1}{2}\!\left(\Omega_{\Lambda_{+}}\!\!+\!\Omega_{\Lambda_{-}}\right)$,
so that the support of the CJ matrix of $\Lambda$, $\Omega_\Lambda\!=\!\sum_{i=1}^r\!\ket{K_i}\bra{K_i}$,
must contain both $\Omega_{\Lambda_\pm}\!=\!\sum_{i=1}^{r_\pm}\ket{K_i^\pm}\bra{K_i^\pm}$.
Thus, we may express any $\ket{K_i^\pm}$ as a superposition of $\{\ket{K_i}\}_{i=1}^r$,
what means that in the channel-picture all Kraus operators $\{K^\pm_i\}_{i=1}^{r_\pm}$ 
can be written as linear compositions of $\{K_i\}_{i=1}^r$, e.g.~$K_{i}^+\!=\!\sum_{j=1}^{r}\!\boldsymbol{\alpha}_{ij}K_{j}$
where $\boldsymbol{\alpha}$ is some (non-unitary) $r^+\!\times\!r$ matrix.
Then, the TP properties of respectively $\Lambda_{+}$ and $\Lambda$ imply 
$\sum_{i,j=1}^{r}\!\left(\boldsymbol{\alpha}^\dagger\boldsymbol{\alpha}\right)_{ij}\!K_{i}^{\dagger}K_{j}\!=\!\mathbb{I}$
and $\sum_{i=1}^{r}\!K_{i}^{\dagger}K_{i}\!=\!\mathbb{I}$,
so that there exists $\boldsymbol{\mu}\!=\!\boldsymbol{\alpha}^\dagger\boldsymbol{\alpha}\!-\!\mathbb{I}\!\ne\!0$
for which $\sum_{i,j=1}^{r}\!\boldsymbol{\mu}_{ij}K_{i}^{\dagger}K_{j}\!=\!0$.
Hence, the set of $\left\{ K_{i}^{\dagger}K_{j}\right\} _{ij}$ is linearly dependent. \hfill {\color{myred} $\blacksquare$}

%% file: Appendices/AppendixB.tex
\chapter{Criterion for $\varphi$-extremality of a quantum channel} 
\label{chap:appPhiExtremCond}

\lhead{Appendix B. \emph{Criterion for $\varphi$-extremality of a quantum channel}} 

Interpreting a family of CPTP maps $\{\Lambda_\varphi\}_\varphi$ parametrised by $\varphi$ as
a curve in the space all quantum channels, we have \emph{geometrically} defined
in \defref{def:ch_extrem} the quantum channel $\Lambda_\varphi$ to be $\varphi$\emph{-extremal} at $\varphi_0$,
if it \emph{cannot} be decomposed there into a probabilistic mixture of two quantum channels $\Lambda_\pm$,
i.e.~$\Lambda_{\varphi_0}\!=\!p_+\Lambda_+\!+\!p_-\Lambda_-$, 
that lie along the tangent to the curve at $\varphi_0$ or, equivalently, along 
the direction defined by the derivative $\left.\partial_\varphi\Lambda_\varphi\right|_{\varphi_0}\!\equiv\!\left.\dot\Lambda_\varphi\right|_{\varphi_0}$,
so that $\Lambda_\pm\!=\!\Lambda_{\varphi_0}\!\pm\!\epsilon\!\left.\dot\Lambda_\varphi\right|_{\varphi_0}$ for some non-zero $\epsilon$.
Due to the CJ-isomorphism described in \secref{sub:CJiso}, the above requirement can be directly translated
onto the density matrix representation of quantum channels, so that it is tantamount to
non-existence of $\epsilon\!>\!0$ such that
\begin{equation}
\Omega_{\Lambda_{\varphi_0}}\pm\epsilon\,\dot\Omega_{\Lambda_{\varphi_0}}\ge0,
\label{eq:phiExtremCond1}
\end{equation}
where $\Omega_{\Lambda_{\varphi_0}}\!=\!\sum_i\ket{K_i}\!\bra{K_i}$ and 
$\dot\Omega_{\Lambda_{\varphi_0}}\!=\!\sum_i\ket{\dot K_i}\!\bra{K_i}\!+\!\ket{K_i}\!\bra{\dot K_i}$
are respectively the CJ matrix of $\Lambda_\varphi$ and its derivative at $\varphi_0$,
with $\ket{K_i}\!=\!K_i\!\otimes\!\mathbb{I}\,\ket{\mathbb{I}}$ (we drop the explicit $\varphi$-dependence
of the Kraus operators for simplicity) corresponding 
to the canonical Kraus representation of $\Lambda_\varphi$ introduced in \secref{sub:CJiso}.
We prove that:
\begin{mytheorem}
$\Lambda_\varphi$ is $\varphi$-extremal at $\varphi_0$ if and only if $\dot{\Omega}_{\Lambda_{\varphi}}$ 
is not contained within the support of $\Omega_{\Lambda_{\varphi}}$ at $\varphi_0$.
\end{mytheorem} 
In other words, there does not exist $\epsilon\!>\!0$ satisfying \eqnref{eq:phiExtremCond1}
if and only if there exists a non-zero Hermitian matrix $\boldsymbol{\mu}$ such that 
\begin{equation}
\dot\Omega_{\Lambda_{\varphi_0}}=\sum_{ij}\boldsymbol{\mu}_{ij}\left|K_{i}\right>\!\left<K_{j}\right|.
\label{eq:phiExtremCond2}
\end{equation}
\textsf{\color{myred} Proof:}~\\
We show the equivalence of \eqnsref{eq:phiExtremCond1}{eq:phiExtremCond2}, what proves
also the equivalence of the above theorem and \defref{def:ch_extrem}.

\eqnref{eq:phiExtremCond1} $\implies$ \eqnref{eq:phiExtremCond2}\\
Assuming \eqnref{eq:phiExtremCond1} to hold, we write explicitly its l.h.s., so that \eqnref{eq:phiExtremCond1} states that
\begin{equation}
\bra{\psi}\left[\sum_{i}\ket{K_{i}}\!\bra{K_{i}}\pm\epsilon\left(\sum_{i}\ket{\dot{K_{i}}}\!\bra{K_{i}}+\ket{K_{i}}\!\bra{\dot{K_{i}}}\right)\right]\ket{\psi}\geq0
\label{eq:poscond}
\end{equation}
for any (even not normalised) $\ket{\psi}$. 
In order to prove \eqnref{eq:phiExtremCond2},
it is enough to show that all $\ket{\dot{K}_{i}}$ can be written as linear
combinations of $\ket{K_{i}}$. If this was not the case for one of them,
e.g.~$\ket{\dot{K}_{\bar{i}}}$, its decomposition would additionally require
a vector $\ket{L_{\bar{i}}}$ which is not 
contained within the support of $\Omega_{\Lambda_{\varphi_0}}$,
i.e.~orthogonal to all $\ket{K_i}$.
Then, taking $\ket{\psi}\!=\!\sqrt{\lambda}\,\ket{K_{\bar{i}}}\!+\!\e^{\ii\phi}\sqrt{1\!-\!\lambda}\,\ket{L_{\bar{i}}}$,
\eqnref{eq:poscond} reads:
\begin{eqnarray}
\lambda\, \braket{K_{\bar{i}}}{K_{\bar{i}}}\left(
\braket{K_{\bar{i}}}{K_{\bar{i}}}\pm\epsilon\left.\frac{\partial\braket{K_{\bar{i}}}{K_{\bar{i}}}}{\partial\varphi}\right|_{\varphi_0}
\right)
\pm2\epsilon\sqrt{\lambda(1-\lambda)}\,\t{Re}\!\left\{\e^{\ii\phi}\braket{L_{\bar{i}}}{\dot{K}_{\bar{i}}}\right\}
&=& \nonumber \\ 
\overset{\lambda\ll1}{=}\;\;\pm2\epsilon\sqrt{\lambda}\,\t{Re}\!\left\{\e^{\ii\phi}\braket{L_{\bar{i}}}{\dot{K}_{\bar{i}}}\right\}\;\;+\;\;O(\lambda)
&\geq&0.
\label{eq:projPsi}
\end{eqnarray}
Thus, for any non-zero $\epsilon$ and $\braket{L_{\hat{i}}}{\dot{K}_{\hat{i}}}$,
we may always set $\lambda$ small enough, so that there exists $\phi$ for which the
l.h.s.~of \eqnref{eq:projPsi} is negative. This leads to a contradiction, hence \eqref{eq:phiExtremCond2}
must hold.

\eqnref{eq:phiExtremCond2} $\implies$ \eqnref{eq:phiExtremCond1}\\
Let us assume the decomposition \eref{eq:phiExtremCond2} to be valid,
so that the condition \eref{eq:phiExtremCond1} now reads
\begin{equation}
\sum_{i}\ket{K_{i}}\!\bra{K_{i}}\pm\epsilon\sum_{ij}\boldsymbol{\mu}_{ij}\ket{K_{j}}\!\bra{K_{i}}\;\geq\;0
\label{eq:PhiExtCond}
\end{equation}
and we must show that it holds for some $\epsilon\!>\!0$.
Yet, we may always define matrices $\boldsymbol{\nu}^{\pm}=\mathbb{I}\!\pm\!\epsilon\boldsymbol{\mu}$ which
are positive semi-definite for small enough $\epsilon$, so that
by taking their square root we can construct $\ket{\tilde{K}_{i}^{\pm}}=\sum_{j}\left[\sqrt{\boldsymbol{\nu}^{\pm}}\right]_{ji}\ket{K_{j}}$.
Now, we recover the l.h.s.~of \eqnref{eq:PhiExtCond} by evaluating adequately 
one of $\sum_{i}\ket{\tilde{K}_{i}^\pm}\!\bra{\tilde{K}_{i}^\pm}$ 
that are clearly positive semi-definite, what completes the proof.
\hfill {\color{myred} $\blacksquare$}

~\\As aside, one should note that in general---by writing explicitly
$\ket{K_i}\!=\!K_i\!\otimes\!\mathbb{I}\,\ket{\mathbb{I}}$ with
$\ket{\mathbb{I}}\!=\!\sum_{i=1}^{\dim\mathcal{H}_\t{in}}{\ket{i}}_\t{\tiny S}\ket{i}_\t{\tiny A}$
and $\varphi$-dependence dropped again for 
simplicity---$\t{Tr}_\t{\tiny S}\!\left\{\dot\Omega_{\Lambda_\varphi}\right\}\!=\!\partial_\varphi\!\left(\sum_i\!K_i^\dagger K_i \right)\!=\!0$,
and thus by tracing out the $\mathcal{H}_\t{in}$ subspace of \eqnref{eq:phiExtremCond2} one obtains:
\begin{equation}
0=\sum_{ij}\boldsymbol{\mu}_{ij}K_{j}^{\dagger}K_{i},
\label{eq:ChoiCondApp}
\end{equation}
what is the negation of \emph{Choi criterion} for channel extremality (\critref{crit:Choi})
discussed in \secref{sub:ExtremCh} and \appref{chap:appChoiCrit}, 
and thus the condition for the \emph{non-extremality} of the channel $\Lambda_\varphi$.
Hence, as \eqnref{eq:phiExtremCond2} implies \eqnref{eq:ChoiCondApp}, an intuitive statement is confirmed 
that: if a channel is $\varphi$-non-extremal, then it cannot be extremal; or equivalently:
\emph{if a channel is extremal, then it must also be $\varphi$-extremal}.

%% file: Appendices/AppendixC.tex
\chapter{Equivalence of the purification-based QFI definitions \eref{eq:QFIPurifEscher} and \eref{eq:QFIPurifFuji}} 
\label{chap:appEquivPurif}

\lhead{Appendix C. \emph{Equivalence of the purification-based QFI definitions \eref{eq:QFIPurifEscher} and \eref{eq:QFIPurifFuji}}} 

For a given state $\varrho_{\varphi}\!=\!\sum_{i}\lambda_{i}(\varphi)\left|e_{i}(\varphi)\right\rangle \!\left\langle e_{i}(\varphi)\right|$
supported by the Hilbert space $\mathcal{H}_\t{\tiny S}$
and a parameter true value $\varphi_{0}$, we unambiguously choose
the system purification $\left|\Psi_{\varphi_{0}}\right\rangle \!=\!\sum_{i}\left|\xi_{i}(\varphi_{0})\right\rangle _\t{\tiny S}\!\left|i\right\rangle _\t{\tiny E}$
defined in the minimal extended Hilbert space $\mathcal{H}_\t{\tiny S}\!\times\!\mathcal{H}_\t{\tiny E}$
with all $\left|\xi_{i}(\varphi_{0})\right\rangle \!=\!\lambda_{i}(\varphi_{0})\!\left|e_{i}(\varphi_{0})\right\rangle $
and $\left\langle i|j\right\rangle \!=\!\delta_{ij}$, so that correctly 
$\varrho_{\varphi_{0}}\!=\!\textrm{Tr}_\t{\tiny E}\!\left\{ \left|\Psi_{\varphi_{0}}\right\rangle \!\left\langle \Psi_{\varphi_{0}}\right|\right\}$.
Then, all the relevant purifications, $\left|\tilde{\Psi}_{\varphi}\right\rangle $,
which differ in their corresponding QFIs \eref{eq:QFI} and lead to a change in the r.h.s.~of \eqnsref{eq:QFIPurifEscher}{eq:QFIPurifFuji},
may be constructed at $\varphi_{0}$ by a unitary rotation of the environment part of $\left|\Psi_{\varphi_{0}}\right\rangle$, i.e.~by applying $u_\varphi^\t{\tiny E}\!=\!\textrm{e}^{-\textrm{i}\hat{h}_\t{\tiny E}(\varphi-\varphi_{0})}$ with any Hermitian $\hat{h}_\t{\tiny E}$ that generates $\left|\tilde{\Psi}_{\varphi}\right\rangle \!=\! u_{\varphi}^\t{\tiny E}\!\left|\Psi_{\varphi}\right\rangle$ and locally shifts only the first derivative:~
$\left|\tilde{\Psi}_{\varphi_{0}}\right\rangle \!=\!\left|\Psi_{\varphi_{0}}\right\rangle$, 
$\left|\dot{\tilde{\Psi}}_{\varphi_{0}}\right\rangle \!=\!\left|\dot{\Psi}_{\varphi_{0}}\right\rangle\!-\!\textrm{i}\hat{h}_\t{\tiny E}\!\left|\Psi_{\varphi_{0}}\right\rangle $.
In \citep{Fujiwara2008}, it has been shown that the purification-based
definition \eref{eq:QFIPurifFuji} is minimised and correctly reproduces
the QFI \emph{if and only if} a purification is chosen that
satisfies the condition 
$\left|\dot{\tilde{\Psi}}_{\textrm{\ensuremath{\varphi}}_{0}}^\t{\tiny opt}\right\rangle \!=\!\frac{1}{2}L_\t{\tiny S}\!\otimes\!\mathbb{I}^\t{\tiny E}\!\left|\tilde{\Psi}_{\varphi_{0}}^\t{\tiny opt}\right\rangle $%
\footnote{
For simplicity, we shorten the notation for the SLD \eref{eq:SLD} of the state $\varrho_\varphi$ at $\varphi_{0}$,
so that $L_\t{\tiny S}\equiv L_\t{\tiny S}\!\left[\varrho_{\varphi_{0}}\right]$.
}.
Firstly, let us note that consistently the second term in the other purification-based QFI definition \eref{eq:QFIPurifEscher}
always vanishes for $\left|\tilde{\Psi}_{\varphi_{0}}^\t{\tiny opt}\right\rangle $, as
\begin{equation}
\left\langle \tilde{\Psi}_{\varphi_{0}}^\t{\tiny opt}\left|\,\dot{\tilde{\Psi}}_{\varphi_{0}}^\t{\tiny opt}\right.\!\right\rangle =\left<\tilde{\Psi}_{\varphi_{0}}^\t{\tiny opt}\right|\frac{1}{2}L_\t{\tiny S}\!\otimes\!\mathbb{I}^\t{\tiny E}\left|\tilde{\Psi}_{\varphi_{0}}^\t{\tiny opt}\right>=\frac{1}{2}\textrm{Tr}\left\{ \varrho_{\varphi_{0}} L_\t{\tiny S}\right\} =\frac{1}{4}\textrm{Tr}\left\{ \varrho_{\varphi_{0}}L_\t{\tiny S}+L_\t{\tiny S}\varrho_{\varphi_{0}}\right\} =\frac{1}{2}\textrm{Tr}\left\{ \dot{\varrho}_{\varphi_{0}}\right\} =0,\label{eq:2TermVanish}
\end{equation}
what follows from the SLD definition \eref{eq:SLD}, so that
\eqnsref{eq:QFIPurifEscher}{eq:QFIPurifFuji} are indeed compatible. 
However, in order to complete the proof and
show their equivalence, we demonstrate that $\left|\tilde{\Psi}_{\textrm{\ensuremath{\varphi}}}^\t{\tiny opt}\right\rangle$
is also the purification that minimises \eqnref{eq:QFIPurifEscher}.

Rewriting the r.h.s.~of the definition \eref{eq:QFIPurifEscher}
at $\varphi_{0}$ for purifications $\left|\tilde{\Psi}_{\textrm{\ensuremath{\varphi}}}\right\rangle$, we obtain
\begin{eqnarray}
 &  & 4\min_{\tilde{\Psi}_{\varphi_{0}}}\left\{ \left\langle \!\left.\dot{\tilde{\Psi}}_{\varphi_{0}}\,\right|\dot{\tilde{\Psi}}_{\varphi_{0}}\right\rangle -\left|\left\langle \tilde{\Psi}_{\varphi_{0}}\left|\,\dot{\tilde{\Psi}}_{\varphi_{0}}\right.\!\right\rangle \right|^{2}\right\} =\label{eq:minHe}\\
 &  & =4\min_{\hat{h}_\t{\tiny E}}\left\{ \left\langle \!\left.\dot{\Psi}_{\varphi_{0}}\,\right|\dot{\Psi}_{\varphi_{0}}\right\rangle +2\,\textrm{Im}\!\left\{ \left\langle \dot{\Psi}_{\varphi_{0}}\right|\hat{h}_\t{\tiny E}\left|\Psi_{\varphi_{0}}\right\rangle \right\} +\left\langle \Psi_{\varphi_{0}}\right|\hat{h}_\t{\tiny E}^{2}\left|\Psi_{\varphi_{0}}\right\rangle -\left|\left\langle\Psi_{\varphi_{0}}\left|\dot{\Psi}_{\varphi_{0}}\!\right.\right\rangle -\textrm{i}\left\langle \Psi_{\varphi_{0}}\right|\hat{h}_\t{\tiny E}\left|\Psi_{\varphi_{0}}\right\rangle \right|^{2}\right\} .\nonumber 
\end{eqnarray}
However, as also $\left|\dot{\Psi}_{\varphi_{0}}\right\rangle \!=\!\sum_{i}\left|\dot{\xi}_{i}(\varphi_{0})\right\rangle _\t{\tiny S}\left|i\right\rangle _\t{\tiny E}$,
we may define the purification derivative with help of a non-Hermitian
operator $D\!=\!\sum_{i}\left|\dot{\xi}_{i}(\varphi_{0})\right\rangle \!\left\langle \xi_{i}(\varphi_{0})\right|$
for which $\left|\dot{\Psi}_{\varphi_{0}}\right\rangle \!=\!\frac{1}{2}D\!\otimes\!\mathbb{I}^\t{\tiny E}\!\left|\Psi_{\varphi_{0}}\right\rangle $.
Notice that $D$ acts only on the system subspace and constitutes a logarithmic derivative, as $\dot{\varrho}_{\varphi_{0}}\!=\!\textrm{Tr}_\t{\tiny E}\!\left\{\left|\Psi_{\varphi_{0}}\right>\!\left<\dot{\Psi}_{\varphi_{0}}\right|+\left|\dot{\Psi}_{\varphi_{0}}\right>\!\left<\Psi_{\varphi_{0}}\right|\right\} \!=\!\frac{1}{2}\!\left(\varrho_{\varphi_{0}}D^{\dagger}+D\varrho_{\varphi_{0}}\right)$.
Hence, we can rewrite \eqnref{eq:minHe} further as 
\begin{equation}
 4\min_{\hat{h}_\t{\tiny E}}\!\left\{ \left\langle \frac{1}{4}D^{\dagger}D+\frac{\textrm{i}}{2}\left(D-D^{\dagger}\right)\otimes\hat{h}_\t{\tiny E}+\hat{h}_\t{\tiny E}^{2}\right\rangle -\left|\left\langle \frac{1}{2}D-\textrm{i}\hat{h}_\t{\tiny E}\right\rangle \right|^{2}\right\},
\end{equation}
where $\left\langle \dots\right\rangle \!\equiv\left\langle \Psi_{\varphi_{0}}\right|\dots\left|\Psi_{\varphi_{0}}\right\rangle$.
Defining the difference of the derivative $D$ from the SLD \eref{eq:SLD} as $\Delta\!=\!D\!-\!L_\t{\tiny S}$,
which satisfies $\varrho_{\varphi_{0}}\Delta^{\dagger}+\Delta\varrho_{\varphi_{0}}\!=\!0$,
and benefiting from:~$\left\langle L_\t{\tiny S}\right\rangle\!=\!0$ that follows from \eqnref{eq:2TermVanish},
$\left\langle\!\left.L_\t{\tiny S}\!\right.^{2}\right\rangle \!=\!\textrm{Tr}\!\left\{ \varrho_{\varphi_{0}}\!\left.L_\t{\tiny S}\!\right.^{2}\right\} \!=\!\left.F_{\textrm{Q}}\!\left[\varrho_{\varphi}\right]\right|_{\varphi_{0}}$, and 
$\left\langle \Delta^{\dagger}L_\t{\tiny S}+L_\t{\tiny S}\Delta\right\rangle \!=\!\textrm{Tr}\!\left\{ \varrho_{\varphi_{0}}\left(\Delta^{\dagger}L_\t{\tiny S}+L_\t{\tiny S}\Delta\right)\right\} \!=\!\textrm{Tr}\!\left\{ \left(\varrho_{\varphi_{0}}\Delta^{\dagger}+\Delta\varrho_{\varphi_{0}}\right)L_\t{\tiny S}\right\} \!=\!0$, 
we arrive at
\begin{align}
 & 4\min_{\hat{h}_\t{\tiny E}}\!\left\{ \left\langle \frac{1}{4}\left(\Delta^{\dagger}+L_\t{\tiny S}\right)\left(\Delta+L_\t{\tiny S}\right)+\frac{\textrm{i}}{2}\left(\Delta-\Delta^{\dagger}\right)\otimes\hat{h}_\t{\tiny E}+\hat{h}_\t{\tiny E}^{2}\right\rangle -\left|\left\langle \frac{1}{2}\Delta-\textrm{i}\hat{h}_\t{\tiny E}\right\rangle \right|^{2}\right\} =\nonumber \\
 & =\min_{\hat{h}_\t{\tiny E}}\!\left\{ \left\langle \left.L_\t{\tiny S}\!\right.^{2}\right\rangle +\left\langle \Delta^{\dagger}\Delta\right\rangle +\left\langle \Delta^{\dagger}L_\t{\tiny S}+L_\t{\tiny S}\Delta\right\rangle +\left\langle 2\textrm{i}\left(\Delta-\Delta^{\dagger}\right)\otimes\hat{h}_\t{\tiny E}+4\hat{h}_\t{\tiny E}^{2}\right\rangle -\left|\left\langle \Delta-2\textrm{i}\hat{h}_\t{\tiny E}\right\rangle \right|^{2}\right\} =\nonumber \\
 & =\left.F_{\textrm{Q}}\!\left[\varrho_{\varphi}\right]\right|_{\varphi_{0}}+\min_{\hat{h}_\t{\tiny E}}\!\left\{ \left\langle \Delta^{\dagger}\Delta\right\rangle +\left\langle 2\textrm{i}\left(\Delta-\Delta^{\dagger}\right)\otimes\hat{h}_\t{\tiny E}+4\hat{h}_\t{\tiny E}^{2}\right\rangle -\left|\left\langle \Delta-2\textrm{i}\hat{h}_\t{\tiny E}\right\rangle \right|^{2}\right\} \nonumber \\
 & =\left.F_{\textrm{Q}}\!\left[\varrho_{\varphi}\right]\right|_{\varphi_{0}}+4\min_{\hat{h}_\t{\tiny E}}\left\langle \left(\frac{1}{2}\Delta^\dagger+\textrm{i}\hat{h}_\t{\tiny E}\right)\left[\mathbb{I}^\t{\tiny SE}-\left|\Psi_{\varphi_{0}}\right\rangle \!\left\langle \Psi_{\varphi_{0}}\right|\right]\left(\frac{1}{2}\Delta-\textrm{i}\hat{h}_\t{\tiny E}\right)\right\rangle.
\label{eq:QFI+min}
\end{align}
The second term in \eqnref{eq:QFI+min}, which should still be minimised over $\hat{h}_\t{\tiny E}$, is importantly non-negative, as it represents
a projection of a non-negative operator, $\mathbb{I}^\t{\tiny SE}-\left|\Psi_{\varphi_{0}}\right\rangle \!\left\langle \Psi_{\varphi_{0}}\right|$,
onto an (unnormalised) state $\left(\frac{1}{2}\Delta-\textrm{i}\hat{h}_\t{\tiny E}\right)\!\left|\Psi_{\varphi_{0}}\right\rangle $.
Importantly, it can be always made zero by choosing $\hat{h}_\t{\tiny E}$
such that $\left(\frac{1}{2}\Delta-\textrm{i}\hat{h}_\t{\tiny E}\right)\!\left|\Psi_{\varphi_{0}}\right\rangle \!=\!0$,
what leads exactly to the previously claimed optimal purification
$\left|\tilde{\Psi}_{\varphi_{0}}\right\rangle \!=\!\left|\tilde{\Psi}_{\varphi_{0}}^\t{\tiny opt}\right\rangle $,
as 
\begin{eqnarray}
 &  & \!\!\!\!\!\!\!\!\! \!\!\!\!\!\!\!\!\! \left(\frac{1}{2}\Delta-\textrm{i}\hat{h}_\t{\tiny E}\right)\!\left|\Psi_{\varphi_{0}}\right\rangle =\left|\dot{\Psi}_{\varphi_{0}}\right\rangle -\left(\frac{1}{2}L_\t{\tiny S}+\textrm{i}\hat{h}_\t{\tiny E}\right)\!\left|\Psi_{\varphi_{0}}\right\rangle =0\\
 & & \!\!\!\!\!\!\!\!\! \!\!\!\!\!\!\!\!\! \quad\qquad\implies \qquad \left|\dot{\Psi}_{\varphi_{0}}\right\rangle -\textrm{i}\hat{h}_\t{\tiny E}\left|\Psi_{\varphi_{0}}\right\rangle =\frac{1}{2}L_\t{\tiny S}\otimes\mathbb{I}^\t{\tiny E}\left|\Psi_{\varphi_{0}}\right\rangle \quad\Longleftrightarrow\quad\left|\dot{\tilde{\Psi}}_{\varphi_{0}}\right\rangle =\frac{1}{2}L_\t{\tiny S}\otimes\mathbb{I}^\t{\tiny E}\left|\tilde{\Psi}_{\varphi_{0}}\right\rangle ,\nonumber 
\end{eqnarray}
and completes an alternative to \citep{Escher2011} proof of the purification-based
QFI definition \eref{eq:QFIPurifEscher}.\\$\t{~}$\hfill {\color{myred} $\blacksquare$}

%% file: Appendices/AppendixD.tex
\chapter{Optimality of covariant POVMs} 
\label{chap:appCovMeas}

\lhead{Appendix D. \emph{Optimality of covariant POVMs}} 

We follow the \emph{covariant POVM} analysis of \citep{Holevo1982}, in order to assess the
optimality of such a measurement class when employed in case of
a single repetition ($\nu\!=\!1$) of the phase estimation 
scheme of \figref{fig:Ph_Est_Scheme}, for which the output state of the
system most generally reads $\rho_{\varphi}^{N}\!=\!\mathcal{U}_{\varphi}^{\otimes N}\!\left[\rho_{0}^{N}\right]$
with $\rho_{0}^{N}\!=\!\mathcal{D}\!\left[\rho_{\textrm{in}}^{N}\right]$.
Let us write the Bayesian, circularly symmetric average cost \eref{eq:QAvCost} for
a particular POVM $M_X$ acting on all the $N$ particles---which satisfies $\forall_{x}\!:M_{x}\!\ge\!0$, 
$\sumint\!\textrm{d}x\, M_{x}\!=\!\mathbb{I}$---and 
an estimator $\tilde{\varphi}(x)\!\equiv\!\tilde{\varphi}_{\nu=1}(x)$:
\begin{eqnarray}
\left\langle \mathcal{C}_{\textrm{H}}(\tilde{\varphi})\right\rangle  & = & \int\!\frac{\textrm{d}\varphi}{2\pi}\sumint\!\textrm{d}x\;\textrm{Tr}\!\left\{ \rho_{\varphi}^{N}\,M_{x}\right\} C_{\textrm{H}}\!\left(\tilde{\varphi}({x})-\varphi\right)\nonumber \\
 & = & \int\!\frac{\textrm{d}\varphi}{2\pi}\sumint\!\textrm{d}x\;\textrm{Tr}\!\left\{\rho_{0}^{N}\; U_{\varphi}^{\dagger\otimes N}\!M_{x}\,U_{\varphi}^{\otimes N}\right\} C_{\textrm{H}}\!\left(\tilde{\varphi}({x})-\varphi\right).
\label{eq:AppB1}
\end{eqnarray}
Interchanging the order of the integrals and freely shifting 
$\varphi\!\rightarrow\!\varphi+\tilde{\varphi}({x})$, what does not affect the $\varphi$-integral limits due to 
invariance of the Haar measure, i.e.~$\forall_\delta\!:\int^{\varphi+\delta}\!\frac{\t{d}\varphi}{2\pi}\dots\!=\!\int^\varphi\!\frac{\t{d}\varphi}{2\pi}\dots$, we may rewrite 
\eqnref{eq:AppB1} as
\begin{eqnarray}
\left\langle \mathcal{C}_{\textrm{H}}(\tilde{\varphi})\right\rangle  & = & \sumint\!\textrm{d}x\int\!\frac{\textrm{d}\varphi}{2\pi}\;\textrm{Tr}\!\left\{\rho_{\varphi}^{N}\;
U_{\tilde{\varphi}({x})}^{\dagger\otimes N}M_{x_{i}}U_{\tilde{\varphi}({x})}^{\otimes N}\right\} C_{\textrm{H}}\!\left(\varphi\right)\nonumber \\
 & = & \int\!\frac{\textrm{d}\varphi}{2\pi}\;\textrm{Tr}\!\left\{\rho_{\varphi}^{N}\,{\Xi^{N}}\right\} C_{\textrm{H}}\!\left(\varphi\right)
\label{eq:AppB2}
\end{eqnarray}
with the \emph{seed element} $\Xi^{N}\!=\!\sumint\!\textrm{d}x\,U_{\tilde{\varphi}({x})}^{\dagger\otimes N}\!M_{x}U_{\tilde{\varphi}({x})}^{\otimes N}$,
which thus $\Xi^{N}\!\!\ge\!0$ corresponding to a mixture of positive semi-definite operators.
Without loss of generality, we may introduce another
parameter $\tilde{\varphi}$ playing now the role of the estimator
that is integrated also with respect to the invariant Haar measure,
so that $\forall_\delta\!:\int^{\tilde\varphi+\delta}\!\frac{\t{d}\tilde\varphi}{2\pi}\dots\!=\!\int^{\tilde\varphi}\!\frac{\t{d}\tilde\varphi}{2\pi}\dots$
 and $\int\!\!\frac{\textrm{d}\tilde{\varphi}}{2\pi}\!=\!1$.
After shifting again $\varphi\!\rightarrow\!\varphi-\tilde{\varphi}$,
\eqnref{eq:AppB2} may finally be written as 
\begin{equation}
\left\langle \mathcal{C}_{\textrm{H}}(\tilde{\varphi})\right\rangle =\int\!\!\frac{\textrm{d}\varphi}{2\pi}\int\!\!\frac{\textrm{d}\tilde{\varphi}}{2\pi}\;\textrm{Tr}\!\left\{ \!\rho_{\varphi}^{N}{M_{\tilde\varphi}}\right\} \; C_{\textrm{H}}\!\left(\tilde{\varphi}-\varphi\right),
\label{eq:AppB3}
\end{equation}
where $M_{\tilde\varphi}\!=\! U_{\tilde{\varphi}}^{\otimes N}\,\Xi^{N} \,U_{\tilde{\varphi}}^{\dagger\otimes N}$
has the interpretation of a \emph{continuously parametrised covariant POVM}. 
Consistently, the elements of ${M_{\tilde\varphi}}$ specified by $\tilde\varphi$ are positive semi-definite
and the overall POVM is complete, as $\int\!\!\frac{\textrm{d}\tilde{\varphi}}{2\pi}{M_{\tilde\varphi}}\!=\!\mathbb{I}$,
what may be verified by:~interchanging the order of the integrals, 
freely shifting the parameter $\tilde\varphi$ to eliminate the $\tilde\varphi({x})$-dependence, 
and acknowledging that $\sumint\!\textrm{d}\!x\,M_{x}\!=\!\mathbb{I}$.
Crucially, \eqnref{eq:AppB3} proves that 
\emph{for any POVM $M_{X}$ assumed
to calculate the average cost \eref{eq:AppB1}, there always exists 
a covariant POVM $M_{\tilde\varphi}$ that also achieves $\left\langle \mathcal{C}_{\textrm{H}}(\tilde{\varphi})\right\rangle$}.
\hfill {\color{myred} $\blacksquare$}

%% file: Appendices/AppendixE.tex
\chapter{RLD bound applies \emph{to and only to} $\varphi$-non-extremal quantum channels} 
\label{chap:appRLDboundCond}

\lhead{Appendix E. \emph{RLD bound applies to and only to $\varphi$-non-extremal quantum channels}} 

In  what follows we adopt the notation of \secref{sub:CJiso} and \appref{chap:appPhiExtremCond},
so that for a given quantum channel $\Lambda_\varphi$:~
$\Omega_{\Lambda_{\varphi}}\!=\!\sum_i\ket{K_i}\!\bra{K_i}$ and 
$\dot\Omega_{\Lambda_{\varphi}}\!=\!\sum_i\ket{\dot K_i}\!\bra{K_i}\!+\!\ket{K_i}\!\bra{\dot K_i}$
respectively represent the corresponding Choi-Jamio\l{}kowski (CJ) matrix and its derivative w.r.t.~$\varphi$,
where $\ket{K_i}\!=\!K_i\!\otimes\!\mathbb{I}\,\ket{\mathbb{I}}$ are defined 
with use of the canonical Kraus operators, which $\varphi$-dependence is dropped for simplicity.

In \citep{Hayashi2011}, it has been proved that the RLD bound \eref{eq:RLDbound} applies to 
a given channel $\Lambda_\varphi$ for a given parameter value $\varphi_0$ if and only if
$\left(\dot{\Omega}_{\Lambda_{\varphi}}\!\right)^{2}$ is contained
within the support of $\Omega_{\Lambda_{\varphi}}$ at $\varphi_0$.
Here, we show that $(\dots)^2$ is unnecessary in the above statement, so 
that---recalling the definition of channel $\varphi$-non-extremality presented in 
\appref{chap:appPhiExtremCond}---it may be rewritten as:
\begin{mytheorem}
The RLD bound \eref{eq:RLDbound} on the extended-channel QFI \eref{eq:ExtChQFI} applies to a given channel $\Lambda_\varphi$ 
at $\varphi_0$ if and only if $\Lambda_\varphi$ is $\varphi$-non-extremal there.
\end{mytheorem}
\textsf{\color{myred} Proof:}~\\
The condition of \citet{Hayashi2011} for the RLD bound \eref{eq:RLDbound} 
to be applicable to the channel $\Lambda_\varphi$ at $\varphi_0$  can be formally written as
\begin{equation}
P_{\perp}\;\left(\dot{\Omega}_{\Lambda_{\varphi_0}}\right)^{2} \,P_{\perp}=0\,,
\label{eq:HayashiCond}
\end{equation}
where we denote by $P_{\parallel}$ and $P_\perp$ projections onto the support and null-space 
of $\Omega_{\Lambda_{\varphi_0}}$ respectively, so that 
$\forall_{i}\!:~P_\parallel\left|K_{i}\right\rangle \!=\!\ket{K_i}$ whereas
$\forall_{i}\!:~P_\perp\left|K_{i}\right\rangle \!=\!0$.
On the other hand, we have shown in \appref{chap:appPhiExtremCond} that 
the definition of the channel $\Lambda_\varphi$ to be  $\varphi$-non-extremal
at $\varphi_0$ is equivalent to the statement that $P_\parallel \,\dot{\Omega}_{\Lambda_{\varphi_0}}\!P_\parallel\!=\!\dot{\Omega}_{\Lambda_{\varphi_0}}$,
or in other words---see \eqnref{eq:phiExtremCond2}---, that there
exists a non-zero Hermitian matrix $\boldsymbol\mu_{ij}$ such that
\begin{equation}
\dot{\Omega}_{\Lambda_{\varphi_0}}=\sum_{ij}\boldsymbol\mu_{ij}\left|K_{i}\right\rangle \!\left\langle K_{j}\right|\!.
\label{eq:CScond}
\end{equation}

\eqnref{eq:CScond} implies \eqnref{eq:HayashiCond}, as by substitution
\begin{equation}
P_\perp\!\left(\!\sum_{ij}\boldsymbol\mu_{ij}\left|K_{i}\right\rangle \!\left\langle K_{j}\right|\!\right)^{2}\!\! P_\perp=\sum_{ij}\!\left(\!\sum_{p}\boldsymbol\mu_{ip}\left\langle K_{p}|K_{p}\right\rangle \boldsymbol\mu_{pj}\!\right)P_\perp\left|K_{i}\right\rangle \!\left\langle K_{j}\right|P_\perp=0\,.
\end{equation}

In order to prove the other direction, we split
the derivatives of each vector $\ket{K_i}$ into components contained
within the support and the null-space of $\Omega_{\Lambda_{\varphi_0}}$,
i.e.~$\left|\dot{K}_{i}\right\rangle \!=\!\sum_{j}\boldsymbol\nu_{ij}\left|K_{j}\right\rangle \!+\!\left|L_{i}^\perp\right\rangle$
with $\forall_{i,j}\!:\,\braket{L_{i}^\perp}{K_{j}}\!=\!0$.
Hence, after substituting for $\dot{\Omega}_{\Lambda_{\varphi_0}}$
the \eqnref{eq:HayashiCond} then simplifies to
\begin{equation}
\left(\sum_{i}\left|L_{i}^{\perp}\right\rangle \!\left\langle K_{i}\right|\right)\left(\sum_{j}\left|K_{j}\right\rangle \!\left\langle L_{j}^{\perp}\right|\right)=0,
\end{equation}
and since $A^{\dagger}A\!=\!0$ implies $A^{\dagger}\!=\! A\!=\!0$
and $\left\{ \left|K_{i}\right\rangle \right\} _{i}$ are orthogonal,
we conclude that all $\left|L_{i}^{\perp}\right\rangle \!=\!0$. Thus,
\eqnref{eq:HayashiCond} implies that $\left|\dot{K}_{i}\right\rangle \!=\!\sum_{j}\boldsymbol\nu_{ij}\left|K_{j}\right\rangle $,
which due to the local ambiguity of Kraus representations \eref{eq:KrausReps}
is equivalent to $\left|\dot{\tilde{K}}_{i}\right\rangle \!=\!\sum_{j}\!\left(\boldsymbol\nu_{ij}\!-\!\textrm{i}\,\mathsf{h}_{ij}\right)\!\left|\tilde{K}_{j}\right\rangle $
for any Hermitian matrix $\mathsf{h}$. Therefore, without loss
of generality, we may set $\mathsf{h}\!=\!-\textrm{i}\,\boldsymbol\nu^\t{\tiny AH}$
after splitting $\boldsymbol\nu$ into its Hermitian and anti-Hermitian parts
$\boldsymbol\nu\!=\!\boldsymbol\nu^\t{\tiny H}\!+\!\boldsymbol\nu^\t{\tiny AH}$, so that 
$\left|\dot{\tilde{K}}_{i}\right\rangle \!=\!\sum_{j}\boldsymbol\nu_{ij}^\t{\tiny H}\left|\tilde{K}_{j}\right\rangle$
with $\boldsymbol\nu^\t{\tiny H}\!\ne\!0$ for any non-trivial channel%
\footnote{Otherwise, there would exist a Kraus representation in which $\forall_i\!:\ket{\dot{\tilde{K}}_{i}}\!=\!0$,
so that the channel QFI \eref{eq:ChQFIPurifK} vanishes independently of the input!}.
Writing the derivative of the CJ matrix in the shifted Kraus representation basis, $\{\ket{\tilde K_i}\}_i$, and
bearing in mind that at $\varphi_0$ all $\ket{\tilde K_i}\!=\!\ket{K_i}$, we obtain
\begin{equation}
\dot{\Omega}_{\Lambda_{\varphi_0}}=\sum_{i}\left|\dot{\tilde{K}}_{i}\right\rangle \!\left\langle \tilde{K}_{i}\right|+\left|\tilde{K}_{i}\right\rangle \!\left\langle \dot{\tilde{K}}_{i}\right|=2\sum_{ij}\boldsymbol\nu_{ji}^\t{\tiny H}\left|\tilde{K}_{i}\right\rangle \!\left\langle \tilde{K}_{j}\right|
=2\sum_{ij}\boldsymbol\nu_{ji}^\t{\tiny H}\left|K_{i}\right\rangle \!\left\langle K_{j}\right|
\label{eq:HayashiCondFin}
\end{equation}
and satisfy the condition \eref{eq:CScond} with $\boldsymbol\mu^T\!=\!2\,\boldsymbol\nu^\t{\tiny H}$. \hfill {\color{myred} $\blacksquare$}

%% file: Appendices/AppendixF.tex
\chapter{Optimal purifications that yield extended-channel QFIs and QS/CE bounds} 
\label{chap:appOptHExtCh}

\lhead{Appendix F. \emph{Optimal purifications that yield extended-channel QFIs and CE bounds}}

We state below the form of the optimal generators $\mathsf{h}$
that shift adequately---see \eqnref{eq:KrausReps}---the first derivatives of the 
canonical Kraus operators
listed in \tabref{tab:noise_models} for each of the noisy-phase--estimation models of \figref{fig:noise_models}.
For each of the noise-types, we provide two versions of optimal 
$\mathsf{h}_\t{\tiny opt}$, first of which minimises \eqnref{eq:ExtChQFIPurif} and thus
yields the correct form of the \emph{extended-channel QFI}, $\mathcal{F}\!\left[\Lambda_{\varphi}\otimes\mathcal{I}\right]$,
or equivalently, the optimal purification $\ket{\tilde\Psi_\varphi^\t{\tiny ext}}$ depicted in
\figref{fig:ch_purif_ext}(\textbf{c}).
The second version of $\mathsf{h}_\t{\tiny opt}$ corresponds in each case
to the optimal generator minimising \eqnref{eq:CEbound}, and thus specifying
the \emph{CE bound}, $\mathcal{F}_\t{as}^\t{\tiny CE}$.
Yet, as for all the channels considered, apart from the spontaneous emission map, such a generator
leads to a Kraus representation satisfying the condition $\alpha_{\tilde K_\t{\tiny opt}}\!\!\!\propto\!\mathbb{I}$
in \eqnref{eq:QSconds}, $\mathsf{h}_\t{\tiny opt}$ minimises also \eqnref{eq:CEbound} and thus yields
also the coinciding \emph{QS bound}, i.e.~$\mathcal{F}_\t{as}^\t{\tiny QS}\!=\!\mathcal{F}_\t{as}^\t{\tiny CE}$.
We adopt the standard notation, in which $\hat\sigma_0$ represents
a $2\!\times\!2$ identity matrix and $\left\{ \hat\sigma_{i}\right\} _{i=1}^{3}$
are the Pauli operators.

\paragraph{Dephasing:}~\\
Optimal $\mathsf{h}$ minimising the expression \eref{eq:ExtChQFIPurif} for the extended-channel QFI, $\mathcal{F}\!\left[\Lambda_{\varphi}\otimes\mathcal{I}\right]$:
\begin{equation}
\mathsf{h}_\t{\tiny opt}=-\frac{\sqrt{1-\eta^{2}}}{2}\;\hat\sigma_{1}\,.
\end{equation}
Optimal $\mathsf{h}$ minimising the expressions \eref{eq:QSbound}/\eref{eq:CEbound} for the QS/CE bounds, $\mathcal{F}_\t{as}^\t{\tiny QS}\!=\!\mathcal{F}_\t{as}^\t{\tiny CE}$:
\begin{equation}
\mathsf{h}_\t{\tiny opt}=-\frac{1}{2\sqrt{1-\eta^{2}}}\;\hat{\sigma}_{1}\,.
\end{equation}

\paragraph{Depolarisation:}~\\
Optimal $\mathsf{h}$ minimising the expression \eref{eq:ExtChQFIPurif} for the extended-channel QFI, $\mathcal{F}\!\left[\Lambda_{\varphi}\otimes\mathcal{I}\right]$:
\begin{equation}
\mathsf{h}_\t{\tiny opt}=-\frac{1}{2}\left(\!\!\!\!\begin{array}{ccc}
0 & \!\!\!\!\!0\;\;0 & \!\!\!\!\!\xi\\
\begin{array}{c}
0\\
0
\end{array} & \!\!\!\!\!\left[\;\overset{\;}{\underset{\;}{\hat\sigma_{2}}\;}\right] & \!\!\!\!\!\begin{array}{c}
0\\
0
\end{array}\\
\xi & \!\!\!\!\!0\;\;0 & \!\!\!\!\!0
\end{array}\!\!\!\!\right)\quad\textrm{with}\quad\xi=\frac{\sqrt{\left(1+3\eta\right)\left(1-\eta\right)}}{1+\eta}.
\end{equation}
Optimal $\mathsf{h}$ minimising the expressions \eref{eq:QSbound}/\eref{eq:CEbound} for the QS/CE bounds, $\mathcal{F}_\t{as}^\t{\tiny QS}\!=\!\mathcal{F}_\t{as}^\t{\tiny CE}$:
\begin{equation}
\mathsf{h}_\t{\tiny opt}=-\frac{c}{2}\left(\!\!\!\!\begin{array}{ccc}
0 & \!\!\!\!\!0\;\;0 & \!\!\!\!\!\xi\\
\begin{array}{c}
0\\
0
\end{array} & \!\!\!\!\!\left[\;\overset{\;}{\underset{\;}{\hat\sigma_{2}}\;}\right] & \!\!\!\!\!\begin{array}{c}
0\\
0
\end{array}\\
\xi & \!\!\!\!\!0\;\;0 & \!\!\!\!\!0
\end{array}\!\!\!\!\right)\quad
\textrm{with}\quad\xi=\frac{\sqrt{\left(1+3\eta\right)\left(1-\eta\right)}}{1+\eta}\quad
\textrm{and}\quad c=\frac{1+\eta}{(1+2\eta)(1-\eta)}
.
\end{equation}

\paragraph{Loss:}~\\
Optimal $\mathsf{h}$ minimising the expression \eref{eq:ExtChQFIPurif} for the extended-channel QFI, $\mathcal{F}\!\left[\Lambda_{\varphi}\otimes\mathcal{I}\right]$:
\begin{equation}
\mathsf{h}_\t{\tiny opt}=
-\frac{1}{2}
\left(\!\!\!
\begin{array}{cc}0 & \!\!\!\!\!0\;\;\;0\\
\begin{array}{c}
0\\
0
\end{array} & \!\!\!\!\!\left[\;\overset{\;}{\underset{\;}{\hat\sigma_{3}}\;}\right]
\end{array}\!
\right)\!.
\end{equation}
Optimal $\mathsf{h}$ minimising the expressions \eref{eq:QSbound}/\eref{eq:CEbound} for the QS/CE bounds, $\mathcal{F}_\t{as}^\t{\tiny QS}\!=\!\mathcal{F}_\t{as}^\t{\tiny CE}$:
\begin{equation}
\mathsf{h}_\t{\tiny opt}=-\frac{1}{2(1-\eta)}
\left(\!\!\!
\begin{array}{cc}0 & \!\!\!\!\!0\;\;\;0\\
\begin{array}{c}
0\\
0
\end{array} & \!\!\!\!\!\left[\;\overset{\;}{\underset{\;}{\hat\sigma_{3}}\;}\right]
\end{array}\!
\right)\!.
\label{eq:hOpt_CE_loss}
\end{equation}

\paragraph{Spontaneous emission (amplitude damping):}~\\
Optimal $\mathsf{h}$ minimising the expression \eref{eq:ExtChQFIPurif} for the extended-channel QFI, $\mathcal{F}\!\left[\Lambda_{\varphi}\otimes\mathcal{I}\right]$:
\begin{equation}
\mathsf{h}_\t{\tiny opt}=\frac{1}{2}\left(\!\!\begin{array}{cc}
\xi & 0\\
0 & 1
\end{array}\!\!\right)\quad\textrm{with}\quad\xi=\frac{1+\sqrt{\eta}}{1-\sqrt{\eta}}\,.
\end{equation}
Optimal $\mathsf{h}$ minimising the expression \eref{eq:CEbound} for the CE bound, $\mathcal{F}_\t{as}^\t{\tiny CE}$:
\begin{equation}
\mathsf{h}_\t{\tiny opt}=\frac{1}{2(1-\eta)}\left(\eta\,\hat{\sigma}_{0}-\hat{\sigma}_{3}\right)
\end{equation}

%% file: Appendices/AppendixG.tex
\chapter{CE bound applies to all $\varphi$-non-extremal maps and is tighter than the RLD bound} 
\label{chap:appCEvsRLD}

\lhead{Appendix G. \emph{CE bound applies to all $\varphi$-non-extremal maps and is tighter than the RLD bound}}

Firstly, let us note that in \appref{chap:appRLDboundCond},
while proving that any channel that admits the RLD bound \eref{eq:RLDbound}
must be $\varphi$-non-extremal,
we chose a Kraus representation generated by $\mathsf{h}\!=\!-\ii\,\boldsymbol\nu^\t{\tiny AH}$
that actually satisfies the necessary \emph{$\beta_{\tilde{K}}\!=\!0$ condition} \eref{eq:BetaKCond}
of the CE method. Recalling that in \appref{chap:appRLDboundCond} the matrix $\boldsymbol\nu$
was specified so that for all $i$: $\left|\dot{K}_{i}\right\rangle \!=\!\sum_{j}\boldsymbol\nu_{ij}\left|K_{j}\right\rangle$,
we obtain exactly \eqnref{eq:BetaKCond} by taking the partial trace over the subspace $\mathcal{H}_\t{\tiny S}$
of both sides of the identity:
\begin{equation}
\sum_{ij}\mathsf{h}_{ij}\left|K_{j}\right\rangle \!\left\langle K_{i}\right| 
=
-\ii\sum_{ij}\frac{1}{2}\left(\boldsymbol\nu_{ij}-\boldsymbol\nu_{ij}^{\dagger}\right)\left|K_{j}\right\rangle \!\left\langle K_{i}\right|
=
\frac{\textrm{i}}{2}\sum_{i}\left|K_{i}\right\rangle \!\left\langle \dot{K}_{i}\right|-\left|\dot{K}_{i}\right\rangle \!\left\langle K_{i}\right|,
\end{equation}
what thus proves that the \emph{CE method applies to all  $\varphi$-non-extremal maps}, which
are really the ones that admit the RLD bound---see \appref{chap:appRLDboundCond}.
This is consistent, as the CE method, being applicable to all the quantum simulable
channels described in  \secref{sub:QSbound}, must also trivially apply to all the classically simulable ones
that are also exactly the $\varphi$-non-extremal maps.

Furthermore, we show that the CE bound
\eref{eq:CEbound} is at least as tight as the RLD bound \eref{eq:RLDbound}. 
We prove this by substituting the representation \eref{eq:HayashiCondFin}
of the CJ matrix derivative into the definition of $\mathcal{F}^\t{\tiny RLD}\!\left[\Lambda_{\varphi}\otimes\mathcal{I}\right]$
in \eqnref{eq:RLDbound}, so that
\begin{equation}
\mathcal{F}^\t{\tiny RLD}\!\left[\Lambda_{\varphi}\otimes\mathcal{I}\right] 
= 
4\left\Vert \textrm{Tr}_{\mathcal{H}_\t{\tiny S}}\!\!\left\{ \sum_{ij}\boldsymbol\nu_{ji}^\t{\tiny H}\left|\tilde{K}_{i}\right\rangle \sum_{pq}\boldsymbol\nu_{pq}^\t{\tiny H}\left\langle \tilde{K}_{q}\right|\right\} \right\Vert =4\left\Vert \sum_{i}\dot{\tilde{K}}_{i}^{\dagger}\dot{\tilde{K}}_{i}\right\Vert ,
\end{equation}
where we have used the fact that $\left\langle \tilde{K}_{j}\right|\Omega_{\varphi}^{-1}\left|\tilde{K}_{p}\right\rangle \!=\!\delta_{jp}$.
Hence, $\mathcal{F}^\t{\tiny RLD}\!\left[\Lambda_{\varphi}\otimes\mathcal{I}\right]$
is an example of the CE bound \eref{eq:CEbound} with a possibly sub-optimal
Kraus operators chosen that satisfy $\forall_{i}\!:\left|\dot{\tilde{K}}_{i}\right\rangle \!=\!\sum_{j}\boldsymbol\nu_{ij}^\t{\tiny H}\left|\tilde{K}_{j}\right\rangle $
and the $\beta_{\tilde{K}}=0$ condition \eref{eq:BetaKCond}, as shown in the paragraph above. \hfill{\color{myred} $\blacksquare$}

%% file: Appendices/AppendixH.tex
\chapter{Optimal local QS of a channel} 
\label{chap:appQSasCE}

\lhead{Appendix H. \emph{Optimal local QS of a channel}} 

A given channel $\Lambda_{\varphi}$ of rank $r$ in order to be locally
\emph{quantum simulable} at $\varphi_0$ must fulfil the condition (see \secref{sub:QSbound}):
\begin{equation}
\Lambda_{\varphi}\!\left[\varrho\right] =  \textrm{Tr}_{\textrm{\tiny E}_{\Phi}\textrm{\tiny E}_{\sigma}}\!\!\left\{\,U\!\left(\varrho\otimes\left|\xi_{\varphi}\right\rangle \!\left\langle \xi_{\varphi}\right|\right)U^{\dagger}\right\} +O(\delta\varphi^{2})=\sum_{i=1}^{r^{\prime}\ge r}\bar{K}_{i}(\varphi)\,\varrho\,\bar{K}_{i}(\varphi)^{\dagger}+O(\delta\varphi^{2}),\label{eq:LocQSCond}
\end{equation}
where $\varphi\!=\!\varphi_0\!+\!\delta\varphi$  and $\bar{K}_{i}(\varphi)\!=\!\left\langle i\right|U\left|\xi_{\varphi}\right\rangle $
and $\left\{ \left|i\right\rangle \right\} _{i=1}^{r^{\prime}}$ form
any basis in the $r^{\prime}$ dimensional $\mathcal{H}_{\textrm{\tiny E}_{\Phi}}\!\!\times\!\mathcal{H}_{\textrm{\tiny E}_{\sigma}}$
space containing $\ket{\xi_{\varphi}}$. Hence, $\Lambda_{\varphi}$ must
admit at $\varphi_0$ a Kraus representation $\{\tilde{K}_{i}\}_{i=1}^{r^{\prime}}$---with 
possibly linearly dependent Kraus operators, as for generality
we assume $r^{\prime}\!\ge\! r$---that coincides with the one of \eqnref{eq:LocQSCond}
up to $O(\delta\varphi^{2})$, i.e.~satisfies $\tilde{K}_{i}\!=\!\bar{K}_{i}$
and $\dot{\tilde{K}}_{i}\!=\!\dot{\bar{K}}_{i}$ for all $i$. 

We construct a valid decomposition of $|\dot{\xi}_{\varphi_0}\rangle$ into its
(normalised) components parallel and perpendicular to 
$\ket{\xi_{\varphi_0}}$:~$\left|\dot{\xi}_{\varphi_0}\right\rangle \!\!=\!\textrm{i}\, a\!\left|\xi_{\varphi_0}\right\rangle \!-\!\textrm{i}\, b\!\left|\xi_{\varphi_0}^{\perp}\right\rangle $,
where we can choose $a,b\!\in\!\mathbb{R}$ because of $\frac{\partial\left\langle \xi_{\varphi}|\xi_{\varphi}\right\rangle}{\partial\varphi} \!=\!0$
and the irrelevance of the global phase. Then, the asymptotic bound
$\mathcal{F}_\t{\tiny as}^\t{\tiny bound}$ of \eqnref{eq:QFIAsBound}
determined by the local QS \eref{eq:LocQSCond} at $\varphi_0$ simply
reads $\left.F_{\textrm{Q}}\!\left[\left|\xi_{\varphi}\right\rangle \right]\right|_{\varphi_0}\!=\!4b^{2}$
and the required Kraus operators $\{\tilde{K}_{i}\}_{i=1}^{r^{\prime}}$
of $\Lambda_{\varphi}$ must fulfil conditions $\tilde{K}_{i}\!=\!\left\langle i\right|U\left|\xi_{\varphi_0}\right\rangle $
and $\dot{\tilde{K}}_{i}\!=\!\left\langle i\right|U\left|\dot{\xi}_{\varphi_0}\right\rangle \!\!=\!\textrm{i}\, a\tilde{K}_{i}\!-\!\textrm{i}\, b\left\langle i\right|U\left|\xi_{\varphi_0}^{\perp}\right\rangle $.
Hence, for the local QS of channel $\Lambda_{\varphi}$ to be valid,
$b$ must be finite and we must always be able  to construct
\begin{equation}
U=\left[\begin{array}{ccccc}
\tilde{K}_{1} & \frac{a}{b}\tilde{K}_{1}+\frac{\textrm{i}}{b}\dot{\tilde{K}}_{1} & \bullet & \ldots & \bullet\\
\tilde{K}_{2} & \frac{a}{b}\tilde{K}_{2}+\frac{\textrm{i}}{b}\dot{\tilde{K}}_{2} & \bullet & \ldots & \bullet\\
\tilde{K}_{3} & \frac{a}{b}\tilde{K}_{3}+\frac{\textrm{i}}{b}\dot{\tilde{K}}_{3} & \vdots & \ddots & \vdots\\
\vdots & \vdots & \bullet & \ldots & \bullet
\end{array}\right]\label{eq:UQS}
\end{equation}
with first two columns adequately fixed to give for all $i$ the correct $\left\langle i\right|U\left|\xi_{\varphi_0}\right\rangle $
and $\left\langle i\right|U\left|\xi_{\varphi_0}^{\perp}\right\rangle $
respectively. Yet, owing to the locality of the condition \eref{eq:LocQSCond}, all entries marked with $\bullet$
in \eqnref{eq:UQS} can be chosen freely, in order to satisfy the unitarity 
condition:~$U^{\dagger}U\!=\! UU^{\dagger}\!=\!\mathbb{I}$, which nevertheless  constraints
the Kraus operators to simultaneously fulfil $\textrm{i}\sum_{i=1}^{r^{\prime}}\!\dot{\tilde{K}}_{i}^{\dagger}\tilde{K}_{i}\!=\! a\,\mathbb{I}$
and $\sum_{i=1}^{r^{\prime}}\!\dot{\tilde{K}}_{i}^{\dagger}\dot{\tilde{K}}_{i}\!=\!\left(b^{2}+a^{2}\right)\mathbb{I}$.
However, as without loss of generality we may shift their phase at $\varphi_0$
via $\tilde{K}_{i}\!\rightarrow\!\textrm{e}^{-\textrm{i}a\varphi_0}\tilde{K}_{i}$,
the above conditions can be made independent of $a$, so that:~$\textrm{i}\sum_{i=1}^{r^{\prime}}\!\dot{\tilde{K}}_{i}^{\dagger}\tilde{K}_{i}\!=0$
and $\sum_{i=1}^{r^{\prime}}\!\dot{\tilde{K}}_{i}^{\dagger}\dot{\tilde{K}}_{i}\!=\! b^{2}\mathbb{I}$.
On the other hand, such constraints do not require $r^{\prime}\!>\! r$,
as rewriting for example the first one as $\textrm{i}\sum_{i=1}^{r^{\prime}}\!\left\langle \dot{\xi}_{\varphi_0}\right|U\left|i\right\rangle \!\left\langle i\right|U\left|\xi_{\varphi_0}\right\rangle \!=0$,
one can always resolve the identity with some basis vectors $\sum_{i=1}^{r^{\prime}}\!\left|i\right\rangle\!\left\langle i\right|=\sum_{i=1}^{r}\!\left|e_{i}\right\rangle\!\left\langle e_{i}\right|$
and define linearly independent Kraus operators $\left\{ K_{i}\!=\!\left\langle e_{i}\right|U\left|\xi_{\varphi_0}\right\rangle \right\} _{i=1}^{r}$
also fulfilling the necessary requirements.

Finally, realising that $b^2\!=\!\frac{1}{4}\!\left.F_{\textrm{Q}}\!\left[\left|\xi_{\varphi}\right\rangle\right]\right|_{\varphi_0}$, 
we may conclude that $\Lambda_{\varphi}$ is locally \emph{quantum
simulable} at $\varphi_0$, if---as stated in the main text in \eqnref{eq:QSconds}---it admits there a Kraus representation
satisfying conditions:
\begin{equation}
\textrm{i}\sum_{i=1}^{r}\dot{K}_{i}(\varphi_0)^{\dagger}K_{i}(\varphi_0)=0\qquad\textrm{and}\qquad\sum_{i=1}^{r}\dot{K}_{i}(\varphi_0)^{\dagger}\dot{K}_{i}(\varphi_0)=\frac{1}{4}\!\left.F_{\textrm{Q}}\!\left[\left|\xi_{\varphi}\right\rangle\right]\right|_{\varphi_0}\,\mathbb{I},
\end{equation}
where $\{K_i(\varphi)\}_i$ should be constructable from any valid linearly independent Kraus operators
via the unitary transformation \eref{eq:KrausReps} generated by some Hermitian $r\!\times\! r$ matrix $\mathsf{h}$.\hfill {\color{myred} $\blacksquare$}

%% file: Appendices/AppendixI.tex
\chapter{Finite-$N$ CE method as an SDP} 
\label{chap:appFinNCEasSDP}

\lhead{Appendix I. \emph{Finite-$N$ CE method as an SDP}} 

The finite-$N$ CE bound has been defined in \eqnref{eq:CEboundN} as
\begin{equation}
\mathcal{F}_{N}^\t{\tiny CE}=4\min_{\mathsf{h}}\left\{ \left\Vert \alpha_{\tilde{K}}\right\Vert +\left(N-1\right)\left\Vert \beta_{\tilde{K}}\right\Vert ^{2}\right\} \!,\label{eq:FqCENApp}
\end{equation}
where $\left\Vert \cdot\right\Vert $ denotes the operator norm, $\alpha_{\tilde{K}}\!=\!\sum_{i}\dot{\tilde{K}}_{i}^{\dagger}\dot{\tilde{K}}_{i}$
and $\beta_{\tilde{K}}\!=\!\mathrm{i}\sum_{i}\dot{\tilde{K}}_{i}^{\dagger}\tilde{K}_{i}$.
Given a channel $\Lambda_{\varphi}$ mapping states between $d_{\textrm{in}}$- and
$d_{\textrm{out}}$-dimensional Hilbert spaces and the set of its
linearly independent Kraus operators $\left\{ K_{i}\right\} _{i=1}^{r}$---corresponding 
to $d_{\textrm{out}}\!\times\! d_{\textrm{in}}$
matrices---in order to compute
$\mathcal{F}_{N}^\t{\tiny CE}$ we must minimise \eqnref{eq:FqCENApp}
over locally equivalent Kraus representations \eref{eq:KrausReps}
of $\Lambda_{\varphi}$ generated by all Hermitian, $r\!\times\! r$
matrices $\mathsf{h}$.

Adopting a concise notation
in which $\mathbf{K}$ is a column vector containing the starting
Kraus operators $K_{i}$ as its elements, we can associate all locally
equivalent Kraus representations $\tilde{\mathbf{K}}$ in \eqnref{eq:FqCENApp}
with those generated by any $\mathsf{h}$ via $\tilde{\mathbf{K}}\!=\!\mathbf{K}$
and $\dot{\tilde{\mathbf{K}}}\!=\!\dot{\mathbf{K}}\!-\!\mathrm{i}\mathsf{h} \mathbf{K}$.
By constructing matrices---$\mathbb{I}_{d}$ represents a $d\!\times\! d$
identity matrix---:
\begin{equation}
\mathbf{A}\!=\!\left[\begin{array}{cc}
\sqrt{\lambda_{a}}\mathbb{I}_{d_{\textrm{in}}} & \dot{\tilde{\mathbf{K}}}^{\dagger}\\
\dot{\tilde{\mathbf{K}}} & \sqrt{\lambda_{a}}\mathbb{I}_{r\cdot d_{\textrm{out}}}
\end{array}\right]\quad\quad\quad\mathbf{B}\!=\!\left[\begin{array}{cc}
\sqrt{\lambda_{b}}\mathbb{I}_{d_{\textrm{in}}} & \left(\mathrm{i}\dot{\tilde{\mathbf{K}}}^{\dagger}\mathbf{\tilde{K}}\right)^{\dagger}\\
\mathrm{i}\dot{\tilde{\mathbf{K}}}^{\dagger}\mathbf{\tilde{K}} & \sqrt{\lambda_{b}}\mathbb{I}_{d_{\textrm{in}}}
\end{array}\right]\!,
\end{equation}
which positive semi-definiteness conditions correspond respectively
to
\begin{equation}
\alpha_{\tilde{K}}=\dot{\tilde{\mathbf{K}}}^{\dagger}\dot{\tilde{\mathbf{K}}}\;\leq\,\;\lambda_{a}\mathbb{I}_{d_{\textrm{in}}}\quad\quad\quad\beta_{\tilde{K}}^{\dagger}\beta_{\tilde{K}}=\mathbf{\tilde{K}}^{\dagger}\dot{\tilde{\mathbf{K}}}\,\dot{\tilde{\mathbf{K}}}^{\dagger}\tilde{\mathbf{K}}\;\leq\;\lambda_{b}\mathbb{I}_{d_{\textrm{in}}},
\end{equation}
we may rewrite \eqnref{eq:FqCENApp} into the desired SDP form as
\begin{eqnarray}
\mathcal{F}_{N}^\t{\tiny CE} & = & 4\min_{\mathsf{h}}\left\{ \lambda_{a}+(N-1)\lambda_{b}\right\} ,\label{eq:FqCENAppSDP}\\
 &  & \textrm{s.t. \; }\mathbf{A}\ge0,\,\mathbf{B}\ge0.\nonumber
\end{eqnarray}
For the purpose of this work we have implemented all the
required SDPs using the CVX package for Matlab \citep{CVX}, which efficiently
evaluates \eqnref{eq:FqCENAppSDP} given the set of Kraus operators
and their derivatives of a generic channel $\Lambda_{\varphi}$. 


Lastly, one should note that by slightly modifying the program in
\eqnref{eq:FqCENAppSDP} we are able to also efficiently evaluate:~the
\emph{extended-channel QFI} \eref{eq:ExtChQFIPurif}, as $\mathcal{F}\!\left[\Lambda_{\varphi}\otimes\mathcal{I}\right]\!=\!\mathcal{F}_{N=1}^\t{\tiny CE}$;
and the \emph{CE bound} \eref{eq:CEbound}, $\mathcal{F}_\t{as}^\t{\tiny CE}$,
by setting $N\!=\!1$ and imposing in \eqnref{eq:FqCENAppSDP}
also the $\beta_{\tilde{K}}\!=\!0$ condition \eref{eq:BetaKCond} 
that is importantly linear in $\tilde{\mathbf{K}}$ and $\dot{\tilde{\mathbf{K}}}$.

%% file: Appendices/AppendixJ.tex
\chapter{Lossy interferometry with distinguishable photons and adaptive measurements} 
\label{chap:appMZinterAdaptM}

\lhead{Appendix J. \emph{Lossy interferometry with distinguishable photons and adaptive measurements}}

\paragraph{Distinguishability of photons}~\\
If the photons travelling through the lossy interferometer of \figref{fig:MZinter_losses} 
are \emph{distinguishable}, e.g.~they are prepared in different time-bins,
we must generalise the description of the input state \eref{eq:Input_Nph} 
in accordance with \secref{sub:sys_indist_part} to:
\begin{equation}
\ket{\psi_\t{in}^N}\; =\sum_{\boldsymbol{n}=\boldsymbol{0}^N}^{\boldsymbol{1}^N}\alpha_{\boldsymbol{n}}\left|\boldsymbol{n}\right\rangle,
\end{equation}
where the sum runs over all $N$-bit sequences $\boldsymbol{n}$ with 
$\left|\boldsymbol{n}\right\rangle\!=\!\left|n_{1}\right\rangle \otimes\dots\left|n_{N}\right\rangle$,
and $\left|n_{i}\right\rangle =\ket{1} (\ket{0})\equiv \ket{a} (\left|{b}\right\rangle )$
denotes the photon in the $i$-th time-bin, propagating in the $a(b)$
arm of the interferometer respectively.

Taking photonic losses into account, we additionally need to track 
the time-slots in which photons were lost.
Therefore, we define
a binary string $\boldsymbol{l}_{a}=l_{a,1}l_{a,2}\dots l_{a,N}$ with 1's
representing the time-bins in which a given photon was lost in arm $a$ and
similarly $\boldsymbol{l}_{b}$ for the arm $b$.
Formally, using the bitwise subtraction, we can thus also define $\boldsymbol{N}^\prime \!=\!\boldsymbol{1} - \boldsymbol{l}_a - \boldsymbol{l}_b$,
where 1's in the binary string $\boldsymbol{N}^\prime$ denote the time-bins in which photons were successfully transmitted.
As a result---introducing the notation in which for any binary sequence $\boldsymbol{x}$ 
we denote by $x\!=\!|\boldsymbol{x}|\!=\!\sum_{i=1}^Nx_i$ the number of appearing 1's---we 
may identify $N'\!=\!|\boldsymbol{N}'|$ as exactly the overall number of surviving photons
introduced in \secref{sec:LossyInter}.

Following the same argumentation as in the indistinguishable photons case, we may assume, due 
to the lack of global-phase reference \citep{Molmer1997,Bartlett2007,Jarzyna2012}, the 
general \emph{seed element} of a covariant POVM to 
still have a block-diagonal structure, but now w.r.t.~different patterns
of surviving photons, i.e.~$\Xi^N = \bigoplus_{\boldsymbol{N}^\prime=\boldsymbol{0}^N}^{\boldsymbol{1}^N} \Xi^{\boldsymbol{N}^\prime}$.
Nevertheless, let us emphasise that---by the reasoning of \appref{chap:appCovMeas}---for 
any general measurement $M_{\boldsymbol{i}}$ on the $N$ photons, possessing now potentially $2^N$ 
outcomes:~$\boldsymbol{i}\!=\!i_1 i_2\dots i_N$, a covariant POVMs constructed with help of the above $\Xi^N$ 
can always be found that achieves the same precision 
of estimation. Moreover, we may write each block---parametrised by a particular pattern---of the seed element in a 
basis:~$\Xi^{\boldsymbol{N}^\prime} = \sum_{\boldsymbol{n}^\prime, \boldsymbol{m^\prime} = \boldsymbol{0}^{N^\prime}}^{\boldsymbol{1}^{N^\prime}}
\Xi^{\boldsymbol{N}^\prime}_{\boldsymbol{n}^\prime,\boldsymbol{m}^\prime} \ket{\boldsymbol{n}^\prime} \bra{\boldsymbol{m}^\prime}$,
in which $\boldsymbol{n^\prime}$ stands for a string with $N^\prime$ bits placed at positions corresponding to 1's in $\boldsymbol{N}^\prime$
with complementary positions left empty (neither $0$ nor $1$).

Adapting \eqnref{eq:AvCost_losses_expl} to the distinguishable-photons case
and, for completeness, assuming a general form of a 
circularly symmetric cost function \eref{eq:CostFun_gen}, we obtain
(after conveniently setting $\boldsymbol{0}\!\equiv\!\boldsymbol{0}^N$, $\boldsymbol{1}\!\equiv\!\boldsymbol{1}^N$):
\begin{equation}
\left<\mathcal{C}\right>  =  c_{0}+ \sum_{\underset{n\ne m}{\boldsymbol{n},\boldsymbol{m}=\boldsymbol{0}}}^{\boldsymbol{1}}
\sum_{\boldsymbol{l}_{a}=\boldsymbol{0}}^{\min(\boldsymbol{n},\boldsymbol{m})}
\sum_{\boldsymbol{l}_{b}=\boldsymbol{0}}^{\boldsymbol{1}-\max(\boldsymbol{n},\boldsymbol{m})}
\frac{c_{\left|n-m\right|}}{2}\;
\sqrt{B_n^{(l_a,l_b)} B_m^{(l_a,l_b)}}\;
\alpha_{\boldsymbol{n}}^*\alpha_{\boldsymbol{m}}\;
\Xi_{\boldsymbol{n}\setminus (\boldsymbol{l}_a+\boldsymbol{l}_b),\boldsymbol{m}\setminus (\boldsymbol{l}_a+\boldsymbol{l}_b)}^{\boldsymbol{1} - (\boldsymbol{l}_a+\boldsymbol{l}_b)}\;, 
\label{eq:AvCost_dist}
\end{equation}
which for the cost function \eref{eq:CostFun_H}, $C_H$, utilised in the
main text must be substituted $c_{0}\!=\!-c_{1}\!=\!2$ and $\forall_{k>1}\!:\, c_{k}\!=\!0$ (see \secref{sub:AvCostCircSymm}).
The $\min$, $\max$ in \eqnref{eq:AvCost_dist} should be understood as bitwise operations and $B_n^{(l_a,l_b)}$ are the non-combinatorial 
constituents of the binomial coefficients \eref{eq:b^(la,lb)_n},
so that $b_{n}^{(l_a,l_b)}\!=\!\binom{n}{l_a}\binom{N-n}{l_b}B_n^{(l_a,l_b)}$.
For compactness, we have introduced above the notation, in which $\boldsymbol{x}\setminus \boldsymbol{y}$ 
represents a binary string $\boldsymbol{x}$ with empty entries at positions
corresponding to 1's in $\boldsymbol{y}$.

We now split each of the sums over $\boldsymbol{l}_{a/b}$ in \eqnref{eq:AvCost_dist} 
into a sum over $l_{a/b}$, i.e.~number of 1's in $\boldsymbol{l}_{a/b}$ respectively,
and a sum over all permutations of 1's within $\boldsymbol{l}_{a/b}$. We proceed analogously
for summations over $\boldsymbol{n}$  and $\boldsymbol{m}$ obtaining
\begin{align}
\left\langle \mathcal{C}\right\rangle \;=\; c_{0}\;+ & 
\sum_{\underset{n\ne m}{n,m=0}}^{N}
\sum_{l_{a}=0}^{\min(n,m)}
\sum_{l_{b}=0}^{N-\max(n,m)}
\frac{c_{\left|n-m\right|}}{2}\;
\sqrt{B_n^{(l_a,l_b)} B_m^{(l_a,l_b)}}\times
\label{eq:fullloss}\\
& \times\sum_{\underset{|\boldsymbol{n}|=n}
{\boldsymbol{n}=\boldsymbol{0}}}^{\boldsymbol{1}}
\sum_{\underset{|\boldsymbol{m}|=m}
{\boldsymbol{m}=\boldsymbol{0}}}^{\boldsymbol{1}}
\alpha_{\boldsymbol{n}}^{*}
\alpha_{\boldsymbol{m}}
\sum_{\underset{|\boldsymbol{l}_{a}|=l_{a}}
{\boldsymbol{l}_{a}=\boldsymbol{0}}}^{\min(\boldsymbol{n},\boldsymbol{m})}
\sum_{\underset{|\boldsymbol{l}_{b}|=l_{b}}{\boldsymbol{l}_{b}=\boldsymbol{0}}}^{\boldsymbol{1}-
\max(\boldsymbol{n},\boldsymbol{m})}\Xi_{\boldsymbol{n}\setminus(\boldsymbol{l}_{a}+\boldsymbol{l}_{b}),\boldsymbol{m}\setminus(\boldsymbol{l}_{a}+\boldsymbol{l}_{b})}^{\boldsymbol{1}-(\boldsymbol{l}_{a}+\boldsymbol{l}_{b})}\nonumber\;.
\end{align}
In order to proceed further let us for the moment return to the lossless scenario ($\eta_a\!=\!\eta_b\!=\!1$), in which the above formula simplifies to:
\begin{equation}
\left\langle \mathcal{C}\right\rangle   \;=\; c_{0}+
\sum_{\underset{n\ne m}{n,m=0}}^{N}
\frac{c_{\left|n-m\right|}}{2}
\sum_{\underset{|\boldsymbol{n}|=n}{\boldsymbol{n}=\boldsymbol{0}}}^{\boldsymbol{1}}
\sum_{\underset{|\boldsymbol{m}|=m}{\boldsymbol{m}=\boldsymbol{0}}}^{\boldsymbol{1}}
\alpha_{\boldsymbol{n}}^*\alpha_{\boldsymbol{m}}
\Xi_{\boldsymbol{n},\boldsymbol{m}}^{\boldsymbol{1}}\;.
\label{eq:noloss}
\end{equation}
Recall (see \secref{sub:QEstGlobal}) that $\Xi^{\boldsymbol{1}}$ needs to be a 
positive semi-definite operator, and by the completeness constraint has to also
fulfil:~$\Xi^{\boldsymbol{1}}_{\boldsymbol{m},\boldsymbol{n}}\!\underset{n=m}{=}\!\delta_{\boldsymbol{m},\boldsymbol{n}}$. 
Since the diagonal blocks of $\Xi^{\boldsymbol{1}}$ (corresponding to $n\!=\!m$) must thus correspond to identity matrices, this implies that none
of the off-diagonal blocks of $\Xi^{\boldsymbol{1}}$ (corresponding to $n\!\neq\!m$) can have a singular value larger than $1$.
Such fact can be proven as follows.

Let us assume that for certain block $[n,m]$ with $n\!\neq\!m$, the largest singular value $\lambda\!>\!1$ and
let $\ket{\boldsymbol{v}_n}$, $\ket{\boldsymbol{w}_m}$ ($|\boldsymbol{v}_n|\!=\!n$, $|\boldsymbol{w}_m|\!=\!m$)
be the normalised left and right singular vectors 
corresponding to the singular value $\lambda$.
Defining $\ket{\boldsymbol{z}}=\ket{\boldsymbol{v}_n}-\ket{\boldsymbol{w}_m}$, we calculate
\begin{equation}
\bra{\boldsymbol{z}} \Xi^{\boldsymbol{1}} \ket{\boldsymbol{z}} = \bra{\boldsymbol{v}_n} \Xi^{\boldsymbol{1}} \ket{\boldsymbol{v}_n} + \bra{\boldsymbol{w}_m} \Xi^{\boldsymbol{1}} \ket{\boldsymbol{w}_m}
- 2\, \t{Re}\!\left\{\bra{\boldsymbol{v}_n} \Xi^{\boldsymbol{1}} \ket{\boldsymbol{w}_m}\right\} = 2(1-\lambda) < 0,
\end{equation}
what contradicts the positivity semi-definiteness of $\Xi^{\boldsymbol{1}}$ and proves that $\lambda$ can never be greater than 1. \hfill {\color{myred} $\blacksquare$}

Crucially, as all the singular values of any $[n,m]$ block of $\Xi^{\boldsymbol{1}}$ cannot exceed one, 
the following inequality must 
hold:~$\sum_{\underset{|\boldsymbol{n}|=n}{\boldsymbol{n}=\boldsymbol{0}}}^{\boldsymbol{1}}
\sum_{\underset{|\boldsymbol{m}|=m}{\boldsymbol{m}=\boldsymbol{0}}}^{\boldsymbol{1}}
\alpha_{\boldsymbol{n}}^*\alpha_{\boldsymbol{m}}
\Xi^{\boldsymbol{1}}_{\boldsymbol{n},\boldsymbol{m}} \leq \alpha_n^* \alpha_m$,
where $\alpha_n = \sqrt{\sum_{\underset{|\boldsymbol{n}|=n}{\boldsymbol{n}=\boldsymbol{0}}}^{\boldsymbol{1}} |\alpha_{\boldsymbol{n}}|^2}$
may then be interpreted as the symmetric coefficients built for a given set of $\alpha_{\boldsymbol{n}}$
that parametrise an input state \eref{eq:Input_Nph}, $\ket{\psi_\t{in}^N}$,
consisting of indistinguishable photons.
Hence, we may always lower-bound the average cost \eref{eq:noloss}
via:
\begin{equation}
\left\langle \mathcal{C}\right\rangle   \;\geq \;c_{0}\;+
\sum_{\underset{n\ne m}{n,m=0}}^{N} \frac{c_{\left|n-m\right|}}{2} \alpha_n^\star \alpha_m,
\label{eq:Ineq_noloss}
\end{equation}
where the r.h.s.~represents the general average cost 
optimised over measurements for the input $\ket{\psi_\t{in}^N}$, i.e.~the generalised
version of the minimal $\left<\mathcal{C}_H\right>$ \eref{eq:QAvCostCovMZ_OptXi} obtained for the 
lossless interferometer assuming the cost function \eref{eq:CostFun_H}.
In other words, \eqnref{eq:Ineq_noloss} proves that in the absence of losses,
for any given input state consisting of distinguishable photons, there 
always exists a state $\ket{\psi_\t{in}^N}$ that performs at least as good, so 
one may always restrict to inputs \eref{eq:Input_Nph} without sacrificing
optimality.

Returning to \eqnref{eq:fullloss}, we realise that a similar argumentation
can be applied after acknowledging the positive semi-definiteness of
the operator:
$\Xi^{(l_a,l_b)}_{\boldsymbol{m},\boldsymbol{n}}\!=\!\sum_{\underset{|\boldsymbol{l}_a|=l_a}{\boldsymbol{l}_a=\boldsymbol{0}}}^{\min(\boldsymbol{n},\boldsymbol{m})}
\sum_{\underset{|\boldsymbol{l}_b|=l_b}{\boldsymbol{l}_b=\boldsymbol{0}}}^{\boldsymbol{1} - \max(\boldsymbol{n},\boldsymbol{m})}
\Xi_{\boldsymbol{m}\setminus (\boldsymbol{l}_a+\boldsymbol{l}_b),\boldsymbol{n}\setminus (\boldsymbol{l}_a+\boldsymbol{l}_b)}^{\boldsymbol{1} - (\boldsymbol{l}_a+\boldsymbol{l}_b)}$. We notice that the completeness constraint again implies a block structure of
 $\Xi^{(l_a,l_b)}$ w.r.t.~$[n,m]$, with the diagonal blocks $[n,n]$ corresponding now 
to $\sum_{\underset{|\boldsymbol{l}_a|=l_a}{\boldsymbol{l}_a=\boldsymbol{0}}}^{\boldsymbol{n}}
\sum_{\underset{|\boldsymbol{l}_b|=l_b}{\boldsymbol{l}_b=\boldsymbol{0}}}^{\boldsymbol{1} - \boldsymbol{n}}1
\!=\!
\binom{n}{l_a} \binom{N-n}{l_b}$ times the identity matrix. Consequently, the maximum 
singular value of any non-diagonal block, $[n,m]$ with $n\!\neq\!m$, of $\Xi^{(l_a,l_b)}$
is upper-limited by $\binom{\min(n,m)}{l_a} \binom{N-\max(n,m)}{l_b}$. As a result, we obtain the following lower bound
on the general average cost \eref{eq:fullloss}:
\begin{align}
\left\langle \mathcal{C}\right\rangle  \;\geq\; & 
c_0\; +
\sum_{\underset{n\ne m}{n,m=0}}^{N}
\sum_{l_{a}=0}^{\min(n,m)}
\sum_{l_{b}=0}^{N-\max(n,m)}
\frac{c_{\left|n-m\right|}}{2} 
\sqrt{B_n^{(l_a,l_b)} B_m^{(l_a,l_b)}}
\binom{\min(n,m)}{l_a} \binom{N-\max(n,m)}{l_b}\;\alpha_n^\star\alpha_m \nonumber\\
\ge\;&c_0\;+
\sum_{\underset{n\ne m}{n,m=0}}^{N}
\sum_{l_{a}=0}^{\min(n,m)}
\sum_{l_{b}=0}^{N-\max(n,m)}
\frac{c_{\left|n-m\right|}}{2} 
\sqrt{b_n^{(l_a,l_b)} b_m^{(l_a,l_b)}}
\;|\alpha_n| |\alpha_m|,
\label{eq:fullloss2}
\end{align}
which is just the generalisation of the minimal $\left<\mathcal{C}_H\right>$ \eref{eq:AvCost_losses_expl}
evaluated for $\ket{\psi_\t{in}^N}$ and the lossy interferometer of \figref{fig:MZinter_losses}
to all valid cost functions \eref{eq:CostFun_gen}%
\footnote{%
Such a fact becomes completely evident after exchanging the order of the sums in \eqnref{eq:fullloss2},
so that 
$\sum_{\underset{n\ne m}{n,m=0}}^{N}
\sum_{l_{a}=0}^{\min(n,m)}
\sum_{l_{b}=0}^{N-\max(n,m)}
\equiv
\sum_{l_{a}=0}^{N}
\sum_{l_{b}=0}^{N-l_a}
\sum_{\underset{n\ne m}{n,m=l_a}}^{N-l_b}$,
and re-parametrising 
$l_b$ to $N'\!=\!N-l_a-l_b$.
}.
Thus, the inequality in \eqnref{eq:fullloss2} completes the proof that
the optimal phase estimation in the presence of losses is indeed achievable within the Bayesian approach
when considering only the input states \eref{eq:Input_Nph} consisting of indistinguishable photons.
\hfill {\color{myred} $\blacksquare$}

\paragraph{Adaptive measurement schemes}~\\
Let us describe the general structure of \emph{adaptive measurement schemes}
performed on $N$ distinguishable photons,
e.g.~arriving in consecutive time-bins.
Let $\{\!M^{(1)}_{i_1}\}_{i_1}$ be a POVM performed on the first photon. Then, depending on the measurement result $i_1$
a POVM $\{\!M^{(2)}_{i_2}(i_1)\}_{i_2}$ is performed on the second one. In general, a POVM performed on the $k$-th photon
$\{\!M^{(k)}_{i_k}(i_1,\dots,i_{k-1})\}_{i_k}$ depends on all previous measurement outcomes. Thus, the adaptive measurement
mathematically corresponds to an overall POVM:
\begin{equation}
M_{\boldsymbol{i}} = M^{(1)}_{i_1} \otimes M^{(2)}_{i_2}(i_1) \otimes M^{(3)}_{i_3}(i_1,i_2)\otimes \dots \otimes M^{(N)}_{i_N}(i_1,\dots,i_{N-1}),
\end{equation}
which, on the other hand, can be treated as an instance of a \emph{global}---acting on all the photons---POVM 
with potentially $2^N$ measurement outcomes indexed by $\boldsymbol{i}\!=\!i_1 i_2\dots i_N$.
Such an observation proves that the optimisation of a given quantum estimation problem
over \emph{all global} POVMs, $M_{\boldsymbol{i}}$, naturally covers also the case of adaptive measurements
described in \citep{Wiseman1997,Wiseman1998,Wiseman2009}.
Therefore, as we have proved above that the ultimate Bayesian bounds on precision,
derived for the lossy interferometry scheme after restricting to bosonic input states, 
apply equivalently to the scenarios employing distinguishable photons and their most general global measurements,
these global bounds on precision must also apply to \emph{all} adaptive measurement strategies.
\hfill {\color{myred} $\blacksquare$}